\documentstyle[12pt,cont,epsfig,twoside]{report}

\newcommand{\textlineskip}{\baselineskip=14pt}
\newcommand{\smalllineskip}{\baselineskip=12pt}

\newcommand{\Ss}{\scriptstyle}

\newtheorem{satz}{\large \bf}
\newtheorem{subsatz}{\bf}[chapter]

\def\contentsname{Contents}

\def\thebibliography#1{\leftline{\large \bf References}\list
  {[\arabic{enumi}]}{\settowidth\labelwidth{[#1]}\leftmargin\labelwidth
    \advance\leftmargin\labelsep
    \usecounter{enumi}}
    \def\newblock{\hskip .11em plus .33em minus .07em}
    \sloppy\clubpenalty4000\widowpenalty4000}

\newcounter{itemlistc}
\newcounter{romanlistc}
\newcounter{alphlistc}
\newcounter{arabiclistc}

\newcounter{tempfigtabc}                        

\newcommand{\be}{\begin{eqnarray}}
\newcommand{\ee}{\end{eqnarray}}
\newcommand{\dslash}{\partial \hskip -0.5em /}
\newcommand{\Dslash}{D \hskip -0.7em /}

\newcommand{\tr}{{\rm tr}}
\newcommand{\Tr}{{\rm Tr}}

\newcommand{\A}{{\cal A}}

\newcommand{\T}{{\cal T}}

\newcommand{\fp}{F^\prime}
\newcommand{\fpt}{F^{\prime2}}
\newcommand{\sF}{{\rm sin}F}
\newcommand{\sFt}{{{\rm sin}^2}F}
\newcommand{\cF}{{\rm cos}F}
\newcommand{\cFt}{{{\rm cos}^2}F}
\newcommand{\ch}{{\rm cos}{F\over2}}
\newcommand{\sh}{{\rm sin}{F\over2}}
\newcommand{\Wp}{W^\prime}
\newcommand{\Wpt}{W^{\prime2}}

\newcommand{\ft}{f_\pi^2}

\newcommand{\Dp}{D^\prime}
\newcommand{\Sp}{S^\prime}
\newcommand{\gp}{G^\prime}
\newcommand{\gpt}{G^{\prime2}}
\newcommand{\go}{\gamma_1}
\newcommand{\gw}{\gamma_2}
\newcommand{\gt}{\gamma_3}
\newcommand{\wo}{\omega}
\newcommand{\wop}{\omega^\prime}
\newcommand{\wpt}{\omega^{\prime2}}
\newcommand{\alp}{\alpha^\prime}

\newcommand{\bep}{\beta^\prime}
\newcommand{\bpp}{\beta^{\prime\prime}}
\newcommand{\app}{\alpha^{\prime\prime}}

\newcommand{\dep}{\delta^\prime}
\newcommand{\ddp}{\delta^{\prime\prime}}

\newcommand{\pmt}{{\mbox{\tiny $\pm$}}}

\def\qed{\hbox{${\vcenter{\vbox{                          
   \hrule height 0.4pt\hbox{\vrule width 0.4pt height 6pt
   \kern5pt\vrule width 0.4pt}\hrule height 0.4pt}}}$}}

\begin{document}
\normalsize\textlineskip
\baselineskip=22pt
\setcounter{page}{1}

\rightline{UNITU-THEP-13/1995}
\rightline{December 1995}

\vskip 4cm

\centerline{\LARGE\bf Baryons as Three Flavor Solitons}
\vskip 2cm
\centerline{\large Herbert Weigel$^{\dag}$}
\vskip 1cm
\centerline{\large Institute for Theoretical Physics}
\centerline{\large T\"ubingen University}
\centerline{\large Auf der Morgenstelle 14}
\centerline{\large D-72076 T\"ubingen}

\vskip 4cm
\centerline{\large \it To appear in Int. J. Mod. Phys. A}

\noindent
\vfil
$^{\dag}$
{\footnotesize{Supported by a Habilitanden--Scholarship of the
Deutschen Forschungsgemeinschaft (DFG).}}
\eject

\newpage
\baselineskip=18 true pt
~
\vskip3cm

\centerline{\Large\bf Abstract}
\vskip2cm
\noindent
The description of baryons as soliton solutions of effective meson 
theories for three flavor (up, down, strange) degrees of freedom 
is reviewed and the phenomenological implications are illuminated. In 
the collective approach the soliton configuration is equipped with 
baryon quantum numbers by canonical quantization of the coordinates
describing the flavor orientation. The baryon spectrum resulting from 
exact diagonalization of the collective Hamiltonian is discussed. 
The prediction of static properties such as the baryon magnetic moments 
and the Cabibbo matrix elements for semi--leptonic hyperon decays
are explored with regard to the influence of flavor symmetry 
breaking. In particular, the role of strange degrees of freedom 
in the nucleon is investigated for both the vector and axial--vector 
current matrix elements. The latter are discussed extensively 
within in the context of the {\it proton spin puzzle}. The influence 
of flavor symmetry breaking on the shape of the soliton is examined 
and observed to cause significant deviations from flavor covariant 
predictions on the baryon magnetic moments. Short range effects are 
incorporated by a chiral invariant inclusion of vector meson fields. 
These extensions are necessary to properly describe the singlet 
axial--vector current and the neutron proton mass difference. The 
effects of the vector meson excitations on baryon properties are 
also considered. The bound state description of hyperons and its 
generalization to baryons containing a heavy quark are illustrated. 
In the case of the Skyrme model a comparison is performed between the 
collective quantization scheme and bound state approach. Finally, the 
Nambu--Jona--Lasinio model is employed to demonstrate that hyperons 
can be described as solitons in a microscopic theory of the quark 
flavor dynamics. This is explained for both the collective and the 
bound state approaches to strangeness.

\newpage
\baselineskip16pt
\setcounter{tocdepth}{3}
\tableofcontents

\normalsize\textlineskip
\newpage
\begin{satz}
\label{chap_intro}{\large \bf \hskip1cm Introduction and Motivation}
\addcontentsline{toc}{chapter}{\protect\ref{chap_intro} 
Introduction and Motivation}
\end{satz}
\stepcounter{chapter}

It is well established that the theory of Quantum Chromodynamics 
(QCD) properly accounts for the strong interaction processes of 
hadrons \cite{Yn93,Mu87}. In this theory hadrons are considered as 
complicated composites of quarks and gluons. The interaction of these 
fields is described within the framework of a non--abelian gauge 
theory, the gauge group being color $SU(3)$. The quark fields are 
represented in the fundamental representation while the gluons, which 
are the gauge bosons mediating the interaction, reside in the adjoint 
representation. Although we are still lacking a rigorous proof, the 
confinement hypothesis is commonly accepted, which states that only 
color singlet objects are observable. These singlet states represent 
the physical hadrons. 

The solution to the renormalization group equation tells us that 
the QCD coupling decreases with increasing momentum transfer 
(asymptotic freedom). In this energy region QCD can therefore been 
treated within perturbation theory. The predictions, which result from 
these analyses of QCD, agree favorably with the experimental data 
obtained {\it e.g.} in deep inelastic scattering (DIS) processes. 
However, the behavior of the solution to the renormalization group 
equation unfortunately prohibits the application of perturbative 
techniques in the low--energy region. It is therefore mandatory to 
consider models which can be deduced or at least motivated from QCD 
in order to describe the low--energy properties of hadrons. 

One such model is the description of baryons as solitons, the 
so--called Skyrme approach. This method emphasizes the role of 
spontaneously broken chiral symmetry and treats the baryons as 
collective excitations of meson fields. In particular, the knowledge 
of the physics of the low--lying mesons provides an exhaustive 
amount of (almost) parameter free predictions on properties of 
baryons. As an additional advantage over many other models the soliton 
description represents a means for studying various aspects 
(spectrum, electromagnetic and axial form factors, meson--baryon 
scattering, baryon--baryon interaction, etc.) within a unique 
framework without making any further assumptions. Before 
explaining the Skyrme approach in detail, it is appropriate the 
straighten up a few misconceptions about this description. The 
Skyrme approach has frequently been criticized as being too crude. 
This criticism is based on the fact that the original Skyrme model, 
which only contains pseudoscalar degrees of freedom, yields incorrect 
predictions on several baryon observables. As will become apparent 
during the course of this report (see chapter \ref{chap_vector}) many 
of these problems are linked to the feature that the pseudoscalar 
fields contain the long--range physics only. A suitable extension of 
the model to account for short--range effects as well, provides an 
appealing solution to these problems. A prominent example is the 
influence of the vector meson fields on the strong interaction piece 
of the neutron--proton mass difference and the axial singlet current 
matrix element which both exactly vanish in the pseudoscalar 
model. A further common criticism concerns the too large predictions 
for the absolute values of the baryon masses. This problem has 
recently been solved by the proper treatment of the quantum 
corrections to the soliton mass. Except for one application in 
the baryon number two sector (see section \ref{sec_further}) this 
topic, however, will not be addressed in the present article.

There exist a couple of other review articles on the soliton picture 
for baryons \cite{Ho86,Za86,Me88,Sch89}. However, these reviews are 
mainly limited to the two flavor version of the Skyrme approach. If 
at all, these articles contain only general aspects of the treatment 
of strange degrees of freedom. In particular, within the Skyrme 
approach a detailed survey of neither the influence of flavor 
symmetry breaking on baryon properties nor on the effects of strange 
quarks in the nucleon are available. It is the main goal of the 
present article to fill this gap. For this purpose several treatments 
of flavor symmetry breaking in the baryon sector will be introduced 
and critically compared. In various aspects a detailed discussion 
must be beyond the scope of this article. However, it is intented to 
provide a background, which should enable the interested reader to 
consult the original literature.

\bigskip

\begin{subsatz}
\label{sec_soliton}{\bf \hskip1cm The Soliton Picture}
\addcontentsline{toc}{section}{\protect\ref{sec_soliton}
The Soliton Picture}
\end{subsatz}
\stepcounter{section}

For the purpose of modeling QCD, ideas originally proposed by 
t` Hooft \cite{tH74} and later pursued by Witten \cite{Wi79} 
have turned out to be very fruitful. In these examinations QCD has been
generalized from the physical value for the numbers of colors $N_C=3$ 
to an arbitrary value. Subsequently its inverse ($1/N_C$) has been
treated as an effective expansion parameter. It was recognized that 
in the limit $N_C\rightarrow\infty$ only a special class of Feynman 
diagrams survived. These are the planar diagrams with quark loops only 
at the edges. Applying crossing symmetry and unitarity, as 
well as assuming confinement, Witten showed \cite{Wi79} that QCD is 
equivalent to an effective theory of weakly interacting mesons (and 
glueballs). In this context weakly refers to the fact that an effective 
four--meson vertex scales like $1/N_C$. Consequently a major goal of 
phenomenological studies is the construction of effective meson 
theories. Such approaches are guided by requiring the symmetries of 
QCD for the meson Lagrangian.

In the case of arbitrary $N_C$ a color singlet baryon consists of $N_C$ 
quarks. As a result it is obvious that the properties of a single 
quark cause the masses of baryons to be of the order $N_C$. Furthermore 
it can be shown that those contributions to the masses, which are due 
to the exchange of gluons, are of this order as well. Since the baryons 
are color singlet states their wave--functions are completely 
anti--symmetric in the color degrees of freedom of the quarks. As a 
consequence of the Pauli Principle the wave--function must be symmetric 
in all other quantum numbers. This allows the quarks to reside in 
$S$--wave states causing the mean square radius of baryons to be of the 
order $N_C^0$ in the limit of large numbers of colors\footnote{This is 
in contrast to an atom. In that case the wave--function needs to be 
anti--symmetric in the spatial quantum numbers and hence the radius 
increases with the number of electrons.}. Obviously the masses of 
baryons scale like the inverse of the coupling constant of the 
effective meson theory associated with QCD while the extension of the 
baryons is essentially independent of this coupling constant. From 
these analogies Witten argued \cite{Wi79} that baryons emerged as the 
soliton solutions in the effective meson theory. The solitons are the 
solutions to the (static) classical Euler--Lagrange equations and may 
be considered as mappings from coordinate space to the configuration
space of the mesons. The latter is commonly given by flavor $SU(2)$ or 
$SU(3)$ in the cases of two (up, down) or three (up, down, strange) 
flavors. From a topological point of view these mapping are 
characterized by the so--called winding number which determines the 
number of coverings of the configuration space when the coordinate 
space is passed through exactly once. Solitons with different winding 
numbers are topologically distinct meaning that there exists no 
continuous deformation connecting solitons of different winding
numbers. Witten conjectured that this winding number be identified 
with the baryon number. Later on we will see that this identification 
can be justified once the effective meson theory for three flavors 
is properly constructed. Witten furthermore analyzed the $N_C$ 
behavior of other baryon observables as {\it e.g.} the scattering 
amplitudes of baryon--baryon and meson--baryon interactions. These 
were all found to comply with the picture that baryons emerge as 
the solitons of the effective meson theory to which QCD is equivalent.

As a matter of fact such a topological soliton was already constructed 
by Skyrme \cite{Sk61} before the notion of quarks and gluons was
invented. However, it was just the above mentioned considerations, 
which enabled the Skyrme model to be established within the context 
of QCD. Subsequent to the first 
application of this model to investigate observables of the 
nucleon and $\Delta$ resonance by Adkins, Nappi and Witten \cite{Ad83} 
the Skyrme model became a major subject of interest. The Skyrme model 
contains as fundamental mesonic degrees of freedom only the isovector of 
the pseudo--scalar pions. As these are by far the lightest (135MeV) mesons
their importance at low energies is obvious. The fact that these mesons 
are the ``would--be" Goldstone bosons of the spontaneous breaking of 
chiral symmetry provides another reason to consider them as the most 
relevant ingredients. The first results of the examinations within the 
Skyrme model are reviewed \cite{Ho86,Za86}. Later the Skyrme model 
was extended (and improved) to also contain the light vector mesons 
$\rho(770)$ and $\omega(783)$ (for a review see ref \cite{Me88}). All 
extensions have been performed such that they satisfy the symmetries 
of QCD. Besides Pioncar\'e invariance the chiral symmetry (see eqs 
(\ref{uchitrans}) and (\ref{qchitrans})) and its spontaneous breaking 
represents a guiding principle for extenting the Skyrme model. Such 
approaches not only allow one to study static properties of the nucleon 
but also dynamics as {\it e.g.} exhibited in pion--nucleon scattering 
\cite{Sch89}. It will be the primary purpose of this review article 
to provide a survey on progress achieved in the three flavor 
generalization of Skyrme type models, this includes those models which 
contain strange vector mesons as well.

\begin{figure}
\centerline{
\setlength{\unitlength}{.46mm}
\begin{picture}(300,150)
\thicklines
\put(30,50){\line(1,0){60}}
\put(0,90){\line(1,0){120}}
\put(0,90){\line(3,4){30}}
\put(90,50){\line(3,4){30}}
\put(30,130){\line(1,0){60}}
\put(90,130){\line(3,-4){30}}
\put(0,90){\line(3,-4){30}}
\put(27,135){$n$}
\put(87,135){$p$}
\put(-10,93){$\Sigma^-$}
\put(49,93){$\Sigma^0$}
\put(120,93){$\Sigma^+$}
\put(63,93){$\Lambda$}
\put(27,40){$\Xi^-$}
\put(87,40){$\Xi^0$}
\thinlines
\put(-10,90){\line(1,0){10}}
\put(120,90){\vector(1,0){20}}
\put(60,40){\line(0,1){90}}
\put(60,130){\vector(0,1){20}}
\put(63,83){{\small 0}}
\put(63,133){{\small 1}}
\put(120,83){{\small 1}}
\put(63,152){{\small $Y$}}
\put(138,80){{\small $I_3$}}
\thicklines
\put(200,50){\line(1,0){60}}
\put(170,90){\line(1,0){120}}
\put(140,130){\line(1,0){180}}
\put(140,130){\line(3,-4){90}}
\put(230,10){\line(3,4){90}}
\put(135,135){$\Delta^{-}$}
\put(195,135){$\Delta^{0}$}
\put(200,128){\line(0,1){4}}
\put(255,135){$\Delta^{+}$}
\put(260,128){\line(0,1){4}}
\put(315,135){$\Delta^{++}$}
\put(169,93){$\Sigma^{*-}$}
\put(215,93){$\Sigma^{*0}$}
\put(278,93){$\Sigma^{*+}$}
\put(190,40){$\Xi^-$}
\put(262,40){$\Xi^0$}
\put(234,5){$\Omega$}
\thinlines
\put(160,90){\line(1,0){10}}
\put(290,90){\vector(1,0){20}}
\put(230,0){\line(0,1){130}}
\put(230,130){\vector(0,1){20}}
\put(233,83){{\small 0}}
\put(233,133){{\small 1}}
\put(292,83){{\small 1}}
\put(233,152){{\small $Y$}}
\put(310,80){{\small $I_3$}}
\end{picture}
}
\caption{\label{fi_su3rep}\tenrm
The low--lying baryons in the octet (left) and decouplet (right) 
representations of flavor ${\Ss SU(3)}$. These baryons are characterized by 
their isopsin--projection (${\Ss I_3}$) and hypercharge 
(${\Ss Y}$) quantum numbers.}
\end{figure}
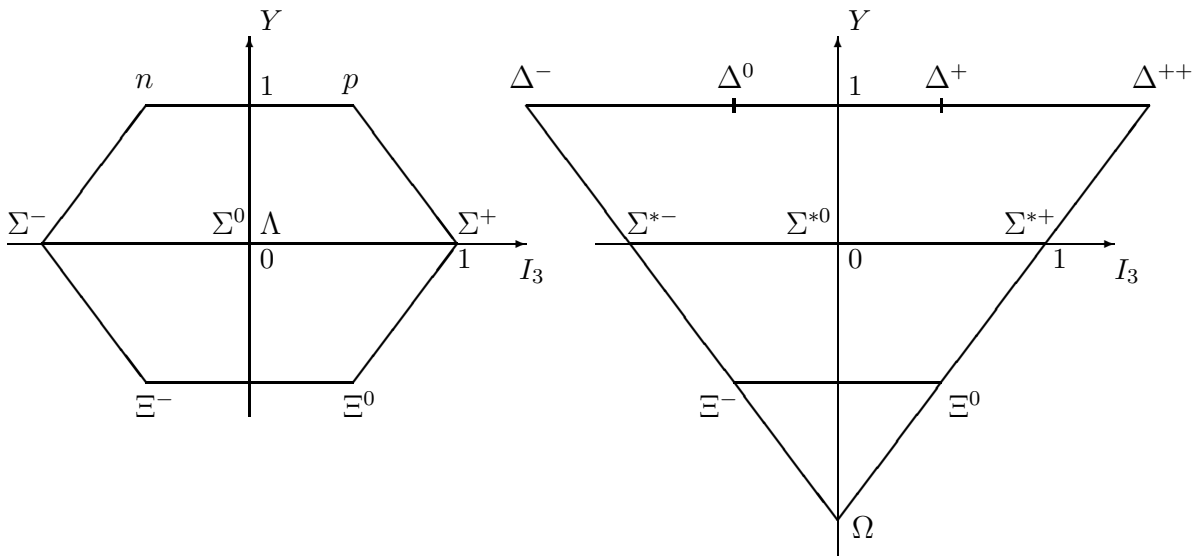

The major focus of these three flavor soliton models is on the 
description of the low--lying $J^\pi=\frac{1}{2}^+$ and 
$\frac{3}{2}^+$ baryons. These objects of interest are displayed in 
figure \ref{fi_su3rep} in form of the popular representations of 
flavor $SU(3)$. The baryons $p,n,\ldots,\Delta^-,\ldots,\Omega$ are 
shown as states in the octet and decouplet representations. Later we 
will see that higher dimensional representations of $SU(3)$ are as 
well relevant for the low--lying baryons. 

The first study within a given model, of course, concerns the 
spectrum of the baryons. Here the mass splittings between the 
baryons shown in figure \ref{fi_su3rep} will represent the main 
issue. For a long time is was believed that soliton models 
considerably overestimated the absolute masses of these baryons. 
However, recently significant progress has been made by including 
the quantum corrections associated with meson loops in the 
background of the soliton \cite{Mo91,Mo93,Ho94,We95,Ho95}.
Adopting a regularization scheme motivated by the chiral 
expansion \cite{Ga84} indeed reproduces the experimental values 
for the absolute masses reasonably well \cite{Mo93,Ho95}. Since 
these corrections effect all baryons approximately equally the 
consideration of mass differences appears to be a suitable measure 
in the semi--classical treatment.  In the next step static properties 
are explored. Although these models are formulated in terms of 
meson fields it is nevertheless possible to study the quark 
structure of baryons. The natural question arises, how to identify 
quark operators within a theory within such a model. Obviously 
that can only be done for objects which are bilinear in the quark 
fields. For operators which are related to symmetry currents this 
is straightforward since imitating the symmetries of QCD within the 
meson theory directly allows one to identify the currents. Phrased 
otherwise, advantage is taken of the fact that the effective meson 
theory imitates the Ward identities of QCD. For operators which do 
not exhibit this feature one commonly refers back to simplified 
models of the quark flavor dynamics like the Nambu--Jona--Lasino (NJL) 
model \cite{Na61}. In contrast to QCD such models can exactly be 
converted into meson theories \cite{Eg74,Eb86}. This also provides the 
quark bilinears in terms of the meson fields. We will make extensive 
use of such identifications. 

Rather than providing a detailed description of the contents of this 
article it is more appropriate to qualitatively describe some of the 
major applications of these three flavor soliton models for recent, 
present or up--comming experiments. The main part of the review 
article will serve to make these discussions quantitative. 
The relation to these experiments will in addition reveal the 
relevance of ``fine--tuned" three flavor soliton models for 
understanding the structure of the low--lying baryons.

\bigskip

\begin{subsatz}
\label{sec_pspin}{\bf \hskip1cm The Proton Spin Puzzle}
\addcontentsline{toc}{section}{\protect\ref{sec_pspin}
The Proton Spin Puzzle}
\end{subsatz}
\stepcounter{section}

The analysis \cite{Br88} of data obtained in the EMC experiments 
\cite{As88} scattering polarized muons off polarized protons 
has led to extensive activities on investigating the 
nucleon matrix elements of axial current. In this analysis 
the separate contributions of the up, down and strange degrees of 
freedom to the axial current have been disentangled yielding two 
surprising results. At first sight, the singlet combination turned 
out to be compatible with zero. This is counter--intuitive because 
in the non--relativistic quark model \cite{Ko79} this quantity is 
identical to twice the spin carried by the quarks and hence should 
be close to unity. In contrast to the non--relativistic quark model
it was soon realized that the Skyrme model indeed predicted an 
identically vanishing matrix element of the axial singlet in nucleon 
states \cite{Br88}. This observation caused a renewed interest in 
soliton models. Second, it turned out that the strange quarks 
contributed a significant amount to the axial current of the 
nucleon, about 30\% of the down quarks. In a na\"\i ve valence quark
picture of the baryon this quantity would identically vanish because
strange degrees of freedom are ignored.  One major input into the 
analyses of the various contributions to the axial current has been 
the assumption of flavor $SU(3)$ symmetry which permits to relate 
nucleon matrix elements of axial current to strangeness changing 
matrix elements of the axial current. The latter can be measured 
in semi--leptonic hyperon decays like 
$\Lambda\rightarrow p e^{-}{\bar \nu}_e$ \cite{Bo83}. Again the 
$SU(3)$ Skyrme model reasonably reproduced this result when the three 
flavor symmetry was assumed as well \cite{Br88}. However, it was soon 
realized \cite{Pa89} that waving this assumption could significantly 
reduce the strange contribution to the axial current, but the singlet 
matrix element turned out to be quite insensitive with respect to 
flavor symmetry breaking \cite{Jo90}. Hence, even a sizable 
breaking of flavor symmetry does not lead to a matrix element of 
the singlet axial current which is close to unity as na\"{\i}vely 
expected. The interested reader may peek ahead to figure 
\ref{fi_hhh} where this feature is illustrated.

As the precision of the experiments improved it became clear 
that the matrix element of the axial singlet current indeed is 
significantly smaller than unity although different from zero. 
In addition refining the soliton models by {\it e.g.} incorporating 
vector mesons or explicit quarks yields also a non--vanishing matrix 
element \cite{Jo90,Bl93c}. The review of these investigations will 
constitute a major part of the present article. Special emphasis 
will be placed on the role of flavor symmetry breaking and the strange 
degrees of freedom. To further elaborate these issues it will also 
be interesting to study the related problem of semi--leptonic 
hyperon decays because for these processes the concept of $SU(3)$ 
symmetry is well established \cite{Jaf90}. It has turned out that 
indeed a large flavor symmetry breaking occurs for the baryon 
wave--functions without contradicting the phenomenology of the 
semi--leptonic hyperon decays. Although this has recently also 
been understood in the context of extended quark models 
\cite{Li94,Li95} the three flavor soliton models provide unique 
frameworks to study these issues comprehensively.

\bigskip

\begin{subsatz}
\label{sec_strange}{\bf \hskip1cm Strangeness in the Nucleon}
\addcontentsline{toc}{section}{\protect\ref{sec_strange}
Strangeness in the Nucleon}
\end{subsatz}
\stepcounter{section}

\begin{figure}
\centerline{
\setlength{\unitlength}{.50mm}
\begin{picture}(300,80)
\thicklines
\put(15.0,40){\vector(-1,1){10}}
\put(5,50){\line(-1,1){8}}
\put(-3.0,22){\vector(1,1){10}}
\put(7,32){\line(1,1){8}}
\put(75.0,40){\vector(1,1){10}}
\put(85,50){\line(1,1){8}}
\put(93.0,22){\vector(-1,1){10}}
\put(83,32){\line(-1,1){8}}
\multiput(17.5,40)(10,0){6}{\oval(5,5)[t]}
\multiput(22.5,40)(10,0){6}{\oval(5,5)[b]}
\put(110,35){\vector(1,0){70}}
\put(110,35){\vector(-1,0){5}}
\put(45,50){$Z^0$}
\put(5,55){$e,\nu$}
\put(5,22){$e,\nu$}
\put(77,55){$q$}
\put(77,22){$q$}
\put(115,45){$\langle N|{\bar q}\gamma_\mu(\gamma_5) q|N\rangle$}
\put(215.0,40){\vector(-1,1){10}}
\put(205,50){\line(-1,1){8}}
\put(197.0,22){\vector(1,1){10}}
\put(207,32){\line(1,1){8}}
\put(275,40){\line(1,1){18}}
\put(274.3,40){\line(1,1){18.3}}
\put(275.7,40){\line(1,1){17.7}}
\put(285,50){\line(-1,0){4}}
\put(285,50){\line(0,-1){4}}
\put(275,40){\line(1,-1){18}}
\put(274.3,40){\line(1,-1){18.3}}
\put(275.7,40){\line(1,-1){17.7}}
\put(283,32){\line(1,0){4}}
\put(283,32){\line(0,-1){4}}
\multiput(217.5,40)(10,0){6}{\oval(5,5)[t]}
\multiput(222.5,40)(10,0){6}{\oval(5,5)[b]}
\put(275,40){\circle*{6}}
\put(245,50){$Z^0$}
\put(205,55){$e,\nu$}
\put(205,22){$e,\nu$}
\put(275,54){$N$}
\put(275,21){$N$}
\end{picture}
}
\caption{\label{fi_elweak}\tenrm
The matrix element
${\Ss \langle N|{\bar q}\gamma_\mu(\gamma_5) q |N\rangle}$ 
relates the electro--weak interaction of the nucleon 
to the elementary processes of the quarks.}
\end{figure}
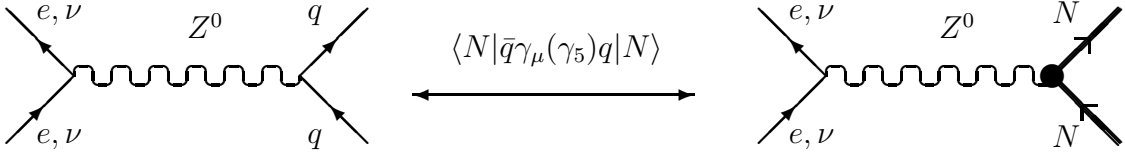

We have just observed in the discussion of the axial current that 
the generalization of soliton models to flavor $SU(3)$ encourages 
one to study effects of strange degrees of freedom in the nucleon. 
As explained above this generalization makes possible the investigation 
of matrix elements of various other strange quark operators. Let us 
consider two prominent examples. The first one concerns the matrix 
element of the strange scalar operator in the nucleon 
$\langle N|{\bar s}s|N\rangle$. Adopting, for the sake of argument, 
the na\"\i ve valence quark picture this quantity is zero. With this 
assumption one may relate the pion nucleon sigma term 
$\sigma_{\pi N}$, which is defined via the double commutator of 
the axial generator with the Hamiltonian, to the spectrum of the 
low--lying $\frac{1}{2}^+$ baryons \cite{Ch88}. This yields 
$\sigma_{\pi N}\approx35{\rm MeV}$. On the other hand 
$\sigma_{\pi N}$ can be extracted from the isospin symmetric part 
of the ${\pi N}$ scattering amplitude resulting in 
$\sigma_{\pi N}\approx45{\rm MeV}$ \cite{Ga91}. Although these 
analyses are somewhat model dependent this discrepancy can 
only be resolved when accepting that
$\langle N|{\bar s}s|N\rangle$ is non--negligible. The second 
example refers to the matrix element of the strange vector current 
$\langle N|{\bar s}\gamma_\mu s|N\rangle$. This quantity is of special 
interest in the context of electro--weak processes where neutral gauge 
bosons are exchanged. In the context of the standard model these bosons 
couple to the quarks and further, the couplings to various quark flavors 
is completely determined in terms of the parameters of the standard 
model. In order to describe the electro--weak interaction of the nucleon 
we require information about the behavior of the quarks inside the 
nucleon. It is exactly this information, which is contained in the 
matrix element $\langle N|{\bar q}\gamma_\mu(\gamma_5) q|N\rangle$ 
where $q$ stands for any quark flavor, see figure \ref{fi_elweak} . 
The axial piece ($\gamma_5$) has 
already been mentioned in the context of the proton spin puzzle.
It is obvious that $q=u,d$ play the dominant role but according to 
the above discussion one expects also the strange vector current 
matrix element $\langle N|{\bar s}\gamma_\mu s|N\rangle$ to be 
non--negligible. In the original studies of electroweak processes, 
this matrix element was set to zero. This permitted one to consider 
reactions like $eN\rightarrow eN$ via $Z^0$--exchange as precision 
measurements for the parameters of the standard model, especially the 
Weinberg angle \cite{Be89} (left arrow in figure \ref{fi_elweak}). 
In the meantime DIS experiments have provided such precise data 
for these parameters\footnote{For a discussion see {\it e.g.} 
chapter 26 of ref \cite{PDG94}.} that one may turn around this 
argument and try to extract a reliable value for 
$\langle N|{\bar s}\gamma_\mu s|N\rangle$ from $eN\rightarrow eN$ or 
$\nu N\rightarrow \nu N$ (right arrow in figure \ref{fi_elweak}). Quite 
a few such experiments are scheduled for the near future (or are already 
taking data). An extensive survey on these experiments may {\it e.g.} be 
found in refs \cite{Mu93,Mu94}. There have already been several 
attempts to estimate $\langle N|{\bar s}\gamma_\mu s|N\rangle$. The 
first one was carried out in ref \cite{Jaf89} performing a three--pole
vector meson fit to dispersion relations \cite{Ho74}. The effect of
$\phi-\omega$ mixing has also been investigated in the framework of 
vector meson dominance \cite{Pa91}. This picture has even been combined 
\cite{Co93} with the kaon loop calculation of ref \cite{Mu93}. 
The study of the matrix element 
$\langle N|{\bar s}\gamma_\mu s|N\rangle$ in various three flavor 
soliton models will be reviewed in the present article. Such 
computations have been performed in the Skyrme model \cite{Pa91}, 
the Skyrme model with vector mesons \cite{Pa92} and the NJL soliton 
approach \cite{We95a}. Although these explorations will be 
discussed in detail later it is appropriate to mention that 
incorporating flavor symmetry breaking into the nucleon wave--function 
reduces the effect of strangeness. This feature has already shown up 
in the discussion of the axial current and can easily be understood 
qualitatively since without symmetry breaking in the baryon
wave--function virtual strange and non--strange quark--antiquark 
pairs are equally probable. However, this feature is almost the only 
one which all these $SU(3)$ soliton models have in common concerning 
their predictions for $\langle N|{\bar s}\gamma_\mu s|N\rangle$. 
Therefore the data accumulated in the experiments measuring this 
quantity might also serve to discriminate between these models.

As already indicated, this review article will primarily be concerned 
with topics related to questions raised in the preceding discussions.
The corresponding predictions obtained in different effective theories 
will be discussed in chapters \ref{chap_coll}, \ref{chap_vector} and
\ref{chap_NJL}. Different approaches to include strange degrees of 
freedom in soliton models will be explained in chapters \ref{chap_coll}, 
\ref{chap_exten} and \ref{chap_bound}. In particular, a comparison of 
these treatments may be found in section \ref{sec_comp}. Furthermore a 
number of other interesting applications of three flavor soliton models 
will at least briefly be exemplified.

\vskip1cm
\begin{satz}
\label{chap_coll}{\large \bf \hskip1cm The Collective Approach to the
SU(3) Skyrme Model}
\addcontentsline{toc}{chapter}{\protect\ref{chap_coll} 
The Collective Approach to the SU(3) Skyrme Model}
\end{satz}
\stepcounter{chapter}

\begin{subsatz}
\label{sec_Skyrme}{\bf \hskip1cm The Skyrme Model}
\addcontentsline{toc}{section}{\protect\ref{sec_Skyrme}
The Skyrme Model}
\end{subsatz}
\stepcounter{section}

Before presenting the details of the three flavor model it is 
appropriate to discuss general aspects of the soliton solution in 
the Skyrme model. These are most conveniently presented in the 
framework of the two flavor reduction. The starting point for 
many of these effective meson theories is the non--linear 
$\sigma$ model. At low energies one expects the most important part 
of an effective theory to only include the lightest mesons ({\it i.e.} 
pions). In order to incorporate these features, chiral symmetry is 
realized by adopting the non--linear representation of the pion 
fields $\mbox{\boldmath $\pi$}(x)$ 
\be
U(x)={\rm exp}\left(\frac{i}{{\tilde f}_\pi}
\mbox{\boldmath $\tau$}\cdot\mbox{\boldmath $\pi$}(x)\right),
\label{chfield}
\ee
where the isovector $\mbox{\boldmath $\tau$}$ contains the Pauli 
matrices. The physical interpretation of the constant ${\tilde f}_\pi$ 
will be explained shortly. The matrix $U(x)$ is commonly referred to 
as the chiral field. The chiral transformations are parametrized by 
the constant matrices $L$ and $R$ via
\be
U(x)\longrightarrow L U(x) R^{\dag}.
\label{uchitrans}
\ee
Chiral invariance is then manifested by the symmetry of the Lagrangian 
under this trans\-formation\footnote{The relation of (\ref{uchitrans}) to
the transformation properties of left-- and right--handed quark fields 
will be explained in chapter \ref{chap_NJL}. See also section 
\ref{sec_su3ext}.}. The fact that the vacuum configuration 
($\mbox{\boldmath $\pi$}=0$, {\it i.e.} $U=1$) is only invariant under 
the coset $L=R$ reflects the spontaneous breaking of chiral symmetry. 
In terms of $U(x)$ the non--linear $\sigma$ model is defined by the 
Lagrangian
\be
{\cal L}_{nl\sigma}=\frac{{\tilde f}_\pi^2}{4}{\rm tr}\ 
\left(\partial_\mu U \partial^\mu U^{\dag}\right).
\label{Lnls}
\ee
It is straightforward to construct the Noether currents associated 
with the symmetry transformation (\ref{uchitrans}). The vector and 
axial--vector currents ($V_\mu$ and $A_\mu$) correspond to $R=L$ and 
$R^{\dag}=L$, respectively. These currents may most conveniently be 
presented by introducing $\alpha_\mu=\partial_\mu UU^{\dag}$
and $\beta_\mu=U^{\dag}\partial_\mu U$
\be
V^a_\mu&=&-i\frac{{\tilde f}_\pi^2}{2}{\rm tr}
\left[\frac{\tau^a}{2}\left(\alpha_\mu-\beta_\mu\right)\right]
=\epsilon_{abc}\pi_b\partial_\mu\pi_c+\cdots \ ,
\label{veccur} \\
A^a_\mu&=&-i\frac{{\tilde f}_\pi^2}{2}{\rm tr}
\left[\frac{\tau^a}{2}\left(\alpha_\mu+\beta_\mu\right)\right]
=\tilde f_\pi\partial_\mu\pi_a+\cdots \ ,
\label{axcur}
\ee
where an expansion in terms of derivatives of the pion fields is 
indicated. In the framework of the electroweak theory the matrix 
element of $A^a_\mu$ between the vacuum and a state containing one 
pion of momentum $p_\mu$
\be
\langle0|A^a_\mu(x)|\pi^b(p)\rangle=
i\delta^{ab}p_\mu f_\pi {\rm e}^{ipx}
\label{deffpi}
\ee
enters the decay width for $\pi\rightarrow\mu{\bar \nu}_\mu$. Its 
measurement then determines the pion decay constant\footnote{In the 
three flavor model the physical pion decay constant will not exactly 
be identical to ${\tilde f}_\pi$.} 
$f_\pi\approx{\tilde f}_\pi=93{\rm MeV}$ \cite{PDG94}.

Simple scaling arguments $U(t,\mbox{\boldmath $r$})
\rightarrow U(t,\lambda\mbox{\boldmath $r$})$ \cite{De64} show that 
the model Lagrangian (\ref{Lnls}) does not possess stable soliton 
solutions which minimize the energy functional. For that reason Skyrme 
\cite{Sk61} added a term which is of fourth order in the derivatives
\be
{\cal L}_{Sk}=\frac{1}{32e^2} {\rm tr}
\left(\left[\alpha_\mu,\alpha_\nu\right]
\left[\alpha^\mu,\alpha^\nu\right]\right)
\label{Skterm}
\ee
but nevertheless only quadratic in the time derivative. This feature 
will be advantageous for quantizing the theory canonically. The 
parameter $e$ remains undetermined at the moment. The Lagrangian 
(\ref{Skterm}) may also be motivated as the remnant of the 
$\rho$--meson exchange in the limit $m_\rho\rightarrow\infty$
\cite{Ig85}.

Static soliton configurations $U(\mbox{\boldmath $r$})$ represent
mappings $U: I\!\!R^3 \to SU (2)$. Demanding the soliton to possess 
a finite energy requires the boundary condition 
$U(\mbox{\boldmath $r$})
{\buildrel{r\rightarrow\infty}\over\longrightarrow}1$. This identifies 
all points at spatial infinity and thus compactifies the 
three--dimensional space to a sphere $S^3$. Consequently
\be
U : S^3 \to S^3,
\label{umap}
\ee
since $SU(2)$ is equivalent to a three--dimensional sphere. The 
mappings (\ref{umap}) are characterized by the integer winding number
\be
\nu [U] = \int d^3 x B^0 (x) 
\quad {\rm with} \quad
B^\mu (x) = \frac 1 {24\pi ^2} \epsilon ^{\mu \nu \rho\sigma}
{\rm tr} \left(\alpha_\nu \alpha_\rho \alpha_\sigma\right),
\label{bnumber}
\ee
which gives the number of complete coverings of the target space 
($SU(2)$) when the configuration space ($I\!\!R^3$) is passed through
exactly once. $B^\mu(x)$ is referred to as the topological or winding 
number current and is conserved independently of the dynamics.
As already mentioned in the introduction, it is the main feature 
of soliton models for baryons to identify the topological current 
with the baryon number current. Hence we are interested in 
configurations with $\nu[U]=B=1$.

\begin{figure}[t]
\centerline{\hskip -1.5cm
\epsfig{figure=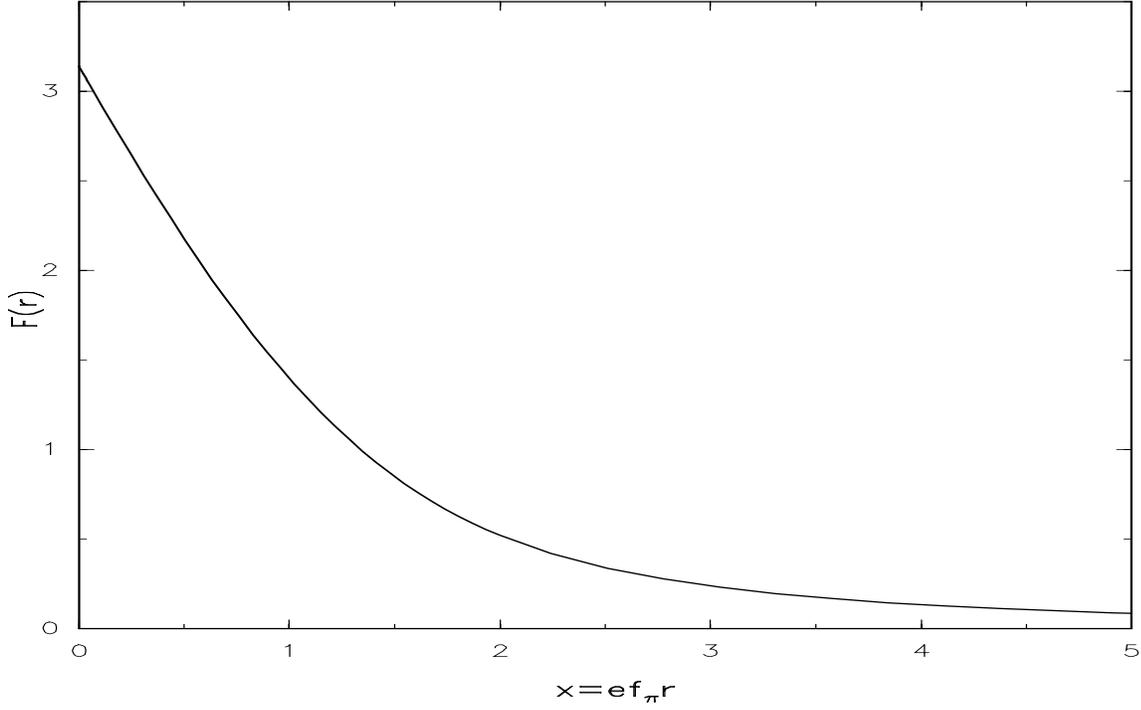,height=10.0cm,width=16.0cm}}
\caption{\label{fi_skyrme}\tenrm
The radial dependence of the chiral angle which 
minimizes (\protect\ref{Skeng}).}
\end{figure}

\begin{figure}[t]
\centerline{\hskip -1.5cm
\epsfig{figure=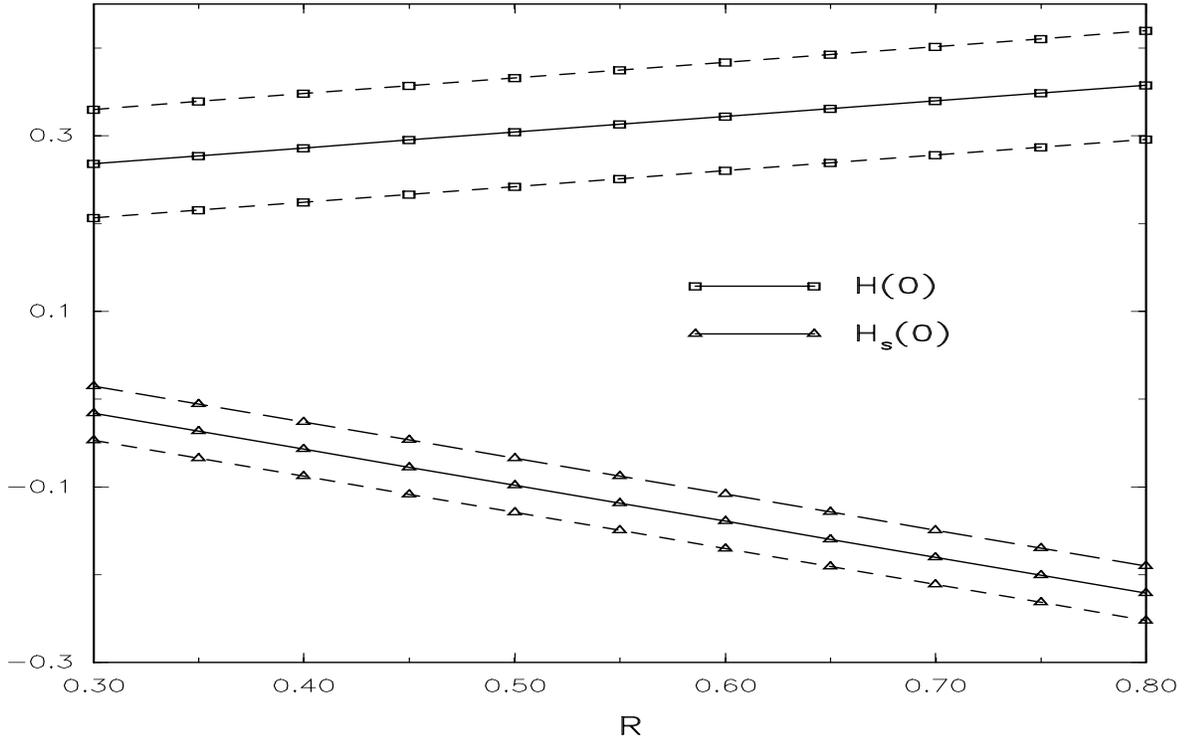,height=9.0cm,width=16.0cm}}
\caption{\label{fi_hhh}\tenrm
The dependence of the matrix elements of the singlet axial 
current ${\Ss H(0)}$ and the strange contribution ${\Ss H_s(0)}$ on 
flavor symmetry breaking. The flavor symmetric case corresponds to 
${\Ss R=0.575\pm0.016}$. The dashed lines indicate errors caused by 
the uncertainty of ${\Ss \Gamma_1^p}$. See also refs \protect\cite{Jo90} 
and \protect\cite{Li95}.}
\end{figure}

In order to construct a soliton solution with unit winding number
Skyrme proposed to adopt the static hedgehog {\it ansatz} 
\cite{Pa46}
\be
U_0(\mbox{\boldmath $r$})=
{\rm exp}\left(i\mbox{\boldmath $\tau$}\cdot
\hat{\mbox{\boldmath $r$}}F(r)\right)
\label{hedgehog}
\ee
which defines the chiral angle $F(r)$. This {\it ansatz} possesses
the famous grand spin symmetry, {\it i.e.} it is invariant under 
the combined spin--isospin transformation generated by 
$\mbox{\boldmath $G$}=\mbox{\boldmath $j$}+
\mbox{\boldmath $\tau$}/2$, where $\mbox{\boldmath $j$}$ is the spin 
operator. Substitution of the {\it ansatz} (\ref{hedgehog}) into the 
Lagrangian ${\cal L}_{nl\sigma}+{\cal L}_{Sk}$ provides the static 
energy functional 
\be
E[F]=\frac{2\pi f_\pi}{e}\int_0^\infty dx\Bigg\{
\left(x^2\fpt+2\sFt\right)+
\sFt\left(2\fpt+\frac{\sFt}{x^2}\right)\Bigg\}
\label{Skeng}
\ee
where a prime indicates a derivative with respect to the dimensionless
coordinate $x=ef_\pi r$. Imposing $F(\infty)$=0 and noting that 
$\nu[U_0]=(F(0)-F(\infty))/\pi$ leads to the boundary condition 
$F(0)=\pi$. The corresponding solution, which minimizes (\ref{Skeng}),
is displayed in figure \ref{fi_skyrme}.
The energy obtained by substituting this solution into (\ref{Skeng})
is found to be $E=23.2\pi f_\pi/e$ \cite{Ad83}. Up to now only 
massless pions have been considered. The inclusion of a pion mass 
term will be discussed together with the mass terms of the 
pseudo--scalar mesons in the $SU(3)$ extension.

\bigskip

\begin{subsatz}
\label{sec_su3ext}{\bf \hskip1cm The SU(3) Extension of the 
Skyrme Lagrangian}
\addcontentsline{toc}{section}{\protect\ref{sec_su3ext}
The SU(3) Extension of the Skyrme Lagrangian}
\end{subsatz}
\stepcounter{section}

For the inclusion of strange degrees of freedom the chiral 
field $U$ is elevated from a $2\times2$ to a $3\times3$ unitary 
unimodular matrix, which in addition to the pions, contains the kaons 
and the octet component of the $\eta$. These fields are most 
conveniently incorporated in the framework of the so--called 
{\it Eightfold Way}
\be
\Phi=\sum_{a=1}^8\frac{\phi^a}{f_a}\lambda^a=\pmatrix{
\frac{1}{\sqrt2}\frac{\pi^0}{f_\pi}+
\frac{1}{\sqrt6}\frac{\eta_8}{f_\eta} &
\frac{\pi^+}{f_\pi} & \frac{K^+}{f_K} \cr
\frac{\pi^-}{f_\pi} & 
-\frac{1}{\sqrt2}\frac{\pi^0}{f_\pi}+
\frac{1}{\sqrt6}\frac{\eta_8}{f_\eta} & \frac{K^0}{f_K} \cr
\frac{K^-}{f_K} & \frac{{\bar K}^0}{f_K} & 
-\frac{2}{\sqrt6}\frac{\eta_8}{f_\eta} \cr},
\label{eightfold}
\ee
where $\lambda^a$ denote the Gell--Mann matrices. The $SU(3)$ chiral
field is then defined as
\be
U(x)={\rm exp}\left(i\Phi\right).
\label{su3chif}
\ee
In eq (\ref{eightfold}) care has been taken of the different 
decay constants. These decay constants are defined via the gradient 
expansion of the axial--vector current analogous to (\ref{axcur})
\be
A_\mu^a=f_a\partial_\mu\phi^a+\ldots \ , 
\qquad
(a=1,\ldots,8),
\label{axsu3}
\ee
{\it i.e.}, $f_{1,2,3}=f_\pi=93{\rm MeV}$, 
$f_{4,\ldots,7}=f_K=113{\rm MeV}$ and 
$f_8=f_\eta=78-94{\rm MeV}$. These data are obtained from the 
leptonic decays of the pseudo--scalar mesons \cite{PDG94}. Since the 
$\eta_8$ meson is of no special importance for the soliton 
calculations\footnote{There is one exception from this statement which 
will be discussed in section \ref{sec_two}.} and $f_\eta$ is not that 
well known, henceforth the approximation $f_\eta=f_\pi$ will be 
adopted.

The difference in the decay constants reflects a feature 
of flavor symmetry breaking. Let us now construct the appropriate 
symmetry breaking terms systematically from the QCD mass term in 
the case of three flavors
\be
{\cal L}_{\rm QCD}^{\rm mass}
= - m_u {\bar u} u - m_d {\bar d} d - m_s {\bar s} s
= - {\hat m}{\bar q}{\cal M}q \ , \qquad
q=\pmatrix{u \cr d \cr s \cr},
\label{QCDmass}
\ee
with ${\hat m}=(m_u+m_d)/2$. The mass matrix is parametrized as 
\cite{SSW93}
\be
{\cal M}=y\lambda^3+T+xS,
\label{massmat}
\ee
where $T={\rm diag}(1,1,0)$ and $S={\rm diag}(0,0,1)$ are the 
projectors on the non--strange and strange subspaces, respectively.
Furthermore $x$ and $y$ are the quark mass ratios
\be
x=\frac{m_s}{{\hat m}}\ , \qquad
y=\frac{1}{2}\frac{m_u-m_d}{{\hat m}}\ .
\label{qmratio}
\ee
The mass term (\ref{QCDmass}) is not invariant under the 
chiral transformation
\be
q_L\rightarrow Lq_L\ ,\quad
q_R\rightarrow Rq_R\ ,\quad
q_{R,L}=\frac{1}{2}\left(1\pm\gamma_5\right)q
\label{qchitrans}
\ee
but rather the matrix $q{\bar q}=q_L{\bar q}_R+q_R{\bar q}_L$ 
transforms like the representation 
$(\mbox{\boldmath $3$},\mbox{\boldmath $3$}^*)+{\rm h.c.}$ of 
$(L,R)$. The symmetry breaking terms of the meson Lagrangian are 
required to imitate this transformation behavior under prospect 
of (\ref{uchitrans}). A minimal set of symmetry breaking terms is 
added to the chirally symmetric Lagrangian discussed above. In order 
to allow for different meson masses and decay constants two 
additional terms are needed
\be
{\cal L}_{\rm SB}={\rm tr}\
\left\{{\cal M}\left[-\beta^\prime\left(
\partial_\mu U \partial^\mu U^{\dag}U
+U^{\dag}\partial_\mu U \partial^\mu U^{\dag}\right)
+\delta^\prime\left(U+U^{\dag}-2\right)\right]\right\}.
\label{LSB}
\ee
The phase conventions have been chosen such that ${\cal M}$ is 
self--adjoint. In general this may not be possible, see {\it e.g.} 
eq (\ref{lambdatheta}). In that case one has to substitute
${\rm tr}\{{\cal M}(U+U^{\dag})\}\rightarrow
{\rm tr}\{{\cal M}U+U^{\dag}{\cal M}^{\dag}\}$, etc. .

For the moment it is appropriate to assume the isospin limit $y=0$.
The new parameters ($x$, $\beta^\prime$ and $\delta^\prime$) are 
determined from the masses and decay constants of the pseudo--scalar 
mesons \cite{Ja89}
\be
m_\pi^2=\frac{4\delta^\prime}{f_\pi^2}\ , \quad
m_K^2=\frac{2(1+x)\delta^\prime}{f_K^2}\ \quad {\rm and}\quad
\left(\frac{f_K}{f_\pi}\right)^2=
1+4\beta^\prime\frac{1-x}{f_\pi^2}.
\label{pspara}
\ee
It has already been remarked that ${\tilde f}_\pi$ in eq (\ref{Lnls}) 
is not exactly identical to the physical pion decay constant 
$f_\pi=93{\rm MeV}$, rather $f_\pi^2={\tilde f}_\pi^2-8\beta^\prime$. 
Substitution of the experimental data \cite{PDG94} yields \cite{We90}
\be
\beta^\prime=-2.64\times10^{-5}{\rm GeV}^2\ , \quad
\delta^\prime=4.15\times10^{-5}{\rm GeV}^4\ 
\quad {\rm and}\quad x=37.3.
\label{pspara1}
\ee
Obviously the difference between ${\tilde f}_\pi$ and $f_\pi$
is only minor and for convenience we will henceforth omit this 
distinction. It is also noteworthy that due to the inclusion of
different decay constants in the present prediction for the quark 
mass ratio, $x$ is considerably larger than the value ($25.0\pm2.5$) 
of ref \cite{Ga82}. Further, it should also be noted that additional 
symmetry breaking terms might be introduced because 
${\cal M}^{{\dag}-1}{\rm det}{\cal M}^{\dag}$ transforms under 
chiral $SU(3)_L\times SU(3)_R$ in the same way as ${\cal M}$
\cite{Ka86}. These terms, however, will not be considered here 
because the minimal set (\ref{LSB}) adequately describes the mesonic 
data.

Now all ingredients necessary to describe the pseudo--scalar mesons 
properly have been presented in eqs (\ref{Lnls}, \ref{Skterm},
\ref{LSB}). However, as compared to QCD (or nature) this Lagrangian 
has still a superfluous symmetry. It is straightforward to verify that 
the above presented pieces of the effective meson Lagrangian are
separately invariant under the transformations 
$U\leftrightarrow U^{\dag}$ and $\mbox{\boldmath $r$}
\leftrightarrow-\mbox{\boldmath $r$}$. The pseudo--scalar character 
of the low--lying mesons requires the effective theory to be 
invariant only when these two transformations are combined.
Unfortunately in four space time dimensions no local term can be 
added such that the separate symmetry is lost. Witten argued 
\cite{Wi83} however, that a suitable  term can be added to the 
equations of motion
\be
\frac{f_\pi^2}{2}\partial_\mu\alpha^\mu+\ldots
+5\lambda\epsilon_{\mu\nu\rho\sigma}
\alpha^\mu\alpha^\nu\alpha^\rho\alpha^\sigma=0\ ,
\label{WZeqm}
\ee
where the dots refer to the contributions from ${\cal L}_{Sk}$ and 
${\cal L}_{\rm SB}$. This equation of motion is only invariant under 
the combined transformation 
$U\leftrightarrow U^{\dag}$ and $\mbox{\boldmath $r$}
\leftrightarrow-\mbox{\boldmath $r$}$. Within the context of the 
action, this term can be obtained by adding a totally antisymmetric 
object on a five dimensional manifold $M_5$ and using Stoke's theorem. 
The important point is that the boundary of $M_5$ is the real Minkowski 
space {\it i.e.} $\partial M_5=M_4$. From the path--integral 
formulation one requires that the action may only change by a multiple 
of $2\pi$ when going from $M_5$ to its complement, which has the 
identical boundary. Therefore $n=240i\pi^2\lambda$ must be 
integer. This actually is completely analogous to Dirac's quantization 
of a magnetic monopole \cite{Di30}. In the next step Witten included 
the photon fields in a gauge invariant way \cite{Wi83,Ka84b}. This 
generates a vertex for the decay $\pi^0\rightarrow\gamma\gamma$ 
\be
\frac{-n}{96\pi^2f_\pi}\pi^0\epsilon_{\mu\nu\rho\sigma}
F^{\mu\nu}F^{\rho\sigma},
\nonumber
\ee
where $F^{\mu\nu}$ is the field strength tensor of the photon field.
Comparison with the triangle anomaly \cite{ABJ69} brings into the game 
the number of colors $n=N_C$. The same result can be obtained from 
considering the process $\gamma\rightarrow\pi^+\pi^0\pi^-$. The 
Wess--Zumino term is now completely determined\footnote{We use the
notation of differential forms $\alpha=\alpha^\mu dx_\mu$.}
\be
\Gamma_{\rm WZ}=-\frac{iN_C}{240\pi^2}
\int_{M_5}(\alpha)^5 \ .
\label{WZterm}
\ee
It should be remarked that as $\Gamma_{\rm WZ}$ is a five--form it 
vanishes in the case of two flavors since the maximal number 
of generators is only four (${\bf 1},\mbox{\boldmath $\tau$}$).
The complete effective action of the pseudo--scalar mesons for flavor 
$SU(3)$ is finally given by
\be
\Gamma=
\int d^4x\left({\cal L}_{nl\sigma}+{\cal L}_{Sk}+
{\cal L}_{\rm SB}\right)+\Gamma_{\rm WZ}.
\label{totact1}
\ee

The Wess--Zumino term (\ref{WZterm}) exhibits one more important 
feature. It provides the only contribution to the Noether current 
associated with the $U_V(1)$ symmetry $L=R={\rm exp}(i\epsilon{\bf 1})$
(\ref{uchitrans}). This current is, of course, nothing but the baryon 
number current, which indeed is recognized to be identical to 
the winding number current $B_\mu$ defined in eq (\ref{bnumber}). 
Obviously the sign of the Wess--Zumino term is fixed by these 
arguments. The opposite sign of $\Gamma_{\rm WZ}$ causes the 
boundary condition $F(0)=-\pi$. Glancing at eq (\ref{Skeng}) we 
see that the substitution $F(r)\rightarrow-F(r)$ leaves the energy 
invariant. As a matter of fact the model can consistently be 
formulated with this opposite sign of the chiral angle.

The first step towards describing baryons in the three flavor model 
consists of constructing the classical soliton. Although various 
embeddings of the hedgehog (\ref{hedgehog}) within the $SU(3)$ matrix 
$U$ are possible demanding a minimal static energy fixes it to 
\be
U_0(\mbox{\boldmath $r$})=\pmatrix{
{\rm exp}\left(i\mbox{\boldmath $\tau$}\cdot
\hat{\mbox{\boldmath $r$}}F(r)\right) & 
{| \atop |}&\hspace{-10pt}{\mbox{\small $0$}
\atop \mbox{\small $0$}}\cr
-------\ -\hspace{-8pt}&-&\hspace{-10pt}-----\cr
0\qquad 0&|&\hspace{-8pt} 1\cr}.
\label{su3hedgehog}
\ee
Other embeddings yield larger energies as a consequence of the 
symmetry breaking terms. Hence as compared to the two flavor model the 
classical mass acquires only a minor correction associated with the 
$\beta^\prime$ and $\delta^\prime$ terms
\be
E_{\rm cl}&=&4\pi\int_0^\infty \ dr\Bigg\{
\left[\frac{f_\pi^2}{2}+4\beta^\prime(1-\cF)\right]
\left(r^2\fpt+2\sFt\right)
\nonumber \\ && \hspace{1.5cm}
+\frac{\sFt}{2e^2}\left(2\fpt+\frac{\sFt}{r^2}\right)
+4\delta^\prime r^2(1-\cF)\Bigg\}.
\label{eclsu3}
\ee
The chiral angle $F(r)$ is again determined as the solution to the 
Euler--Lagrange equations extremizing (\ref{eclsu3}) subject to the 
boundary conditions $F(0)=\pi$ and $F(\infty)=0$. The resulting profile 
is comparable to the one shown in figure \ref{fi_skyrme}.

\bigskip

\begin{subsatz}
\label{sec_quant}{\bf \hskip1cm Quantization of 
the Collective Coordinates}
\addcontentsline{toc}{section}{\protect\ref{sec_quant}
Quantization of the Collective Coordinates}
\end{subsatz}
\stepcounter{section}

It can easily be verified that the static field configuration 
(\ref{su3hedgehog}) does not yield states of good spin and/or flavor 
because $U_0$ does not commute with the corresponding generators, 
$\mbox{\boldmath $r$}\times\mbox{\boldmath $\partial$}$ and
$\lambda^a/2$. States with good spin and isospin 
quantum numbers are generated by the well--known cranking 
procedure \cite{In54}.

The generators for the flavor transformations and spatial 
rotations are constructed as Noether charges 
\be
{\cal Q}^a=\int d^3r \left\{
\frac{\partial{\cal L}}{\partial{\dot U}}
\left(\delta^a U\right)\ + \ {\rm h. c.}\right\} ,
\label{defNoe}
\ee
associated with the field configuration which solves the full 
Euler--Lagrange equations. The infinitesimal form of the symmetry 
transformation is denoted by $\delta^a U$. It is therefore obvious that 
time--dependent solutions are required, not only static solutions.
Unfortunately no such time--dependent solution is known at present
and approximations have to be imposed. Ignoring (for the time being) 
symmetry breaking effects one recognizes that the energy remains 
unchanged by global rotations of the hedgehog solution (\ref{su3hedgehog}) 
in flavor and coordinate spaces. Hence, as a first step a reasonable 
approximation to the time--dependent solutions is presumably given by 
assuming these transformations to vary in time,
\be
U(\mbox{\boldmath $r$},t)= {\tilde A}(t) 
U_0\left(R^{-1}(t)\mbox{\boldmath $r$}\right) \tilde A^{\dag}(t)
=A(t) U_0\left(\mbox{\boldmath $r$}\right) A^{\dag}(t) \ .
\label{su3rot}
\ee
Later more elaborate {\it ans{\"a}tze} for the time--dependent 
configuration will be introduced. In eq (\ref{su3rot}) the hedgehog 
symmetry of $U_0$ has been employed to express the spatial rotation 
as a flavor transformation. The $SU(3)$ matrix $A$ is referred to as 
the collective rotation and contains collective coordinates. It may 
conveniently be parametrized in terms of eight ``Euler angles", 
{\it cf.} eq (\ref{Apara}). The time--dependence of $A$ is measured 
by eight angular velocities $\Omega_a$, which are defined via
\be
A^{\dag}(t)\frac{d A(t)}{dt}=\frac{i}{2}\sum_{a=1}^8
\lambda^a\Omega_a \ . 
\label{defomega}
\ee
It is easy to verify that the angular velocities are invariant under
global left transformations $A(t)\rightarrow LA(t)$ while under 
global right transformations $A(t)\rightarrow A(t)R^{\dag}$ they 
behave like a member of the adjoint representation
\be
\Omega_a\rightarrow \sum_{a=1}^8D_{ab}(R)\Omega_b \ , \quad
D_{ab}(R)=\frac{1}{2}{\rm tr}
\left(\lambda^a R \lambda^b R^{\dag}\right) \ .
\label{omtrans}
\ee
The $8\times8$ rotation matrix $D_{ab}(A)$ behaves like 
a vector under both left-- and right transformations
\be
D_{ab}(A)\rightarrow \sum_{a^\prime, b^\prime=1}^8
D_{aa^\prime}(L^{\dag})
D_{a^\prime b^\prime}(A) D_{b^\prime b}(R^{\dag})\ .
\label{Dtrans}
\ee
For that reason $D_{ab}(A)$ is frequently called the adjoint 
representation of the rotation $A$. The combination 
$\Omega_a^\prime=D_{ab}(A)\Omega_b$ obviously transforms like a left 
vector $\Omega_a^\prime\rightarrow D_{ab}(L^{\dag})\Omega_b^\prime$.

In general the Lagrange function of the collective coordinates 
(the so--called collective Lagrangian) contains all combinations 
of $D_{ab}(A)$ and $\Omega_a$ which are consistent with the effective 
meson action as {\it e.g.} (\ref{totact1}). From the {\it ansatz} 
(\ref{su3rot}) we infer that the flavor rotations correspond to the 
left transformation ($L$) of $A$. The flavor symmetric part of the 
action can therefore only contribute terms to the collective 
Lagrangian with all left indices saturated, {\it e.g.} 
$\Omega_a^\prime\Omega_a^\prime=\Omega_a\Omega_a$. On the other hand 
the flavor symmetry breaking term (\ref{LSB}) transforms like the 
eighth component of a flavor vector\footnote{The isospin breaking, 
which is omitted here corresponds to the third component of a flavor 
vector.}. Hence it contributes terms of the structure 
$\Omega_8^\prime=D_{8b}(A)\Omega_b$. In order to gain restrictions of 
the right indices one recognizes that the spatial rotations may be 
identified with the ``isospin--subset" of the right transformation 
($R$) as a consequence of the hedgehog symmetry. The symmetry under 
spatial rotations thus permits contributions which separately saturate 
the subsets\footnote{From now on a vector will always refer to the
first three components.} $i=1,2,3$, $\alpha=4,\ldots,7$ and $b=8$ as 
{\it e.g.} $\Omega_i\Omega_i=\mbox{\boldmath $\Omega$}\cdot
\mbox{\boldmath $\Omega$}$ or $D_{8\alpha}(A)\Omega_\alpha$. The latter, 
of course, stems from a term which breaks the flavor symmetry. Employing 
these arguments it is straightforward to verify that up to quadratic 
order in both the time derivatives and symmetry breaking the general 
form of the collective Lagrangian is given by
\be
L(A,\Omega_a)&=&-E+\frac{1}{2}\alpha^2\sum_{i=1}^3\Omega_i^2
+\frac{1}{2}\beta^2\sum_{\alpha=4}^7\Omega_\alpha^2
-\frac{N_CB}{2\sqrt3}\Omega_8+\alpha_1\sum_{i=1}^3D_{8i}\Omega_i
+\beta_1\sum_{\alpha=4}^7D_{8\alpha}\Omega_\alpha
\nonumber \\ && \hspace{-1.0cm}
-\frac{1}{2}\gamma \left(1-D_{88}\right)
-\frac{1}{2}\gamma_S\left(1-D_{88}^2\right)
-\frac{1}{2}\gamma_T\sum_{i=1}^3D_{8i}D_{8i}
-\frac{1}{2}\gamma_{TS}\sum_{\alpha=4}^7D_{8\alpha}D_{8\alpha}.
\label{colllag}
\ee
Here $D_{ab}=D_{ab}(A)$ is implied. The coefficients 
$\alpha^2,\ldots,\gamma_{TS}$ are functionals of the soliton profile(s) 
and hence are model dependent. It is the main objective of 
the various models to determine these coefficients. The quantities 
$\alpha^2$ and $\beta^2$ denote the moments of inertia for rotations 
in coordinate space and flavor rotations in direction of the strange 
degrees of freedom, respectively. From symmetry 
arguments one might have also expected a term proportional to 
$\Omega_8^2$. Such a contribution, however, is absent because the 
hedgehog configuration (\ref{su3hedgehog}) commutes with $\lambda_8$. 
An expression of the form $D_{88}\Omega_8$ does not appear because of
the same reason. Nevertheless a term linear in $\Omega_8$ has shown 
up. It is a surface term which arises when applying Stoke's theorem 
to the Wess--Zumino term (\ref{WZterm}) and its coefficient is 
uniquely related to the baryon number $B$. Shortly we will see that 
this term provides an important restriction on the allowed baryon 
states. It should be remarked that the quantities involving the 
coefficients $\gamma_S$, $\gamma_T$ and $\gamma_{TS}$ only appear when
terms of the order ${\cal M}^2$ are included in the meson Lagrangian.

In order to quantize this ``classical" theory we require the operators 
for spin and flavor from eq (\ref{defNoe}). As a consequence of the 
hedgehog structure the infinitesimal change under spatial rotations can 
be written as a derivative with respect to $\mbox{\boldmath $\Omega$}$
\be
\left[\mbox{\boldmath $r$}\times\mbox{\boldmath $\partial$},
U(\mbox{\boldmath $r$},t)\right]=
\frac{\partial \dot U(\mbox{\boldmath $r$},t)}
{\partial \mbox{\boldmath $\Omega$}}\ .
\label{infrot}
\ee
Upon substitution of this relation into the defining equation 
(\ref{defNoe}) one observes for the spin operator $\mbox{\boldmath $J$}
=\partial L(A,\Omega_a)/\partial\mbox{\boldmath $\Omega$}$.
The quantization now proceeds along the lines of an $SU(3)$ rigid 
top by generalizing this result to the right generators 
\be
R_a=-{{\partial L}\over{\partial\Omega_a}}=\cases{
-(\alpha^2\Omega_a+\alpha_1D_{8a})=-J_a,&a=1,2,3\cr
-(\beta^2\Omega_a+\beta_1D_{8a}),&a=4,..,7\cr
\frac{N_CB}{2\sqrt3},&a=8} \ .
\label{Rgen}
\ee
The quantization prescription then demands the commutation 
relation $[R_a,R_b]=-if_{abc}R_c$ with $f_{abc}$ being the 
antisymmetric structure constants of $SU(3)$. Of course, the 
identification of 
$\partial L(A,\Omega_a)/\partial\mbox{\boldmath $\Omega$}$
as the right generators is consistent with the transformation 
behavior of a vector (\ref{omtrans}). The explicit forms of these
generators in terms of an ``Euler--angle" parametrization of 
$A$ is presented in appendix A. 

The generator $R_8$ is linearly connected to the so--called right 
hypercharge $Y_R=2R_8/\sqrt3=1$ for $B=1$ and $N_C=3$. In analogy 
to the Gell--Mann Nishijima relation \cite{Ge53,Ni53} a right charge 
\be
Q_R=-J_3+\frac{Y_R}{2}
\label{GMN}
\ee
may be defined for the right generators. Completing the analogy we 
note that the eigenvalues 
of $Q_R$ are $0,\pm1/3$, $\pm2/3$, $\pm1$, $\ldots$. Hence for 
$Y_R=1$ the relation (\ref{GMN}) can only be fulfilled when the 
eigenvalue of $J_3$ is half--integer. This yields the important 
conclusion that the $SU(3)$ model describes 
fermions. This is {\it a priori} not expected since the starting 
point has been an effective model of bosons. Arguing from a 
path--integral point of view Witten has even shown that 
these solitons describe fermions when $N_C$ is odd and bosons
when $N_C$ is even \cite{Wi83}. This, of course, is expected from 
considering baryons as being composed of $N_C$ quarks.
It should be stressed that this result could only be gained by 
generalizing the Skyrme model to $SU(3)$ since in $SU(2)$ the 
Wess--Zumino term (\ref{WZterm}) vanishes. We therefore conclude
that the proper incorporation of the anomaly structure of QCD leads 
to the desired spin--statistics relation. 

The left generators, which are defined by the rotation 
$L_a=D_{ab}R_b$, satisfy the commutation relations 
$[L_a,L_b]=if_{abc}L_c$. They provide the isospin, 
$I_i=L_i\ (i=1,2,3)$ and hypercharge, $Y=2L_8/\sqrt3$ operators.

Finally the collective Hamiltonian is conventionally obtained 
as the Legendre transformation $H=-\sum_{a=1}^8R_a\Omega_a-L$
\be
H(A,R_a)&=&E
+\frac{1}{2}\left[\frac{1}{\alpha^2}-\frac{1}{\beta^2}\right]
\mbox{\boldmath $J$}^2+\frac{1}{2\beta^2}C_2(SU(3))
-\frac{3}{8\beta^2}
\nonumber \\ && 
+\frac{\alpha_1}{2\alpha^2}\sum_{i=1}^3D_{8i}
\left(2R_i+\alpha_1D_{8i}\right)
+\frac{\beta_1}{2\beta^2}\sum_{\alpha=4}^7D_{8\alpha}
\left(2R_\alpha+\beta_1D_{8\alpha}\right)
+\frac{1}{2}\gamma \left(1-D_{88}\right)
\nonumber \\ &&
+\frac{1}{2}\gamma_S\left(1-D_{88}^2\right)
+\frac{1}{2}\gamma_T\sum_{i=1}^3D_{8i}D_{8i}
+\frac{1}{2}\gamma_{TS}\sum_{\alpha=4}^7D_{8\alpha}D_{8\alpha}
\label{collham}
\ee
for $B=1$ and $N_C=3$. The constraint $R_8=\frac{\sqrt3}{2}$, 
which yielded the spin--statistics relation, commutes with $H$ 
permitting one to substitute this value. The term involving 
$\sum_{\alpha=4}^7R_\alpha^2$ has been expressed by introducing the 
quadratic Casimir operator of $SU(3)$, $C_2(SU(3))=\sum_{a=1}^8R_a^2$. 
The standard $SU(3)$ representations are eigenstates of $C_2(SU(3))$ 
with eigenvalues $\mu$. For the representations displayed in figure 
\ref{fi_su3rep} one finds $\mu({\bf 8})=3$ and $\mu({\bf 10})=3$. 
The eigenstates of $H(A,R_a)$ have to satisfy an additional 
condition. This arises from the hedgehog structure of the static 
field configuration. The eigenstates of the flavor symmetric 
part of $H(A,R_a)$ are $SU(3)$ representations constituting the 
basis for diagonalizing the whole Hamiltonian.  As outlined in
appendix A, the allowed $SU(3)$ representations must contain
at least one state which has identical spin and isospin \cite{Ma84}, 
like the nucleon or the $\Delta$ resonance.

Under flavor transformations the symmetry breaking parts linear in 
${\cal M}$ behave like the eighth component of an octet \cite{Gu84} as 
can be seen from $[L_a,D_{8b}]=if_{8ac}D_{cb}$. Hence the Gell--Mann 
Okubo mass formulae \cite{Ge64,Ok62}
\be
2\left(M_N+M_\Xi\right)&=&M_\Sigma+3M_\Lambda
\label{GMOmass1} \\
M_\Omega-M_{\Xi^*}&=&M_{\Xi^*}-M_{\Sigma^*}=
M_{\Sigma^*}-M_\Delta
\label{GMOmass2}
\ee
will automatically hold at first order. 
Eqs (\ref{GMOmass2}) are referred to as the equal 
spacing relation for the $\frac{3}{2}^+$ baryons. However, 
there is no guarantee that calculations at leading order in 
symmetry breaking are reliable and indeed higher order 
terms may play a significant role. To investigate these one first 
notes that the symmetry breaking parts mix various $SU(3)$ 
representations, while they are diagonal in the physical quantum 
numbers of the baryons like spin and flavor. As an example a 
matrix element of the nucleon in the octet (${\bf 8}$) and 
anti--decouplet (${\overline {\bf 10}}$) reads
$\langle N, {\bf 8}|D_{88}|N {\overline {\bf 10}}\rangle
=\sqrt5 / 10$. Such matrix elements can be computed by means of 
$SU(3)$ Clebsch--Gordon coefficients \cite{Sw63,Pa89}. 
Up to third order in the perturbation expansion only the 
$SU(3)$ representations ${\bf 8}$, ${\overline {\bf 10}}$ 
and ${\bf 27}$ contribute for the $\frac{1}{2}^+$ baryons.
As an example the shifts in energy, which are associated with 
the dominant symmetry breaking term, are given by \cite{Pa89}
\be
\delta M_N&=&\frac{1}{2\beta^2}\left\{
-0.3\gamma\beta^2-0.0287\left(\gamma\beta^2\right)^2
+0.0006\left(\gamma\beta^2\right)^3+\ldots\right\}
\nonumber \\
\delta M_\Lambda&=&\frac{1}{2\beta^2}\left\{
-0.1\gamma\beta^2-0.0180\left(\gamma\beta^2\right)^2
-0.0003\left(\gamma\beta^2\right)^3+\ldots\right\}
\nonumber \\
\delta M_\Sigma&=&\frac{1}{2\beta^2}\left\{
0.1\gamma\beta^2-0.0247\left(\gamma\beta^2\right)^2
+0.0002\left(\gamma\beta^2\right)^3+\ldots\right\}
\nonumber \\
\delta M_\Xi&=&\frac{1}{2\beta^2}\left\{
0.2\gamma\beta^2-0.0120\left(\gamma\beta^2\right)^2
-0.0006\left(\gamma\beta^2\right)^3+\ldots\right\}\ .
\label{delE3}
\ee
The leading order obviously satisfies (\ref{GMOmass1}). Since the 
second order correction to the ground states is always negative
the deviation from (\ref{GMOmass1}) remains moderate at this order.
Apparently the series converges sufficiently fast because the products 
of Clebsch--Gordon coefficients are always significantly smaller than 
unity. There is one more important fact that can be extracted from the 
expansion (\ref{GMOmass1}). Obviously the effective symmetry breaking 
parameter is the product $\gamma\beta^2$ rather than only the 
coefficient $\gamma$. This can easily be understood because the 
probability for rotations into strange directions not only depends the 
on the repulsive potential (measured by $\gamma$) but also the inertia 
parameter. This feature can also be observed when considering the 
expansion of the nucleon wave--function in terms of $SU(3)$ 
representations
\be
|N\rangle=|N,{\bf 8}\rangle
+0.0745\gamma\beta^2|N, {\overline {\bf 10}}\rangle
+0.0490\gamma\beta^2|N, {\bf 27}\rangle\ldots\ .
\label{nwfexp}
\ee
As in eq (\ref{GMOmass1}) the coefficients are computed from 
$SU(3)$ Clebsch--Gordon coefficients \cite{Sw63}. These admixtures 
of higher dimensional $SU(3)$ representations can be interpreted 
as additional quark--antiquark excitations in the nucleon.

In appendix A it is indicated how the collective Hamiltonian 
(\ref{collham}) can be diagonalized exactly by adopting 
an ``Euler-angle" parametrization of $A$ (\ref{Apara}). Upon
canonical quantization of these ``Euler-angle" the generators 
$R_a$ become differential operators (\ref{Rexpl}) and the 
eigenvalue problem $H\Psi=\epsilon\Psi$ turns into coupled 
partial differential equations. Fortunately the symmetry breaking 
terms acquire quite simple forms as {\it e.g.} 
$1-D_{88}=\frac{3}{2}{\rm sin}^2\nu$. The angle $\nu$ interpolates 
between strange and non--strange directions\footnote{The angle $\nu$ 
will henceforth be denoted as strangeness changing angle.}, {\it cf.} 
eq (\ref{Apara}). Therefore the partial differential equations 
simplify considerably to a set of ordinary coupled
differential equations in the angle $\nu$ when an appropriate 
parametrization for the baryon wave function is adopted 
(\ref{Dpsi}). This finally yields the mass formula
\be
M_B=E+\frac{1}{2}\left(\frac{1}{\alpha^2}-\frac{1}{\beta^2}\right)
J(J+1)-\frac{3}{8\beta^2}+\frac{1}{2\beta^2}\epsilon_{\rm SB}\ ,
\label{bmass}
\ee
where $J$ denotes the spin of the baryon $B$ and $\epsilon_{\rm SB}$
is the eigenvalue of
\be
&&C_2+\beta^2\gamma\left(1-D_{88}\right)
+\beta^2\gamma_S\left(1-D_{88}^2\right)
+\beta^2\gamma_T\sum_{i=1}^3D_{8i}D_{8i}
+\beta^2\gamma_{TS}\sum_{\alpha=4}^7D_{8\alpha}D_{8\alpha}
\nonumber  \\
&&\qquad\qquad
+\beta^2\frac{\alpha_1}{\alpha^2}\sum_{i=1}^3D_{8i}
\left(2R_i+\alpha_1D_{8i}\right)
+\beta_1\sum_{\alpha=4}^7D_{8\alpha}
\left(2R_\alpha+\beta_1D_{8\alpha}\right) \ .
\label{c2problem}
\ee
In ref \cite{Pa89} it has been shown for the simplest case where only 
the term proportional to $\gamma$ is included, that the perturbative 
expansion up to third order deviates from the exact result by a 
negligible amount even for large values of $\gamma\beta^2$.

In the literature the exact diagonalization using the ``Euler--angle" 
parametrization is known as the Yabu--Ando approach \cite{Ya88} to 
three flavor soliton models. In their original approach only the 
simplest symmetry breaking term ($\gamma(1-D_{88})$) had been 
considered because it is the only one appearing in the Skyrme 
model of pseudo--scalar as will be seen in the proceeding section.
Later this treatment was extended to include all types 
of symmetry breaking terms listed in eq (\ref{c2problem}). These 
terms come into the game when vector mesons \cite{Pa92} or explicit 
quark degrees of freedom \cite{We92,Bl93} are included. These models 
will be discussed in chapters \ref{chap_vector} and \ref{chap_NJL}, 
respectively. Of course, the quality of such approaches is
as good as the approximation (\ref{su3rot}) to the exact 
time--dependent solution to the Euler--Lagrange equations,
which should be reasonable for small symmetry breaking. In chapter 
\ref{chap_bound} an approach will be introduced which starts from 
considering flavor symmetry to be large by not treating the 
kaon fields being collective excitations of the hedgehog but rather 
as small amplitude fluctuations in the background of the soliton. 
{\it A posterior} the comparison of both treatments should permit 
the justification of at least one of them. 

\bigskip

\begin{subsatz}
\label{sec_masses}{\bf \hskip1cm Spectrum and Form Factors}
\addcontentsline{toc}{section}{\protect\ref{sec_masses}
Spectrum and Form Factors}
\end{subsatz}
\stepcounter{section}

We now return to the special form of the $SU(3)$ Skyrme model 
defined in eq (\ref{totact1}). In that case the collective Lagrangian 
(\ref{colllag}) simplifies considerably since the coefficients 
$\alpha_1$, $\beta_1$, $\gamma_S$, $\gamma_{T}$ and $\gamma_{TS}$ 
vanish. Substituting the {\it ansatz} (\ref{su3rot}) gives the 
moment of inertia for rotations in coordinate space
\be
\alpha^2=\frac{8\pi}{3}\int dr r^2 \sFt
\left\{\ft+\frac{1}{e^2}\left[\fpt+\frac{\sFt}{r^2}\right]
+8\beta^\prime(1-\cF)\right\} ,
\label{alsky}
\ee
while the only remaining symmetry breaking parameter is obtained to 
be, {\it cf.} eq (\ref{qmratio})
\be
\gamma=\frac{32\pi}{3}(x-1)\int dr \left\{
\delta^\prime r^2(1-\cF)-\beta^\prime\cF
\left(\fpt r^2+2\sFt\right)\right\}.
\label{gasky}
\ee
Although the direct influence of the $\beta^\prime$ type 
symmetry breaker is small its impact on $\gamma$ is very important 
because it accounts for $f_K\ne f_\pi$. As a consequence 
$\delta^\prime(x-1)\propto f_K^2m_K^2-f_\pi^2m_\pi^2\approx
1.5f_\pi^2m_K^2$ and the symmetry breaker $\gamma$ is significantly
increased as compared to the case when the $\beta^\prime$ term is 
omitted \cite{We90}.

Unfortunately the strange moment of inertia is not as straightforwardly 
obtained from the {\it ansatz} (\ref{su3rot}) \cite{Go88}. As already 
remarked the Wess--Zumino term (\ref{WZterm}) is linear in the time 
derivatives. As a consequence static kaon fluctuations, which are 
represented in the form of a two--component isospinor 
$K(\mbox{\boldmath $r$})$,
\be
U(\mbox{\boldmath $r$},t)= A^{\dag}(t)\
{\rm e}^{iZ(\mbox{\boldmath $r$})}\ U_0 \
{\rm e}^{iZ(\mbox{\boldmath $r$})}\ A(t)\ ,
\qquad
Z(\mbox{\boldmath $r$})=\pmatrix{\mbox{\large $0$} &
{| \atop |}&\hspace{-10pt}K(\mbox{\boldmath $r$}) \cr
-----\hspace{-8pt}&-&\hspace{-10pt}---\cr
\ K^{\dag}(\mbox{\boldmath $r$})&|&\hspace{-8pt} 0\cr},
\label{kaonind}
\ee
appear in the action with a linear coupling to the time derivative of 
the rigidly rotating hedgehog (\ref{su3hedgehog}). Stated otherwise, 
the Wess--Zumino term provides the source for induced kaon fields
$K(\mbox{\boldmath $r$})$. As this source is linear in the angular 
velocities $\Omega_4,\ldots, \Omega_7$ the induced kaon fields are 
as well. Hence a suitable {\it ansatz}, which also takes care of the 
pseudo--scalar nature of the kaon fields is \cite{Go88,We90}
\be
K(\mbox{\boldmath $r$})=W(r)\mbox{\boldmath $\tau$}\cdot
\hat{\mbox{\boldmath $r$}}\Omega_K\ , \quad
\Omega_K=
\pmatrix{\Omega_4-i\Omega_5 \cr\Omega_6-i\Omega_7}.
\label{kaonansatz}
\ee
Expanding the Lagrangian up to quadratic order in $\Omega_\alpha$ 
yields additional contributions for the strange moment of inertia
$\beta^2$ which then is a functional of the radial function $W(r)$.
The explicit form of this functional is displayed in appendix B,
eqs (\ref{betasks},\ref{betasksb}). In principle $W(r)$ is complex, 
however, it turns out that only the real part becomes excited by the 
Wess--Zumino term. The radial dependence of $W(r)$ is finally 
determined by extremizing $\beta^2$. Although this completes the 
calculation of the collective Lagrangian a note should be added 
concerning possible double counting of kaonic fields. Denoting 
those kaon fields which are already contained in the original 
{\it ansatz} (\ref{su3hedgehog}) by $K_0$ it turns out that the 
overlap
\be
\langle K_0 | K \rangle \propto
\int_0^\infty r^2 dr W(r)\ \sh
\label{overlap}
\ee
indeed is non--zero. Although an overall normalization of the 
overlap is missing it turns out that (\ref{overlap}) decreases 
rapidly when increasing the parameter representing the kaon mass 
$m_K$. In ref \cite{We90} one order of magnitude of decrease was 
obtained for (\ref{overlap}) by changing $m_k$ from $200{\rm MeV}$ 
to the physical value of $495{\rm MeV}$. This behavior indicates that 
for physically relevant parameters the overlap is actually 
negligible\footnote{A more elaborate metric in (\ref{overlap}) might 
alter this conclusion \cite{Wapriv}.}. It is worthwhile to mention 
that at least in principle also the symmetry breaking will induce 
kaon fields \cite{Pa92}. At present, however, these have not been 
considered.

Except for the Skyrme parameter $e$ all parameters are determined 
in the meson sector (\ref{pspara1}). Choosing $e=4.0$ finally 
provides\footnote{In the main part of ref \cite{We90} a simplified 
version of ${\cal L}_{SB}$ was employed to compute $\beta^2$, which
yielded a somewhat larger value, see however appendix C of that 
reference.} 
\be
E&=&1.744{\rm GeV}\ , \qquad
\gamma= 1.374{\rm GeV}\ , 
\nonumber \\
\alpha^2&=& 6.04 {\rm GeV}^{-1}\ ,\qquad 
\beta^2= 4.53 {\rm GeV}^{-1}\ ,
\label{pscopara}
\ee
while all other parameters in the collective Hamiltonian 
(\ref{collham}) vanish. Although the strange moment of inertia 
$\beta^2$ is dominated by the hedgehog contribution the share due to 
the induced components (\ref{kaonind}) is about 35\% . One might be 
tempted to enforce a vanishing matrix element (\ref{overlap})
by including an appropriate Lagrange multiplier. This reduces 
$\beta^2$ somewhat to $3.52{\rm GeV}^{-1}$. Although such a treatment 
would be mandatory if one wanted to construct $W(r)$ in the flavor 
symmetric case, there is no apparent need to do so when this symmetry 
is broken because then the homogeneous part of the differential 
equation does not possess a regular solution. In other words, the 
constraint represents the driving term in the equation of motion only 
when flavor symmetry breaking is small. We will therefore stick to the 
value in eq (\ref{pscopara}). In any event, the appearance of the 
induced components are an indication that the collective treatment is 
merely an approximation to the physical case when flavor symmetry is 
broken. This is especially reflected by the fact that the radial 
function $W(r)$ decays like ${\rm exp}(-m_K r)$ in contrast to the 
hedgehog configuration rotated into strange direction according to 
(\ref{su3rot}). The latter behaves like ${\rm exp}(-m_\pi r)$ at large 
$r$. Furthermore the inclusion of induced kaon fields is mandatory to 
obtain the proper divergence of the axial current. This proof is 
indicated in appendix C.

This value $e=4.0$ has been chosen since it leads to a reasonable 
description of the baryon mass differences as can be seen from 
table \ref{ta_mdiffps}. 
\begin{table}
\caption{\label{ta_mdiffps}\tenrm
The mass differences, which are obtained by exact diagonalization 
of the collective Hamiltonian (\protect\ref{collham}), of the 
neighboring ${\Ss \frac{1}{2}^+}$ and ${\Ss \frac{3}{2}^+}$ in the 
pseudo--scalar model for ${\Ss e=4.0}$ are compared to the 
experimental data. Also the mass differences with respect to the 
nucleon are listed. The values in parentheses are obtained with a 
Lagrange multiplier included to guarantee a vanishing overlap 
(\protect\ref{overlap}) of the induced kaon components with the 
would--be zero mode of the hedgehog configuration. In that case the 
Skyrme parameter has slightly been readjusted to ${\Ss e=3.9}$.
All data are in MeV.}
~
\newline
\centerline{\tenrm\smalllineskip
\begin{tabular}{c | c  c || c | c c}
Baryons & Model & Expt. &
Baryons & Model & Expt. \\
\hline
$\Lambda-N$        & 154 (163) & 177 &
$\Lambda-N$        & 154 (163) & 177 \\
$\Sigma-\Lambda$   & 88  (101)  & 77  &
$\Sigma-N$         & 242 (264) & 254 \\
$\Xi-\Sigma$       & 124 (122) & 125 &
$\Xi-N$            & 366 (388) & 379 \\
$\Delta-\Xi$       &-88 (-120) & -86 &
$\Delta-N$         & 278 (268) & 293 \\
$\Sigma^*-\Delta$  & 132 (138) & 153 &
$\Sigma^*-N$       & 410 (406) & 446 \\
$\Xi^*-\Sigma^*$   & 134 (139) & 145 &
$\Xi^*-N$          & 544 (545) & 591 \\
$\Omega-\Xi^*$     & 133 (135) & 142 &
$\Omega-N$         & 677 (680) & 733 \\
\end{tabular}}
\end{table}
A major reason for this result is the fact that $\gamma$ is 
significantly enlarged by including the effects associated with 
$f_K\ne f_\pi$. These effects were omitted in the original 
studies \cite{Gu84,Pr83,Ch85,Ya88} yielding far too low mass 
splittings between baryons of different strangeness for physically 
motivated parameters of the effective Lagrangian\footnote{Many 
of these authors used to consider $f_\pi$ as a free parameter fitted 
to the absolute values of the baryon masses . Without the 
$\beta^\prime$ term this yielded $f_\pi$ as low as 25MeV 
\cite{Pr83}.}. It is also apparent from table \ref{ta_mdiffps}
that the inclusion of a constraint to ensure a vanishing 
overlap (\ref{overlap}) can be compensated by a small variation of 
the Skyrme parameter, $e$. This indicates that double counting 
effects play only a subleading role. It is interesting to remark that 
the mass differences for the $\frac{1}{2}^+$ baryons deviate strongly 
from the prediction in leading order of the flavor symmetry breaking. 
This can easily be observed from the ratios
\be
\left(M_\Lambda-M_N\right):
\left(M_\Sigma-M_\Lambda\right):
\left(M_\Xi-M_\Sigma\right)=1:0.52:0.85
\label{mratio}
\ee
which are in much better agreement with the experimental data 
(1:0.43:0.69) than the leading order result (1:1:0.5) of eq 
(\ref{delE3}). Obviously the higher order contributions are 
important. This also indicates that the baryon wave--functions 
contain sizable admixtures of higher dimensional $SU(3)$ 
representations, {\it cf.} eq (\ref{nwfexp}). Nevertheless the 
deviation from the Gell--Mann--Okubo relations (\ref{GMOmass1}) is 
only moderate, in particular the equal spacing among the 
$\frac{3}{2}^+$ baryons is well reproduced.

The value for the Skyrme parameter $e=4.0$ obtained from this
best fit to the baryon mass differences is then employed to evaluate 
static properties of baryons within this model. In order to do 
so one first constructs the Noether currents associated 
with the symmetry transformation (\ref{uchitrans}). A convenient 
method is to extend these global symmetries to local ones by 
introducing external gauge fields ({\it e.g.} the gauge fields 
of the electroweak interactions) into the action (\ref{totact1}). 
The Noether currents are then read off as the expressions which 
couple linearly to these gauge fields. This procedure is especially 
appropriate for the Wess--Zumino term (\ref{WZterm}) because this 
non--local term can only be made gauge invariant by a trial and 
error type procedure \cite{Wi83,Ka84b}. The final form of the nonet 
($a=0,\ldots,8$) vector ($V_\mu^a$) and axial--vector ($A_\mu^a$) 
currents reads \cite{Pa91} (for $N_C=3$)
\be
V_\mu^a (A_\mu^a) & = & -\frac{i}{2}f_\pi^2\ 
{\rm tr}\left\{Q^a\left(\alpha_\mu\mp\beta_\mu\right)\right\}
-\frac{i}{8e^2}{\rm tr}\left\{Q^a\left(
\left[\alpha_\nu,\left[\alpha_\mu,\alpha_\nu\right]\right]\mp
\left[\beta_\nu,\left[\beta_\mu,\beta_\nu\right]\right]
\right)\right\}
\nonumber \\ &&
-\frac{1}{16\pi^2}\epsilon^{\mu\nu\rho\sigma}
{\rm tr}\left\{Q^a\left(\alpha_\nu\alpha_\rho\alpha_\sigma
\pm\beta_\nu\beta_\rho\beta_\sigma\right)\right\}
\nonumber \\ &&
-i\beta^\prime{\rm tr}\left\{Q^a\left(
\{{U{\cal M}+\cal M}U^{\dag},\alpha_\mu\}\mp
\{{\cal M}U+U^{\dag}{\cal M},\beta_\mu\}\right)\right\},
\label{currps}
\ee
where $Q^a=(\frac{1}{3},\frac{\lambda^1}{2},\ldots,
\frac{\lambda^8}{2})$ denote the Hermitian nonet generators.
The combination
\be
Q^{\rm e.m.}={\rm diag}
\left(\frac{2}{3},-\frac{1}{3},-\frac{1}{3}\right)
=Q^3+\frac{1}{\sqrt3}Q^8
\label{emgen}
\ee
is of special interest because it enters the computation of the 
electromagnetic properties. The associated form factors of the 
$\frac{1}{2}^+$ baryons ($B$) are defined by
\be
\langle B(\mbox{\boldmath $p$}^\prime)|V_\mu^{\rm e.m.}|
B(\mbox{\boldmath $p$})\rangle=
{\overline u}(\mbox{\boldmath $p$}^\prime)\left[
\gamma_\mu F^B_1(q^2)+
\frac{\sigma_{\mu\nu}q^\nu}{2M_B}F^B_2(q^2)\right]
u(\mbox{\boldmath $p$})\ , \qquad
q_\mu=p_\mu-p_\mu^\prime \ .
\label{defff}
\ee
Frequently it is convenient to introduce ``electric" and 
``magnetic" form factors
\be
G_E^B(q^2)=F^B_1(q^2)-\frac{q^2}{4M_B^2}F^B_2(q^2)\ , \qquad
G_E^B(q^2)=F^B_1(q^2)+F^B_2(q^2)\ .
\label{emff}
\ee
Further relevant form factors are the matrix elements of the 
diagonal vector currents of the individual quarks
in the proton state
\be
\langle P(\mbox{\boldmath $p$}^\prime)|
{\overline q}_i\gamma_\mu q_i|
P(\mbox{\boldmath $p$})\rangle=
{\overline u}(\mbox{\boldmath $p$}^\prime)\left[
\gamma_\mu F_i(q^2)+
\frac{\sigma_{\mu\nu}q^\nu}{2M_p}\tilde F_i(q^2)\right]
u(\mbox{\boldmath $p$})\ , \qquad
i={\rm u},{\rm d},{\rm s} \ ,
\label{pvecmat}
\ee
as well the analogous expressions for the axial--vector currents
\be
\langle P(\mbox{\boldmath $p$}^\prime)|
{\overline q}_i\gamma_\mu\gamma_5 q_i|
P(\mbox{\boldmath $p$})\rangle=
{\overline u}(\mbox{\boldmath $p$}^\prime)\left[
\gamma_\mu H_i(q^2)+
\frac{q_\mu}{2M_p}\tilde H_i(q^2)\right]\gamma_5
u(\mbox{\boldmath $p$})\ .
\label{paxmat}
\ee
The list of relevant form factors is completed by the matrix 
elements of strangeness changing components of the currents 
(\ref{currps}) between different baryon states. These form 
factors are relevant for the description of the semi--leptonic 
decays of the hyperons. Their generic form is
\be
\langle B^\prime(\mbox{\boldmath $p$}^\prime)|V_\mu^\alpha|
B(\mbox{\boldmath $p$})\rangle & = &
{\overline u}(\mbox{\boldmath $p$}^\prime)\left[
\gamma_\mu g_V(q^2)+\ldots\right]u(\mbox{\boldmath $p$})\ ,
\nonumber \\
\langle B^\prime(\mbox{\boldmath $p$}^\prime)|A_\mu^\alpha|
B(\mbox{\boldmath $p$})\rangle & = &
{\overline u}(\mbox{\boldmath $p$}^\prime)\left[
\gamma_\mu\gamma_5 g_A(q^2)+\ldots\right]
u(\mbox{\boldmath $p$})\ .
\label{semlepmat}
\ee
Here the ellipses represent contributions involving the momentum 
transfer $q_\mu$, which need not be considered in the present context. 
To be precise, separate $g_V(q^2)$ and $g_A(q^2)$ must be introduced
for each baryon pair ($B^\prime,B$) and flavor index 
$\alpha=4,\ldots,7$.

Having collected these definitions of the form factors it is 
straightforward (although tedious) to compute the corresponding 
predictions of the $SU(3)$ Skyrme model. In the first step the 
parametrization (\ref{kaonind}) is substituted into the defining 
equation of the currents (\ref{currps}). This yields for the 
spatial components of the vector current\footnote{The conventions 
are $i,j,k=1,2,3$ and $\alpha,\beta=4,\ldots,7$.}
\be
V_i^a&=&V_1(r)\epsilon_{ijk}x_jD_{ak}
+\frac{\sqrt3}{2}B(r)\epsilon_{ijk}\Omega_jx_kD_{a8}
+V_2(r)\epsilon_{ijk}x_jd_{d\alpha\beta}D_{a\alpha}\Omega_\beta
\nonumber \\ &&
+V_3(r)\epsilon_{ijk}x_jD_{88}D_{ak}
+V_4(r)\epsilon_{ijk}x_jd_{d\alpha\beta}D_{8\alpha}D_{a\beta}
+\ldots \ ,
\label{spcurrps}
\ee
where 
\be
B(r)=\frac{-1}{2\pi^2}\fp\frac{\sFt}{r^2}
\label{bdensity}
\ee
is the baryon number density (\ref{bnumber}). The explicit form of 
the radial functions $V_1(r),\ldots,V_4(r)$ is given in appendix B of 
ref \cite{Pa91}. The ellipsis in eq (\ref{spcurrps}) represent 
terms, which vanish when sandwiched between baryon 
states. According to the quantization prescription (\ref{Rgen}) 
the angular velocities $\Omega_a$ are substituted by the 
right generators $R_a$ of $SU(3)$. Taking the Fourier transform 
of the resulting matrix elements allows one to identify the 
magnetic form factor in the Breit frame \cite{Br86,Me87}
\be
G_M^B(\mbox{\boldmath $q$}^2)&=&-8\pi M_B
\int_0^\infty \hspace{-0.2cm} r^2 dr
\frac{r}{|\mbox{\boldmath $q$}|}j_1(r|\mbox{\boldmath $q$}|)
\Bigg\{V_1(r)\langle D_{e3}\rangle_B
-\frac{1}{2\alpha^2}B(r)\langle D_{e8}R_8\rangle_B
\label{Gmps} \\ && \hspace{1cm}
-\frac{1}{\beta^2}V_2(r)\langle
d_{3\alpha\beta}D_{e\alpha}R_\beta\rangle_B
+V_3(r)\langle D_{88}D_{e3}\rangle_B
+V_4(r)\langle d_{3\alpha\beta}D_{e\alpha}D_{8\beta}\rangle_B
\Bigg\} \ .
\nonumber
\ee
Here the flavor index $e$ refers to the ``electromagnetic"
direction (\ref{emgen}). The magnetic moment corresponds to the 
magnetic form factor at zero momentum transfer $\mu_B=G_M^B(0)$.
Similarly the electric form factor is given by Fourier 
transforming the time component of the electromagnetic current
\be
G_E^B=4\pi\int_0^\infty \hspace{-0.2cm}
r^2 dr j_0(r|\mbox{\boldmath $q$}|)
\left\{\frac{\sqrt3}{2}B(r)\langle D_{e3}\rangle_B
+\frac{1}{\alpha^2}V_7(r)\langle D_{ei}R_i\rangle_B
+\frac{1}{\beta^2}V_8(r)\langle D_{e\alpha}R_\alpha\rangle_B
\right\} \hspace{-0.1cm} .
\label{Geps}
\ee
The two new radial functions $V_7(r)$ and $V_8(r)$ are listed in 
appendix B of ref \cite{Pa91} as well. Integrating $V_7$ and $V_8$ 
yields the moments of inertia, $\alpha^2$ and $\beta^2$, respectively. 
Hence the electric charges are properly normalized. It should be 
remarked that the baryon matrix elements in the space of the collective 
coordinates are computed using the exact eigenstates (\ref{Dpsi})
of (\ref{c2problem}) and adopting the representations (\ref{Rexpl})
for the $SU(3)$ generators. The results for the magnetic 
moments and the radii
\be
r_M^2=-\frac{6}{\mu_B}\frac{d G_M^B(\mbox{\boldmath $q$}^2)}
{d\mbox{\boldmath $q$}^2}\Bigg|_{\mbox{\boldmath $q$}^2=0}\ ,
\quad
r_E^2=-6\frac{d G_E^B(\mbox{\boldmath $q$}^2)}
{d\mbox{\boldmath $q$}^2}\Bigg|_{\mbox{\boldmath $q$}^2=0}\ .
\label{radii}
\ee
are shown in table \ref{ta_emps}.
\begin{table}
\caption{\label{ta_emps}\tenrm
The electromagnetic properties of the baryons compared to the 
experimental data. The predictions of the Skyrme model are taken 
from ref \protect\cite{Pa91}.}
~
\newline
\centerline{\tenrm\smalllineskip
\begin{tabular}{c| c c | c c | c c}
 & \multicolumn{2}{c|}{$\mu_B({\rm n.m.})$} &
\multicolumn{2}{c|}{$r_M^2({\rm fm}^2)$} &
\multicolumn{2}{c}{$r_E^2({\rm fm}^2)$} \\
$B$ & $e=4.0$ & Expt. & $e=4.0$ & Expt. & $e=4.0$ & Expt. \\
\hline
$p$         & 2.03 & 2.79 & 0.43 & 0.74 & 0.59 & 0.74 \\
$n$         &-1.58 &-1.91 & 0.46 & 0.77 &-0.22 &-0.12 \\
$\Lambda$   &-0.71 &-0.61 & 0.36 & ---  &-0.08 & ---  \\
$\Sigma^+$  & 1.99 & 2.42 & 0.45 & ---  & 0.59 & ---  \\
$\Sigma^0$  & 0.60 & ---  & 0.36 & ---  &-0.02 & ---  \\
$\Sigma^-$  &-0.79 &-1.16 & 0.58 & ---  &-0.63 & ---  \\
$\Xi^0$     &-1.55 &-1.25 & 0.38 & ---  &-0.15 & ---  \\
$\Xi^-$     &-0.64 &-0.69 & 0.43 & ---  &-0.49 & ---  \\
$\Sigma^0\rightarrow\Lambda$
            &-1.39 &-1.61 & 0.48 & ---  & ---  & ---  \\
\end{tabular}
}
\end{table}
As in the two flavor model \cite{Ad83} the isovector part of the 
magnetic moments is underestimated while the isoscalar part is 
reasonably well reproduced. Despite the fact that the flavor 
symmetry breaking is large for the baryon wave--functions, the 
predicted magnetic moments do not strongly deviate from the 
$SU(3)$ relations \cite{Ad85}
\be
\mu_{\Sigma^+}&=&\mu_p\ ,\quad \mu_{\Sigma^0}=\frac{1}{2}
(\mu_{\Sigma^+}+\mu_{\Sigma^-})\ ,\quad 
\mu_{\Sigma^-}=\mu_{\Xi^-}\ ,
\nonumber \\
2\mu_\Lambda&=&-(\mu_{\Sigma^+}+\mu_{\Sigma^-})
=-2\mu_{\Sigma^0}=\mu_n=\mu_{\Xi^0}=
{2\over{\sqrt3}}\mu_{\Sigma^0\Lambda}\ .
\label{su3mag}
\ee
Later it will become clear that an even more elaborate treatment of 
the flavor symmetry breaking is necessary in order to accommodate for 
the experimental breaking of the $U$--spin symmetry which {\it e.g.} 
causes the approximate identity $\mu_{\Sigma^+}\approx\mu_p$.
The moderate difference between the various magnetic radii 
$r_M^2$ is a further hint that symmetry breaking effects are 
mitigated. The comparison with the available empirical data for 
the radii shows that the predictions turn out too small in 
magnitude (except of the neutron electric radius). This is a strong 
indication that essential ingredients are still missing in the 
model. In chapter \ref{chap_vector} it will be explained that 
the effects, which are associated with vector meson dominance 
(VMD), will account for this deficiency. Nevertheless the overall 
picture gained for the electromagnetic properties of the 
$\frac{1}{2}^+$ and $\frac{3}{2}^+$ can at least be characterized 
as satisfactory, especially in view of the fact that 
the only free parameter of the model has been fixed beforehand.
As a side remark it should be mentioned that the direct contribution 
of the induced fields (\ref{kaonind}) to the magnetic moments is 
as small as 5--10\%. For example, when supplementing the model by 
a Lagrange multiplier such that the overlap (\ref{overlap}) vanishes,
the proton and neutron magnetic moments are changed to 1.88 and -1.49,
respectively. The results displayed in table \ref{ta_emps} might 
be regained by a minor change of the Skyrme parameter, $e$. The 
relevance of these fields appears merely to be of formal nature.

From here it is straightforward to compute the strange vector 
form factors $F_s(q^2)$ and $\tilde F_s(q^2)$, which are defined in 
eq (\ref{pvecmat}). Up to now no precise measurement 
of the associated form factors has been performed. As already indicated 
in section \ref{sec_strange}, these form factors are currently under 
intensive experimental investigation, {\it cf.} refs \cite{Mu93,Mu94}.
These form factors have been estimated in various models. They range 
from vector--meson--pole fits \cite{Jaf89} of dispersion relations
\cite{Ho74} through vector meson dominance approaches \cite{Pa91}
and kaon--loop calculations with \cite{Mu93} and without \cite{Fo94}
vector meson dominance contributions to soliton model calculations
\cite{Pa91,Pa92,We95a}. The numerical results for the strange 
magnetic moment $\mu_S=\tilde F_s(0)\approx
-0.31\pm0.09\ \ldots \ 0.25$ are quite diverse. The predictions 
for the strange charge radius $r_S^2=-6dF_s(q^2)/dq^2|_{q=0}$
are almost equally scattered $r_S^2\approx -0.20\ \ldots \ 
0.14 {\rm fm}^2$. See table \ref{ta_strange} for a comprehensive list 
of predictions on $\mu_S$ and $r_S^2$. In order to evaluate these 
objects in the three flavor Skyrme model one requires the matrix 
elements of the ``strange" combination
\be
Q^s=\frac{1}{3}{\bf 1}-\frac{1}{\sqrt3}\lambda_8=
Q^0-\frac{2}{\sqrt3}Q^8
\label{strgen}
\ee
between proton states rather than the electromagnetic ones 
(\ref{emgen}). Using the above established value $e=4.0$ yields
\be
\mu_S=-0.13 \ , \qquad
r_S^2=-0.10 {\rm fm}^2 \ .
\label{stresps}
\ee
As usual, magnetic moments are in units of nuclear magnetons.
It should be stressed that these results are obtained within 
the Yabu--Ando approach, {\it i.e.} the proton wave--function
contains sizable admixtures of higher dimensional representations.
If a pure octet wave--function were employed to compute the 
matrix elements of the collective operators the strange magnetic 
moment would have been $\mu_S=-0.33$. The proper inclusion of 
symmetry breaking into the nucleon wave--function apparently reduces 
the effect of the strange degrees of freedom in the nucleon. This 
is also intuitively clear, since for a flavor symmetric 
wave--function virtual ${\overline s}s$ pairs are as probable as 
virtual ${\overline u}u$ or ${\overline d}d$ pairs. However, 
as the strange quarks within the nucleon become more massive 
(effect of symmetry breaking) their excitation is less likely.

Before discussing the predictions of the three flavor Skyrme 
model on the axial properties of the $\frac{1}{2}^+$ baryons it 
is illuminating to discuss a few general aspects of the 
so--called proton spin puzzle. The purpose of these studies is to 
disentangle the three form factors $H_i(q^2=0)$ of the proton matrix 
element (\ref{paxmat}). Assuming isospin invariance the analysis of 
the neutron--beta decay provides the linear combination 
\be
H_u(0)-H_d(0)=g_A=1.257 \ .
\label{gaexp}
\ee
More recently extensive measurements have been carried out to 
gain data for the combination
\be
\frac{1}{9}H_u(0)\left(C_{ns}(q^2)+C_s(q^2)\right)
-\frac{1}{18}\left(H_d(0)+H_s(0)\right)
\left(C_{ns}(q^2)+2C_s(q^2)\right) 
= \Gamma_1^p(q^2)\ .
\label{EMCcomb}
\ee
The momentum dependent coefficients $C_s,\ C_{ns}$ are computed 
in perturbative QCD and may {\it e.g.} be taken from ref \cite{La94}.
The combined analysis of the EMC \cite{As88}, SLAC \cite{An93} 
and SMC \cite{Ad94} data may be summarized as $\Gamma_1^p(q^2=
(10.7{\rm GeV})^2)=0.129\pm0.010$. Defining $R$ as the linear 
combination, which corresponds to the eighth component of the 
axial--vector current matrix element
\be
H_u(0)+H_d(0)-2H_s(0)=R\ ,
\label{defR}
\ee
flavor symmetry may provide the final ingredient to extract all 
three $H_i(0)$. Taking this symmetry for granted allows one to relate 
$R$ to the data obtained from the semi--leptonic hyperon decays 
such as $\Lambda\rightarrow pe^-{\overline{\nu}}_e$ yielding
$R=0.575\pm0.016$ \cite{Bo83,Hs88,Jaf90,Cl93}. For the time being
let us, however, treat $R$ as an independent parameter. The 
dependences of $H_s(0)$ as well as of the singlet combination
\be
H(q^2)=H_u(q^2)+H_d(q^2)+H_s(q^2)
\label{defsing}
\ee
at $q^2=0$ on $R$ are shown in figure \ref{fi_hhh}.
One observes that $H(0)\approx0.3\pm0.1$ within a wide range of $R$. 
The issue of $H(0)$ being considerably smaller than unity has become 
known as the proton spin puzzle (or crisis) because in a quark parton 
model interpretation this quantity is identical to twice the quarks' 
contribution to the proton spin. The puzzle is caused by the fact that 
$H(0)$ is unity in a na\"{\i}ve quark model. On the other hand the 
strange quark contribution varies strongly with symmetry breaking. For 
the flavor symmetric value of $R$ one finds $H_s(0)\approx-0.11\pm0.05$. 
Originally this sizable value caused an additional point of confusion
because it implies the unexpectedly large ratio
$|H_s(0)/H_d(0)|\approx0.3$. However, as the role of 
symmetry breaking became clearer it was soon recognized that a 
smaller $H_s(0)$ may well be in agreement with the experimental data. 

Simultaneously with the publication of the first data on $\Gamma_1^P$
\cite{As88}, the three flavor Skyrme model was observed to be an 
excellent candidate to explain the smallness of $H(0)$ \cite{Br88}.
From eq (\ref{currps}) the axial singlet current ($Q^0=1/3$) is 
found to be
\be
A_\mu^0=-\frac{i}{3}f_\pi^2{\rm tr}
\left(\partial_\mu U U^{\dag}\right)
-4i\beta^\prime\partial_\mu{\rm tr}
\left({\cal M}\left[U-U^{\dag}\right]\right) \ .
\label{A0ps}
\ee
Separating the singlet piece of the nonet matrix $U$ by introducing 
the unitary--unimodular matrix $\tilde U={\rm e}^{-i\chi}U$ leads to
\be
A_\mu^0=f_\pi^2\partial_\mu\chi
-4i\beta^\prime\partial_\mu{\rm tr}
\left({\cal M}\left[U-U^{\dag}\right]\right)
\label{A0ps1}
\ee
which obviously is a total derivative \cite{Pa89a}. Hence only the 
induced form factor 
$\tilde H(q^2)=\tilde H_u(q^2)+\tilde H_d(q^2)+\tilde H_s(q^2)$ in 
eq (\ref{paxmat}) acquires a non--zero value. This implies that
in the Skyrme model which contains only pseudo--scalar fields
\be
H(q^2) \equiv 0\ ,
\label{Hpsmodel}
\ee
even in the presence of flavor symmetry breaking terms. This is in 
contrast to the original estimates \cite{Br88} of the effect of 
symmetry breaking on $H(q^2)$. Later it will be argued that short 
range effects are responsible for a non--vanishing (although small) 
value for $H(0)$. There have been several attempts to obtain 
a non--vanishing $H(0)$ in the pseudo--scalar Skyrme model. 
Adding the symmetric fourth order term 
${\rm tr}(\alpha_\mu\alpha^\mu\alpha_\nu\alpha^\nu)$ gives a 
contribution to axial the singlet current \cite{Ry89}
\be
\delta A_\mu^0 = ic{\rm tr}
\left(\alpha_\nu\alpha_\mu\alpha^\nu\right)\ ,
\label{Ryzak}
\ee
where $c$ is an arbitrary constant. This expression contains terms 
which are quadratic in the angular velocities $\Omega_a$ causing 
ordering ambiguities when applying the quantization  prescription 
(\ref{Rgen}). Although originally a non--vanishing contribution to 
$H(0)$ was obtained \cite{Ry89} it has recently been shown that this 
result was almost completely due to those ordering ambiguities. 
In the two flavor case one can easily demand the proper behavior 
under the particle conjugation transformation. This requires one to 
choose an ordering such that $H(0)$, associated with (\ref{Ryzak}, 
vanishes. In the three flavor model, however, the proper ordering of 
the collective coordinates might give $H(0)\ne0$ \cite{Ka94}. From a 
physical point of view it is, nevertheless, more appealing to find 
already a non--vanishing $H(0)$ in the two flavor version of the model 
under consideration. Another way to achieve $H(0)\ne0$ is to augment 
the Lagrangian by 
\be
\delta{\cal L}=
{\tilde c}\ \epsilon_{\mu\nu\rho\sigma}
\partial^\mu({\rm ln \ det}U)B^\nu\partial^\rho B^\sigma \ ,
\label{locapp}
\ee
which is motivated by the local approximation to vector mesons
\cite{Wa95a} and indeed leads to a non--trivial contribution to 
the axial singlet current \cite{Co89}
\be
\delta A_\mu^0 = 3{\tilde c}\ \epsilon_{\mu\nu\rho\sigma}
B^\nu\partial^\rho B^\sigma \ .
\label{a0maryl}
\ee
However, as (\ref{locapp}) does not constitute a single trace, this 
type of singlet current will suffer additional suppression from the 
quark line (or OZI \cite{OZI63}) rule. Therefore the explicit 
consideration of vector meson seems to be more appropriate (see 
chapter \ref{chap_vector}). Aiming towards completeness (without 
possibly gaining it) on this subject, it should be mentioned that 
there are other studies of the axial singlet matrix elements in 
related models. These are {\it e.g.} a linear chiral model with 
quark and gluon confinement \cite{St89} or a hybrid chiral bag model 
\cite{Dr89}. It is interesting to note that in this bag model the 
axial singlet matrix element is quite sensitive to the bag 
radius\footnote{See ref \cite{Rho94} for a review on bag models
and the {\it Cheshire Cat} principle.}, $R$, in contrast to $g_A$ 
\cite{Ka86a}. The authors of ref \cite{Dr89} have employed this 
sensitivity to constrain the radius $R{_<\atop^\sim}0.5{\rm fm}$
from the data on the axial singlet matrix element.

Due to the vanishing singlet matrix element in the Skyrme model 
of pseudo--scalars it is more appropriate to discuss the general 
behavior of matrix elements of the axial current under symmetry 
breaking rather than the detailed results for these quantities.
The leading order term (in $1/N_C$) of the spatial components of 
the axial current is straightforwardly obtained to be
\be
\int d^3r A_i^a = {\cal C}D_{ai}(A) \ .
\label{aiagen}
\ee
The constant ${\cal C}$ denotes an integral over the chiral angle.
However, since its actual value will be of no importance for the 
following discussions we refer the interested reader to refs 
\cite{Pa90,Pa91} for the explicit expression. The semi--leptonic 
decays of the hyperons\footnote{In ref \cite{Pr86} non--leptonic 
decays have been studied using flavor symmetric wave--functions.} are 
described by the ratios $g_A/g_V$ defined in eq (\ref{semlepmat}). 
The numerator can easily be obtained from (\ref{aiagen})
\be
g_A^a(B^\prime,B)=
{\cal C}\langle B^\prime | D_{a3} | B \rangle \ .
\label{calgaa}
\ee
The flavor index $a$ has to be chosen according to whether 
strangeness conserving ($a=1,2,3,8$) or strangeness changing
($a=4,\ldots,7$) processes are considered.
Although for the current studies the actual value of ${\cal C}$ 
is not important, it should be admitted that the corresponding 
result for the axial charge of the nucleon 
$g_A=g_A^{1+i2}(n,p)$, as measured in neutron--beta decay, is 
predicted too low in many soliton models\footnote{Note, however, 
that in the chiral quark model $g_A$ is somewhat overestimated
\cite{Ja88a}.}. This problem is already encountered in the two 
flavor model and gets worse in $SU(3)$ as the Clebsch--Gordon 
coefficient associated with $D_{1+i2\ 3}$ changes by a factor 
of 7/10. As symmetry breaking is included the $SU(3)$ prediction 
for $g_A$becomes larger \cite{Pa89}
\be
g_A(SU(3))=\frac{7}{10}\left[1
+0.0514\gamma\beta^2+\ldots\right]g_A(SU(2)) \ .
\label{expga}
\ee
Actually the exact treatment shows that with increasing symmetry 
breaking the two flavor result is approached, although only slowly.
Taking everything together, including subleading terms in 
(\ref{calgaa}), finally gives $g_A=0.98$ for $e=4.0$ \cite{Pa91}
which is about 4/5 of the experimental value 
$g_A({\rm expt.})=1.26$.

Returning to the more general studies it should first be remarked 
that flavor symmetry relates the matrix elements between various 
baryons. Most conveniently they are expressed in terms of two 
unknown constants $F$ and $D$. One has to use models to determine 
these constants. In the flavor symmetric Skyrme model one finds 
\cite{Ad85} $D/F=9/5$ and $D+F=7{\cal C}/15=g_A$. In table 
\ref{ta_hyperon} the flavor symmetric dependencies of the matrix 
elements on $F$ and $D$ are displayed. In addition one finds $R=3F-D$.
\begin{table}
\caption{\label{ta_hyperon}\tenrm
The matrix elements of the axial--vector current \protect\ref{calgaa}
between different baryon states in the flavor symmetric limit.
Displayed are both the strangeness conserving (a) and strangeness 
changing (b) processes. The first column gives the relevant 
flavor component of the axial current.}
~
\newline
\centerline{\tenrm\smalllineskip
\begin{tabular}{c| c c c c}
& \multicolumn{4}{c}{(a)} \\
& $n\rightarrow p$ & $\Sigma^-\rightarrow\Lambda$ &
$\Sigma^-\rightarrow\Sigma^0$ & $\Xi^-\rightarrow\Xi^0$ \\
$A^{\pi^-}$ & $F+D$ & $\frac{2}{\sqrt6}D$ & $\sqrt2 F$ & $D-F$ \\
\hline
& \multicolumn{4}{c}{(b)} \\
& $\Lambda\rightarrow p$ & $\Sigma^-\rightarrow n$ &
$\Xi^-\rightarrow\Lambda$ & $\Xi^-\rightarrow\Sigma^0$ \\
$A^{K^-}$ & $\frac{1}{\sqrt6}(3F+D)$ & $D-F$ & 
$\frac{1}{\sqrt6}(3F-D)$ & $\frac{1}{\sqrt2}(F+D)$ \\
\end{tabular}}
\end{table}
As one departs from the flavor symmetric case the baryon 
wave--functions acquire admixtures from higher dimensional $SU(3)$ 
representations making the notion of $F$ and $D$ obsolete. 

\begin{figure}[t]
\centerline{\hskip -1.5cm
\epsfig{figure=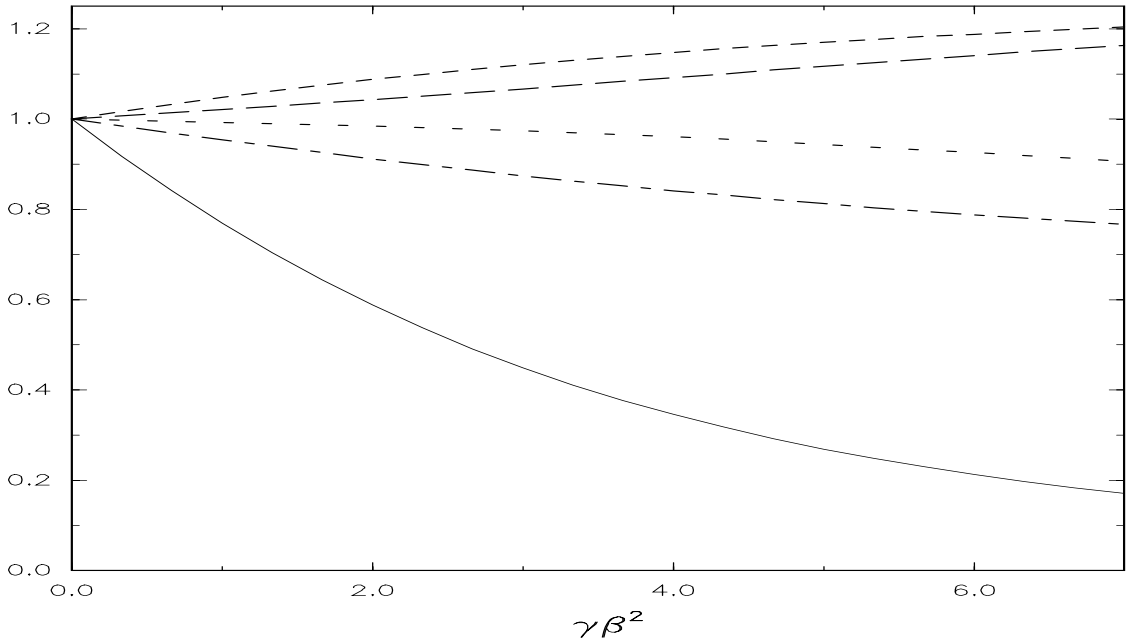,height=9.0cm,width=16.0cm}}
\caption{\label{fi_axial}\tenrm
The variation of the matrix elements describing the semi--leptonic 
hyperon decays with the effective symmetry breaking parameter 
${\Ss \gamma\beta^2}$. Full line:
${\Ss \langle p|{\overline{s}}\gamma_3\gamma_5s| p\rangle}$; 
dashed dotted line:
${\Ss \langle p|{\overline{u}}\gamma_3\gamma_5s| \Lambda\rangle}$; 
dotted line:
${\Ss \langle n|{\overline{u}}\gamma_3\gamma_5s|
\Sigma^-\rangle}$; 
long dashed line:
${\Ss \langle\Lambda|{\overline{u}}\gamma_3\gamma_5s|
\Xi^-\rangle}$; 
dashed line:
${\Ss \langle p|{\overline{u}}\gamma_3\gamma_5d|
n\rangle}$; These matrix elements, which are taken from refs 
\protect\cite{Pa89a} and \protect\cite{Pa90}, are normalized to the 
flavor symmetric values.}
\end{figure}

In addition to the expansion (\ref{nwfexp}) one 
{\it e.g.} finds for the $\Lambda$ hyperon
\be
|\Lambda\rangle=|\Lambda, {\bf 8}\rangle 
+\frac{3}{50}\gamma\beta^2|\Lambda, {\bf 27}\rangle+\ldots \ .
\label{lamwfexp}
\ee
Such expansions may be used to evaluate the behavior of the 
matrix elements entering the Cabibbo scheme \cite{Ca63} for the 
semi--leptonic hyperon decays. Noting that the $D$--functions 
mix the $SU(3)$ representations, a typical example is
\be
\langle p \uparrow | D_{K^-3} |\Lambda \uparrow\rangle =
\frac{2}{5\sqrt3}-\frac{7\sqrt3}{1125}\gamma\beta^2+\ldots \ , 
\qquad
D_{K^-3}=\frac{1}{\sqrt2}\left(D_{43}-iD_{53}\right) \ .
\label{lampexp}
\ee 
Of course, this expansion just provides a first approximation to 
the symmetry breaking dependence of the Cabibbo matrix elements. 
Using the exact treatment initiated by Yabu and Ando \cite{Ya88} 
this dependence can be computed numerically as shown in figure
\ref{fi_axial} for the processes of interest. Those results 
are normalized to the $SU(3)$ symmetric values (\ref{ta_hyperon})
to illuminate that these matrix elements vary in a completely 
different way with symmetry breaking.
In this figure also the variation of the strange quarks' contribution
to the nucleon matrix element of the axial singlet current 
$H_3(0)=\langle N|{\overline s}\gamma_3\gamma_5s|N\rangle$ is 
displayed. Obviously $H_3(0)$ decreases very rapidly with increasing
symmetry breaking. On the contrary the Cabibbo matrix elements 
exhibit only a moderate dependence on $\gamma\beta^2$.
It is this different behavior of the matrix elements that makes the
application of flavor symmetry to the analyses of the 
EMC--SLAC--SMC experiments suspicious \cite{Li88}. Stated otherwise, 
the strange quark contribution to the proton spin may be decreased 
significantly as a consequence of symmetry breaking without 
contradicting the successful Cabibbo scheme for the semi--leptonic 
decays of the hyperons. 

As the ratio $g_A/g_V$ enters the description of the hyperon decays 
one may worry whether the dependence of $g_V$ on the symmetry 
breaking changes this conclusion. The dominating contribution is 
given by the matrix elements of the flavor generators
\be
g_V^a(B^\prime,B)=\langle B^\prime | L_a | B \rangle \ .
\label{calgva}
\ee
From a general point of view it is important to remark that these 
matrix elements are protected by the Ademollo--Gatto theorem
\cite{Ad64}. This theorem states that the leading term in the 
difference of $g_V^a(B^\prime,B)$ from its symmetric value must 
be at least quadratic in symmetry breaking. In the framework of 
the Skyrme model this can easily be understood because the matrix 
elements of the generators $L_a$ vanish between different $SU(3)$ 
representations. Using the Yabu--Ando scheme this has numerically been 
confirmed \cite{Pa90} resulting in, an at most, 10\% deviation from 
the symmetric values, even for large symmetry breaking, {\it e.g.}
$\gamma\beta^2\approx7$. 

A reduction of the strangeness in the nucleon is also observed 
for the scalar strange content fraction of the proton
\be
X_s=\frac{\langle p|{\overline s}s|p\rangle
-\langle0|{\overline s}s|0\rangle}
{\langle p|{\overline u}u+{\overline d}d+{\overline s}s|p\rangle-
\langle 0|{\overline u}u+{\overline d}d+{\overline s}s|0\rangle} \ .
\label{defxs}
\ee
Here the state $|0\rangle$ refers to the soliton being absent.
As will become apparent from NJL--model studies ({\it cf.}
chapter \ref{chap_NJL}) $X_S$ is related to
\be
X_s=\frac{1}{3}\langle p |1-D_{88}| p \rangle
\approx\frac{7}{30}-\frac{43}{2250}\gamma\beta^2+\ldots \ .
\label{xsskyrme}
\ee
In this case, however, the deviation from the flavor symmetric 
result \cite{Do86} ($X_s=7/30$) is considerably mitigated \cite{Ya89} 
as compared to the variation of $H_s$. The symmetry breaking has to 
be as large as $\gamma\beta^2\approx4.5$ to obtain a reduction of the 
order of 50\%. In the case of $H_s$ this was already achieved for 
$\gamma\beta^2\approx2.5$. In any event, the additional 
quark--antiquark excitations in the nucleon, which are 
parametrized by the admixtures of higher dimensional $SU(3)$ 
representations (\ref{nwfexp}), apparently tend to cancel the 
virtual strange quarks of the octet nucleon.

To summarize this section on baryon spectroscopy and form 
factors it should be mentioned that the effects of higher orders 
in the flavor symmetry breaking are important. Furthermore the 
soliton approach provides a unique framework to demonstrate that 
flavor symmetry breaking may effect various physical quantities 
in a completely different manner. Also the conjecture \cite{El93}
is confirmed that some operators see a fairly large amount of 
strangeness in the nucleon while others do not. In this section we 
have concentrated on general aspects of flavor symmetry breaking of 
baryon properties in the soliton approach rather than gathering 
numerical results of the model. The reason is that the Skyrme model, 
which contains only pseudo--scalar degrees of freedom, lacks important 
short range effects. Therefore more reliable predictions are expected 
in extensions of the model containing meson fields whose profile 
functions have their major support at small distances, say $r\le0.5fm$. 
An excellent candidate is the vector meson model to be described in 
chapter \ref{chap_vector}.

In ref \cite{Pa91a} the use of an alternative Skyrme term 
\be
{\cal L}^{\rm alt.}_{Sk}=\frac{1}{16e^2}
\left[{\rm tr}\left(\alpha_\mu\alpha_\nu\right)
{\rm tr}\left(\alpha^\mu\alpha^\nu\right)
-\left({\rm tr}\left(\alpha_\mu\alpha^\mu\right)\right)^2\right]
\label{Skalt}
\ee
has been considered together with the meson configuration (\ref{su3rot}). 
Although the resulting Lagrange function is identical to (\ref{Skterm}) 
in the two flavor reduction its contribution to the moment of inertia 
$\beta^2$ is four times as large. According to eqs (\ref{delE3}) and 
(\ref{nwfexp}) the effect of symmetry breaking is increased yielding a 
slightly improved pattern for the baryon spectrum. However, the 
predictions for the static properties of the baryons exhibit almost no 
variation as compared to the ordinary Skyrme term. The effects of the 
induced components (\ref{kaonansatz}) have not been investigated in 
the context of (\ref{Skalt}).

\bigskip

\begin{subsatz}
\label{sec_further}{\bf \hskip1cm Further Developments}
\addcontentsline{toc}{section}{\protect\ref{sec_further}
Further Developments}
\end{subsatz}
\stepcounter{section}

In this section studies will briefly be discussed, which do not 
directly effect static properties of the strange baryons but are 
special beyond the two flavor model and should be mentioned 
to achieve a higher level of completeness.

By adding a term to the Lagrangian which represents the $U_A(1)$
anomaly of QCD \cite{Ro80}
\be
{\cal L}_A=\frac{\kappa}{48}\left[{\rm ln}
\left({\rm det}U-{\rm det}U^{\dag}\right)\right]^2
\label{ua1ano}
\ee
and extending the symmetry breaking matrix (\ref{massmat})
\be
{\cal M}\rightarrow {\cal M}+i\frac{\lambda_\Theta}{4f_\pi^2} \ ,
\label{lambdatheta}
\ee 
it is possible to imitate the effects of the QCD $\Theta$ angle
\cite{Hu82} since a non--zero $\Theta$ is equivalent to ${\cal M}$ 
having a non--vanishing phase \cite{Ba79}. The constant $\kappa$ can 
be determined from the mass of the $\eta^\prime$ meson, {\it cf.} 
section \ref{sec_two}. Substituting the experimental values one 
obtains \cite{Di91} $\lambda_\Theta=9\times10^{-3}\Theta{\rm GeV}^2$. 
This extended Lagrangian contains a strong $CP$ violating term
\be
\delta {\cal L}_{CP}\approx \frac{\lambda_\Theta}{\sqrt3}f_\pi
\left(1-\cF\right)\left(\phi^8+\sqrt2\phi_0\right)
\label{lcp}
\ee 
where $\phi_8$ and $\phi_0$ denote the flavor octet and singlet 
components (\ref{eightfold}) of the $\eta$--fields, respectively.
Including the contribution of $\delta {\cal L}_{CP}$ to the 
Euler Lagrange equations causes non--trivial profiles for 
$\phi_8$ and $\phi_0$. As a consequence the expression for 
the neutron electric dipole moment
\be
D_n=\langle n|\int d^3r \mbox{\boldmath $r$} J_0^{\rm e.m.}
| n \rangle \ ,
\label{defdn}
\ee
which is linear in these $\eta$ fields, allows one to relate the 
neutron electric dipole moment to the QCD $\Theta$--angle. In ref 
\cite{Di91} the numerical result $D_n=2\times10^{-6}\Theta e{\rm cm}$ 
was obtained which is somewhat smaller than the current algebra result 
($3.6\times10^{-6}\Theta e{\rm cm}$)
of ref \cite{Cr79}. Future experiments on the neutron electric dipole
moment will provide a means to estimate QCD $\Theta$--angle.

A prominent issue to be discussed within the $SU(3)$ Skyrme model 
is the question of whether or not the $H$--dibaryon is bound. 
Originally the $H$--dibaryon was considered as a six quark 
state, which has strangeness -2, in the MIT bag model \cite{Jaf77}.
It was found to be bound by about $80{\rm MeV}$ against a decay 
into two $\Lambda$ hyperons. For the Skyrme model calculation a 
corresponding {\it ansatz} for a soliton configuration 
\be
V_H(\mbox{\boldmath $r$}) = {\rm e}^{i\Psi(r)}
+i\mbox{\boldmath $\Lambda$}\cdot\hat{\mbox{\boldmath $r$}}
{\rm sin}\chi(r)
+\left(\mbox{\boldmath $\Lambda$}\cdot
\hat{\mbox{\boldmath $r$}}\right)^2
{\rm e}^{-i\Psi(r)/2}{\rm cos}\chi(r){\rm e}^{i\Psi(r)}
\label{balansatz}
\ee
was proposed by Balachandran {\it et al.} \cite{Ba84,Ba85}.
Here $\Lambda_1=\lambda^7,\ \Lambda_2=-\lambda^5$ and 
$\Lambda_3=\lambda^2$ denote the generators of an $SO(3)$
subgroup of $SU(3)$. Minimizing the static energy by variation 
of the radial functions $\Psi(r)$ and $\chi(r)$ and subsequent 
canonical quantization yielded a binding energy of about 
$130{\rm MeV}$ \cite{Ba85}. More recently the order $N_C^0$ 
corrections to the mass of the $H$--dibaryon were 
estimated \cite{SSG93}. These estimates are based on the 
observation that zero mode channels provide the dominant 
contribution to the Casimir energy of the soliton
\cite{Ca76,Ho94}
\be
E_{\rm Cas}=-\frac{1}{4}{\rm tr}
\left[\frac{\left(H-H_0\right)^2}{H_0}\right]
\le-\frac{1}{4}\sum_{k,a,\zeta}\omega(k)
\left|\langle \mbox{\boldmath $k$}, a| \zeta \rangle\right|^2\ .
\label{ecas}
\ee
Here $H$ and $H_0$ denote the Hamilton operator for the meson fields
in the background and absence of the soliton, respectively.
$|\mbox{\boldmath $k$}, a\rangle$ is a plane wave state of momentum
$\mbox{\boldmath $k$}$ and flavor $a$, while $\zeta$ refers to the 
zero modes of the soliton. Including these corrections it is 
found \cite{SSG93} that $E[V_H]\approx 3 E_{B=1}$, where $E_{B=1}$ 
is the hedgehog energy with the associated Casimir energy included.
Obviously the quantum corrections, which the configuration $V_H$ 
suffers, are significantly lower than twice the corrections of the 
hedgehog.  This is intuitively clear from eq (\ref{ecas}) because the 
number of zero modes does not depend on the baryon number but rather 
on the symmetries of the model and {\it ans{\"a}tze}. As the 
$1/N_C$--rotational corrections are included a slightly bound 
$H$--dibaryon is found. However, as (\ref{ecas}) just represents a 
bound for the quantum corrections, the authors of ref \cite{SSG93}
conjecture that the $H$--dibaryon is most likely unbound.

In ref \cite{SSG95} further dibaryon configurations have been 
considered by generalizing the product {\it ansatz} 
\cite{Ja85} to flavor $SU(3)$
\be
U(\mbox{\boldmath $r$})=
A U_0\left(x,y,z-\frac{R}{2}\right)C
U_0\left(x,y,z+\frac{R}{2}\right)C^{\dag}A \ .
\label{su3prod} 
\ee
Here $A$ and $C=A^{\dag}B$ denote constant flavor rotations which 
are defined such that $A$ and $B$ describe the individual flavor 
orientations of the single $SU(3)$ Skyrmions (\ref{su3hedgehog}). 
Furthermore $R$ measures the distance between the two Skyrmions. 
Substitution of (\ref{su3prod}) into the action gives the typical 
interaction term \cite{SSG95}
\be
V(R;A,B)\ {\stackrel{{\scriptstyle R\to\infty}}
{\textstyle\longrightarrow}}\ 4\pi f_\pi^2\partial_i\partial_j
\left\{D_{ki}(A)D_{kj}\frac{{\rm e}^{-m_\pi R}}{R}+
\ldots\right\} \ .
\label{typicalterms}
\ee
This permits one to extract the potential for the baryon--baryon 
interaction in certain spin and flavor channels\footnote{The 
$SU(3)$--central potential has also been studied in ref 
\cite{Ka92a}.} since, according to the Wigner--Eckart theorem, 
$D_{ab}\sim L_a R_b$ when taking matrix elements. In this case
bound dibaryon states were found for the channels
$N-N (I=1/2,J=1)$, $N-\Sigma (I=1/2,J=1)$, $N-\Xi (I=,J=0)$,
$\Lambda-\Lambda (I=0,J=0)$, $\Sigma-\Sigma (I=2,J=2)$ and 
$\Xi-\Xi (I=0,J=1)$. These results are in agreement with the 
general expectations from boson exchange models. See ref \cite{Do89}
for a review on these models and a compilation of references.

It is also interesting to note that the collective treatment of 
section \ref{sec_quant} can be generalized to flavor $SU(N_f)$ 
\cite{Wa92a}, with $N_f$ being the number of flavor degrees of 
freedom. As a generalization of the {\it ansatz} (\ref{su3hedgehog}) 
the hedgehog is embedded in the isospin subgroup and collective 
rotations are parametrized by the $SU(N_f)$ matrix $A$ as in eq 
(\ref{su3rot}). Here the case of $N_f=4$ will briefly be discussed. 
Then, eq (\ref{defomega}) defines 15 angular velocities $\Omega_a$. 
Further in analogy to the quantization prescription, (\ref{Rgen}), 
15 right generators $R_a$ are introduced. The flavor symmetric 
part of the collective Hamiltonian is similar to the symmetric part 
in the three flavor case (\ref{collham})
\be
H_{\rm sym.}=E+\frac{1}{2\alpha^2}\sum_{i=1}^3R_i^2
+\frac{1}{2\beta^2}\left(\sum_{\alpha=4}^7R_\alpha^2
+\sum_{m=9}^{12}R_m^2\right).
\label{clham4}
\ee
The moments of inertia are identical to the three flavor expressions. 
First class constraints \cite{Di64} are obtained for all generators 
associated with those $\lambda$ matrices of the group $SU(N_f)$ which 
commute with the static hedgehog configuration
\be
R_8=\frac{N_C B}{2\sqrt3}\ , \qquad
R_{13}=R_{14}=0\ , \qquad
R_{15}=\frac{N_C B}{2\sqrt6}\ . \qquad 
\label{su4const}
\ee
The first one, which again requires half--integer spin for the 
allowed eigenstates, is the same as in $SU(3)$, {\it cf.} eq 
(\ref{Rgen}). Also the last one stems from the Wess--Zumino term 
(\ref{WZterm}). These constraints uniquely select (for $N_C=3$ and 
$B=1$) the ground state representations of $SU(4)$
\be
{\bf 20}\ , \quad J=\frac{1}{2}
\qquad {\rm and} \qquad
{\bf 20}^\prime\ , \quad J=\frac{3}{2} \ .
\label{su4ground}
\ee
A diagrammatic representation of these multiplets is {\it e.g.} given 
in figure 30.2 of ref \cite{PDG94}. In the same way the isospin 
multiplets are subsets of the $SU(3)$ multiplets in figure 
\ref{fi_su3rep}, the $SU(3)$ multiplets are subsets of the $SU(4)$ 
multiplets (\ref{su4ground}). The energy difference between the lowest 
lying states in these multiplets (nucleon and $\Delta$) is found to be 
independent of $N_f$: $M_\Delta-M_N=3/2\alpha^2$.
Adding symmetry breaking terms to account for different meson 
masses $m_D=1.9{\rm GeV}$ and decay constants $f_D\approx1.7f_\pi$
shows that the relevant symmetry breaking parameters in the baryon
sector are about an order of magnitude larger than in the 
case of flavor $SU(3)$ in reasonable agreement with the experimental 
data. Unfortunately the complete symmetry breaking 
pattern for the $SU(4)$ effective Lagrangian is still unknown. 
Furthermore, due to the lack of a comprehensive set of $SU(4)$ 
Clebsch--Gordon coefficients, the important expansion in symmetry 
breaking, as in eq (\ref{delE3}), has not been carried out. Due to the 
large symmetry breaking in the charm sector, the consideration of the 
$D$--meson as a ``would--be" Goldstone boson and hence the application 
of the collective approach is doubtful. Nevertheless it is interesting 
that this approach yields the same ground states as the quark model. 
As in the three flavor model this is caused by the constraints 
originating from Wess--Zumino term (\ref{WZterm}), which represents 
the anomaly structure of QCD.

Another generalization of the Skyrme model wave--functions is to 
consider an arbitrary, odd number of colors, $N_C$, in particular
the limit $N_C\rightarrow\infty$ is of interest. This has been 
studied for the three flavor symmetric model \cite{Ma84}. The main 
result is that for $N_C\rightarrow\infty$ the matrix elements 
of collective space operators are (up to an overall factor) identical 
to their analogous matrix elements obtained in the non--relativistic 
quark model in the same limit \cite{Ka84c}. In the three flavor 
description the generalization to arbitrary $N_C$ is more involved as 
in the two flavor case because the tensor structure of the baryon 
wave--functions is more complicated \cite{Bi86,Da94}.

In what comes we are exclusively interested in the predictions of the 
soliton picture for physical observables. Henceforth we will only 
consider the case $N_C=3$ in three flavor models.
\vskip1cm
\begin{satz}
\label{chap_exten}{\large \bf \hskip1cm 
Symmetry Breaking and the Size of the Skyrmion}
\addcontentsline{toc}{chapter}{\protect\ref{chap_exten}
Symmetry Breaking and the Size of the Skyrmion}
\end{satz}
\stepcounter{chapter}

Up to now non--vanishing strange fields have been constructed 
simply by rigidly rotating the hedgehog configuration, which 
is embedded in the isospin subgroup, into strange directions
(\ref{su3rot}). Except for the induced components (\ref{kaonind})
the effect of symmetry breaking on the meson configuration has 
completely been ignored. In this chapter two approaches will be 
presented, which go beyond this approximation by allowing the 
radial dependence of the chiral angle $F(r)$ to depend on the 
flavor orientation. This approach is intuitive because for the 
kaon fields one expects the asymptotic behavior ${\rm exp}(-m_Kr)$ 
rather than ${\rm exp}(-m_\pi r)$ as is implied by the configuration 
(\ref{su3rot}). This change in the profile will also cause 
favorable deviations from the $SU(3)$ relations (\ref{su3mag}) 
which could not be achieved by any form of the Yabu--Ando method,
meaning {\it e.g.} that whatever complicated structure of the 
symmetry breaking is assumed for the collective Hamiltonian 
(\ref{collham}), the ratio $\mu_{\Sigma^+}/\mu_p$ stays close 
to unity rather than 0.85 as demanded by the experimental data.

\bigskip

\begin{subsatz}
\label{sec_slowrot}{\bf \hskip1cm The ``Slow--Rotator'' Approach}
\addcontentsline{toc}{section}{\protect\ref{sec_slowrot}
The ``Slow--Rotator'' Approach}
\end{subsatz}
\stepcounter{section}

Let us, for the moment, ignore the time dependence of the 
collective rotations. Then the angles describing the spatial and 
isospin orientations of the hedgehog may be absorbed by symmetry 
transformations. This, however, is not the case for the whole flavor 
space due to $SU(3)$ symmetry breaking. Denoting the angle which 
describes the strangeness changing orientation by $\nu=0\ldots \pi/2$ 
a suitable parametrization for (\ref{su3rot}) is therefore given by 
(see also appendix A)
\be
U(\mbox{\boldmath $r$},\nu)={\rm e}^{-i\nu\lambda_4}
U_0(\mbox{\boldmath $r$}){\rm e}^{i\nu\lambda_4} \ .
\label{slowrot}
\ee
Substituting this {\it ansatz} into the action (\ref{totact1}) yields 
an energy functional which is not only a functional of the chiral 
angle $F$, but also a function of the strangeness changing angle 
$\nu$ \cite{Sch92b}
\be
E(\nu)[F]&\hspace{-0.15cm}=&\hspace{-0.15cm}\int d^3r 
\Bigg[\frac{f_{\pi}^2}{2}(\fpt +
2\frac{\sFt}{r^2})+\frac{\sFt}{e^2r^2}(\fpt+\frac{\sFt}{2r^2})
+\epsilon_6^2\frac{\fpt{\rm sin}^4F}{8\pi^4r^4}
+m^2_{\pi}f^2_{\pi}(1-\cF)
\nonumber \\ && \hspace{-0.5cm}
+{\rm sin}^2 \nu \left(\frac{1}{2}(f^2_{K}-f^2_{\pi})(\fpt +
\frac{2\sFt}{r^2})\cF+(m^2_{K}f^2_{K}-m^2_{\pi}f^2_{\pi})(1-\cF)\right)
\Bigg] \ .
\label{nuemass}
\ee
Here the relations (\ref{pspara}) have been employed to make the 
dependence on the physical parameters more transparent. Also an 
additional parameter, $\epsilon_6$, has been introduced via the sixth 
order term
\be
{\cal L}_6=-\frac{\epsilon_6^2}{2}B_\mu B^\mu \ ,
\label{sixorder}
\ee
with the baryon number current being defined in eq (\ref{bnumber}). 
Since this term may be considered as the approximation to the exchange 
of an infinitely heavy $\omega$--meson \cite{Ja85} one obtains at 
least an estimate $\epsilon_6\approx 0.015{\rm MeV}^{-1}$ \cite{La86}.
Nevertheless, $\epsilon_6$ has been considered to be an adjustable 
parameter in ref \cite{Sch92b}.

\begin{figure}[t]
\centerline{\hskip -0.5cm 
\epsfig{figure=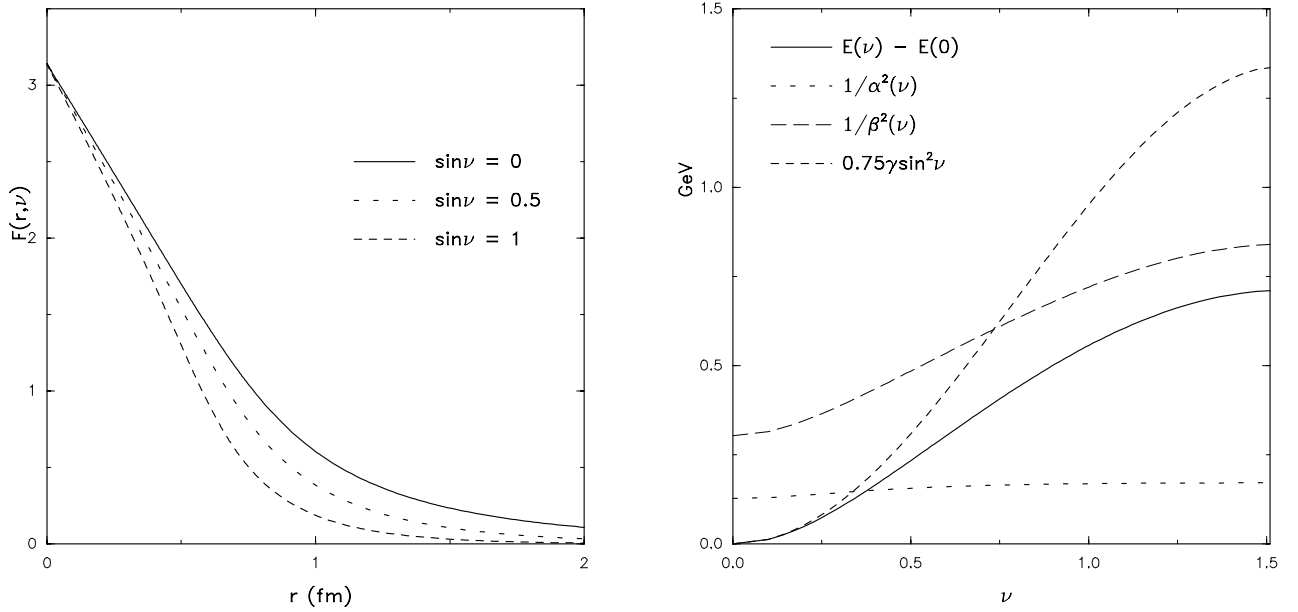,height=8.0cm,width=16.0cm}}
\caption{\label{fi_slow}\tenrm
Left panel: The chiral angle as functions of the radial 
distance ${\Ss r}$ and the ${\Ss SU(3)}$ angle ${\Ss \nu}$.
Right panel: The quantities in the eigenvalue problem 
(\protect\ref{slowham}) as functions of the strangenesss changing 
angle ${\Ss \nu}$. For comparison also the behavior of the 
symmetry breaker in the ``rigid rotator" approach is shown. The 
results correspond to the model with the sixth order term 
(\protect\ref{sixorder}) included.}
\end{figure}

For a given value of $\nu\in [0,\pi/2]$ the minimum of the energy 
functional (\ref{nuemass}) is obtained. Hence the chiral angle 
is not only a function of the radial distance but also of the 
$SU(3)$ ``Euler angle" $\nu$: $F=F(r,\nu)$. The philosophy is that 
the rotation (\ref{su3rot}) proceeds slowly enough that $F$ can 
adjust itself to the momentary orientation in flavor space. This 
causes the notion of ``slow rotator" in contrast to the ``rigid 
rotator" which is characterized by the chiral angle confined to 
two--flavor reduction as discussed in the preceding chapter. From 
(\ref{nuemass}) it is furthermore transparent that the symmetry 
breaking forces yield $F(r,\pi/2)\rightarrow {\rm exp}(-m_K r)$ for 
$r\rightarrow\infty$, {\it i.e.} the kaon field has the proper 
long--distance behavior. This is indicated in the left part of 
figure \ref{fi_slow}.
In the next step the time--independence of the flavor rotations 
is waived. This yields kinetic terms in the collective 
Lagrangian as in eq (\ref{colllag}) 
\be
L=-E(\nu)+\frac{1}{2}\alpha^2(\nu)\sum^3_{i=1}\Omega^2_i
+\frac{1}{2}\beta^2(\nu)\sum^7_{\alpha=4} \Omega^2_\alpha-
\frac{N_C}{2\sqrt{3}}\Omega_8+\dot\nu^2\Delta(\nu) \ .
\label{slowlag}
\ee
Since the moments of inertia $\alpha^2$ and $\beta^2$ are 
functionals of the chiral angle ({\it cf.} eq (\ref{alsky})) 
these quantities parametrically depend on the strangeness changing 
angle $\nu$. For the expressions of
$\alpha^2(\nu)$ and $\beta^2(\nu)$ the reader may consult ref 
\cite{Sch92b}. The static part $E(\nu)$ not only has an explicit 
dependence on $\nu$ via $1-D_{88}$, but also an implicit one implied by 
the chiral angle. The $\nu$--dependencies of these quantities is shown 
in the right part of figure \ref{fi_slow}. Obviously the symmetry 
breaking exhibited by $E(\nu)$ is much more moderate than in the 
case of the ``rigid rotator". The positivity of $E(\nu)-E(0)$ 
also verifies the earlier statement that the hedgehog in the isospin 
subgroup indeed minimizes the classical energy functional in the 
presence of flavor symmetry breaking.
The implicit time--dependence of the chiral 
angle causes the appearance of the additional term $\dot\nu^2\Delta(\nu)$ 
in eq (\ref{slowlag}). It has been argued \cite{Sch92b} that its 
influence on baryon properties is only moderate and may easily be 
compensated by a slight change of the model parameters. Due to the 
$\nu$--dependence of the moment of inertia for rotations into 
strange directions, $\beta^2$, the quantization is plagued by ordering 
ambiguities. The most straightforward procedure to obtain an 
Hermitian Hamiltonian is to substitute the expression 
$(1/\beta^2)C_2$ by the anti--commutator which leads to the 
collective Hamiltonian
\be
H = E(\nu)&+&\left(\frac{1}{2\alpha^2(\nu)}-
\frac{1}{2\beta^2(\nu)}\right) {\bf J}^2
+\frac{1}{2}\left\{\frac{1}{2\beta^2(\nu)}
,C_2\left[ SU(3)\right]\right\} 
-\frac{3}{8\beta^2(\nu)}\ .
\label{slowham}
\ee
From figure \ref{fi_slow} it is observed that $1/\beta^2$ depends 
on $\nu$ only moderately. Hence a different ordering of the collective 
operators, as {\it e.g.} suggested by the Pauli--prescription
\cite{Pa33}, does not cause substantial changes. The eigenvalues and 
eigenstates of (\ref{slowham}) are constructed in analogy to the 
Yabu--Ando approach \cite{Ya88} by separating the $\nu$--dependence of
the baryon wave--functions and writing the quadratic Casimir 
operator $C_2$ as a differential operator with respect to the $SU(3)$ 
``Euler angles". The details of this treatment may be extracted from 
appendix A. 

\begin{table}
\caption{\label{ta_diffslow}\tenrm
The mass differences of the low--lying ${\Ss \frac{1}{2}^+}$ and
${\Ss \frac{3}{2}^+}$ with respect to the nucleon in the 
``slow--rotator" approach. The models SK4 and SK46 are explained 
in the text. All data (from refs \protect\cite{Sch91,Sch92b}) are 
in MeV.}
~
\newline
\centerline{\tenrm\smalllineskip
\begin{tabular}{c | c  c | c}
Baryons & SK4 & SK46 & Expt. \\
\hline
$\Lambda$   & 177 & 180 & 177 \\
$\Sigma$    & 285 & 299 & 254 \\
$\Xi$       & 381 & 400 & 379 \\
$\Delta$    & 298 & 296 & 293 \\
$\Sigma^*$  & 477 & 457 & 446 \\
$\Xi^*$     & 619 & 590 & 591 \\
$\Omega$    & 731 & 699 & 733 \\
\end{tabular}}
\end{table}
In table \ref{ta_diffslow} the resulting mass differences 
are listed for two sets of parameters
\be
\begin{array}{c c}
SK4 & SK46 \\
e=3.46 & e=5.61 \\
\epsilon_6=0 & \epsilon_6=0.0118{\rm MeV}^{-1} \\
f_K=118{\rm MeV} & f_K=108{\rm MeV}\ . \\ 
\end{array}
\label{slowpara}
\ee
In the presence of the repulsive sixth order term (\ref{sixorder}) 
the coefficient of the Skyrme term may be decreased without strongly 
effecting the soliton profile. The predictions of both models favorably 
agree with the experimental data. Although $f_K$ has been fine--tuned 
to better reproduce the baryon mass differences it is only a little 
different from the experimental value (113MeV).

Having established the slow--rotator approach from the baryon mass 
differences it is, of course, suggestive to apply this treatment for 
a study of the static properties of the baryons. This investigation
proceeds analogously to those for the rigid rotator by calculating 
the form factors defined in section \ref{sec_masses}. Although an
extensive study of static properties of baryons has been carried 
out \cite{Sch92b,Kr94} we will focus on the magnetic moments. Here 
care has to be taken of the fact that the radial functions, $V_i$, 
defined in (\ref{spcurrps}) not only depend on the radial coordinate 
$r$, but also on the strangeness changing angle $\nu$ as a consequence 
of the implicit dependence of the chiral angle. Hence we write 
$V_i=V_i(r,\nu)$. A typical term contributing to the magnetic moments 
is then 
given by
\be
G_M^B(\mbox{\boldmath $q$}^2)&=&-8\pi M_B
\Bigg\langle\int_0^\infty \hspace{-0.2cm} r^2 dr
\frac{r}{|\mbox{\boldmath $q$}|}j_1(r|\mbox{\boldmath $q$}|)
\Bigg\{V_1(r,\nu) D_{e3}+\ldots\Bigg\}\Bigg\rangle_B\ .
\label{magslow}
\ee
Numerically these matrix elements may be evaluated with the 
techniques explained at the end of appendix A. The results 
are displayed in the lower part of table \ref{ta_uspin}.
\begin{table}[t]
\caption{\label{ta_uspin}\tenrm
The U--spin symmetry for the magnetic moments (in units of the 
proton magnetic moment) compared to the experimental data. 
The ${\Ss \frac{1}{2}^+}$ baryons are arranged as in figure 
\protect\ref{fi_su3rep}. The rigid rotator results correspond 
to table \protect\ref{ta_emps} while the data for the 
non--relativistic quark model are taken from ref
\protect\cite{Wa95}. The lower part gives the predictions 
of the slow--rotator approach \protect\cite{Sch92b}.}
~
\newline
{\footnotesize 
\begin{tabbing}
123456789012345678901234567890\=1234\=1234\=1234\=1234\=1234\=\kill
12345\=12345\=12345\=12345\=12345\=12345\=12345678901234567890\=
12345\=12345\=12345\=12345\=12345\=\kill
\>\>     \>-0.68   \>    \>1.0   \>     \>        \>       \>b     \>
\>1.0\\
\>\>     \>     \>-0.22    \> \>     \>        \>       \>      \>-b/2
\\
\>\>-0.42     \>\> ~~?   \>\>0.87      \>   \>c      \>
\>\,b/2\>\>1.0\\
\>\>     \>-0.25 \>  \>-0.45\>  \>          \>       \>c     \>
\>b\\
\\
\>\>     \>Experiment \>\>\>\>\>\quad U-spin symmetric case\\
\\ \\
\>\>     \>-0.78\>     \>1.0   \>     \>          \>      \>-0.60  \>
\>1.0  \\
\>\>    \>     \>-0.35 \>     \>     \>          \>      \>     \> -0.20 \\
\>\>-0.39  \>  \>0.30  \>     \>0.98 \>          \> -0.41\>
\>0.29 \>\>0.99\\
\>\>     \>-0.31 \>     \>-0.76 \>     \>          \>      \> -0.24\>
\>-0.46\\
\\
\>\>    \>rigid rotator\> \> \> \> \> \>nr quark model\\
\\ \\
\>\>     \>-0.83   \>    \>1.0   \>     \>        \>       \>-0.79     \>
\>1.0\\
\>\>     \>     \>-0.25    \> \>     \>        \>       \>      \>-0.29
\\
\>\>-0.40     \>\> 0.23  \>\>0.85      \>   \>-0.40      \>
\>0.27\>\>0.94\\
\>\>     \>-0.20 \>  \>-0.54\>  \>          \>       \>-0.24     \>
\>-0.63\\
\\
\>\>\>     \>Sk4\>\>\>\>\>\>Sk46
\\
\end{tabbing}}
\end{table}
As already mentioned in the previous chapter the rigid rotator 
approach is not capable of reproducing the experimentally observed
deviations from the $SU(3)$ predictions (\ref{su3mag}). Actually 
these relations just reflect the U--spin symmetry of the magnetic 
moments in the flavor octet, which is also indicated in table 
\ref{ta_uspin}. This feature occurs in the rigid rotator approach 
although the U--spin operation does not relate 
various eigenstates of the collective Hamiltonian (\ref{collham}) 
since these eigenstates represent are no pure octet states; {\it cf} 
eq (\ref{nwfexp}). Amusingly, the non--relativistic quark model, 
which commonly is celebrated for its successful determination of the 
baryon magnetic moments, does not depart from this symmetry either,
especially for the ratio $\mu_{\Sigma^+}/\mu_p$ \cite{Wa95}. One 
immediately observes that for the pattern of the magnetic moments 
the slow rotator approach gives the desired improvement compared 
to the rigid rotator. Stated otherwise, the influence of flavor 
symmetry breaking on the soliton profile is responsible for breaking 
the U--spin symmetry. Obviously this effect is stronger when the 
soliton is stabilized by the Skyrme term than in the case when also 
a sixth order term is present. It should, however, be admitted that
the absolute values for the magnetic moments are predicted to be 
too small for both sets of parameters (\ref{slowpara})
\be
{\rm SK4:}\quad \mu_p=1.78 \qquad
{\rm SK46:}\quad \mu_p=1.90 \ .
\label{magprtslow}
\ee

Also in ref \cite{Sch92b} the magnetic moments of the 
$\frac{3}{2}^+$ baryons and the electric radii have been calculated. 
Since experimentally little is known about these quantities 
we will not go into further detail but only mention that again 
sizable deviations from the flavor symmetric predictions were observed. 
Again these turned out to be larger for SK4 than SK46. In addition the 
strangeness matrix elements of the (axial)vector currents, which are 
relevant for the description of the semi--leptonic hyperon decays were 
studied. Apart from the the too small result for $g_a\approx 1$,  
reasonable agreement with the experimentally known data was obtained
despite of the large deviation of the wave--functions from their 
flavor symmetric forms. In particular the strange quark contribution 
to the nucleon matrix element of the axial current was found to 
be reduced to about a quarter of the flavor symmetric value 
in the model SK4. For SK46 the reduction is mitigated to 40\%.

In the framework of the the slow rotator approach the electric 
quadrupole moments of the $\frac{3}{2}^+$ baryons have been studied 
in ref \cite{Kr94}. In particular the influence of flavor symmetry 
breaking has been explored. Various strengths of symmetry breaking 
are parametrized by a different value of the quark mass ration $x$ 
(\ref{qmratio}). As a change in $x$ gives different results for $X_s$ 
(\ref{defxs},\ref{xsskyrme}) these strengths can be translated into 
the strangeness content of the proton. As already explained a 
large strangeness content $X_s=0.23$ is found in the flavor symmetric 
case while $X_s$ vanishes for infinite symmetry breaking. It turns out 
that the electric quadrupole moments are rather sensitive to the 
assumed symmetry breaking. While the use of flavor symmetric 
wave--functions causes the electric quadrupole moments to be 
proportional to the charge $Q$ of the considered baryon, an infinite 
symmetry breaking only yields a proportionality to the isospin 
projection $I_3$ (the constant of proportionality depends on the isospin 
multiplet under consideration). These two cases are very interesting 
from a conceptual point of view. In a quenched lattice gauge calculation 
\cite{Le92} a proportionality to the charges was obtained. This is in 
contrast to the recent chiral perturbation calculation of ref \cite{Bu94} 
which found the quadrupole moments to be proportional to the isospin.
In the sense that symmetry breaking may be considered an adjustable 
quantity measured by the strangeness content of the proton, the three 
flavor soliton models permit a smooth interpolation between the 
limiting cases. However, some care has be taken since we have already 
observed that $X_s$ decreases only slowly as $x$ increases. Other 
quantities like $H_s$ exhibit a stronger dependence on $x$. For the 
above established model SK4, a charge dominated pattern for the 
quadrupole moments is actually found. For example, the quadrupole 
moment of the $\Omega^-$ is obtained as $0.024{\rm efm}^2$, while 
that of the $\Xi^{*0}$ almost vanishes ($-0.007{\rm efm}^2$). This 
pattern agrees with the lattice result, although it is an order of 
magnitude larger \cite{Le92}.

Summarizing the slow rotator approach, it has to be concluded that 
the force exerted by symmetry breaking on the soliton configuration 
has significant effects which, in particular, provide the proper 
deviation from the U--spin symmetric relations.

\bigskip
\begin{subsatz}
\label{sec_mixing}{\bf \hskip1cm Radial Excitations}
\addcontentsline{toc}{section}{\protect\ref{sec_mixing}
Radial Excitations}
\end{subsatz}
\stepcounter{section}

In section \ref{sec_quant} the admixtures of higher dimensional 
$SU(3)$ representations to the standard octet and decouplet
wave--functions have been shown to account for a proper description 
of the flavor symmetry breaking pattern in the spectrum of the 
low--lying baryons. Naturally the question arises as to what extent 
the states in these representations are related to experimentally
observed baryons. Ignoring symmetry breaking the mass difference 
between the state with nucleon quantum numbers in the 
${\overline{\bf 10}}$ and the nucleon itself is given by
\be
M_{N({\overline{\bf 10}})}-M_{N({\bf 8})}
=\frac{1}{2\beta^2}\left\{
C_2\left({\overline{\bf 10}}\right)-C_2\left({\bf 8}\right)
\right\}\approx350{\rm MeV}
\label{m10bar}
\ee
for typical values of $\beta^2$ (\ref{pscopara}0). Since this is not 
too different from the position of the  $\Delta$ resonance the physical 
significance of such states needs to explored. As long as 
$\beta^2<\alpha^2$, to which no exception has been obtained in any 
known soliton model, this state lies above the $\Delta$ resonance. The 
quantum numbers of the $N({\overline{\bf 10}})$ state suggest its 
identification as the Roper resonance (1440). As has been mentioned 
earlier, the admixture of higher dimensional $SU(3)$ representations 
can be interpreted as additional quark--antiquark excitations. Hence 
the identification of the $N({\overline{\bf 10}})$ state as the Roper 
resonance would imply that the Roper represents a quark--antiquark 
rather than a radial excitation of the nucleon. In general one, 
however, expects a complicated interplay between states of $SU(3)$ 
representations and radial excitations. In order to study such 
questions the field configuration
\be
U(\mbox{\boldmath $r$,t})=A(t)
U\left(\mu(t)\mbox{\boldmath $r$,t}\right)A^{\dag}(t)
\label{su3breath}
\ee
has been considered in refs \cite{Sch91a,Sch91b}. While $A(t)$ 
again describes the flavor orientation of the hedgehog, the collective
coordinate $\mu(t)$ is introduced to describe radial excitations
\cite{Ha84a}. Commonly this degree of freedom it is referred to 
as the breathing mode. As will be explained shortly the resulting 
baryon spectrum exhibits too strong a flavor symmetry breaking. For 
that reason a model has been considered in ref \cite{Sch91b} which 
contains an additional isoscalar scalar meson field $\sigma$. This 
degree of freedom is incorporated as an order parameter for the 
gluon condensate to imitate the QCD trace anomaly \cite{Ni77}. For 
the details of this effective meson Lagrangian the reader may consult 
ref \cite{Go86}. Here it suffices to display the flavor symmetry 
breaking part
\be
{\cal L}_{SB}&=&{\rm tr}\left\{
{\cal M}\left[-\bep {\rm e}^{2\sigma}
\left(\partial_\mu U\partial^\mu U^{\dag}U
+U^{\dag}\partial_\mu U\partial^\mu U^{\dag}\right)
+\dep {\rm e}^{3\sigma}\left(U+U^{\dag}\right)\right]\right\} \ .
\label{lsbglue}
\ee
As a further change of the Lagrangian the non--linear 
$\sigma$ term (\ref{Lnls}) acquires a factor ${\rm e}^{2\sigma}$. 
There are also kinetic and mass terms for the $\sigma$--field.
These involve an additional parameter which can be determined 
from the empirical value of the gluon condensate \cite{Go86}.
Since the profile $\sigma(r)\le0$, it is obvious that the extension
by this scalar fields mitigates the effects of flavor symmetry
breaking. Only configurations, which correspond to shallow 
bags, {\it e.g.} $\sigma(r=0)$ being close to zero, 
have been considered in ref \cite{Sch91b}.

Substitution of the parametrization (\ref{su3breath}) into the 
mesonic action (\ref{totact1},\ref{lsbglue}) yields a Lagrangian 
for the collective coordinates $A(t)$ as well as
$x(t)=[\mu(t)]^{-3/2}$
\be
L(x,\dot x,A,\dot A)&=&{4\over9}
\left(a_1+a_2x^{-{4\over3}}\right){\dot x^2}-
\left(b_1x^{2\over3}+b_2x^{-{2\over3}}+b_3x^2\right)
+{1\over2}\left(\alpha_1x^2+\alpha_2x^{2\over3}\right)
\sum^3_{a=1}\Omega^2_a
\nonumber \\ && \hspace{-1.5cm}
+{1\over2}\left(\beta_1x^2+\beta_2x^{2\over3}\right)
\sum^7_{a=4}\Omega^2_a +{\sqrt3\over2}\Omega_8
-\left(s_1x^2+s_2x^{2\over3}+{4\over9}s_3{\dot x^2}\right)
\left(1-D_{88}\right) \ .
\label{lagbreath}
\ee
A term linear in $\dot x$, which would originate from flavor symmetry 
breaking terms, has been omitted because the matrix elements of the 
associated $SU(3)$ operators vanishes when properly accounting for 
Hermiticity in the process of quantization \cite{Pa91}. The expressions 
for the constants $a_1,\ldots,s_3$ as functionals of the chiral angle 
as well as numerical values may be extracted from refs 
\cite{Sch91a,Sch91b}.
 
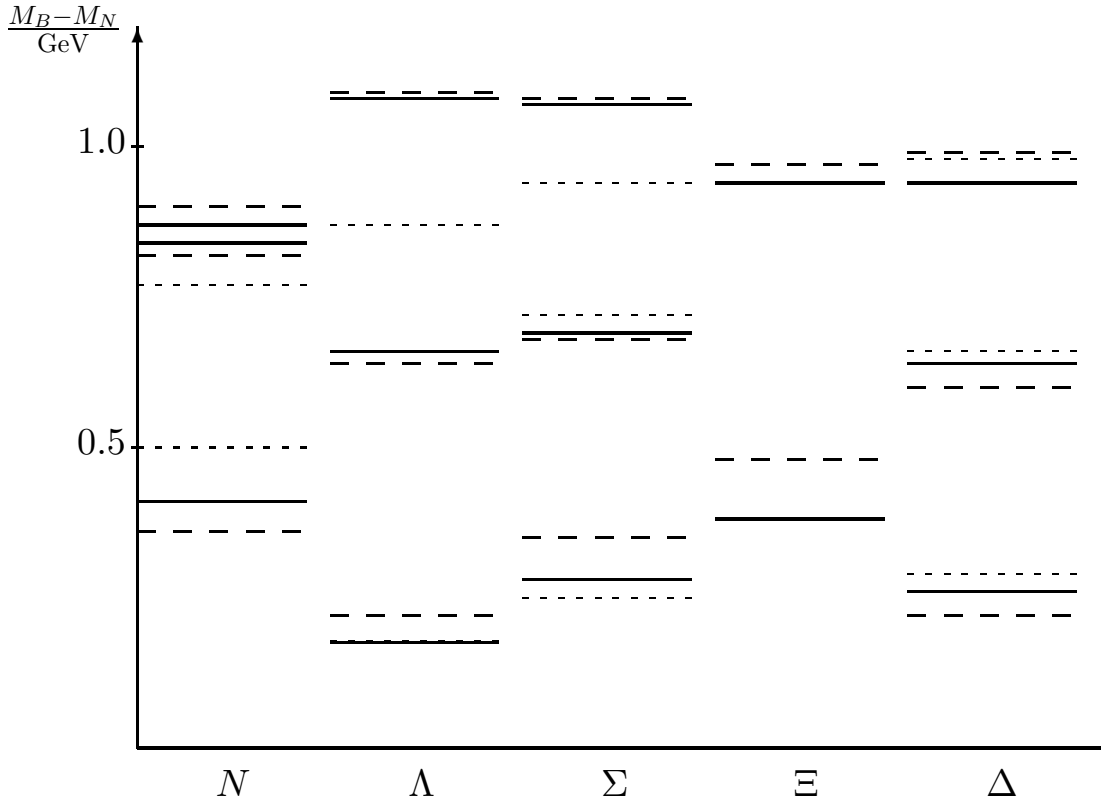
\begin{figure}[t]
\centerline{
\setlength{\unitlength}{0.8mm}
\begin{picture}(160,120)
\thicklines
\put(0,0){\line(1,0){160}}
\put(0,0){\line(0,1){115}}
\put(0,115){\vector(0,1){5}}
\put(-1,50){\line(1,0){2}}
\put(-1,100){\line(1,0){2}}
\put(-22,118){\large {${\frac{M_B-M_N}{{\rm GeV}}}$}}
\put(-10,49){\large $0.5$}
\put(-10,99){\large $1.0$}
\put(13,-8){\large $N$}
\put(45,-8){\large $\Lambda$}
\put(77,-8){\large $\Sigma$}
\put(109,-8){\large $\Xi$}
\put(141,-8){\large $\Delta$}
\multiput(0,36)(6,0){5}{\line(1,0){3}}
\multiput(0,82)(6,0){5}{\line(1,0){3}}
\multiput(0,90)(6,0){5}{\line(1,0){3}}
\multiput(32,22)(6,0){5}{\line(1,0){3}}
\multiput(32,64)(6,0){5}{\line(1,0){3}}
\multiput(32,109)(6,0){5}{\line(1,0){3}}
\multiput(64,35)(6,0){5}{\line(1,0){3}}
\multiput(64,68)(6,0){5}{\line(1,0){3}}
\multiput(64,108)(6,0){5}{\line(1,0){3}}
\multiput(96,48)(6,0){5}{\line(1,0){3}}
\multiput(96,97)(6,0){5}{\line(1,0){3}}
\multiput(128,22)(6,0){5}{\line(1,0){3}}
\multiput(128,60)(6,0){5}{\line(1,0){3}}
\multiput(128,99)(6,0){5}{\line(1,0){3}}
\thinlines
\multiput(0,50)(3,0){10}{\line(1,0){1}}
\multiput(0,77)(3,0){10}{\line(1,0){1}}
\multiput(32,17.7)(3,0){10}{\line(1,0){1}}
\multiput(32,66)(3,0){10}{\line(1,0){1}}
\multiput(32,87)(3,0){10}{\line(1,0){1}}
\multiput(64,25)(3,0){10}{\line(1,0){1}}
\multiput(64,72)(3,0){10}{\line(1,0){1}}
\multiput(64,94)(3,0){10}{\line(1,0){1}}
\multiput(96,37.9)(3,0){10}{\line(1,0){1}}
\multiput(128,29)(3,0){10}{\line(1,0){1}}
\multiput(128,66)(3,0){10}{\line(1,0){1}}
\multiput(128,98)(3,0){10}{\line(1,0){1}}
\thicklines
\put(0,41){\line(1,0){28}}
\put(0,84){\line(1,0){28}}
\put(0,87){\line(1,0){28}}
\put(32,17.5){\line(1,0){28}}
\put(32,66){\line(1,0){28}}
\put(32,108){\line(1,0){28}}
\put(64,28){\line(1,0){28}}
\put(64,69){\line(1,0){28}}
\put(64,107){\line(1,0){28}}
\put(96,38.1){\line(1,0){28}}
\put(96,94){\line(1,0){28}}
\put(128,26){\line(1,0){28}}
\put(128,64){\line(1,0){28}}
\put(128,94){\line(1,0){28}}
\end{picture}
}
~
\vskip0.2cm
\caption{\label{fi_breath}\tenrm
The mass differences of the predicted baryons in the breathing mode 
treatment of the three flavor Skyrme model with ${\Ss e=5.0}$. Full 
lines: scalar field included \protect\cite{Sch91b}; dashed lines: 
pure Skyrme model \protect\cite{Sch91a}; dotted lines: experimentally 
observed states \protect\cite{PDG94}. In this presentation the ground 
states of the ${\Ss \Lambda}$ and ${\Ss \Xi}$ channels in the 
model with the scalar field and the experimentally observed 
counterparts are (almost) indistinguishable.}
\end{figure}

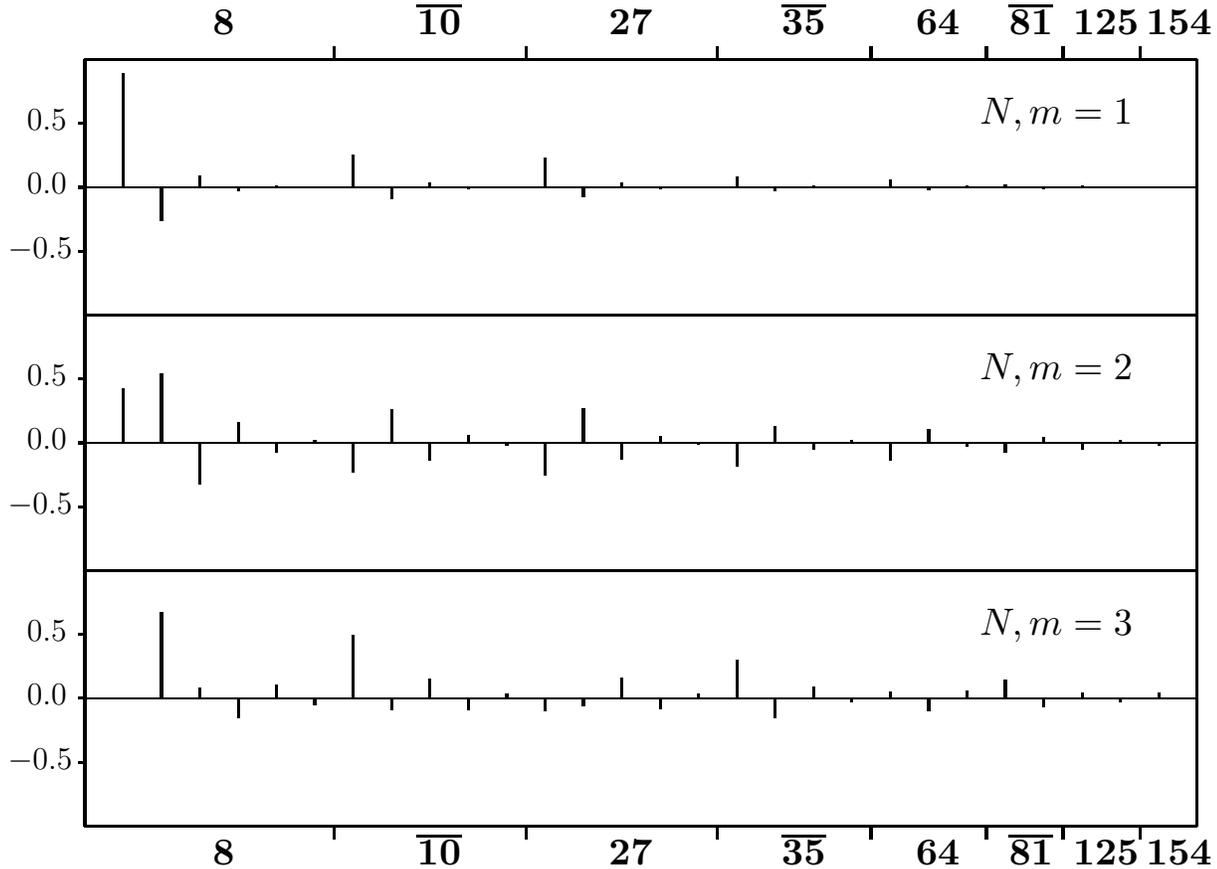
\begin{figure}[t]
\centerline{
\setlength{\unitlength}{0.17mm}
\begin{picture}(870,600)
\thicklines
\put(0,0){\line(1,0){870}}
\put(0,0){\line(0,1){600}}
\put(0,600){\line(1,0){870}}
\put(870,0){\line(0,1){600}}
\put(0,200){\line(1,0){870}}
\put(0,400){\line(1,0){870}}
\thinlines
\put(0,100){\line(1,0){870}}
\put(0,300){\line(1,0){870}}
\put(0,500){\line(1,0){870}}
\thicklines
\put(-5,100){\line(1,0){5}}
\put(-5,300){\line(1,0){5}}
\put(-5,500){\line(1,0){5}}
\put(-5,150){\line(1,0){5}}
\put(-5,350){\line(1,0){5}}
\put(-5,550){\line(1,0){5}}
\put(-5,50){\line(1,0){5}}
\put(-5,250){\line(1,0){5}}
\put(-5,450){\line(1,0){5}}
\put(-60,45){$-0.5$}
\put(-45,95){$0.0$}
\put(-45,145){$0.5$}
\put(-60,245){$-0.5$}
\put(-45,295){$0.0$}
\put(-45,345){$0.5$}
\put(-60,445){$-0.5$}
\put(-45,495){$0.0$}
\put(-45,545){$0.5$}
\put(100,-30){\large ${\bf 8}$}
\put(100,620){\large ${\bf 8}$}
\put(195,-10){\line(0,1){10}}
\put(195,600){\line(0,1){10}}
\put(260,-30){\large ${\bf{\overline{10}}}$}
\put(260,620){\large ${\bf{\overline{10}}}$}
\put(345,-10){\line(0,1){10}}
\put(345,600){\line(0,1){10}}
\put(410,-30){\large ${\bf 27}$}
\put(410,620){\large ${\bf 27}$}
\put(495,-10){\line(0,1){10}}
\put(495,600){\line(0,1){10}}
\put(545,-30){\large ${\bf{\overline{35}}}$}
\put(545,620){\large ${\bf{\overline{35}}}$}
\put(615,-10){\line(0,1){10}}
\put(615,600){\line(0,1){10}}
\put(650,-30){\large ${\bf 64}$}
\put(650,620){\large ${\bf 64}$}
\put(705,-10){\line(0,1){10}}
\put(705,600){\line(0,1){10}}
\put(723,-30){\large ${\bf{\overline{81}}}$}
\put(723,620){\large ${\bf{\overline{81}}}$}
\put(765,-10){\line(0,1){10}}
\put(765,600){\line(0,1){10}}
\put(773,-30){\large ${\bf 125}$}
\put(773,620){\large ${\bf 125}$}
\put(825,-10){\line(0,1){10}}
\put(825,600){\line(0,1){10}}
\put(830,-30){\large ${\bf 154}$}
\put(830,620){\large ${\bf 154}$}
\put(700,550){\large $N,m=1$}
\put(30,500){\line(0,1){89}}
\put(60,500){\line(0,-1){26}}
\put(90,500){\line(0,1){9}}
\put(120,500){\line(0,-1){3}}
\put(150,500){\line(0,1){1}}
\put(180,500){\line(0,1){0}}
\put(210,500){\line(0,1){25}}
\put(240,500){\line(0,-1){9}}
\put(270,500){\line(0,1){3}}
\put(300,500){\line(0,-1){1}}
\put(330,500){\line(0,1){0}}
\put(360,500){\line(0,1){23}}
\put(390,500){\line(0,-1){7}}
\put(420,500){\line(0,1){3}}
\put(450,500){\line(0,-1){1}}
\put(480,500){\line(0,1){0}}
\put(510,500){\line(0,1){8}}
\put(540,500){\line(0,-1){3}}
\put(570,500){\line(0,1){1}}
\put(600,500){\line(0,1){0}}
\put(630,500){\line(0,1){6}}
\put(660,500){\line(0,-1){2}}
\put(690,500){\line(0,1){1}}
\put(720,500){\line(0,1){2}}
\put(750,500){\line(0,-1){1}}
\put(780,500){\line(0,1){1}}
\put(810,500){\line(0,-1){0}}
\put(840,500){\line(0,1){0}}
\put(700,350){\large $N,m=2$}
\put(30,300){\line(0,1){42}}
\put(60,300){\line(0,1){54}}
\put(90,300){\line(0,-1){32}}
\put(120,300){\line(0,1){16}}
\put(150,300){\line(0,-1){7}}
\put(180,300){\line(0,1){2}}
\put(210,300){\line(0,-1){23}}
\put(240,300){\line(0,1){26}}
\put(270,300){\line(0,-1){14}}
\put(300,300){\line(0,1){6}}
\put(330,300){\line(0,-1){2}}
\put(360,300){\line(0,-1){25}}
\put(390,300){\line(0,1){27}}
\put(420,300){\line(0,-1){13}}
\put(450,300){\line(0,1){5}}
\put(480,300){\line(0,-1){1}}
\put(510,300){\line(0,-1){18}}
\put(540,300){\line(0,1){13}}
\put(570,300){\line(0,-1){5}}
\put(600,300){\line(0,1){2}}
\put(630,300){\line(0,-1){14}}
\put(660,300){\line(0,1){10}}
\put(690,300){\line(0,-1){3}}
\put(720,300){\line(0,-1){7}}
\put(750,300){\line(0,1){4}}
\put(780,300){\line(0,-1){5}}
\put(810,300){\line(0,1){2}}
\put(840,300){\line(0,-1){2}}
\put(700,150){\large $N,m=3$}
\put(30,100){\line(0,1){0}}
\put(60,100){\line(0,1){67}}
\put(90,100){\line(0,1){8}}
\put(120,100){\line(0,-1){15}}
\put(150,100){\line(0,1){10}}
\put(180,100){\line(0,-1){5}}
\put(210,100){\line(0,1){49}}
\put(240,100){\line(0,-1){9}}
\put(270,100){\line(0,1){15}}
\put(300,100){\line(0,-1){9}}
\put(330,100){\line(0,1){3}}
\put(360,100){\line(0,-1){10}}
\put(390,100){\line(0,-1){6}}
\put(420,100){\line(0,1){16}}
\put(450,100){\line(0,-1){8}}
\put(480,100){\line(0,1){3}}
\put(510,100){\line(0,1){30}}
\put(540,100){\line(0,-1){15}}
\put(570,100){\line(0,1){9}}
\put(600,100){\line(0,-1){3}}
\put(630,100){\line(0,1){5}}
\put(660,100){\line(0,-1){10}}
\put(690,100){\line(0,1){6}}
\put(720,100){\line(0,1){14}}
\put(750,100){\line(0,-1){7}}
\put(780,100){\line(0,1){4}}
\put(810,100){\line(0,-1){3}}
\put(840,100){\line(0,1){4}}
\end{picture}
}
~
\vskip0.2cm
\caption{\label{fi_amplitudes}\tenrm
The amplitudes ${\Ss C_{\mu,n_\mu}^{(N,m)}}$ of the three lowest 
states with nucleon quantum numbers. The intervals indicated along 
the abscissa refer to the radially excited states within an 
${\Ss SU(3)}$ representation ${\Ss\mu}$. Here the model including 
the scalar field has been used with ${\Ss e=5.0}$.}
\end{figure}

The baryon states corresponding to the Lagrangian (\ref{lagbreath}) 
are obtained in a two--step procedure. For convenience one defines 
\be
m=m(x)=\frac{8}{9}(a_1+a_2x^2)\ , 
\alpha=\alpha(x)=\alpha_1x^2+\alpha_2 x^{\frac{2}{3}}\ , 
\ldots \ , s=s(x)=s_1x^2+sx^{\frac{2}{3}} \ .
\label{defbreath}
\ee
Then the flavor symmetric part of the collective Hamiltonian
\be
H=-\frac{1}{2\sqrt{m\alpha^2\beta^4}}\frac{\partial}{\partial x}
\sqrt{\frac{\alpha^3\beta^4}{m}}\frac{\partial}{\partial x}
+b+\left(\frac{1}{2\alpha}-\frac{1}{2\beta}\right)J(J+1)
+\frac{1}{2\beta}C_2(\mu)-\frac{3}{8\beta}+s \ ,
\label{freebreath}
\ee
is diagonalized for a definite $SU(3)$ representation $\mu$. Denote 
the eigenvalues by ${\cal E}_{\mu,n_\mu}$ and the corresponding 
eigenstates by $|\mu,n_\mu\rangle$, where $n_\mu$ labels the radial 
excitations. Actually the eigenstates factorize 
$|\mu,n_\mu\rangle=|\mu\rangle|n_\mu\rangle$. In this language the 
nucleon corresponds to $|{\bf 8},1\rangle$ while the first excited 
state, which commonly is identified with the Roper (1440) resonance,
would be $|{\bf 8},2\rangle$. Of course, we are interested in the 
role of states like 
$|{\overline {\bf 10}},n_{\overline {\bf 10}}\rangle$. In order to 
answer this question the symmetry breaking part has to be taken into 
account as well. This is done by employing the states $|\mu,n_\mu\rangle$
as a basis to diagonalize the complete Hamiltonian matrix
\be
H_{\mu,n_\mu;\mu^\prime,n^\prime_{\mu^\prime}}=
{\cal E}_{\mu,n_\mu}\delta_{\mu,\mu^\prime}
-\langle\mu|D_{88}|\mu^\prime\rangle
\langle n_\mu|s(x)|n^\prime_{\mu^\prime}\rangle \ .
\label{hammatr}
\ee
The flavor part of these matrix elements is computed using 
$SU(3)$ Clebsch--Gordon coefficients while the radial part 
is calculated using the appropriate eigenstates of 
(\ref{freebreath}). Of course, this can be done for each isospin 
multiplet separately, {\it i.e.} flavor quantum numbers are not 
mixed. The physical baryon states $|B,m\rangle$ are finally 
expressed as linear combinations of the eigenstates of 
the symmetric part 
\be
|B,m\rangle=\sum_{\mu,n_\mu}C_{\mu,n_\mu}^{(B,m)}
|\mu,n_\mu\rangle \ .
\label{bsbreath}
\ee
The corresponding eigen energies are denoted by $E_{B,m}$.
The nucleon $|N,1\rangle$ is then identified as the lowest energy 
solution with the associated quantum numbers, while the Roper 
is defined as the next state ($|N,2\rangle$) in the same spin -- 
isospin channel. Turning to the quantum numbers of the $\Lambda$ 
provides not only the energy $E_{\Lambda,1}$ and 
wave--function $|\Lambda,1\rangle$ of this hyperon but also the 
analogous quantities for the radially excited $\Lambda$'s:
$E_{\Lambda,n}$ and $|\Lambda,n\rangle$ with $n\ge2$. These 
calculations are repeated for the other spin -- isospin channels 
yielding the spectrum not only of the ground state $\frac{1}{2}^+$ 
and $\frac{3}{2}^+$ baryons but also their radial excitations.
For both models (with and without scalar field) the predicted 
eigen energies of the Hamiltonian (\ref{hammatr}) are compared to those 
of the experimentally observed baryons in figure \ref{fi_breath}.
As remarked previously this treatment of the pure Skyrme model 
yields mass differences of the hyperons which are considerably larger 
than the empirical data. Including, however, the scalar field according 
to eq (\ref{lsbglue}) provides an excellent agreement for the 
mass differences for the ground states of each spin--isospin channel.
The inclusion of the scalar field also leads to an improved prediction 
for the Roper resonance although its position is still underestimated 
by about 80MeV. Nevertheless, the overall picture obtained for the 
mass differences suggests that the model gives quite reliable 
information about the structure of the baryon wave--functions. This 
structure is parametrized by the coefficients $C_{\mu,n_\mu}^{(B,m)}$ 
defined in eq (\ref{bsbreath}). For the lowest energy states with 
nucleon quantum numbers these are displayed in figure 
\ref{fi_amplitudes}.
As expected the nucleon state $|N,1\rangle$ is dominated by the 
lowest octet excitation. In addition to the admixtures of the 
higher dimensional $SU(3)$ representations, which are expected 
from the expansion (\ref{nwfexp}), the second octet excitation
has a sizable amplitude. More surprisingly, the structure of the 
Roper resonance is very complicated. Although the assumption 
that this state is dominanted by $|{\bf 8},2\rangle$ is 
confirmed in so far that this component has the largest 
amplitude, other components (especially $|{\bf 8},1\rangle$) 
are equally important. The picture of the Roper being a radial
excited nucleon is partially verified because it is always the 
second radial state, which dominates in the higher dimensional 
$SU(3)$ representations. In the case of the pure Skyrme model
one actually finds 
$|C^{\rm Roper}_{{\bf 8},1}|>|C^{\rm Roper}_{{\bf 8},2}|$
\cite{Sch91a}. Amusingly the $|{\bf 8},1\rangle$ 
component is completely absent in the wave--function of the next state 
above the Roper with nucleon quantum numbers. This state should be 
identified with the $P_{11}(1710)$ resonance. The analogous studies 
for the $\Lambda$, $\Sigma$ and $\Xi$ channels are presented in 
ref \cite{Sch91a} for the case of the pure Skyrme model. 

In the framework of the scaling treatment various static properties of 
the low--lying baryons have been computed \cite{Sch91a,Sch91b}. As an 
example the predicted magnetic moments of the $\frac{1}{2}^+$ 
baryons are shown in table \ref{ta_magscal}.

\begin{table}
\caption{\label{ta_magscal}\tenrm
The magnetic moments in the scaling treatment of the Skyrme model 
(I). Model (II) contains the additional scalar field 
(\protect\ref{lsbglue}). Results are given in nucleon magnetons 
as well as ratios of the proton magnetic moment.}

~
\newline
\centerline{\tenrm\smalllineskip
\begin{tabular}{c | c c | c c | c c}
& \multicolumn{2}{c|}{I} & \multicolumn{2}{c|}{II}
& \multicolumn{2}{c}{Expt.} \\
\hline
Baryon & $\mu_B$ & $\mu_B/\mu_p$ &
$\mu_B$ & $\mu_B/\mu_p$ & $\mu_B$ & $\mu_B/\mu_p$ \\
\hline
$p$        & 2.58 & 1.00 & 2.21 & 1.00 & 2.79 & 1.00 \\
$n$        &-2.26 &-0.86 &-1.84 &-0.83 &-1.91 &-0.68 \\
$\Lambda$  &-0.50 &-0.20 &-0.52 &-0.24 &-0.61 &-0.22 \\
$\Sigma^+$ & 2.01 & 0.78 & 1.82 & 0.82 & 2.42 & 0.87 \\
$\Sigma^0$ & 0.40 & 0.16 & 0.44 & 0.20 & ---  & ---  \\
$\Sigma^-$ &-1.21 &-0.47 &-0.94 &-0.43 &-1.16 &-0.42 \\
$\Xi^0   $ &-1.06 &-0.41 &-1.06 &-0.48 &-1.25 &-0.45 \\
$\Xi^-   $ &-0.37 &-0.14 &-0.41 &-0.19 &-0.69 &-0.25 \\
$\Sigma^0\rightarrow\Lambda$ 
&-1.55 &-0.60 &-1.37 &-0.62 &-1.61 &-0.58 \\
\end{tabular}}
\end{table}
As for the mass differences in the pure Skyrme model, the scaling 
treatment over--estimates the symmetry breaking: The deviation from the 
$SU(3)$ relations (\ref{su3mag}) is larger than experimentally found. 
On the other hand the additional scalar field mitigates the flavor 
symmetry breaking effects properly. Again the experimentally observed 
deviation from the U--spin relations ({\it cf.} table \ref{ta_uspin}) is 
reproduced, comparable to the slow rotator approach. It is especially 
this result for the magnetic moments which indicates that the scaling 
and slow rotator approaches are quite similar. As a matter of fact 
both treatments attempt to incorporate the effects of flavor symmetry 
breaking on the soliton profile. While the slow rotator does this 
on a classical level, the admixture of radially excited states 
may be regarded as a quantum description of the same problem.

For completeness it should also be remarked that the axial 
properties of the hyperons have also been studied in the scaling 
treatment. As in the rigid and slow rotator approaches, consistency 
with the Cabibbo model \cite{Ca63} was obtained together with a strong 
reduction of strange quark contribution to the axial current of the 
nucleon. This appears to be a common feature of models, which 
incorporate flavor symmetry breaking in the baryon wave--function.
To be specific, the matrix element of the eighth component of 
the axialvector current between nucleon states (see eq (\ref{defR}))
is only one third of its flavor symmetric value while the 
predictions for the matrix elements characterizing the 
semi--leptonic hyperons decays agree with the experimental values 
within 10\% after normalizing with respect to the axial charge 
of the nucleon.

In order to summarize the studies compiled in this chapter it is 
important to remark that one has to go beyond the rigid rotator 
treatment in order to reproduce the detailed structure of the baryon
properties. In particular, this has become obvious for the magnetic 
moments of the $\frac{1}{2}^+$ baryons. As will be discussed in 
chapter \ref{chap_bound} the bound state approach to include strange 
degrees of freedom \cite{Ca85,Ca88} can be compared to the rigid 
rotator in the limit of very large flavor symmetry breaking. Hence 
the bound state approach result $\mu_{\Sigma^+}/\mu_p\approx1.1$ 
\cite{Ku89,Oh91} (see also table \ref{ta_magbs}) has to be considered 
as an additional support of the above statement that the rigid 
rotator is too crude an approximation for a detailed prescription 
of baryon properties.

Unfortunately the two treatments introduced in the present chapter 
cannot straightforwardly be applied to more complicated but also 
more realistic effective meson Lagrangians. This is also the case for 
the vector meson model to be discussed next. The reason being that the 
time components of the vector meson fields commonly have to satisfy 
special constraints which are fulfilled by the classical soliton 
configuration. However, the introduction of additional time dependence 
(either via the strangeness changing angle $\nu(t)$ as in section 
\ref{sec_slowrot} or the breathing mode coordinate $\mu(t)$ as in section 
\ref {sec_mixing}) would spoil these constraints making impossible 
the quantum prescription.

Furthermore, all the above described modifications of the rigid rotator 
treatment predict $\mu_{\Xi^-}/\mu_\Lambda<1$ in contrast to the 
experimental observation. This indicates that further refinements 
are necessary.

\vskip1cm
\newpage
\begin{satz}
\label{chap_vector}{\large \bf \hskip1cm 
Inclusion of Vector Mesons}
\addcontentsline{toc}{chapter}{\protect\ref{chap_vector}
Inclusion of Vector Mesons}
\end{satz}
\stepcounter{chapter}

As already mentioned earlier, short range effects are missing 
within the pseudo--scalar Skyrme model. Such effects may elegantly 
been incorporated by extending the model to contain vector meson 
fields. The associated profile functions have their major support 
in the vicinity of the origin. From the studies of two nucleon 
reactions in meson exchange models \cite{La80} it is well known that 
these fields play an important role in the description of low 
energy processes. Also the phenomenologically successful concept of 
vector meson dominance (VMD) \cite{Sa69} supports the explicit 
consideration of these fields. The VMD hypothesis states that a photon 
does not interact directly with the nucleon but rather it switches 
to a heavy vector meson, which then couples to the nucleon. On the 
other hand it has already been remarked in the preceding chapters that 
the stabilizing terms (\ref{Skterm}) and (\ref{sixorder}) can be 
motivated within the infinite mass limit of the interaction of the 
pions with the $\rho$ and $\omega$ vector mesons, respectively. In that 
limit the Skyrme parameter is identical to the coupling constant of the 
$\rho$ meson to two pions, $g_{\rho\pi\pi}$ while $\epsilon_6$ is 
proportional to $g_{\omega\pi\pi\pi}$, the coupling of the $\omega$ 
meson to three pions.  Of course, it is more appealing to keep these 
mesons explicitly in the effective meson Lagrangian rather than to 
assume the infinite mass approximation. The explicit incorporation of 
these fields yields important improvements for the description of 
baryon properties. Some of them were already encountered soon after 
the soliton picture for baryons had become popular. Before going into 
the details of the vector meson model for three flavors, it is 
illuminating to briefly comment on a few of these improvements. The 
presence of a finite mass $\omega$ meson provides an increase of 
the isoscalar radius \cite{Me87}
\be
\langle r^2\rangle _{I=0} \approx
\langle r^2\rangle _B +\frac{6}{m_V^2} \ ,
\label{changeR0}
\ee
where $\langle r^2\rangle _B$ is the radius associated with baryon 
number current (\ref{bnumber}). The additional piece in eq 
(\ref{changeR0}) is a consequence of (approximate) 
VMD, which indeed  is observed when including the vector mesons in a 
chirally invariant manner. As can be seen from table \ref{ta_emps} this 
increase of about $0.35{\rm fm}^2$ will significantly improve the 
prediction for the radii. Another interesting feature of finite mass 
vector mesons concerns the $\pi N$ scattering amplitudes. In this 
context it was soon realized \cite{Sch87} that keeping the mass, $m_V$, 
of the vector mesons finite solves the problem of the ever--rising 
phase shifts observed in the Skyrme model description of $\pi N$ 
scattering \cite{Ec86}. Actually, this effect is similar to the 
transition from the Fermi model for weak interactions to the standard 
model. It just implies a reduction of the contact interaction at large 
momentum
\be
g_{\rho\pi\pi}^2\longrightarrow
\frac{g_{\rho\pi\pi}^2m_V^2}{m_V^2-q^2} \ .
\label{changeVMprop}
\ee
Other interesting features of the vector mesons are that these fields 
are mandatory to obtain a finite matrix element of the singlet axial 
current \cite{Jo90} as well as a non--vanishing prediction for the 
strong interaction piece of the neutron proton mass difference 
\cite{Ja89}. As a matter of fact, these two issues are strongly 
related as will be discussed in section \ref{sec_two}.

\bigskip

\begin{subsatz}
\label{sec_veclag}{\bf \hskip1cm The Vector Meson Lagrangian}
\addcontentsline{toc}{section}{\protect\ref{sec_veclag}
The Vector Meson Lagrangian}
\end{subsatz}
\stepcounter{section}

In this section a realistic effective Lagrangian, which contains vector 
meson degrees of freedom as well as the pseudo--scalar fields, will be 
constructed. Here the language of the massive Yang--Mills approach 
will be used although the resulting Lagrangian is identical to the 
one obtained in the hidden symmetry approach \cite{Tu85} when 
terms, which do not transform properly under charge conjugation, are 
omitted \cite{Ja88}.

In order to avoid the appearance of explicit axial--vector fields 
without violating global chiral symmetry it is helpful to introduce 
auxiliary ``gauge fields" $A_\mu^L$ and $A_\mu^R$ \cite{Ka84b}. Under 
a local chiral transformation (\ref{uchitrans}) these fields are 
assumed to behave like
\be
A_\mu^L\rightarrow 
L\left(A_\mu^L+\frac{i}{g}\partial_\mu\right)L^{\dag}
\quad {\rm and} \quad
A_\mu^R\rightarrow
R\left(A_\mu^R+\frac{i}{g}\partial_\mu\right)R^{\dag} .
\label{vectransf}
\ee
The physical relevance of the gauge coupling constant $g$ will 
be explained shortly. Under parity one demands 
$A_\mu^L\leftrightarrow A^{R\mu}$. The axial--vector content of 
$A_\mu^L$ and $A_\mu^R$ is then eliminated by imposing the condition 
\be
A_\mu^L=U\left(A_\mu^R+\frac{i}{g}\partial_\mu\right)U^{\dag}
\label{chiconst}
\ee
on these auxiliary fields. Here $U$ denotes the chiral field 
(\ref{su3chif}) which transforms according to (\ref{uchitrans}).
Since the $RHS$ of eq (\ref{chiconst}) transforms like a ``left gauge 
field" chiral invariance is not violated when replacing 
$A_\mu^L$ accordingly. A convenient realization of the 
condition (\ref{chiconst}) is given by
\be
A_\mu^L=
\xi\left(\rho_\mu+\frac{i}{g}\partial_\mu\right)\xi^{\dag}
\quad {\rm and} \quad
A_\mu^R=
\xi^{\dag}\left(\rho_\mu+\frac{i}{g}\partial_\mu\right)\xi ,
\label{rho1}
\ee
where the root of the chiral field has been introduced, $\xi=U^{1/2}$.
Since under parity $\xi\leftrightarrow\xi^{\dag}$ it is easy to 
verify that $\rho_\mu$ contains vector meson degrees of freedom only. 
This field is therefore identified with the physical vector meson nonet
\be
\rho=\sum_{a=1}^8\frac{\rho^a}{2}\lambda^a=\pmatrix{
\frac{1}{2}\rho^0+\frac{1}{2}\omega &
\frac{1}{\sqrt2}\rho^+ & \frac{1}{\sqrt2}K^{*+} \cr 
\frac{1}{\sqrt2}\rho^- 
&-\frac{1}{2}\rho^0+\frac{1}{2}\omega &\frac{1}{\sqrt2}K^{*0}\cr
\frac{1}{\sqrt2}K^{*-} & \frac{1}{\sqrt2}{\bar K}^{*0}& 
\frac{1}{\sqrt2}\varphi \cr},
\label{eightfoldvec}
\ee
where the Lorentz indices have been suppressed. Under the chiral 
transformation this vector field also behaves like a gauge 
field \cite{Ca69}
\be
\rho_\mu\rightarrow
K\left(\rho_\mu+\frac{i}{g}\partial_\mu\right)K^{\dag} 
\quad {\rm and} \quad
\xi\rightarrow L\xi K^{\dag}=K\xi R^{\dag} .
\label{defK}
\ee
For vector transformations the last equation, which defines the 
matrix $K$, is trivially solved by $L=R=K$ while for axial 
transformations, $L=R^{\dag}$, the matrix $K$ depends on the meson 
field configuration $\xi$.

The effective vector meson pseudo--scalar Lagrangian is developed in two
steps. First a chirally invariant Lagrangian is constructed in terms 
of the auxiliary fields $A_\mu^{L,R}$. Subsequently the axial--vector 
degrees of freedom are eliminated by the realization (\ref{rho1}). As 
this realization merely is a gauge transformation for the fields 
$A_\mu^{L,R}$, the kinetic part for the vector mesons is uniquely given 
by
\be
-\frac{1}{2}{\rm tr}\left[F_{\mu\nu}(\rho)F^{\mu\nu}(\rho)\right] ,
\label{rhokin}
\ee
where $F_{\mu\nu}(\rho)=\partial_\mu\rho_\nu-\partial_\nu\rho_\mu
-ig[\rho_\mu , \rho_\nu]$ refers to the field strength tensor of 
the $\rho$ meson. This field acquires a mass by adding terms 
quadratic in $A_\mu^{L,R}$ which are invariant under global 
chiral transformations only
\be
\frac{m_0^2}{2}{\rm tr}\left[A_\mu^L A^{L\mu}+A_\mu^R A^{R\mu}\right]
-\frac{B}{2}{\rm tr}\left[A_\mu^L U A^{R\mu} U^{\dag}\right] .
\label{ALRmass}
\ee
The identical coefficients of the first two terms are demanded by 
parity invariance. Employing the realization (\ref{rho1}) one 
observes that the expression (\ref{ALRmass}) not only contains the 
vector meson mass term but also the non--linear $\sigma$ model when 
identifying $4m_0^2-2B=m_V^2$ and $4m_0^2+2B=g^2\tilde f_\pi^2$. Here 
$m_V$ denotes an average mass of the vector meson nonet
(\ref{eightfoldvec}). Next, one has to account for the mass 
splittings of the vector mesons. To leading order in the symmetry 
breaking, an appropriate term which properly behaves under chiral 
transformations, can be constructed from the last expression in 
(\ref{ALRmass})
\be
-\alpha^\prime{\rm tr}\left[{\cal M}\left(A_\mu^L U A^{R\mu}+
A_\mu^R U^{\dag} A^{L\mu}\right)\right] .
\label{vmsymbr}
\ee
The reader may consult ref \cite{Ha95} for a very recent and 
elaborate study of symmetry breaking in the meson sector.
Introducing, for convenience 
\be
p_\mu&=&\partial_\mu\xi\xi^{\dag}+\xi^{\dag}\partial_\mu\xi\ , \quad
R_\mu=\rho_\mu-\frac{i}{2g}
\left(\partial_\mu\xi\xi^{\dag}+\xi^{\dag}\partial_\mu\xi\right)
\nonumber \\
T^\pm&=&\xi T \xi\pm\xi^{\dag}T\xi^{\dag}\ , \quad
S^\pm=\xi S \xi\pm\xi^{\dag}S\xi^{\dag}
\label{defpR}
\ee
(see eq (\ref{massmat}) for the relevant definitions)
the explicit form of the Lagrangian may compactly be written as
\be
\hspace{-1cm}
{\cal L}_{\rm non-an}&=&{\rm tr}\Bigg[
-\frac{1}{4}\tilde f_\pi^2p_\mu p^\mu
-\frac{1}{2}F_{\mu\nu}(\rho)F^{\mu\nu}(\rho)
+m_V^2R_\mu R^\mu
\nonumber \\ &&
-2\alpha^\prime R_\mu R^\mu\left(T^+ +xS^+\right)
-\frac{i}{g}[R_\mu,p^\mu]\left(T^- +xS^-\right)
\nonumber \\ &&
+\left(\beta^\prime-\frac{\alpha^\prime}{2g^2}\right)
p_\mu p^\mu\left(T^+ +xS^+\right)
+\delta^\prime\left(T^+-2T +x(S^+-2S)\right)\Bigg] ,
\label{Lnanvm}
\ee
where the index indicates that the anomalous terms (involving the 
anti--symmetric tensor $\epsilon_{\mu\nu\rho\sigma}$) have not yet 
been included. Expanding this Lagrangian up to quadratic order in 
the meson fields allows one to express the parameters of the model
in terms of the physical decay constants and masses. Typical examples 
are (see also eq (\ref{pspara}))
\be
f_\pi^2=\tilde f_\pi^2 +\alpha^\prime-8\beta^\prime
\ , \quad
m^2_{K^*}=m_V^2-2\alpha^\prime\left(1+x\right)\ .
\label{vmpara}
\ee
The complete set of equations may be found in ref \cite{Ja89}. The 
third term in eq (\ref{Lnanvm}) contains the $\rho\pi\pi$ vertex
\be
m_V^2{\rm tr}\left[R_\mu R^\mu\right]=
\frac{m_V^2}{2}\left\{\rho_\mu^a\rho^{\mu a}+
\frac{1}{g \tilde f_\pi^2}\mbox{\boldmath $\rho$}_\mu\cdot
\left(\mbox{\boldmath $\pi$}\times\partial^\mu
\mbox{\boldmath $\pi$}\right)+\ldots\right\} .
\label{expRR}
\ee
Utilizing the experimental data ($f_\pi=93{\rm MeV}$,
$g_{\rho\pi\pi}=m_V^2/\sqrt2 g\tilde f_\pi^2=8.66$\footnote{The 
coupling constant $g_{\rho\pi\pi}$ is computed from the 
width $\Gamma_{\rho\pi\pi}=151{\rm MeV}$ \cite{PDG94}.}) yields 
the parameters \cite{Pa92}
\be
&&\tilde f_\pi\approx0.092{\rm GeV}\ , \qquad
m_V\approx0.766{\rm GeV}\ , \qquad
g\approx5.57 \ , \qquad x\approx36\ ,
\nonumber \\ && \hspace{-1cm}
\alpha^\prime\approx-2.8\times10^{-3}{\rm GeV}^2\ , \quad
\beta^\prime\approx-7.14\times10^{-5}{\rm GeV}^2\ , \quad
\delta^\prime\approx4.15\times10^{-5}{\rm GeV}^2\ .
\label{vmpara1}
\ee
Due to the presence of the $\alpha^\prime$ type symmetry breaker 
the values of $\beta^\prime$ and $x$ are slightly changed compared to 
the pure pseudo--scalar model (\ref{pspara1}). One might want to 
include a symmetry breaker for the kinetic part of the vector meson 
fields \cite{Ja89}. As it is the goal to describe the baryons with a 
minimal set of terms in the effective meson Lagrangian, such a term 
is omitted unless some fine--tuning of the model is required.

In order to complete the vector meson model Lagrangian the anomalous 
terms have to be added. For their presentation it is most useful to 
introduce the notation of differential forms: $A^R=A_\mu^R dx^\mu,\
d=\partial_\mu dx^\mu$, etc. . Since the left and right ``gauge 
fields" are related via the chiral constraint (\ref{chiconst}) the
number of linear independent terms, which transform properly under 
the chiral transformation as well as parity and charge conjugation, 
is quite limited \cite{Ka84b,Ja88}:
\be
A^L\alpha^3 \ , \quad
dA^L\alpha A^L-A^L\alpha dA^L+A^L\alpha A^L\alpha \ , \quad
2\left(A^L\right)^3\alpha+\frac{i}{g}A^L\alpha A^L\alpha\ .
\label{anom1}
\ee
Of course, including these terms in the model Lagrangian will 
introduce three more parameters: $\gamma_1,\gamma_2$ and $\gamma_3$.
Noting that $\alpha=\xi p \xi^{\dag}$ a suitable presentation of 
the anomalous part of the action is given in terms of the 
quantities defined in eq (\ref{defpR})
\be
\hspace{-1cm}
\Gamma_{\rm an}&=&
\frac{iN_C}{240\pi^2}\int_{M_5} {\rm tr}\left(p^5\right)
\nonumber \\ &&
+\int_{M_4}{\rm tr}\left(
\frac{1}{6}\left[\gamma_1+\frac{3}{2}\gamma_2\right]Rp^3
-\frac{i}{4}g\gamma_2 F(\rho)\left[pR-Rp\right]
-g^2\left[\gamma_2+2\gamma_3\right]R^3p\right) ,
\label{anom2}
\ee
where also the Wess--Zumino term (\ref{WZterm}) is included. In 
ref \cite{Ja88} two of the three unknown constants, $\gamma_{1,2,3}$ 
were determined from purely strong interaction processes like 
$\omega\rightarrow 3\pi$. Defining $\tilde h=-2\sqrt2\gamma_1/3$, 
$\tilde g_{VV\phi}=g\gamma_2$ and $\kappa=\gamma_3/\gamma_2$ 
the central values $\tilde h=\pm0.4$ and $\tilde g_{VV\phi}=\pm1.9$ were 
found. Within experimental uncertainties (stemming from the errors 
in the $\omega - \phi$ mixing angle) these may vary in the range 
$\tilde h=-0.15,\ldots,0.7$ and $\tilde g_{VV\phi}=1.3,\ldots, 2.2$ 
subject to the condition 
$\vert\tilde g_{VV\phi}-\tilde h\vert\approx 1.5$. The third parameter, 
$\kappa$ could not be fixed in the meson sector. From studies 
\cite{Me89} of nucleon properties in the two flavor model it was 
argued that $\kappa\approx1$ represents a reasonable choice. These 
studies also allowed one to fix the overall signs of the parameters 
$\gamma_{1,2,3}$ (which cannot be determined in the meson sector) to 
be $\tilde g_{VV\phi}\approx+1.9$ when $F(0)=+\pi$. The model can 
equivalently be formulated for $F(0)=-\pi$. Then the signs of the 
coefficients $\gamma_{1,2,3}$ as well as that of the Wess--Zumino 
term (\ref{WZterm}) have to be changed.

\bigskip

\begin{subsatz}
\label{sec_inertia}{\bf \hskip1cm Baryon Masses}
\addcontentsline{toc}{section}{\protect\ref{sec_inertia}
Baryon Masses}
\end{subsatz}
\stepcounter{section}

\begin{figure}[t]
\centerline{\hskip -2cm
\epsfig{figure=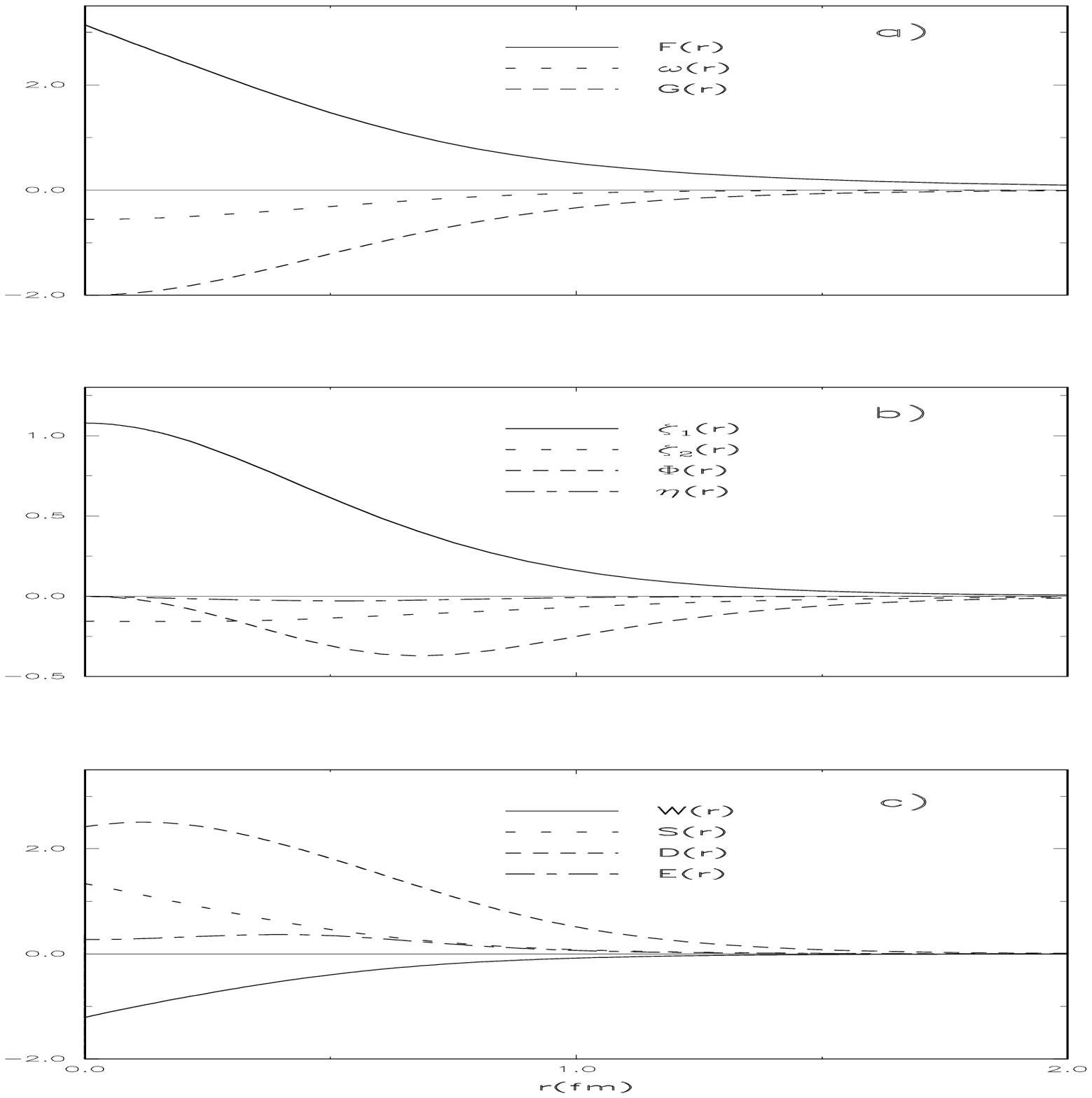,height=14.0cm,width=16.0cm}}
\caption{\label{fi_vm}\tenrm
The radial dependencies  of the profile functions in the 
vector meson model. a) classical fields minimizing the static 
energy ${\Ss M}$, b) induced fields for the moment of inertia 
${\Ss \alpha^2}$, c) induced fields for the moment of inertia
${\Ss \beta^2}$. ${\Ss \omega}$ is in ${\Ss {\rm GeV}}$, 
${\Ss \phi, W}$ and ${\Ss D}$ are in ${\Ss {\rm GeV}^{-1}}$ while 
all other fields are dimensionless. The profile functions 
correspond to the parameter set (\protect\ref{vmpara1}) and
(\protect\ref{parabf}).}
\end{figure}

In order to compute the spectrum of the low--lying baryons in the 
vector meson model the coefficients in the collective Hamiltonian 
(\ref{collham}) have to be evaluated from the action
\be
\Gamma=\int d^4x {\cal L}_{\rm non-an}\ +\ \Gamma_{\rm an} \ .
\label{Atotvm}
\ee
The generalization of the hedgehog {\it ansatz} (\ref{hedgehog}) to 
the vector meson model requires the time component of the 
$\omega$ field and the space components of the $\rho$ field to
be different from zero. All static fields are embedded in the 
isospin subgroup of $SU(3)$. Parity and grand spin symmetry 
allow for three radial functions
\be
\xi_\pi=
{\rm exp}\left(\frac{i}{2}\hat{\mbox{\boldmath $r$}}\cdot
{\mbox{\boldmath $\tau$}} F(r)\right)\ ,\quad
\omega_0={{\omega(r)}\over{2g}}\ ,\quad
\rho_i^a={{G(r)}\over{gr}}\epsilon_{ija}\hat r_j\ .
\label{vmhedgehog}
\ee
Substituting these {\it ans\"atze} into the action (\ref{Atotvm})
yields the classical mass $E$. Its functional form is displayed in 
eq (\ref{mcl}). Application of the variational principle to this 
functional leads to second order coupled non--linear differential 
equations for the radial functions $F(r)$, $\omega(r)$ and $G(r)$. The 
boundary conditions for the chiral angle $F(r)=\pi$ and $F(\infty)=0$, 
which correspond to a unit baryon number, also determine the boundary 
conditions of the  vector meson profiles via the differential equations,
{\it e.g.} $G(0)=-2$. A typical set of resulting profile functions 
is shown in figure \ref{fi_vm}a. These radial functions are subsequently
employed to evaluate the symmetry breaking parameters 
$\gamma$, $\alpha_1$ and $\beta_1$, {\it cf.} eqs (\ref{gammavm}) and 
(\ref{ab1vm}).

The computation of the moments of inertias $\alpha^2$ and $\beta^2$ is 
more involved due to the appearance of induced components. In contrast 
to the pure pseudo--scalar model also (mainly vector meson) fields are 
also excited by the collective rotation in coordinate space. In the 
context of the collective approach to the three flavor model the 
collectively rotating vector meson fields are parametrized as
\be
\rho_\mu(\mbox{\boldmath $r$},t)=A(t)
\pmatrix{\mbox{\boldmath $\rho$}_\mu\cdot\mbox{\boldmath $\tau$}
+\omega_\mu & K^*_\mu \cr
K^{*\dag}_\mu & 0 \cr} A^{\dag}(t),
\label{rotvm}
\ee
where $A(t)$ again is an $SU(3)$ matrix. In addition to the static
vector meson profiles $\rho(\mbox{\boldmath $r$})$ and 
$\omega(\mbox{\boldmath $r$})$ (\ref{vmhedgehog}) three 
radial fields are induced by the spatial rotation \cite{Me87}
\be
\mbox{\boldmath $\rho$}_0
=\frac{1}{2g}\left[\xi_1(r)\mbox{\boldmath $\Omega$}
+\xi_2(r)(\hat{\mbox{\boldmath $r$}}\cdot\mbox{\boldmath $\Omega$})
\hat{\mbox{\boldmath $r$}}\right]\ , \quad
\omega_i=\frac{\Phi(r)}{2g}\epsilon_{ijk}\Omega_j\hat r_k\ .
\label{vmind1}
\ee
Suitable {\it ans\"atze} for the vector meson fields\footnote{In 
ref \cite{Ma88} a simplified treatment has been considered where 
the time component of the vector field, $S(r)$ represents the 
only non--trivial excitation due to the rotation into the direction 
of strangeness.} which are excited by the flavor rotation into 
the direction of strangeness are given by \cite{Pa91b}
\be
K^*_0=\frac{S(r)}{g}\Omega_K,\quad
K^*_i=\frac{1}{2g}\left[iE(r)\hat r_i +
{D(r)\over r}\epsilon_{ijk}\hat r_j \tau_k\right]\Omega_K.
\label{vmind2}
\ee
Again the flavor decomposition of the angular velocity matrix 
$\sum_{a=1}^8\Omega_a\lambda^a$, as in eqs (\ref{defomega}) 
and (\ref{kaonansatz}), has been employed. The parametrization 
(\ref{kaonind}) for the pseudo--scalar fields has to be augmented 
by an $\eta$ field component
\be
Z(\mbox{\boldmath $r$})=\pmatrix{\mbox{\large $\eta$}_T 
(\mbox{\boldmath $r$}) &
{| \atop |}&\hspace{-10pt}K(\mbox{\boldmath $r$}) \cr
-----\hspace{-8pt}&-&\hspace{-10pt}---\cr
\ K^{\dag}(\mbox{\boldmath $r$})&|&\hspace{-8pt} 0\cr} \ .
\label{kaonetaind}
\ee
The pseudo--scalar nature of the $\eta$ meson requires the 
{\it ansatz} $\eta_T=\eta(r)\hat{\mbox{\boldmath $r$}}\cdot
\mbox{\boldmath $\Omega$}/f_\pi$. The kaon field is again given 
by (\ref{kaonind}). As explained in the context of the pseudo--scalar 
model (section \ref{sec_quant}) those terms in the effective action, 
which are linear in the time derivative, provide the sources for these 
induced radial functions. These sources are functions of the classical 
fields (\ref{vmhedgehog}). The variational treatment of the spatial 
moment of inertia, $\alpha^2$ then yields the radial functions 
$\xi_1,\xi_2,\Phi$ and $\eta$, which are displayed in figure 
\ref{fi_vm}b. Similarly the radial functions $S,E,D$ and $W$ 
drawn in figure \ref{fi_vm}c, are obtained from the strange moment of 
inertia, $\beta^2$. In an attempt to freighten the reader, the 
explicit expressions for these moments of inertia are presented 
in appendix B.

The remaining coefficients $\gamma_S,\gamma_T$ and $\gamma_{TS}$ of 
the collective Hamiltonian (\ref{collham}) vanish for the action 
(\ref{Atotvm}).

For a given set of parameters of the effective action the coefficients 
of the collective Hamiltonian (\ref{collham}) may now be evaluated.
Subsequently its eigenvalues, which are identified as the baryon masses, 
are computed using the exact diagonalization procedure described in 
section \ref{sec_quant} and appendix A. The remaining freedom in 
choosing the parameters $\tilde h$, $\tilde g_{VV\phi}$ and 
$\kappa$ is used to optimize the predictions for the baryon mass
differences. This yields
\be
\tilde h=0.3\ , \quad \tilde g_{VV\phi}=1.3
\ ,\quad \kappa=1.2 \ ,
\label{parabf}
\ee
which actually is close to the central values discussed above. 
These parameters together with (\ref{vmpara1}) result in 
\be
E&=&1.628GeV,\ \alpha^2=5.144GeV^{-1},\ \beta^2=4.302GeV^{-1},
\nonumber \\
\gamma&=&1.755GeV,\ \alpha_1=2\beta_1=0.454
\label{numbf}
\ee
for the non--vanishing coefficients of the collective Hamiltonian 
(\ref{collham}). The induced meson fields $\eta,\ldots,D$ actually 
cause the moments of inertia to decrease compared to the case when 
these fields are omitted. While the contribution of the induced 
components to the spatial moment of inertia $\alpha^2$ is sizable,
it is almost negligible \cite{Pa91b} for the strange moment of 
inertia $\beta^2$.

In table \ref{ta_vmdiff} the mass differences obtained for the set 
(\ref{numbf}) are compared with the experimental data. Obviously 
a reasonable agreement with experimental data can be gained.
\begin{table}
\caption{\label{ta_vmdiff}\tenrm
The predictions for the mass differences of various baryons with 
respect to the nucleon in the vector meson model (VM) compared to 
the experimental data (all in MeV).}
~
\newline
\centerline{\tenrm\smalllineskip
\begin{tabular}{l| c c c c c c c}
& $\Lambda$  &  $\Sigma$   & $\Xi$      & $\Delta$   &
  $\Sigma^*$ &  $\Xi^*$    & $\Omega$   \\
\hline
VM    & 159 & 270 & 398 & 311 & 448 & 592 & 718 \\
Expt. & 177 & 254 & 379 & 293 & 446 & 591 & 733 \\
\end{tabular}}
\end{table}
These mass differences certainly represent an improvement over 
the pseudo--scalar model (see table \ref{ta_mdiffps}). In particular,
the $\frac{3}{2}^+$ baryons lie higher in energy (relative to the 
nucleon) than in the pseudo--scalar model. This is caused by the 
decrease of the spatial moment of inertia $\alpha^2$, which is due 
to quite an extended soliton, see figure \ref{fi_vm}. As will be
discussed in the proceeding section such large extension of the 
soliton is also reflected in the predictions for the baryon radii.
Simultaneously the larger soliton leads to a sizable symmetry breaking 
coefficient $\gamma$. Nevertheless the ratios (\ref{mratio}) of 
the mass differences are closer (1:0.70:0.81) to the leading order 
(in symmetry breaking) prediction than in the pseudo--scalar model. 
The reason being that the additional coefficients $\alpha_1$ and 
$\beta_1$ mitigate the effect of symmetry breaking. In case 
$\alpha_1$ and $\beta_1$ had turned out to have the opposite sign, 
the effects of the symmetry breaking terms would have added 
coherently to $\gamma$. 

Up to now isospin breaking effects have been ignored, {\it i.e.}
$y=0$ in eq (\ref{qmratio}). Substituting the above described 
meson field configurations into the relevant pieces of the action 
yields
\be
{\cal L}_{\triangle I}=
\left(\frac{2y\delta^\prime}{3f_\pi}\eta\sF+\ldots\right)
D_{3i}\Omega_i
-D_{38}\left(\frac{2y\delta^\prime}{\sqrt3}(1-\cF)+\ldots\right),
\label{isospinbreak}
\ee
where only typical two and three flavor expression involving 
respectively the collective expressions $D_{3i}\Omega_i$ and $D_{38}$ 
are shown\footnote{Obviously ${\cal L}_{\triangle I}$ yields
the strong interaction parts to the mass differences within an 
isospin multiplet when first computing the integrals over the 
radial functions and subsequently evaluating the matrix elements of 
the collective operators.}. For the nucleon the matrix elements of 
$D_{38}$ and $D_{83}$ only differ in sign. From figure \ref{fi_axial} 
one therefore deduces that $\langle N |D_{38}| N\rangle$ suffers a 
large reduction for actual amount of symmetry breaking. Hence the 
proper description of the neutron proton mass difference $\triangle_N$ 
requires the major contribution from the two flavor piece. This, of 
course, is expected because this mass difference is nothing but a 
consequence of different current up and down quark masses and should 
therefore already be present at the two flavor level. However, in a 
purely pseudo--scalar model one finds $\eta(r)\equiv0$. Hence there is no 
two flavor contribution to $\triangle_N$ in purely pseudo--scalar models. 
This conclusion also holds for the terms not explicitly shown in eq 
(\ref{isospinbreak}). The $\eta$ meson is only induced in the presence 
of vector mesons. Hence a two flavor contribution to $\triangle_N$ 
only occurs when the pseudo--scalar model is augmented by short range 
effects, which are imitated by the vector mesons, The actual 
calculation \cite{Ja89} in the vector meson model gives about 
1.3MeV for the two flavor contribution to $\triangle_N$. Three flavor 
effects may account for additional 0.5MeV. Hence the sum (1.8MeV) 
favorably agrees with the experimental value (1.3MeV) when the effects 
of the electromagnetic interaction \cite{Ga82}, 
$\triangle_N^{\rm e.m}=-0.76\pm0.30$MeV are taken into account. In 
section \ref{sec_two} we will see that fine--tuning $\triangle_N$ 
provides a handle to disentangle the matter and glue contributions to 
the axial singlet matrix element of the nucleon.

\bigskip

\begin{subsatz}
\label{sec_static}{\bf \hskip1cm Static Properties}
\addcontentsline{toc}{section}{\protect\ref{sec_static}
Static Properties}
\end{subsatz}
\stepcounter{section}

The appropriate definitions of the observables and form factors have 
already been provided in section \ref{sec_masses}. The covariant
expression for the currents associated with the action is presented 
in eq (2.12) of ref \cite{Pa92}. Rather than going into detail it 
is interesting to consider the expansion of the vector current in 
terms of the meson fields (\ref{su3chif},\ref{defpR})
\be
V_\mu^a=\frac{m_V^2}{g}{\rm tr}\left\{Q^a\left[
2\rho_\mu-i\left(\frac{2g^2f_\pi^2}{m_V^2}-1\right)
\Phi\partial_\mu\Phi\right]\right\}+\ldots .
\label{vcurexp}
\ee
The first term in this expression reflects exact VMD while the 
second one approximately vanishes\footnote{It vanishes identically 
when parameters satisfying the KSRF relation \cite{Ka66} are used.} 
for the parameters (\ref{vmpara1}). Hence the model contains the 
VMD concept, at least approximately \cite{Sch86}.

When substituting the parametrizations 
(\ref{vmhedgehog},\ref{vmind1},\ref{vmind2}), the formal structure of 
the currents remains unaltered compared to that of the pseudo--scalar 
model. Here, however, the radial functions $V_1(r)$, etc. in eq 
(\ref{spcurrps}) are given in terms of the profiles $F(r)$, 
$\omega(r)$, $G(r)$, $\xi_1(r),\ldots,D(r)$, whose computation has been
described in the preceding section. The explicit expressions, which are 
quite awkward, are given in appendix B of ref \cite{Pa92}. Only for 
the case of the axial singlet current (which in the pseudo--scalar model
is a total derivative (\ref{A0ps1})) conceptually important changes 
will appear. This will be considered at the pertinent occasion.

The numerical results for electromagnetic observables obtained in the 
vector meson model, which are computed in analogy to the pseudo--scalar 
model, are presented in table \ref{ta_emvm}.
\begin{table}
\caption{\label{ta_emvm}\tenrm
The electromagnetic properties of the baryons compared to the
experimental data. The predictions of the vector meson model are 
taken from ref \protect\cite{Pa92} and correspond to the parameter 
set (\protect\ref{vmpara1}) and (\protect\ref{parabf}).}
~
\newline
\centerline{\tenrm\smalllineskip
\begin{tabular}{c| c c | c c | c c}
 & \multicolumn{2}{c|}{$\mu_B({\rm n.m.})$} &
\multicolumn{2}{c|}{$r_M^2({\rm fm}^2)$} &
\multicolumn{2}{c}{$r_E^2({\rm fm}^2)$} \\
$B$ & VM & Expt. & VM & Expt. & VM & Expt. \\
\hline
$p$         & 2.36 & 2.79 & 0.94 & 0.74 & 1.20 & 0.74 \\
$n$         &-1.87 &-1.91 & 0.94 & 0.77 &-0.15 &-0.12 \\
$\Lambda$   &-0.60 &-0.61 & 0.78 & ---  &-0.06 & ---  \\
$\Sigma^+$  & 2.41 & 2.42 & 0.96 & ---  & 1.20 & ---  \\
$\Sigma^0$  & 0.66 & ---  & 0.86 & ---  &-0.01 & ---  \\
$\Sigma^-$  &-1.10 &-1.16 & 1.07 & ---  &-1.21 & ---  \\
$\Xi^0$     &-1.96 &-1.25 & 0.90 & ---  &-0.10 & ---  \\
$\Xi^-$     &-0.84 &-0.69 & 0.84 & ---  &-1.21 & ---  \\
$\Sigma^0\rightarrow\Lambda$
            &-1.74 &-1.61 & 0.97 & ---  & ---  & ---  \\
\end{tabular} }
\end{table}
Like in the case of the mass differences an improvement over the 
pseudo--scalar model is obtained. Most apparently, the absolute values 
of the magnetic moments have increased significantly as a consequence 
of the more extended soliton. Accept for the hyperon $\Xi^0$ this 
represents a desired result. The prediction is about 40\% off the 
experimental value in this special case. This is again linked to the 
fact that only minor deviations from the $SU(3)$ relations 
(\ref{su3mag}) occur.  It has already been discussed in chapter 
\ref{chap_exten} that the rigid rotator approach, which has been 
applied here, cannot accommodate the experimentally observed 
deviations. However, the vector meson model correctly describes the 
ratio $\mu_{\Xi^-}/\mu_\Lambda$, which was not possible with the 
pseudo--scalar model, even when the rigid rotator approach was modified. 
Statements on this ratio are apparently model dependent, in particular
because the $SU(3)$ relations (\ref{su3mag}) give no information. One 
further recognizes much larger magnetic and isoscalar electric radii 
than in the pseudo--scalar model. Of course, this is partially due to the 
bigger soliton but mainly this result reflects the effects of VMD as 
discussed in the introduction to this chapter, see eq (\ref{changeR0}). 
In the case of the electric radius of the proton these effects are too 
strong and some improvement appears to be needed. This can also be 
observed from the full momentum dependence of the proton electric form 
factor (see {\it e.g.}  figure 3a of ref \cite{Pa92}), which drops off 
more rapidly than the dipole fit, $(1+Q^2/0.71{\rm GeV}^2)^{-2}$. On 
the other hand the momentum dependence of the neutron electric form 
factor agrees reasonably well with the empirical data.

When adopting the suitable generator (\ref{strgen}) the vector currents,
which have provided the electromagnetic properties, may also be 
employed to compute the strangeness form factors in the nucleon. 
The vector meson model predicts \cite{Pa92}
\be
\mu_S=-0.05 \ , \qquad
r_S^2=0.05 {\rm fm}^2 \ .
\label{stresvm}
\ee
The absolute value of the strange magnetic moment $\mu_S$ is even 
smaller than in the pseudo--scalar model (\ref{stresps}), which already 
represented quite a reduction from Jaffe's result ($-0.31\pm0.09$)
\cite{Jaf89}. The latter was obtained by using a three pole vector meson 
fit to dispersion relations suggested by H\"ohler et al. \cite{Ho74}. 
In ref \cite{Pa92} this has been explained as a consequence of flavor 
symmetry breaking because the use of an $SU(3)$ symmetric nucleon 
wave--function (\ref{su3dexap}) resulted in $\mu_S\approx -0.54$. 
Amusingly, the prediction for the squared strange radius $r_S^2$ has 
the opposite sign as the one in the pseudo--scalar model (\ref{stresps}). 
It has been conjectured that the positiveness of $r_s^2$ is linked 
to VMD. The argument is that a one pole VMD approach, together with 
$\omega-\phi$ mixing, leads to $r_s^2=0.01{\rm fm}^2$ \cite{Pa91}; 
which, however, is much smaller in magnitude. Also Jaffe \cite{Jaf89} 
obtained a positive value $r_s^2=0.16\pm0.06{\rm fm}^2$. In table 
\ref{ta_strange} the predictions of various models for $\mu_S$ and 
$r_s^2$ are summarized.
\begin{table}
\caption{\label{ta_strange}\tenrm
Predictions for the strange magnetic moment ${\Ss \mu_S}$ and the 
strangeness radius ${\Ss r_S^2}$:
a. pole fit \protect\cite{Jaf89},
b. ${\Ss SU(3)}$ symmetric treatment of kaon loops 
\protect\cite{Mu93},
c. kaon loop calculation \protect\cite{Co93},
d. single pole VMD \protect\cite{Pa91},
e. VMD combined with kaon loops \protect\cite{Co93},
f. pseudo--scalar Skyrme model (\protect\ref{stresps}),
g. vector meson Skyrme model (\protect\ref{stresvm}),
h. NJL soliton model \protect\cite{We95a}. }
~
\newline
\centerline{\tenrm\smalllineskip
\begin{tabular}{c| c c c c c c c c}
& a. & b. & c. & d. & e. & f. & g. & h. \\
\hline
$\mu_S$(n.m.) & -0.22...-0.40 & -0.30...-0.40& 
-0.24...-0.32& -0.003 & -0.24...-0.32& -0.13 &
-0.05 & -0.05...0.25\\
$r^2_S$(fm$^{\mbox{\rm\tiny $2$}}$)& 
0.07...0.21&-0.03&-0.02...-0.03&
0.01&-0.04...-0.05&-0.10&0.05&-0.15...-0.25\\
\end{tabular} }
\end{table}
Obviously these predictions are quite diverse and precision 
experiments should soon provide some resolution.

When substituting the meson configurations of section
\ref{sec_inertia} into the covariant expression for the axial current,
one obtains for the octet part ($a=1,\ldots,8$) the general structure 
\be
A_i^a&=&{\cal A}^{(1)}_{ik}D_{ak}
+{\cal A}^{(2)}_{ik}d_{k\alpha\beta}D_{a\alpha}\Omega_\beta
+{\cal A}^{(3)}_{ik}D_{ak}\Omega_k
\nonumber \\ && \hspace{1cm}
+{\cal A}^{(4)}_{ik}D_{ak}\left(D_{88}-1\right)
+{\cal A}^{(5)}_{ik}d_{k\alpha\beta}D_{a\alpha}D_{8\beta}\ ,
\label{axcurrvm}
\ee
with 
${\cal A}^{(l)}_{ik}=A^{(l)}(r)\delta_{ik}+B^{(l)}(r)\hat r_i\hat r_j$.
The radial functions $A^{(l)}$ and $B^{(l)}$ may be extracted from 
appendix B of ref \cite{Pa92}. It is important to remark that 
${\cal A}^{(3)}_{ik}$ is solely due to the anomalous part of the 
action (\ref{anom2}) and would vanish identically if no vector meson
fields were present. This also causes the axial singlet current
to be different from a total derivative
\be
A_i^0=2\sqrt3 f_\pi\partial_i \eta^0 + \tilde A_i^0
\qquad {\rm with} \qquad
\tilde A_i^0=\frac{2}{\sqrt3}{\cal A}^{(3)}_{ik}\Omega_k \ .
\label{axsingl}
\ee 
Here $\eta^0$ refers to the singlet component of the pseudo--scalar field 
$\Phi$ in the decomposition $U={\rm exp}(i(\Phi+\eta^0/\sqrt3f_\pi))$, 
with ${\rm tr}\Phi=0$. Now a non--trivial result may 
be obtained for the axial singel matrix element $H(q^2)$ defined in 
eq (\ref{defsing}). Obviously, the vector mesons provide a conceptual 
difference over pure pseudo--scalar models. Actually the anomaly 
equation $-\partial_iA_i^0=f_\eta m_\eta^2 \eta$ is identical to 
the equation of motion for the profile function $\eta(r)$ obtained 
from varying the spatial moment of inertia $\alpha^2$. This is  
another manifestation of the fact that the $\eta$ field is only 
excited when vector meson fields (more generally short range 
effects) are incorporated because ${\cal A}^{(3)}$ is the source term. 
Apparently, the non--trivial result for $H(q^2)$ is subject to 
the semi--classical treatment (\ref{vmind1}), while $H(q^2)$ vanishes 
for the static configuration (\ref{vmhedgehog}). A similar feature was 
already encountered when studying the neutron proton mass difference. 
This strongly supports the above conjecture that the issue of the 
proton spin puzzle is closely connected to this mass difference.

\begin{figure}[t]
\centerline{\hskip -1.0cm
\epsfig{figure=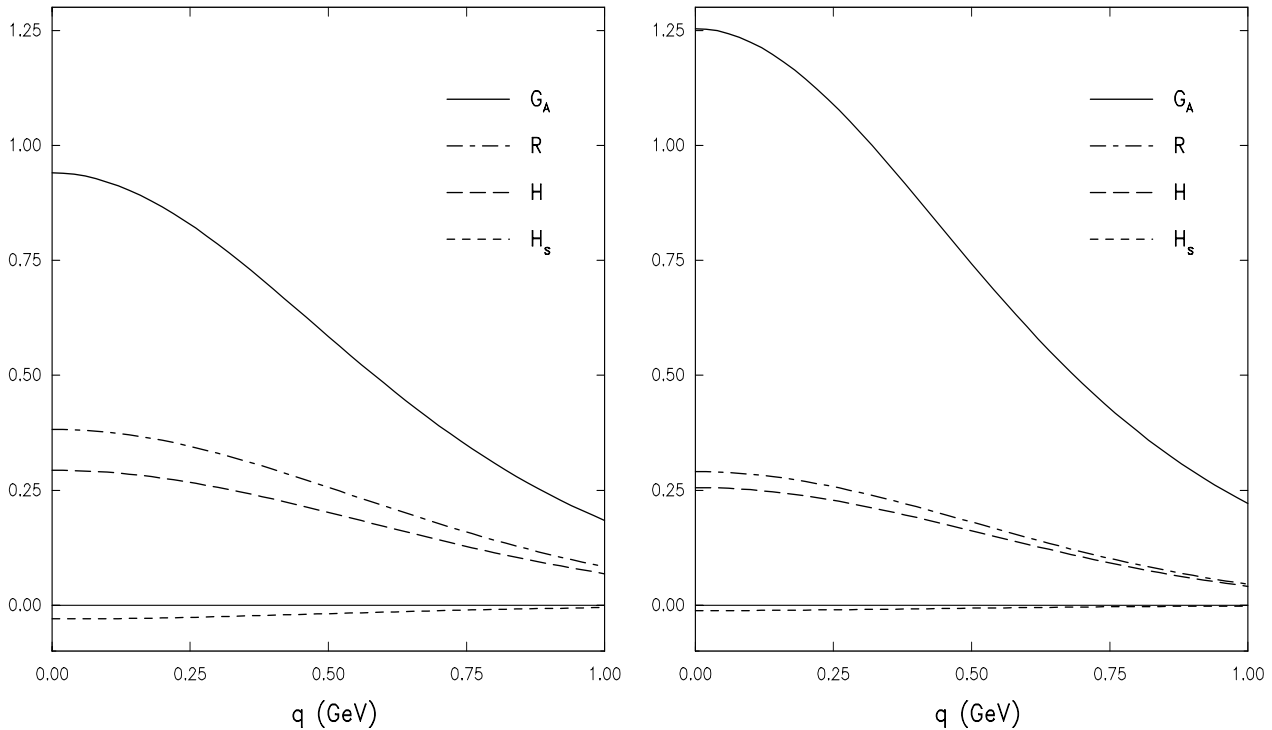,height=9.0cm,width=16.0cm}}
\caption{\label{fi_axialvm}\tenrm
The momentum dependence of the nucleon matrix elements of the 
axial--vector current, which are associated with the strangeness 
conserving components. For the definition of the displayed 
quantities see eqs (\protect\ref{paxmat}) and (\protect\ref{naxff}).
Left panel (\protect\ref{parabf}), right panel (\protect\ref{paraax}). 
In both cases the parameters of the non--anomalous part of the action 
are given by (\protect\ref{vmpara1}).}
\end{figure}

Due to isospin invariance the Fourier transforms of the components 
$2A_3^3$, $2\sqrt3A_3^8$ and $3\tilde A_3^0$ respectively provide the 
form factors $G_A(q^2)$, $R(Q^2)$ and $H(q^2)$, of the nucleon 
after computing the matrix elements of the collective operators between 
the nucleon eigenstate of the collective Hamiltonian. These form 
factors are related to the individual quark contributions defined in eq 
(\ref{paxmat}) via
\be
G_A(q^2)&=&H_u(q^2)-H_d(q^2)\ , 
\nonumber \\
R(q^2)&=&H_u(q^2)+H_d(q^2)-2H_s(q^2)\ ,
\nonumber \\
H(q^2)&=&H_u(q^2)+H_d(q^2)+H_s(q^2)\ .
\label{naxff}
\ee
The predicted momentum dependencies\footnote{Similar studies for the 
two flavor reduction of this models are presented in ref \cite{Be90}.} 
of these form factors as well as the strange quark contribution to 
the axial current matrix element of the proton, $H_s(q^2)$ are 
shown in figure \ref{fi_axialvm}.
It is a long--standing problem of mesonic soliton models that the 
axial charge of the nucleon $g_A=G_A(q^2=0)$ is underestimated when 
parameters are adopted which are suitable for the description of 
the baryon spectrum (see, however, ref \cite{Ja88a}). It is hence 
not surprising that for the set (\ref{parabf}) $g_A$ is predicted 
too small by about 30\%
\be
g_A=0.93\ , \quad R(0)=0.38 \ , \quad H(0)=0.29.
\label{axpred}
\ee
Amusingly these results approximately satisfy the $SU(3)$ relation
$R(0)=3g_A/7$. However, it is important to establish that the 
isosinglet form factors $R(0)$ and $H(0)$ show almost no variation 
within the allowed parameter space of $\tilde h=$, $\tilde g_{VV\phi}$ 
and $\kappa$ \cite{Jo90}, while $g_A$ exhibits quite a strong parameter 
dependence. Stated otherwise, the above given values for $R(0)$ and 
$H(0)$ represent firm predictions of the vector meson model. 
Furthermore these form factors are dominated by the contribution of 
the ${\cal A}^{(3)}_{ik}$ in eq (\ref{axcurrvm}) in contrast to $G_A$, 
which is dominated by ${\cal A}^{(1)}_{ik}$. Hence the above quoted 
$SU(3)$ relation between $R(0)$ and $g_A$ is purely accidental for the
parameters (\ref{parabf}). The reason actually is the too small 
prediction for $g_A$. The correlations between $R(0)$ and $H(0)$ 
furthermore cause the strange contribution to axial current of the 
nucleon, $H_s(q^2)$ always to be tiny as can also be seen from figure 
\ref{fi_axialvm}. In order to confirm these statements a parameter set 
\be
\tilde h=-0.1\ , \quad \tilde g_{VV\phi}=1.4
\ ,\quad \kappa=1.0 \ ,
\label{paraax}
\ee
which properly reproduces $g_A$ has been considered. This set is 
neither in the allowed parameter space nor does it reasonably describe 
the baryon spectrum. Nevertheless it is helpful for the discussion 
of axial properties of the nucleon. As can be observed from the right 
panel in figure \ref{fi_axialvm} the isoscalar form factors are indeed 
strongly correlated and neither of them scales with $G_A$. For this set
of parameters a significant deviation from the $SU(3)$ relation 
between $g_A$ and $R(0)$ is obtained, namely $R(0)=0.23g_A$.
 
Of course, it is very interesting to put these results into perspective
with the EMC, SLAC and SMC measurements \cite{As88,An93,Ad94} of the 
axial form factors of the nucleon. Although the (firm) prediction 
$H(0)\approx0.29$ favorably agrees with the value extracted from these 
experiments ($0.24\pm0.09$ \cite{El93}) neither $R(0)$ nor $H(0)$ are 
suited to carry out this comparison because the present model is 
formulated to incorporate flavor symmetry breaking effects but no 
independent information on $H_s(0)$ is currently available. Rather 
the vector meson model prediction for the structure function
$\Gamma_1^p(q^2)$ (\ref{EMCcomb}), which depends on the individual 
quark pieces $H_{u,d,s}(0)$, should be compared with the ``world
average" $0.129\pm0.010$ at $q^2=(10.7{\rm GeV})^2$. Using the 
$q^2$--dependent coefficients $C_{ns}(q^2)$ and $C_{s}(q^2)$ as 
given in ref \cite{La94} yields\footnote{The variation of 
$\Gamma_1^p(q^2)$ with the scale $q^2$ is quite moderate as can 
be seen from table 1 of ref \cite{El93}.}
\be
\Gamma_1^p\left(q^2=10.7{\rm GeV})^2\right)
=\cases{0.11& {\rm set} (\protect\ref{parabf})\cr
0.13& {\rm set} (\protect\ref{paraax})} .
\label{Ga1pred}
\ee 
Obviously, the vector meson model excellently explains what commonly 
is quoted as the proton spin puzzle. Once the model parameters are 
adjusted to give the correct value for $g_A$, the integral of the 
polarized structure function is exactly reproduced. The na{\"\i}vely
expected matrix element of the axial singlet current, $H(0)$=1, 
together with $g_A=1.25$ and the assumption of flavor symmetry yields 
$0.22$, which is much bigger.

In ref \cite{Pa92} the matrix elements describing the semi--leptonic 
hyperon decays have also been computed. As in the pseudo--scalar model 
these matrix elements are found to be consistent with the 
empirical data of the Cabibbo scheme after accounting for the too 
small value of the nucleon axial charge $g_A$. The 
deviation of these strangeness changing matrix elements from the 
$SU(3)$ symmetry relations is very moderate compared to the change 
which may be observed for the matrix element of strangeness conserving 
component $R$.

This section can be summarized by stating that for many aspects the 
vector meson model represents an improvement over the pseudo--scalar 
model, at least for the treatment within the rigid rotator approach. 
The explicit appearance of the vector mesons incorporates short 
range effects, which are not present in the pseudo--scalar model. As a 
consequence the two flavor contribution to the neutron proton mass 
difference as well as the axial singlet current matrix element are 
different from zero. Especially, since the model nicely describes the 
measurements on polarized lepton--nucleon scattering 
\cite{As88,An93,Ad94} one wonders whether more information on the 
proton spin structure can be extracted from this model. Related 
studies will be the major issue of the following section.

\bigskip

\begin{subsatz}
\label{sec_two}{\bf \hskip1cm Two Component Approach to the 
Proton Spin Puzzle}
\addcontentsline{toc}{section}{\protect\ref{sec_two}
Two Component Approach to the Proton Spin Puzzle}
\end{subsatz}
\stepcounter{section}

Soon after the advent of the proton spin puzzle, explanations
\cite{Al88} for the smallness of $H(0)$ 
were attempted via the axial anomaly of QCD
\be
\partial^\mu A_\mu^0=2i\sum_{i=u,d,s}m_i{\bar q}_i\gamma_5 q_i
+\partial^\mu K_\mu \ ,
\label{QCDan1}
\ee
where
\be
K_\mu=-\frac{3ig_{QCD}^2}{4\pi^2}\epsilon_{\mu\nu\rho\sigma}
{\rm tr}\left\{A^\nu\partial^\rho A^\sigma-
\frac{2ig_{QCD}}{3}A^\nu A^\rho A^\sigma\right\}
\label{QCDan2}
\ee
is the Chern--Simons current of QCD. Here $A_\mu$ refers to the 
gauge fields of QCD. It has been suggested to associate 
the matrix element of $A_\mu^0-K_\mu$ as the ordinary ``matter" 
contribution to the axial singlet current. Hence one would suspect 
$\langle A_3^0-K_3\rangle\approx1$. Essentially it is assumed 
that quark spin is cancelled by the gluonic contribution leading 
to a small $H(0)$. However, this treatment is suspicious because 
the matrix element of $K_\mu$ is not gauge invariant \cite{Ma91},
in contrast to $\partial^\mu K_\mu$. To nevertheless firmly 
disentangle the ``matter" and ``glue" contributions to $H(0)$, it has 
been suggested to take the matrix elements of eq (\ref{QCDan1}) and use 
pole saturation \cite{Ve89,Sch90} for the evaluation of the $RHS$
\be
\langle P(\mbox{\boldmath $p$})|
\partial^\mu A_\mu^0|P(\mbox{\boldmath $p$})\rangle&=&
2i \langle P(\mbox{\boldmath $p$})
|\sum_{i=u,d,s}m_i{\bar q}_i\gamma_5 q_i
|P(\mbox{\boldmath $p$})\rangle
+\langle P(\mbox{\boldmath $p$})|\partial^\mu K_\mu
|P(\mbox{\boldmath $p$})\rangle
\nonumber \\  &=& 2M_N H(0)
{\overline u}(\mbox{\boldmath $p$})\gamma_5
u(\mbox{\boldmath $p$})
\label{QCDan3}
\ee
The ``matter" and ``glue" components of $H(q^2)$ are obtained by 
identifying the appropriate terms on the $RHS$. The ``matter" and 
``glue" pieces correspond to coupling constants $g_{\eta^\prime NN}$ 
of the singlet pseudo--scalar field $\eta^\prime$ and $g_{GNN}$ of an 
effective gluon field $G=\partial_\mu K^\mu$ to the nucleon. In this 
section the Skyrme model analogue \cite{Sch90a} to this investigation, 
which is known as the two component approach to the proton 
spin\footnote{Ref. \cite{Sch92c} provides a brief review and a list 
of references on the gauge invariant two component decomposition to 
the proton spin.}, will be described. For this analysis it is important 
to note the effects of the $\eta$ mesons on baryon properties are 
commonly very moderate \cite{Me89}. There seems to be only one exception, 
the strong interaction part of the neutron proton mass difference 
$\triangle_N$ \cite{Ja89}.

As a first step the anomaly (\ref{QCDan1}) has to be incorporated 
into the effective Lagrangian \cite{Ro80}
\be
{\cal L}_{ax-an}=\frac{1}{\kappa}G^2+\frac{i}{12}G
\left({\rm ln\ det}U-{\rm ln\ det}U^{\dag}\right)
+\ldots
\label{u1aeff}
\ee
where the ellipses indicate flavor symmetry breaking terms associated 
with $G$. These terms are not specified here. The interested reader 
may consult ref \cite{SSW93} for details. In the context of effective 
meson Lagrangians the pseudo--scalar gluon field $G$ is the equivalent 
of the scalar glueball field ${\rm exp}(\sigma)$, which was introduced 
in section \ref{sec_mixing} to imitate the scale anomaly of QCD.
Since there is no kinetic term for $G$, it may be eliminated by 
its equation of motion. This gives rise to $\eta$ meson masses
\be
m^2_{\eta,\eta^\prime}=\frac{1}{4f_\pi^2}\left\{
8\delta^\prime(1+x)+\frac{\kappa}{6}
\pm\sqrt{\left(8\delta^\prime(1-x)+\frac{\kappa}{18}\right)^2
+\frac{4}{9}\kappa^2}\right\}+\ldots \ ,
\label{meta}
\ee
thereby relating the constant $\kappa$ to physical quantities. The 
coupling of the pseudo--scalar glueball field to the nucleon has to be 
put in explicitly \cite{Ve89,Sch90,Ko93}. In the soliton picture this 
is achieved by adding the chirally invariant term 
\be
{\cal L}_{GNN}=\frac{2t}{m_{\eta^\prime}^4}
\tilde A_\mu^0\partial^\mu G
\label{Lgnn}
\ee
to the effective Lagrangian. Finally one may consider a deviation 
from the nonet structure of the pseudo--scalar and vector meson fields 
by introducing the additional parameter $s$ into eq (\ref{axsingl})
\be
A_i^0=2\sqrt3 f_\pi\partial_i \eta^0 + s \tilde A_i^0 \ .
\label{fudge}
\ee
This alters the prediction for the singlet matrix element
\be
H(0)=(0.30\pm0.03)s \ ,
\label{H0s}
\ee
wherein an estimate for the ``theoretical uncertainty" was made. 

Sandwiching the anomaly equation, which essentially is the 
equation of motion of the $\eta^\prime$ field
\be
\left(\partial_\mu\partial^\mu+m_{\eta^\prime}^2\right)\eta^\prime
=\frac{s-t}{2\sqrt3 f_\pi}\partial^\mu\tilde A_\mu^0+\ldots 
\label{etq0eqm}
\ee
between nucleon states, gives the Goldberg--Treiman relation for 
the axial singlet current \cite{Ve89}
\be
g_{\eta^\prime NN}=\frac{s-t}{s}\frac{M_p}{\sqrt3 f_\pi} H(0)\ .
\label{getaNN}
\ee
From eq (\ref{Lgnn}) the gluon coupling constant is extracted 
to be
\be
g_{GNN}=\frac{4tM_N}{sm_{\eta^\prime}^4}H(0)=\frac{t}{t-s}
\frac{2\sqrt3 F_\pi}{m_{\eta^\prime}^4}g_{\eta^\prime NN}\ .
\label{gGNN}
\ee
Solving these two equations for $H(0)$ finally gives the 
``matter"+``glue" decomposition of the axial singlet matrix element
\be
H(0)=\frac{\sqrt3 F_\pi}{2M_p}\left(g_{\eta^\prime NN}
-\frac{m_{\eta^\prime}^4}{2\sqrt3 F_\pi}g_{GNN}\right)
={\rm ``matter"}+{\rm ``glue"} \ .
\label{twocomp}
\ee

The incorporation of the axial anomaly of QCD into the effective 
meson theory and abandoning the nonet structure has introduced two 
dimensionless parameters, $s$ and $t$.  For a given value of $H(0)$ 
the model determines the ``fudge factor" $s$, while the strong 
interaction piece of the neutron proton mass difference,
$\triangle_N\approx2{\rm MeV}$ can be employed to fix $t$. In order 
to explain the latter point it is illustrative to once again consider 
a typical contribution to $\triangle_N$, which has already been
indicated in eq (\ref{isospinbreak}),
\be
\triangle_N=\frac{2y\delta^\prime}{3\alpha^2}
\int d^3r \ {\rm sin}F(r)\eta_T(r)+\ldots ,
\label{mnmp}
\ee
where $\eta_T(r)$ refers to profile function of the non--strange 
combination of the $\eta$ fields, $\eta_T=\eta_T(r)
\hat{\mbox{\boldmath $r$}}\cdot\mbox{\boldmath $\Omega$}$.
In ref \cite{SSW93} a detailed analysis of isospin breaking in the 
meson sector has been performed yielding $y\approx0.42$. This 
analysis also includes second order symmetry breaking terms like 
${\rm tr}({\cal M}U^{\dag}{\cal M}^{\dag}U)$, which in the soliton 
sectors give rise to non--vanishing coefficients $\Gamma_T$, $\Gamma_S$ 
and $\Gamma_{TS}$ in (\ref{collham}). From the equation of motion
(\ref{etq0eqm}) it is apparent that the radial function $\eta_T(r)$, 
which is obtained by extremizing the spatial moment of inertia 
$\alpha^2$, strongly depends on the parameters $s$ and $t$.
It is then obvious that for a given value of $H(0)$ the parameters $s$ 
and $t$ can be determined from eq (\ref{H0s}) and reproducing the 
strong interaction part of the neutron proton mass difference. For $s$ 
of the order unity, $t$ varies in the range $-0.5\ldots-1.5$, the 
uncertainty being caused by the one in $\triangle_N=(2.0\pm0.3)$MeV 
\cite{Ga82}. In any event, $g_{\eta^\prime NN}$ and $g_{GNN}$ have 
opposite signs. For two sets of parameters\footnote{The detailed 
analysis \cite{SSW93} of the symmetry breaking leads to minor changes 
in the parameters (\ref{vmpara1}). A more recent investigation 
including electroweak interactions favors a smaller value $x\approx25$ 
\cite{Ha95}.} the resulting two component decomposition is given in 
table \ref{ta_twocomp}.
\begin{table}
\caption{\label{ta_twocomp}\tenrm
The decomposition of the axial singlet matrix element
${\Ss H(0)}$ into ``matter" and ``glue" contributions for two 
parameter sets. I: ${\Ss \tilde h=0.36}$,
${\Ss \tilde g_{VV\Phi}=1.88}$, ${\Ss \kappa =1.0}$ and
${\Ss x=28}$.  II: ${\Ss \tilde h=0.4}$,
${\Ss \tilde g_{VV\Phi}=1.9}$, ${\Ss \kappa =1.0}$ and
${\Ss x=31.5}$. Results are taken from refs 
\protect\cite{SSW93,We93b}.}
~
\newline
\vspace{-0.6cm}
\centerline{\tenrm\smalllineskip
\begin{tabular}{l|c c|c c}
&\multicolumn{2}{c|}{\rm Set I}
&\multicolumn{2}{c}{\rm Set II} \\
\hline
$H(0)$ & {\rm ``matter"} & {\rm ``glue"}
& {\rm ``matter"} & {\rm ``glue"} \\
\hline
~0.0&0.54$\pmt$0.26&-0.54$\pmt$0.26&0.32$\pmt$0.26&-0.32$\pmt$0.26 \\
~0.1&0.57$\pmt$0.30&-0.47$\pmt$0.30&0.31$\pmt$0.30&-0.21$\pmt$0.27 \\
~0.2&0.58$\pmt$0.30&-0.38$\pmt$0.30&0.31$\pmt$0.26&-0.11$\pmt$0.28 \\
~0.3&0.58$\pmt$0.30&-0.28$\pmt$0.30&0.31$\pmt$0.30&-0.01$\pmt$0.28 \\
~0.4&0.58$\pmt$0.28&-0.18$\pmt$0.28&0.32$\pmt$0.27&~0.08$\pmt$0.27 \\
\end{tabular}}
\end{table}
The important message to be extracted from these investigations is 
that soliton models for baryons indeed indicate that the 
smallness of the singlet matrix element is due to a cancellation 
mechanism between quark and gluon contribution. However, the pure 
matter component of $H(0)$ is still significantly smaller than unity, 
in contrast to the original speculation. Unfortunately, the predictions 
for the individual components of $H(0)$ possess uncertainties, which 
are quite large. Their accuracy could be improved if a more accurate 
value of the photon exchange contribution to $M_n-M_p$ were available.

Let us finish this chapter by mentioning that a similar 
analysis of $\triangle_N$ can be used to study the Gottfried sum 
rule 
\be
S_G=\frac{1}{3}\int_0^1 dx\left\{u(x)+
{\bar u}(x)-d(x)-{\bar d}(x)\right\}
\label{Gsum}
\ee
in soliton models. Here $u(x)$ denotes the up--quark distribution
function in the nucleon, ${\bar u}(x)$ that of the anti up--quark, 
etc.. Assuming that all antiquarks are equally distributed predicts,
$S_G=1/3$ \cite{Go67}. The evaluation based on the recent NMC data, 
however, gives a smaller value $S_G=0.240\pm0.016$ \cite{NMC91}. 
Sandwiching the symmetry breaking term (\ref{QCDmass}) between 
nucleon states and assuming isospin invariance yields 
\be
S_G=\frac{\triangle_N}{3(m_d-m_u)} 
\label{Gsum1}
\ee
to leading order symmetry breaking. In ref \cite{Wa93a} the effects 
of higher order corrections to this relation have been estimated to 
be negligibly small in Skyrme type models. As all available data on 
the difference $m_d-m_u$ are larger than $\triangle_N$ (see {\it e.g.} 
ref \cite{PDG94}), the soliton models provide an interesting 
explanation of the unexpected smallness of $S_G$.

\vskip1cm
\begin{satz}
\label{chap_bound}{\large \bf \hskip1cm 
The Bound State Approach}
\addcontentsline{toc}{chapter}{\protect\ref{chap_bound}
The Bound State Approach}
\end{satz}
\stepcounter{chapter}

In the previous chapters several treatments of the three flavor
soliton have been discussed. These have been based on the assumption
that the time--dependent solution to the Euler Lagrange equations
is reasonably approximated by elevating the coordinates, which 
parametrize the flavor orientation, to time dependent quantities. Of 
course, such treatments are motivated by considering the flavor 
symmetry to be approximately realized, allowing for large amplitude 
fluctuations in the direction of symmetry breaking. Subsequently 
the symmetry breaking effects are treated within the space of these
collective coordinates. The exact diagonalization of the resulting 
collective Hamiltonian is possible according to the Yabu--Ando approach 
\cite{Ya88} and yields the baryon masses and wave--functions. Adopting, 
however, the opposite point of view that only small amplitude 
fluctuations are permitted, implying that the restoring forces exerted
by the symmetry breaking are sizable, leads to the treatment, which has 
become known as the bound state approach. The reason for this notion
is that hyperons are constructed out of the soliton and a kaon mode, 
which is bound in the background field of the soliton. This treatment 
has been initiated by Callan and Klebanov \cite{Ca85} and later been 
employed in many aspects as will be pointed out in the present chapter. 
Actually this approach is comparable to the old compound hypothesis, 
where hyperons are considered as molecules consisting of a nucleon 
and a kaon \cite{Go56}.

\bigskip

\begin{subsatz}
\label{sec_fluc}{\bf \hskip1cm
Small Fluctuations off the Soliton and the Kaon Bound State}
\addcontentsline{toc}{section}{\protect\ref{sec_fluc}
Small Fluctuations off the Soliton and the Kaon Bound State}
\end{subsatz}
\stepcounter{section}

The time dependent, small amplitude fluctuations are conveniently 
described by adopting the parametrization \cite{Ca85,Oh91}
\be
U(\mbox{\boldmath $r$},t)=\xi_\pi(\mbox{\boldmath $r$})
U_K(\mbox{\boldmath $r$},t)\xi_\pi(\mbox{\boldmath $r$})\ , \
U_K(\mbox{\boldmath $r$},t)={\rm exp}\left\{\frac{i\sqrt2}{f_K}
\pmatrix{\mbox{\large $0$} &
{| \atop |}&\hspace{-10pt}K(\mbox{\boldmath $r$},t) \cr
-----\hspace{-8pt}&-&\hspace{-10pt}---\cr
\ K^{\dag}(\mbox{\boldmath $r$},t)&|&\hspace{-8pt} 0\cr}\right\}\ ,
\label{bsansatz}
\ee
where $\xi_\pi$ represents the hedgehog configuration
(\ref{vmhedgehog}).  The general decomposition of the kaon field 
reads
\be
K(\mbox{\boldmath $r$},t)=\int \frac{d\omega}{2\pi}\sum_{GML}
{\rm e}^{-i\omega t}k_{GL}(r,\omega)
{\cal Y}_{GML}(\hat{\mbox{\boldmath $r$}})\ ,
\label{kaondecom}
\ee
where ${\cal Y}_{GML}$ are spinor spherical harmonics. These arise
by coupling the orbital angular momentum $\mbox{\boldmath $L$}$ 
together with isospin $t=1/2$ to the grand spin, $G$. Since the 
soliton background, $\xi_\pi$ has vanishing grand spin the radial 
functions $k_{GL}$ are independent of the grand spin projection 
$M$ and channels with different grand spins, $G$ and $G^\prime$
decouple.

Although the general case (\ref{kaondecom}) is of physical 
relevance ({\it cf.} section \ref{sec_scattering}) the distinct 
channel $L=1$, $G=1/2$ is of special interest for examining the 
spectrum and static properties of the low--lying baryons. This 
channel contains a bound state, evolving from the zero--mode, 
which, in the flavor symmetric case, is associated with the rotation 
into the strange flavor direction. {\it I.e.} this bound state is 
the ``would--be" Goldstone boson of the flavor transformations. A 
suitable {\it ansatz} for this $P$--wave mode reads
\be
K_P(\mbox{\boldmath $r$},t)=\int \frac{d\omega}{2\pi}\
{\rm e}^{-i\omega t}k_P(r,\omega)
\hat{\mbox{\boldmath $r$}}\cdot\mbox{\boldmath $\tau$}
\pmatrix{a_1(\omega) \cr a_2(\omega)\cr}\ .
\label{pwansatz}
\ee
The spectral functions $a_i(\omega)$ become creation 
and annihilation operators in the process of quantization.

\begin{figure}[t]
\centerline{\hskip -2.0cm
\epsfig{figure=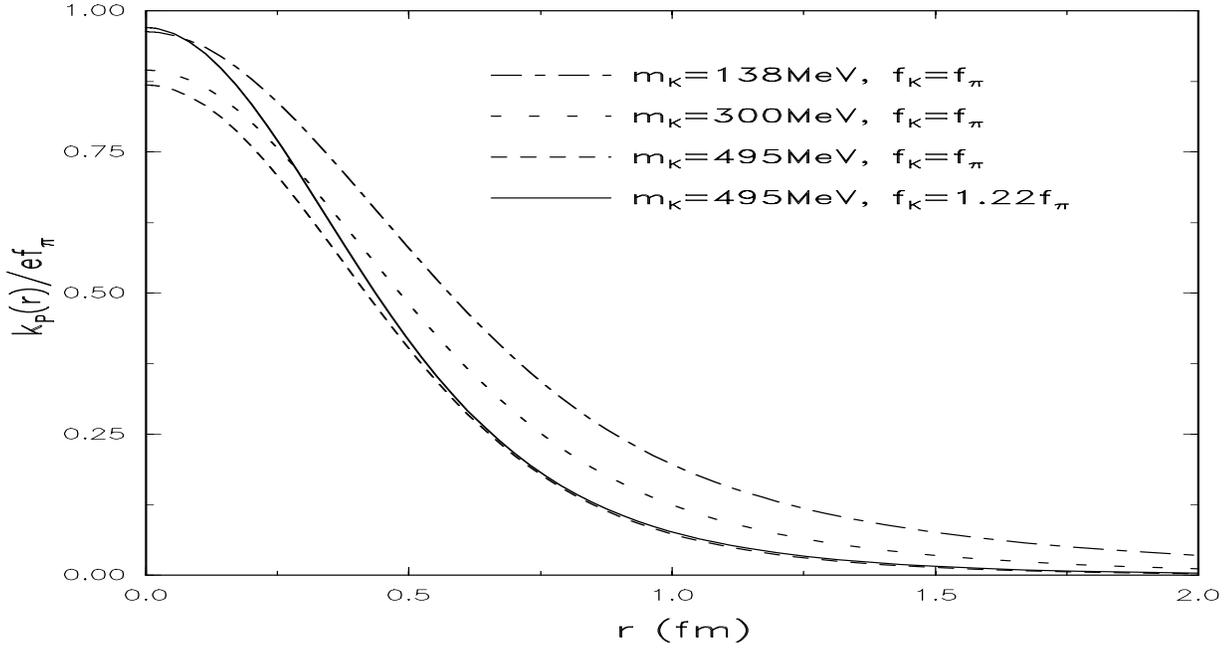,height=9.0cm,width=16.0cm}}
\caption{\label{fi_bound}\tenrm
The radial dependence of the bound state wave function. The
normalization is according to eq (\protect\ref{kbsnorm}). The 
parameters in the non--strange sector are ${\Ss f_\pi=93{\rm MeV}}$,
${\Ss e=4.25}$ and ${\Ss m_\pi=138{\rm MeV}}$.}
\end{figure}

Expanding the Skyrme model action up to quadratic order in the 
fluctuating kaon field, $K$ yields the Lagrangian explicitly 
given in eq (\ref{L2CK}). From this Lagrangian a differential 
equation for $k_P(r,\omega)$ is obtained, which can be cast in 
the generic form \cite{Ca88}
\be
\Bigg\{-\frac{1}{r^2}\frac{d}{dr}\left(r^2h(r)\frac{d}{dr}\right)
+m_K^2+V_P(r)-f(r)\omega^2-2\lambda(r)\omega\Bigg\}k_P(r,\omega)=0\ .
\label{eqmkbs}
\ee
The explicit expressions of the radial functions $h$, $V_P$, $f$ and 
$\lambda$ are also summarized in appendix D. The term linear in the 
frequency $\omega$ originates from the Wess--Zumino action 
(\ref{WZterm}) and removes the degeneracy between solutions 
of positive and negative $\omega$. From the orthogonality condition 
associated with the above differential equation the normalization 
\cite{Ca88}
\be
2\int dr r^2 k_P(r,\omega)\left(f(r)\omega+\lambda(r)\right)
k_P(r,\omega)={\rm sgn}(\omega)
\label{kbsnorm}
\ee
is derived. The above normalized solutions to the bound state 
equation (\ref{eqmkbs}) are drawn in figure \ref{fi_bound}. In 
particular, it is illustrated how the bound state wave--function 
evolves from the zero mode, which is proportional to ${\rm sin}(F/2)$, 
when the parameters are tuned from those describing the flavor 
symmetric case to physical ones.
It can also be observed from figure \ref{fi_bound} that the effect 
of $f_K\ne f_\pi$ is mainly a change in the profile at small $r$. 
This can be traced back to the bound state equation (\ref{eqmkbs}) 
because this inequality only shows up in the Skyrme term, which 
describes the short range behavior, as can be seen from eqs 
(\ref{hrbst}--\ref{vrbst}). The bound state energy computed
with the physical parameters ($m_K=495{\rm MeV}$, $f_K=1.22f_\pi$) 
is $\omega_P=252{\rm MeV}$ for the Skyrme parameter $e=4.25$.
Note that the asymptotic form of the bound state wave--function 
is ${\rm exp}(-\sqrt{m^2_K-\omega_P^2}r)$ rather than 
${\rm exp}(-m_Kr)$ as one would expect for a kaon field.

The infinitesimal symmetry transformation for the variation of 
strangeness\footnote{The generator for strangeness is defined to be
${\rm diag}(0,0,-1)$.} may be para\-metrized as $K(\mbox{\boldmath $r$},t)
\rightarrow K(\mbox{\boldmath $r$},t)-i\epsilon$. It is then
straightforward to obtain the strange\-ness charge from the 
associated Noether current \cite{Ca88}
\be
S=\int \frac{d\omega}{2\pi}\ s(\omega)
\ , \
s(\omega)=-{\rm sgn}(\omega)
\left(a_1^{\dag}(\omega)a_1(\omega)
+a_2^{\dag}(\omega)a_2(\omega)\right)\ ,
\label{strcharge}
\ee
where the condition (\ref{kbsnorm}) has been imposed.
Obviously the $a_i(\omega)$ annihilate modes of strange\-ness $-1$ 
and $+1$ for positive and negative $\omega$, respectively. 
Furthermore the bound state constructed above, carries strangeness
$S=-1$ suitable for the description of physical hyperons. The 
energy operator in a Fock space based on these operators is
$\int (d\omega/2\pi)|\omega s(\omega)|$. Hence each occupation 
of the bound state not only lowers the strangeness by one unit
but also increases the energy by $\omega$. The mass of the $\Lambda$ 
hyperon may then be identified with the sum 
$E+\omega$, where $E$ refers to the soliton mass (\ref{Skeng}).

At present only the degeneracy of the baryons with respect to 
strangeness has been removed. In order to generate states 
corresponding to physical baryons and to obtain the mass differences 
of baryons with identical strangeness but different spin and/or 
isospin, collective coordinates are introduced into the field 
configuration (\ref{bsansatz}) \cite{Ca85}
\be
U(\mbox{\boldmath $r$},t)=
\pmatrix{\hspace{10pt}A(t) &{| \atop |} & \hspace{-10pt} 0\cr 
----\hspace{-10pt}&-&\hspace{-10pt}-\cr 0 &|& \hspace{-10pt} 1\cr}
\xi_\pi(\mbox{\boldmath $r$})
U_K(\mbox{\boldmath $r$},t)\xi_\pi(\mbox{\boldmath $r$})
\pmatrix{\hspace{10pt}A^{\dag}(t) &{| \atop |} & \hspace{-10pt} 0\cr
----\hspace{-10pt}&-&\hspace{-10pt}-\cr 0 &|& \hspace{-10pt} 1\cr}
\label{bsrot}
\ee
where $A(t)$ is an $SU(2)$ matrix. Substituting this {\it ansatz}
into the action augments the collective Lagrangian by terms involving
the time derivative $A^{\dag}\dot A=(i/2)\mbox{\boldmath $\tau$}
\cdot\mbox{\boldmath $\Omega$}$, {\it cf.} eq (\ref{defomega}) 
\be
L_{\mbox{\boldmath $\Omega$}}=
\frac{1}{2}\alpha^2{\mbox{\boldmath $\Omega$}}^2
-\frac{1}{2} \int \frac{d\omega}{2\pi} c(\omega)
{\mbox{\boldmath $\Omega$}}\cdot
\left(\sum_{i,j=1}^2 a_i^{\dag}(\omega)
{\mbox{\boldmath $\tau$}}_{ij}
a_j(\omega)\right)\ .
\label{lbscoll}
\ee
For the P--wave channel the explicit form of the spectral function 
$c(\omega)$ is given in eq (\ref{cpara}). An infinitesimal isospin 
transformation of the meson configuration (\ref{bsrot}) may be related 
to its time derivative
\be
\left[\frac{i}{2}\tau_i , U\right] =
-D_{ij}\frac{\partial \dot U}{\partial \Omega_j} \ ,
\label{isofl1}
\ee
where $D_{ij}$ denotes the adjoint representation of the two flavor 
rotation $A$, {\it cf.} eq (\ref{omtrans}). As a consequence the 
total isospin is completely carried by the collective coordinates 
\be
I_i=-D_{ij}\frac{\partial L_{\mbox{\boldmath $\Omega$}}}
{\partial \Omega_j}\ .
\label{isofl2}
\ee
This identity motivates the decomposition of the total spin 
$\mbox{\boldmath $J$}$ into soliton and bound state pieces, 
$\mbox{\boldmath $J$}^F$ and $\mbox{\boldmath $J$}^K$, respectively. 
The former is defined as
\be
\mbox{\boldmath $J$}^F=
\frac{\partial L_{\mbox {\boldmath $\Omega$}}}
{\partial {\mbox {\boldmath $\Omega$}}}
=\alpha^2{\mbox {\boldmath $\Omega$}}
-\frac{1}{2} \int \frac{d\omega}{2\pi} c(\omega)
\left(\sum_{i,j=1}^2 a_i^\dagger(\omega)
{\mbox{\boldmath $\tau$}}_{ij}
a_j(\omega)\right)\ .
\label{jsol}
\ee
By construction this piece satisfies the relation 
$(\mbox{\boldmath $J$}^F)^2=\mbox{\boldmath $I$}^2$.
The remainder of the total spin can be cast into the form
\be
\mbox{\boldmath $J$}^K=
- \frac{1}{2} \int \frac{d\omega}{2\pi} d(\omega)
\left(\sum_{i,j=1}^2 a_i^{\dag}(\omega)
{\mbox{\boldmath $\tau$}}_{ij}
a_j(\omega)\right).
\label{jk}
\ee
In the Skyrme model the spectral function $d(\omega)$ turns out 
to be unity when the fluctuating fields are normalized according 
to eq (\ref{kbsnorm}). This, however, is not necessarily the case
(see {\it e.g.} section \ref{sec_ckmasses}). The above described 
decomposition, $\mbox{\boldmath $J$}=\mbox{\boldmath $J$}^F+
\mbox{\boldmath $J$}^K$, is further motivated by re--arranging 
\cite{Ca85} the collective rotation (\ref{bsrot})
\be
\xi_\pi(\mbox{\boldmath $r$},t)=
A(t)\xi_\pi(\mbox{\boldmath $r$})A^{\dag}(t)\ , \quad
K_P(\mbox{\boldmath $r$},t)=
A(t)\int \frac{d\omega}{2\pi}k_P(r,\omega)
\hat{\mbox{\boldmath $r$}}\cdot\mbox{\boldmath $\tau$}
\pmatrix{a_1(\omega) \cr a_2(\omega)\cr}\ .
\label{bsrot2}
\ee
Due to the hedgehog structure a spatial rotation can be re--phrased 
as a right multiplication of $A(t)$ and an isospin rotation of 
the isospinor $(a_1(\omega),a_2(\omega))$. The former leads to 
$\mbox{\boldmath $J$}^F$, while $\mbox{\boldmath $J$}^K$ is the 
Noether charge associated with the latter. It is thus clear that
$\mbox{\boldmath $J$}^K$ formally is the isospin associated with 
the isospinor $(a_1(\omega),a_2(\omega))$.

In what follows, only the kaon bound state with $S=-1$ is supposed 
to be occupied. This projects out 
\be
c_P=c(\omega_P)\ ,\ d_P=d(\omega_P)
\label{bspara}
\ee
from the spectral integrals. With this restriction to the kaon bound 
state, the collective Hamiltonian associated with
$L_{\mbox{\boldmath $\Omega$}}$ becomes
\be
H_{\mbox{\boldmath $\Omega$}}=
\frac{1}{2\alpha^2}\left(\mbox{\boldmath $J$}^F +
\frac{c_P}{2}\sum_{i,j=1}^2 a_i^{\dag}
{\mbox{\boldmath $\tau$}}_{ij} a_j\right)^2
=\frac{1}{2\alpha^2}\left(\mbox{\boldmath $J$}^F
+\chi\mbox{\boldmath $J$}^K\right)^2\ ,
\label{homega}
\ee
where the parameter $\chi=-c_P/d_P$ has been introduced.
It is then straightforward to obtain the baryon mass formula 
\cite{Ca88}
\be
M_B=E+|S\omega_P|+\frac{1}{2\alpha^2}
\left(\chi J(J+1)+(1-\chi)I(I+1)\right) . 
\label{bsmass}
\ee
It should be remarked that the contribution\footnote{Note that
$\left(\sum_{i,j=1}^2 a_i^{\dag}
{\mbox{\boldmath $\tau$}}_{ij} a_j\right)^2=S(S-2)$, when canonical 
commutation relations are assumed for the $a_i$.}, 
$c_P(c_P+1)/4S(S-2)$ \cite{Kl89}, which is quartic in the kaon field, 
has been omitted because this order has not consistently been treated.

This approach to quantize the kaon bound state may be phrased in 
terms of slow ($\mbox{\boldmath $\Omega$}$) and fast ($\omega$) 
moving degrees of freedom. The former is of the order $1/N_C$ while 
the latter are ${\cal O}(N_C^0)$. This distinction allows one to 
introduce the notion of Berry phases. For the kaon bound state this 
idea has extensively been discussed in ref \cite{Lee93}.

Substituting the bound state solution of the differential equation
(\ref{eqmkbs}) into (\ref{cpara}) provides the mass differences 
displayed in table \ref{ta_bsdiff}. In these computations the 
physical values for the masses and decay constants of the 
pseudo--scalar mesons have been adopted, $m_\pi=138{\rm MeV}$, 
$m_K=495{\rm MeV}$, $f_\pi=93{\rm MeV}$ and $f_K=113{\rm MeV}$. 
It was common, however, in previous calculations to employ 
parameters, which in the two flavor model reproduce the nucleon and 
$\Delta$ masses
\be
f_\pi&=&64.5{\rm MeV},\ e=5.45,\ m_\pi=0\quad
 \protect\cite{Ad83} \ ;
\nonumber \\
f_\pi&=&54.0{\rm MeV},\ e=4.84,\ m_\pi=138{\rm MeV}\quad
\protect\cite{Ad84} \ .
\label{ANWpara}
\ee
These parameter sets have much smaller decay constants. In the 
meantime it has, however, become clear that the absolute mass of the 
soliton is strongly reduced by quantum corrections \cite{Mo93,Ho94}. 
Then the absolute masses are reasonably reproduced using the physical 
decay constants. It is therefore more appealing to use  the physical 
decay constants and rather consider mass differences than absolute 
masses (This point of view has also been assumed in the previous 
chapters). Then the Skyrme parameter $e$ is the only one which is 
not determined in the meson sector. It may, {\it e.g.} be fixed by 
demanding the experimental $\Delta$--nucleon splitting leading to 
$e=4.25$.
\begin{table}
\caption{\label{ta_bsdiff}\tenrm
The parameters of the mass formula (\protect\ref{bsmass}) and the 
predictions for the mass differences (in MeV) of the low--lying baryons 
with respect to the nucleon in the bound state approach. Various 
values of the Skyrme parameter ${\Ss e}$ are used. The data in 
parenthesis include the quartic term. For completeness also the results
corresponding to the reduced decay constants (\protect\ref{ANWpara}) 
are included with ${\Ss m_K=495}$ MeV and ${\Ss f_K=1.22f_\pi}$ 
\protect\cite{Ri91}.}
~
\newline
\centerline{\tenrm\smalllineskip
\begin{tabular}{l| c | c | c | c | c c}
& $e=$4.00 & $e=$4.25 & $e=$4.50 & Expt. &
\protect\cite{Ad83} &\protect\cite{Ad84} \\
\hline
$\alpha^2({\rm GeV}^{-1})$ & 6.013 & 5.115 & 4.389  & --- &
5.050 & 5.133 \\
$\omega_P({\rm MeV})$      & 265   & 253   & 240    & --- &
221 & 209  \\
$\chi$                     & 0.261 & 0.332 & 0.401  & --- &
0.500 & 0.394 \\
\hline
$\Lambda$          & 218(207) & 205(187) & 189(168) & 177 &
183(165) & 165(147) \\
$\Sigma$           & 342(330) & 334(318) & 325(305) & 254 & 
283(264) & 283(265) \\
$\Xi$              & 530(498) & 505(462) & 479(424) & 379 &
442(392) & 418(371) \\
$\Delta$           & 249(249) & 293(293) & 342(342) & 293 &
293(293) & 292(292) \\
$\Sigma^*$         & 407(394) & 431(415) & 462(442) & 446 &
432(413) & 398(380) \\
$\Xi^*$            & 595(563) & 602(559) & 616(562) & 591 &
591(541) & 533(486) \\
$\Omega$           & 814(754) & 805(724) & 805(702) & 733 &
774(681) & 697(610) \\
\end{tabular}}
\end{table}
From table \ref{ta_bsdiff} we see that the resulting bound state 
energy is somewhat larger that the empirical value 
$\omega_{\rm emp}=(M_\Xi-M_N)/2\approx189{\rm MeV}$ causing the 
flavor symmetry breaking in the baryon spectrum to be overestimated. 
This is in particular the case for the $\frac{1}{2}^+$ baryons. The 
inclusion of the (presumably inconsistent) quartic term provides a 
slight improvement.  Except for the $\Omega$ baryon, which has $S=-3$, 
this term has, however, only minor influence. Similarly, an empirical 
value for the hyperfine parameter $\chi$ can be extracted from the 
mass formula (\ref{bsmass})
$3\chi/2(1-\chi)=(M_{\Sigma^*}-M_\Sigma)/(M_\Sigma-M_\Lambda)$, 
{\it i.e.} $\chi\approx0.62$, which is larger than the model 
prediction. In case the parameters (\ref{ANWpara}) are used together 
with $m_K=495$ MeV and $f_K=1.22f_\pi$, the agreement with 
the experimental mass difference is significantly improved. This is 
not surprising because the energy scale is reduced. However, there is 
no deeper reason to take the experimental ratio between the decay 
constant once $f_\pi$ is moved away from $93{\rm MeV}$.
Originally, the bound state approach was assumed to underestimate 
the flavor symmetry breaking in the baryon spectrum 
\cite{Ca88,Sc88,Bl89}. A better agreement with experiment was 
claimed \cite{Ri91} to be due to the kinetic type symmetry breakers
in eq (\ref{LSB}), {\it cf.} the discussion in the paragraph after eq 
(\ref{gasky}). When taking the physical values for the decay constants 
we are confronted with the situation that the bound state approach 
actually overestimates the flavor symmetry breaking in the baryon 
spectrum. One wonders whether kinematical corrections change this 
conclusion. As a very crude estimate one might replace the kaon mass 
by the reduced mass of the kaon--soliton system into the bound state 
equation (\ref{eqmkbs}). For $e=4.25$ the bound state energy the 
decreases to $191{\rm MeV}$, which, of course, improves on the baryon 
mass differences.

Alternative parametrizations of the fluctuating kaon fields essentially 
reach similar results for the mass differences when identical parameters
are used, see {\it e.g.} ref \cite{Bl89}.

The bound state approach corresponds to filling up levels of a quantum 
system, without taking into account the correlations between the 
fluctuating meson fields. Two heavy meson correlations \cite{Ba95}, 
as they could appear in the case of the $\Xi$ baryon, might alter the 
description.

In order to make statements about baryon properties in the bound 
state approach, the appropriate wave functions have to be constructed. 
These baryon states are products of eigenstates associated with the 
collective coordinates and elements of the Fock space built from the 
kaon bound state. The former are $SU(2)$ D--functions, while the 
latter are not only eigenstates of the strangeness operator 
(\ref{strcharge}) but also of 
$(\mbox{\boldmath $J$}^K)^2=S(S-2)/4$ and the associated projection
$\mbox{\boldmath $J$}^K_3$, defined in eq (\ref{jk}). These 
states are coupled to good total spin, ($\mbox{\boldmath $J$}$),
isospin, ($\mbox{\boldmath $I$}$) and hypercharge, $Y=S+1$
\be
|J,J_3;I=J_F,I_3;Y\rangle=
{\cal N}C^{JJ_3}_{IJ^F_3,J^KJ^K_3}
D_{I^3,-J^F_3}^{I=J^F}(A)|S;J^K=|S|/2,J^K_3\rangle \ . 
\label{bsbaryon}
\ee
The identity, $I=J^F$ results from eq (\ref{isofl2}). Furthermore 
$C$ refers to a Clebsch--Gordon coefficient and ${\cal N}$ is a 
normalization constant. An extensive list of explicit baryon 
wave--functions is provided in table 2 of ref \cite{Oh91}. Apparently 
the collective coordinates, $A$, have to be quantized like a fermion 
if the occupation number of the kaon bound state is odd and like a 
boson if it is even \cite{Ca85}.

The meson configuration (\ref{bsrot}) is then substituted into the 
covariant expression for the vector current $V_\mu^a$ (\ref{currps}) 
and the matrix elements between the baryon states (\ref{bsbaryon}) 
are computed. The magnetic moment operator is related to the spatial 
components of the electromagnetic combination (\ref{emgen}),
\be
\mu=\frac{M_N}{2}\int d^3r \epsilon_{3ij}x_iV_j^{\rm e.m.}
=\mu_{S,0}J^K_3+\mu_{S,1}J^K_3
+\frac{2I_3J^F_3}{I(I+1)}\left(\mu_{V,0}+\mu_{V,1}|S|\right) \ .
\label{magbs}
\ee
where the angular velocities have been eliminated according to the
quantization prescription (\ref{jsol}). The isovector part arises 
from matrix elements of $D_{33}=D_{00}^1(A)$ in the space of 
the collective coordinates, $A$. The coefficients $\mu_{S,V;0,1}$ 
(taken from ref \cite{Oh91}) are summarized in eq (\ref{magbspara}) of 
appendix D. The expressions not involving the fluctuating kaon field 
($\mu_{S,V;0}$) are identical to their two flavor counterparts 
\cite{Ad83}. Hence the effects of virtual strange quark excitations in 
the nucleon are ignored. Substituting the bound state wave--function 
and computing the matrix elements of the operators in eq (\ref{magbs}) 
between the baryon states (\ref{bsbaryon}) then provides the baryon 
magnetic moments in the bound state approach. For $e=4.25$ these are 
listed in table \ref{ta_magbs}.
\begin{table}
\caption{\label{ta_magbs}\tenrm
The magnetic moments in the bound state (BS) approach to the Skyrme 
model.  Results are given in nucleon magnetons
as well as ratios of the proton magnetic moment.}
~
\newline
\centerline{\tenrm\smalllineskip
\begin{tabular}{c | c c | c c}
& \multicolumn{2}{c|}{BS} & \multicolumn{2}{c}{Expt.} \\
\hline
Baryon & $\mu_B$ & $\mu_B/\mu_p$ & $\mu_B$ & $\mu_B/\mu_p$ \\
\hline
$p$        & 1.78 & 1.00 & 2.79 & 1.00 \\
$n$        &-1.42 &-0.80 &-1.91 &-0.68 \\
$\Lambda$  &-0.56 &-0.31 &-0.61 &-0.22 \\
$\Sigma^+$ & 1.97 & 1.10 & 2.42 & 0.87 \\
$\Sigma^0$ & 0.43 & 0.24 & ---  & ---  \\
$\Sigma^-$ &-1.07 &-0.60 &-1.16 &-0.42 \\
$\Xi^0   $ &-1.31 &-0.74 &-1.25 &-0.45 \\
$\Xi^-   $ &-0.33 &-0.18 &-0.69 &-0.25 \\
$\Sigma^0\rightarrow\Lambda$
&-1.54 &-0.86 &-1.61 &-0.58 \\
\end{tabular}}
\end{table}
Although the physical value $f_\pi=93{\rm MeV}$ has been used, these 
results do not differ significantly from those obtained with the sets 
(\ref{ANWpara}) \cite{Ku89,Oh91}. As in the collective approach 
(chapter \ref{chap_coll}) the absolute values of the magnetic moments 
come out too small. The only exception is the transition matrix 
element for $\Sigma^0\rightarrow\Lambda$. Considering the bound state 
results for the magnetic moments in the light of the U--spin relations 
(table \ref{ta_uspin}), sizable deviations from these relations are 
observed. However, these deviations are either in the wrong direction 
($\mu_{\Sigma^+}$/$\mu_p$) or are too strong 
($\mu_{\Sigma^-}$/$\mu_{\Xi^-}$) when compared with the 
experimental situation. Also the prediction for the ratio 
$\mu_\Lambda/\mu_{\Xi^-}\approx1.7$ is in drastic disagreement 
with the physical value ($\sim0.9$). Obviously ample space for
improvement for predictions of magnetic moments is left within the 
bound state description of baryons.

Generalizing the operator (\ref{magbs}) to depend on the momentum 
transfer (analogous to eq (\ref{Gmps})) allows one to compute the 
magnetic dipole ($M1$) transitions for the electromagnetic decay of 
$\frac{3}{2}^+$ baryons to $\frac{1}{2}^+$. Similarly the 
orbital angular momentum $l=2$ component of the electric charge 
operator may be employed to compute the quadrupole transition 
($E2$). The calculated \cite{Sch95b} decay widths exhibit the same 
pattern as in the non--relativistic quark model \cite{Da83}
and the bag model \cite{Bu93}; {\it i.e.} the transitions 
$\Sigma^{*0}\rightarrow \Lambda\gamma$, $\Xi^{*0}\rightarrow
\Xi^0\gamma$ and $\Sigma^{*+}\rightarrow \Sigma^+\gamma$ have 
large widths (100--200keV), while all others are about an order 
of magnitude smaller. In particular the width of the reaction
$\Sigma^{*-}\rightarrow \Sigma^-\gamma$ is predicted to be almost 
zero. All ratios $E2/M1$ for the transition matrix elements 
are predicted to be negative. As a consequence of delicate 
cancellations these ratios are largest for 
$\Sigma^{*-}\rightarrow \Sigma^-\gamma$ and 
$\Xi^{*0}\rightarrow \Xi^0\gamma$ (about -0.5 and -0.2, 
respectively) \cite{Sch95b}. For the other transitions the ratios 
are again down by an order of magnitude. The $E2$ contribution 
to the reaction $\Sigma^{*0}\rightarrow \Sigma^0\gamma$ even 
vanshes identically. 

For all decay widths and most of the $E2/M1$--ratios similar results 
are obtained in the collective model \cite{Ab95a}. For the 
reactions $\Sigma^{*-}\rightarrow\Sigma^-\gamma$ and 
$\Xi^{*0}\rightarrow \Xi^0\gamma$ the collective approach predicts much 
smaller ($\sim -0.05$) $E2/M1$--ratios. In any event, the widths of 
these decays are very tiny. Hence the corresponding $E2/M1$--ratios 
are subject to delicate cancellations. In the collective approach 
the symmetry breaking parameter can continuously be varied. This 
feature has been employed to verify the $U$--spin selection rule 
within the Skyrme model \cite{Ab95a}. This rule states that 
the transition matrix elements associated with the decays of the 
negatively charged $\frac{3}{2}^+$ baryons vanish in the flavor symmetric 
case \cite{Li73}.

It has been indicated above that in the bound state approach the 
nucleon wave--function is identical to the one of the two flavor 
model. Hence one expects the strangeness content fraction $X_s$ 
(\ref{defxs}) to vanish.  However, this is not the case. The 
non--trivial strangeness in the nucleon is caused by the fact 
that the Fock space built from the solutions to the bound state 
equation (\ref{eqmkbs}) is different from the one based on free kaon 
modes, {\it i.e.} in the absence of the soliton \cite{Bl88a}. In 
particular, the nucleon state is annihilated by the operator 
$a(\omega)$ of the Fock space in the presence of the soliton. The 
creation and annihilation operators $\tilde a^{({\dag})}(\omega)$ 
in the Fock space of free kaons are related to those (\ref{pwansatz}) 
for the bound state problem
\be
\pmatrix{\tilde a(\omega) \cr \tilde a^{\dag}(\omega)\cr}=
\int \frac{d\omega^\prime}{2\pi}
\pmatrix{{\cal A}_{++}(\omega,\omega^\prime)&
{\cal A}_{-+}(\omega,\omega^\prime)\cr
{\cal A}_{+-}(\omega,\omega^\prime)&
{\cal A}_{--}(\omega,\omega^\prime)\cr}
\pmatrix{a(\omega) \cr a^{\dag}(\omega)\cr} \ ,
\label{opertrans}
\ee
where the transformation matrix ${\cal A}$ is computed from the 
overlaps between the solutions of the bound state equation
(\ref{eqmkbs}) and free kaon modes in a spherical wave decomposition
\cite{Bl88a,Wo90}. As a consequence of the mixing between the 
operators in the two distinct Fock spaces the nucleon state is not 
annihilated by $\tilde a(\omega)$. The calculations based on the 
transformation (\ref{opertrans}) are essentially equivalent to a 
treatment within the random phase approximation (RPA \cite{Ri80}), 
introducing correlations between the different vacua. Hence the 
nucleon matrix element of the ``free" strangeness charge 
\be
S_{\rm free}=
-\int\frac{d\omega}{2\pi}{\rm sgn}(\omega)
\tilde a^{\dag}(\omega)\tilde a(\omega)
\label{frstrcharge}
\ee
is different from zero. This causes a non--vanishing strangeness 
content fraction $X_s$, which can be computed from the momentum 
space integrals of the off--diagonal elements of ${\cal A}$. This 
yields $X_s\approx3\%$ \cite{Bl88a,Wo90}, which is significantly lower 
than the corresponding result in the collective approach, {\it cf.} 
section \ref{sec_masses}. It should, however, be noted that in general 
these RPA--type integrals diverge; $X_s$ represents one of the few 
exceptions. Thus, in principle, a regularization scheme would be 
required, which might also alter the prediction for $X_s$.

We would like to finish this section by remarking that in addition 
to the P--wave bound state discussed above, the equation (\ref{eqmkbs})
contains a solution of orbital angular momentum $l=0$
\be
K_S(\mbox{\boldmath $r$},t)=\int \frac{d\omega}{2\pi}\
{\rm e}^{-i\omega t}k_S(r,\omega)
\pmatrix{a_1(\omega) \cr a_2(\omega)\cr}\ ,
\label{swansatz}
\ee
with $|\omega|<m_K$ \cite{Ca88}. The appearance of the bound state 
is completely due to dynamics exposed by the background soliton. This 
is in contrast to the P--wave bound state, which arises from the 
existence of a zero--mode in the flavor symmetric case. The S--wave
bound state is well suited to describe the odd parity $\Lambda(1405)$. 
For the {\it ansatz} (\ref{swansatz}) a right transformation of the 
collective coordinates $A$ (which represents the spatial rotation of
the soliton) can also be rephrased as an isospin transformation of 
the isospinor $(a_1(\omega), a_2(\omega))$. Hence the quantization is 
identical to the one for the P--wave. For the parameters established 
above ($f_\pi=93{\rm MeV},e=4.25$, etc.) the mass difference of the 
odd parity $\Lambda$ to the nucleon is predicted to be $450{\rm MeV}$, 
about $40{\rm MeV}$ over the experimental value. Not surprisingly a 
deficit compared to the experimental mass difference of about 
$100{\rm MeV}$ is obtained when the smaller energy scale 
(\ref{ANWpara}) is used. In ref \cite{Sch95a} the magnetic moment 
of this hyperon has been calculated to be about 10\% of the magnetic 
moment of the proton.

\bigskip

\begin{subsatz}
\label{sec_scattering}{\bf \hskip1cm
Remarks on Meson Baryon Scattering}
\addcontentsline{toc}{section}{\protect\ref{sec_scattering}
Remarks on Meson Baryon Scattering}
\end{subsatz}
\stepcounter{section}

Shortly after the rediscovery of the Skyrme model by Adkins, Nappi 
and Witten \cite{Ad83} the model was employed to compute the amplitudes 
of meson baryon scattering. For an extensive list of references the 
reader is referred to the review article \cite{Sch89}. Similar to the 
bound state approach (section \ref{sec_fluc}) such calculations involve
time--dependent meson fluctuations $\eta_a(\mbox{\boldmath $r$},t)$ off
the soliton. A possible parametrization in the flavor rotating frame is 
given by
\be
U(\mbox{\boldmath $r$},t)=
{\rm e}^{i\lambda^a\eta_a({\bf r},t)}
U_0(\mbox{\boldmath $r$})
{\rm e}^{i\lambda^a\eta_a({\bf r},t)} \ ,
\label{deffluc}
\ee
where the static configuration $U_0(\mbox{\boldmath $r$})$ is 
defined in eq (\ref{su3hedgehog}). In the next step the action 
is expanded up to quadratic order in $\eta_a$ in the background 
of the static soliton. From the solutions to the corresponding 
equations of motions the $T$--matrix elements,
${\tau}^{IY}_{Kl^\prime l}$ for meson--soliton scattering may be 
extracted \cite{Wa84,Kr86}. The $T$--matrix is diagonal in the 
intrinsic isospin ($I$), hypercharge ($Y$) and grand spin ($K$) 
(the latter represents the vector sum of total spin and isospin).
Furthermore $l$ and $l^\prime$ denote the orbital angular 
momentum of the in-- and outgoing meson, respectively. For 
example, ${\tau}^{1/2\ 1}_{1/2\ 11}$ represents the $T$--matrix 
in the bound state channel discussed in section \ref{sec_fluc}.
Assuming adiabaticity of the collective rotation and ignoring 
symmetry breaking effects, the $T$--matrix for physical channels 
may be obtained within the geometrical coupling scheme
\cite{Ha84,Kr86}
\be
T_{\rm phys}&=&(-1)^{s^\prime-s}
\frac{dim R dim R^\prime}{dim R^{\rm tot}}
\sum_{IY}\sum_i\sum_K (2i+1) (2K+1)
\left\{\begin{array}{ccc}
K & i & J \\ s^\prime  & l^\prime & I 
\end{array}\right\}
\left\{\begin{array}{ccc}
K & i & J \\ s & l & I
\end{array}\right\}
\nonumber \\ && \hspace{1cm}
\left[\begin{array}{cc|cc}
R_{tot} & \gamma^\prime & R^\prime & {\bf 8} \\
i & Y+1 & s^\prime 1 & IY \end{array}\right]
\left[\begin{array}{cc|cc}
R & {\bf 8} & R_{tot} & \gamma \\
s1 & IY & i & Y+1Y \end{array}\right]
{\tau}^{IY}_{Kl^\prime l}\ .
\label{tphys}
\ee
The objects in curly and squared brackets are $6$--$j$ symbols and
$SU(3)$ isoscalar factors \cite{Sw63}, respectively. Furthermore 
$s$ and $R$ refer to spin and $SU(3)$ representations {\it e.g.}
$s=\frac{1}{2}$ and $R={\bf 8}$ for the nucleon.
The index $tot$ finally denotes the total $SU(3)$ representation 
of the meson baryon system. For further details we refer to the 
original article by Karliner and Mattis \cite{Kr86}. It should 
be stressed that this coupling scheme is purely geometrical, and 
dynamic contributions from the collective rotations have completely 
been ignored.  The so--obtained physical $T$--matrix allows to 
reasonably describe a huge number of scattering processes from 
$\pi N\rightarrow\pi N$ to ${\overline{K}}N\rightarrow \pi\Sigma^*$. 
The authors of ref \cite{Kr86} characterize the agreement with 
experiment to be good in the case that the initial states are 
$\pi N$, mixed for ${\overline{K}}N$ but poor for $KN$. 

Although appealing, this approach suffers from the shortcoming that 
flavor symmetry breaking effects are not accounted for. Dominantly this
concerns the thresholds, which should account for
different baryon masses. Furthermore the baryon wave--functions have 
been assumed to be members of a certain $SU(3)$ representation $R$, 
which is also reflected in the appearance of the isoscalar factors in 
eq (\ref{tphys}). These problems have recently been studied 
\cite{Sch92a} by parametrizing the time--dependent chiral field 
\be
U(\mbox{\boldmath $r$},t)=
A(t)\left(U_0(\mbox{\boldmath $r$})\right)^{\frac{1}{2}}A^{\dag}(t)
\ {\rm exp}\left(i\lambda^a\phi^a(\mbox{\boldmath $r$},t)\right)
A(t)\left(U_0(\mbox{\boldmath $r$})\right)^{\frac{1}{2}}A^{\dag}(t) 
\ .
\label{berndpara}
\ee
Obviously the $\phi_a$ are the meson fluctuations in the 
laboratory frame {\it i.e.} they are defined in the background 
of the collectively rotating hedgehog. Separating the radial 
dependence of the meson fluctuations allows one to define target 
wave--functions 
\be
\phi_{ss_3,ii_3,Y}^{t_a y_a \choose m_a}
(\mbox{\boldmath $r$},\omega)
=\sum_{l,m \atop \tau_\alpha j_\alpha\mu_\alpha \nu_\alpha}
C^{i i_3}_{t_a\tau_\alpha, m_a \mu_\alpha}
C^{s s_3}_{l m, j_\alpha \nu_\alpha}
\phi_{t_a y_a}^{l\tau_\alpha j_\alpha}(r,\omega)
\Psi\left(\tau_\alpha, \mu_\alpha y_\alpha; j_\alpha, \nu_\alpha,1\right)
Y_{lm}(\hat{\mbox{\boldmath $r$}}) \ ,
\label{traget}
\ee
where a Fourier transformation with respect to the time coordinate 
has been performed. Furthermore, $\Psi$ represents the solution to 
the Yabu--Ando eigenvalue problem (\ref{Dpsi}). By varying the 
strength of the symmetry breaking parameters, it 
is possible to interpolate between the flavor symmetric approach 
described above and the two flavor computations for $\pi N$ 
scattering \cite{Ha84}. Furthermore the adiabatic approximation has 
been abandoned, {\it i.e.} time derivatives of the collective 
coordinates are (at least approximately) included, in this approach. 
This dominantly effects the $P$--wave channels. Finally the 
scattering amplitudes are extracted from the solutions of the 
differential equations for the radial fields 
$\phi_{t_a y_a}^{l\tau_\alpha j_\alpha}(r)$. It should 
be stressed that these calculations are rather tedious, in particular 
because the differential equations also contain the flavor 
orientation, $A$, of the hedgehog. As can been seen from the Argand 
diagrams displayed in ref \cite{Sch92a} the agreement with experimental 
data is impressive for the long wavelength behavior of the 
scattering amplitudes, in particular for the number and approximate 
positions of the main decay channels of baryon resonances.
Hence this extensive study of meson baryon scattering represents 
another interesting application of the collective treatment of 
symmetry breaking. As a side product also the bound state equation 
of Callan and Klebanov is obtained. 

An examination of scattering processes in the bound state approach 
appears to be technically unfeasible. Considering for example for 
the description of a  reaction like $KN\rightarrow\pi\Sigma$, the 
action has to be expanded up to third order in the meson fluctuations 
because the scattering meson fields ($K,\pi$) as well as the hyperon 
($\Sigma$) are constructed as small amplitude fluctuations in the 
background of the soliton.

\bigskip

\begin{subsatz}
\label{sec_comp}{\bf \hskip1cm 
Comparison: Collective Approach versus Bound State Approach}
\addcontentsline{toc}{section}{\protect\ref{sec_comp}
Comparison: Collective Approach versus Bound State Approach}
\end{subsatz}
\stepcounter{section}

In this section the interesting question will be addressed as to 
how far the collective and bound state approaches to flavor symmetry 
breaking can be related. In the case of the collective treatment this 
discussion will be restricted to the rigid rotator. To further 
simplify this comparison, the symmetry breaking operator proportional 
to $\gamma$ will be assumed to be the only one contained in the 
collective Hamiltonian (\ref{collham}). Although the conceptual ideas 
of these approaches have already been discussed earlier they are 
compactly summarized and compared in table \ref{ta_comp}.
\begin{table}
\caption{\label{ta_comp}\tenrm 
Comparison of the collective and the bound state 
approaches to strangeness.}
~\newline
\vspace{-0.7cm}
\centerline{\tenrm\smalllineskip
\begin{tabular}{|c|c|}
\hline
Collective approach (YA) & Bound state approach (CK) \\
\hline
Symmetry breaking small & Symmetry breaking large \\
(light strange quark) & (heavy strange quark)\\
$\downarrow$ & $\downarrow$ \\
Strange components as & Restoring force for \\
{\bf collective coordinates} & {\bf strange fluctuations} \\
(analogous to zero modes)    & (``harmonic'' potential) \\
$\downarrow$ & $\downarrow$ \\
Collective Hamiltonian & Bound State Energy and \\
in flavor SU(3) including & Wave Function \\
symmetry breaking & $\downarrow$ \\
$\downarrow$ & Collective quantization of \\
Exact diagonalization & spin and isospin \\
$\downarrow$ & $\downarrow$ \\
{\it HYPERONS} &{\it HYPERONS} \\
\hline
\end{tabular} }
\end{table}

The numerical results for the strangeness  content fraction, $X_s$, 
discussed in the previous section as well as those obtained in the 
context of the rigid rotator suggest that the two approaches have a 
common limit for large symmetry breaking. For this purpose, Yabu and 
Ando \cite{Ya88} have expanded the eigenvalue problem (see appendix 
A for the relevant definitions)
\be
\left\{C_2(SU(3))+\omega^2{\rm sin}^2\nu\right\}\Psi(A)
=\epsilon_{\rm SB}\Psi(A)\ , \quad
\omega^2=\frac{3}{2}\gamma\beta^2
\label{YAprob}
\ee 
in inverse powers of the effective symmetry breaking parameter, {\it
i.e.}, $1/\omega\rightarrow0$. According to the discussion of chapter 
\ref{chap_coll}, the isoscalar functions $f_{M_L,M_R}(\nu)$ contained 
in the collective space wave--function $\Psi(A)$ (\ref{Dpsi}) are 
strongly pronounced at $\nu=0$ in the limit of large symmetry 
breaking. The trigonometrical functions appearing in the differential 
equation (\ref{YAprob},\ref{C2psi}) may thus be linearized in the 
limit $1/\omega\rightarrow0$. Introducing the scaled ``Euler angle"
\be
\xi=\sqrt\omega \nu
\label{scalangle}
\ee
then makes possible the above mentioned expansion \cite{Ya88}
in $1/\omega$
\be
&&\hspace{-1cm}
\left\{\frac{-1}{4\xi^2}\frac{d}{d\xi}\left(\xi^3\frac{d}{d\xi}\right)
+\xi^2+\left(\frac{1}{\xi^2}-\frac{1}{6}\right)\mbox{\boldmath $T$}^2
+\frac{1}{4}\left(Y^2+Y_R^2+YY_R\right)
+\frac{1}{2}\left(\mbox{\boldmath $J$}^2+
\mbox{\boldmath $I$}^2\right)\right\}_
{M_L,M_L^\prime}^{M_R,M_R^\prime}
\hspace{-0.2cm}
f_{M_L^\prime,M_R^\prime}(\nu)
\nonumber \\ && \hspace{3cm}
-\frac{1}{\omega}\left[-\frac{1}{2}\xi\frac{d}{d\xi}
+\frac{1}{3}\xi^4\right]f_{M_L,M_R}(\nu)=
\frac{1}{\omega}\epsilon_{\rm SB}f_{M_L,M_R}(\nu)
+{\cal O}\left(\frac{1}{\omega^2}\right) .
\label{YAexp1}
\ee
In this expression the projection of the quadratic Casimir operator 
onto different components ($M_L,M_R$) of the intrinsic isospin and 
spin has been indicated. The only operator in eq (\ref{YAexp1}), which 
is non--diagonal in these indices, is the ``intrinsic grand spin"
$\mbox{\boldmath $T$}=
\tilde{\mbox{\boldmath $J$}}-\tilde{\mbox{\boldmath $I$}}$, where 
$\tilde{\mbox{\boldmath $J$}}$ and $\tilde{\mbox{\boldmath $I$}}$ 
are the intrinsic spin and isospin operators which act on the 
indices $M_L$ and $M_R$ respectively. Hence the combinations 
\be
F(L,M)(\nu)=\sum_{M_L,M_R}C^{LM}_{I-M_L,JM_R}f_{M_L,M_R}(\nu)
\label{diagT1}
\ee 
diagonalize $\mbox{\boldmath $T$}^2$. The eigenvalues are 
$L(L+1)$ with  \cite{Ya88}
\be
{\rm max}\left(|I-J|,|S|/2\right)\le L \le I+J \ ,
\label{diagT2}
\ee
where the $LHS$ is due to the constraint $S=Y-Y_R=2(M_L-M_R)$. For the 
baryons we are interested in ($N,\Lambda,\ldots,\Omega$), the 
value $L=|S|/2$ is always permitted. Obviously, the ``intrinsic grand
spin" plays the role of the kaon spin operator $\mbox{\boldmath $J$}^K$
in the bound state approach. This can also be argued from the 
decomposition (\ref{bsrot2}) because in the intrinsic frame ($A=1$) 
the grand spin of $\xi_\pi$ vanishes, while the grand spin of 
$K$ is nothing but the isospin of the kaon. In leading order of 
the $1/\omega$ expansion, the eigenvalue problem for $F(L,M)(\nu)$ 
can be formulated as that of a four dimensional harmonic oscillator.
Hence the eigenvalues may readily be evaluated. The state of lowest 
energy corresponds to the case when the wave--function $F(L,M)(\nu)$ 
does not possess a node. The associated contribution to the mass 
formula (\ref{bmass}) is \cite{Ya88}
\be
\frac{1}{2\beta^2}\epsilon_{\rm SB}=\frac{\omega}{2\beta^2}|S|
+\frac{1}{4\beta^2}\left[I(I+1)+J(J+1)\right]
+\frac{1}{8\beta^2}\left[3S-2|S|+\frac{1}{2}S^2\right]
+{\cal O}\left(\frac{1}{\omega}\right) ,
\label{YAexp3}
\ee
where a constant contribution has been ignored because we are only 
interested in mass differences. Comparison with the mass formula of 
the bound state approach (\ref{bsmass}) suggests the identifications
\be
\omega_P=\frac{1}{2\beta^2}\sqrt{\frac{3}{2}\beta^2\gamma}
\approx337{\rm MeV}\ , \quad
\chi=1-\frac{\alpha^2}{2\beta^2}\approx0.33 \ .
\label{compnum}
\ee
These data are obtained from eq (\ref{pscopara}), which have 
been computed for $e=4.0$.  Although the analytical identifications 
(\ref{compnum}) have been obtained in the large symmetry breaking 
limit, the numerical values reasonably agree with the exact results 
in the bound state approach even for the physical case (see table 
\ref{ta_bsdiff}). It should be remarked that the inclusion of induced 
components (\ref{kaonind}) is important to arrive at this result. 
Otherwise it can generally shown that $2\beta^2<\alpha^2$ \cite{Ya88} 
causing the estimate for $\chi$ to be negative. 

Another way to establish the correspondence between the two approaches 
in the large symmetry breaking limit, is to expand \cite{Kl89,West94} 
the Lagrangian (\ref{colllag}) in powers of the strange components 
contained in the three flavor collective matrix $A(t)$ 
(\ref{su3hedgehog}). This procedure provides approximations for 
the parameters in the mass formula (\ref{bsmass}). Unfortunately,
this approximation overestimates the bound state energy $\omega_P$, 
while the hyperfine parameter $\chi$ comes out too small.

In the flavor symmetric case one has $\omega_P=0$ and $\chi=1$ due 
to the appearance of the zero mode. Hence the bound state mass formula 
reproduces the degeneracy of the baryons within a given $SU(3)$
multiplet. One may therefore ask the question whether or not the two 
approaches can be related in the small symmetry breaking limit as 
well. However, it can easily be shown that this is not the case. In 
the first non--trivial order of flavor symmetry breaking the 
Gell--Mann--Okubo mass formula (\ref{GMOmass2}) is not 
recovered in the bound state approach. That is,
\be
2\left(M_N+M_\Xi\right)-M_\Sigma-3M_\Lambda=
\frac{1}{2\alpha^2}\left(1-\chi+\frac{3}{4}\chi(\chi-1)\right),
\label{GMObs}
\ee
which is of the order of 50MeV. The second term is due to the quartic 
term which has been omitted in eq (\ref{bsmass}). Hence the equivalence
between the bound state and collective approaches can only be
established in the large symmetry breaking limit. This statement is 
further supported by comparing the recent computations of the 
electric quadrupole moments of the $\frac{3}{2}^+$ baryons. In the 
bound state approach all these moments are found to be proportional 
to the baryon isospin \cite{Oh95a}. In the collective approach this 
proportionality only appears in the large symmetry breaking limit, 
while for small breaking the quadrupole moments happen to be linked 
to the baryon charge \cite{Kr94}, see also section \ref{sec_slowrot}.
In any event, comparing the results displayed in tables \ref{ta_mdiffps}
and \ref{ta_bsdiff} unambiguously shows that the collective approach 
provides better agreement with the experimental mass differences 
than the bound state approach when the parameters of the Skyrme model
are determined from meson data.

\bigskip

\begin{subsatz}
\label{sec_heavy}{\bf \hskip1cm Baryons with a Heavy Quark}
\addcontentsline{toc}{section}{\protect\ref{sec_heavy}
Baryons with a Heavy Quark}
\end{subsatz}
\stepcounter{section}

In this section a brief remark on an additional application of the 
bound state approach will be given. This application concerns the 
description of baryons containing one (very) heavy quark like charm 
or bottom. For a more detailed survey article the reader may consult 
ref \cite{Min95}. Although these investigations are outside the three 
flavor regime they are worth mentioning here because they are based on 
the bound state approach described in section \ref{sec_fluc}.

The heavy baryons are constructed from a $D$ or $B$ meson bound 
in the background field of the soliton. The straightforward 
generalization of this treatment consists of substituting the 
masses and decay constants of such heavy mesons into the bound 
state equation (\ref{eqmkbs}) \cite{Rho92}. However, this treatment 
does not incorporate the heavy quark symmetry \cite{Ei81,Ne94} which 
(among other features) states that pseudo--scalar and vector mesons 
containing one heavy quark are degenerate. Hence an effective meson
Lagrangian suitable for the soliton picture of heavy baryons should 
contain heavy vector meson fields ($Q_\mu$) in addition to the heavy 
pseudo--scalar fields ($P$). The interaction among the light degrees of 
freedom should be described in accordance to chiral symmetry. 
Since the heavy mesons are composed of a heavy quark ($c,b$)
and a light (anti)quark (up, down or strange) this concept describes 
the interaction of the heavy and light mesons. The latter fields 
support the soliton.

When constructing the part of the action, which describes the 
interaction of the heavy and light mesons, one minimally extents the 
chirally and heavy quark symmetric action to finite masses, $M$, of 
the heavy meson \cite{Sch93b}
\begin{eqnarray}
{\cal L}_H&=&D_\mu P\left(D^\mu P\right)^{\dag}
-\frac{1}{2}Q_{\mu\nu}\left(Q^{\mu\nu}\right)^{\dag}
-M^2PP^{\dag}+M^{*2}Q_\mu Q^{\mu{\dag}}
\nonumber \\ &&
+2iMd\left(Pp_\mu Q^{\mu{\dag}}-Q_\mu p^\mu P^{\dag}\right)
-\frac{d}{2}\epsilon^{\alpha\beta\mu\nu}
\left[Q_{\nu\alpha}p_\mu Q_\beta^{\dag}+
Q_\beta p_\mu \left(Q_{\nu\alpha}\right)^{\dag}\right]
\label{lagheavy} \\ &&
-\frac{2\sqrt{2}icM}{m_V}\left\{
2Q_\mu F^{\mu\nu}\left(\rho\right)Q_\nu^{\dag}
+\frac{i}{M}\epsilon^{\alpha\beta\mu\nu}\left[
D_\beta PF_{\mu\nu}\left(\rho\right)Q_\alpha^{\dag}
+Q_\alpha F_{\mu\nu}\left(\rho\right)\left(D_\beta P\right)^{\dag}
\right]\right\}.
\nonumber
\end{eqnarray}
Here the mass $M$ of the heavy pseudo--scalar meson $P$ is allowed 
to differ from the mass $M^*$ of the heavy vector meson, $Q_{\mu}$.
Note that the heavy mesons are conventionally defined as {\it row}
vectors in isospace. When light vector mesons are included,
the covariant derivative introduces the additional parameter
$\alpha$:
\begin{eqnarray}
D_\mu P^{\dag}=\left\{\partial_\mu-i\alpha g \rho_\mu
+\frac{1-\alpha}{2}\left(\xi\partial_\mu\xi^{\dag}+
\xi^{\dag}\partial_\mu\xi\right)\right\} P^{\dag} \ .
\label{covder}
\end{eqnarray}
The covariant field tensor of the heavy vector
meson is defined as
\begin{eqnarray}
\left(Q_{\mu\nu}\right)^{\dag}=
D_\mu Q_\nu^{\dag}-D_\nu Q_\mu^{\dag}.
\label{heavyft}
\end{eqnarray}
The parameters $d$, $c$ and $\alpha$ have still not been very 
accurately determined. Commonly $d\approx0.53$ and $c\approx1.60$ 
are used \cite{Sch93b}. The assumption of vector meson dominance for 
the electromagnetic form factors of the heavy mesons suggests 
$\alpha\approx1$ \cite{Ja95}, although other values are allowed 
as well.

\begin{figure}[t]
\centerline{\hskip -2.0cm
\epsfig{figure=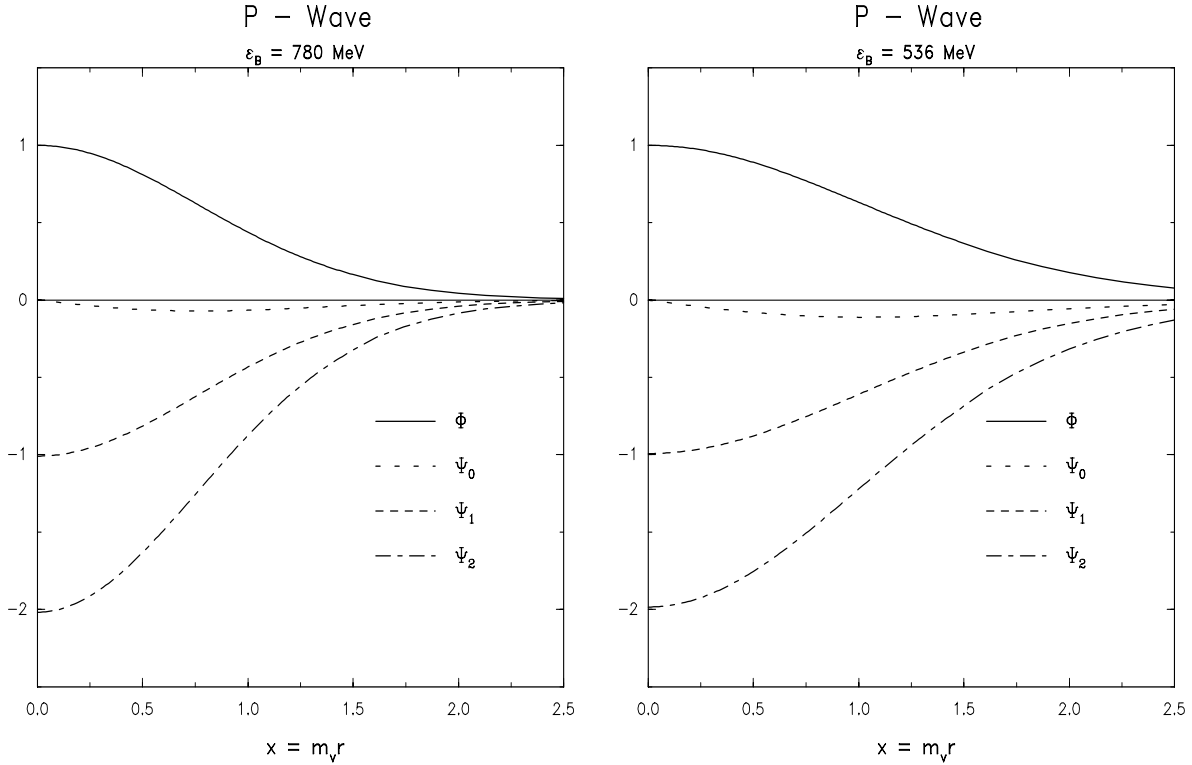,height=9.0cm,width=16.0cm}}
\caption{\label{fi_heavy}\tenrm
The radial dependence of the heavy meson bound state wave 
functions in both the bottom (left) and charm (right) sectors. 
Here ${\Ss m_V}$ is the light vector meson mass
(\protect\ref{vmpara1}). This figure is taken from ref
\protect\cite{Sch95c}.}
\end{figure}

Suitable {\it ans\"atze} for the P--wave heavy mesons are
\begin{eqnarray}
P^{\dag}&=&\frac{\Phi(r)}{\sqrt{4\pi}}\
\hat{\mbox{\boldmath $r$}}\cdot
\mbox{\boldmath $\tau$}\chi e^{i\omega t}, \qquad
Q_0^{\dag}=\frac{\Psi_0(r)}{\sqrt{4\pi}}\chi e^{i\omega t},
\nonumber \\
Q_i^{\dag}&=&\frac{1}{\sqrt{4\pi}}\left[
i\Psi_1(r)\hat r_i+\frac{1}{2}\Psi_2(r)\epsilon_{ijk}
\hat r_j\tau_k\right]\chi e^{i\omega t}
\label{pwavean}
\end{eqnarray}
which actually are motivated by the form of the induced fields for 
the collective rotation into the strange flavor direction 
(\ref{vmind2}). The isospinor $\chi$ contains spectral functions as in 
eq (\ref{pwansatz}). The bound state solutions for the heavy mesons 
have been constructed for the cases that the light soliton only contains 
the pseudo--scalar fields ({\it i.e.} the Skyrmion) \cite{Oh95,Sch95c}
as well as the inclusion of the light vector mesons in the soliton 
configuration as in eq (\ref{vmhedgehog}) \cite{Sch95c}. A consistent 
description of both the light and heavy baryon masses can only be 
obtained when the light vector mesons were included. This improvement
is due to the additional freedom in choosing the parameter $\alpha$.
For $\alpha\approx0$ the binding energy $\epsilon_b=780{\rm MeV}$ 
of the heavy baryon $\Lambda_b$ is reproduced for the experimental 
masses $M=5.28{\rm GeV}, M^*=5.33{\rm GeV}$. Using 
$M=1.87{\rm GeV}, M^*=2.01{\rm GeV}$ in the charm sector then 
predicts $\epsilon_c=536{\rm MeV}$ for the binding of $\Lambda_c$, 
which reasonably agrees with the experimental value of $630{\rm MeV}$.
The associated heavy meson wave--functions are displayed in 
figure \ref{fi_heavy}

Here just a brief introduction has been provided to the description of 
heavy quark baryons as composite systems of solitons and heavy mesons 
in order to demonstrate that the bound state approach can suitably 
be extented. Many aspects of this approach have not been addressed
here. In particular the issue of kinematical corrections has been
ignored \cite{Sch95d}. These corrections are supposedly sizable as a 
consequence of the large meson masses involved.

\vskip1cm
\begin{satz}
\label{chap_NJL}
{\large \bf \hskip1cm 
Explicit Quark Degrees of Freedom: The NJL Model}
\addcontentsline{toc}{chapter}{\protect\ref{chap_NJL}
Explicit Quark Degrees of Freedom: The NJL Model}
\end{satz}
\stepcounter{chapter}

The meson \cite{Vo91,Ha94} as well as the soliton \cite{Al95} sectors of
the Nambu--Jona--Lasinio \cite{Na61} (NJL) model have already been 
reviewed elsewhere. It is therefore beyond the scope of the present 
article to describe the this model in detail. Nevertheless a short 
discussion on the generals aspects of the model is in order to put 
the NJL model in perspective. The main interest in the NJL model is 
due to the fact that it constitutes a simple, chirally symmetric 
theory of the quark flavor dynamics while the flavor symmetry breaking 
is completely located in the current quark mass terms. This feature 
is analogous to QCD. Otherwise the NJL model is much simpler in
structure making possible the derivation of the effective meson theory, 
which is equivalent to the NJL Lagrangian. The derivation of this meson 
theory is commonly referred to as bosonization. The main issue of the 
present chapter will be to demonstrate that the previously discussed 
aspects of soliton models for three flavors can be obtained not only
in purely mesonic models, but also from such a microscopic theory of 
the quark flavor dynamics. For the sake of argument the NJL model 
will be employed although similar studies have been performed 
in bag models \cite{Pa93}.

\bigskip

\begin{subsatz}
\label{sec_quarkmot}{\bf \hskip1cm The NJL Soliton}
\addcontentsline{toc}{section}{\protect\ref{sec_quarkmot}
The NJL Soliton}
\end{subsatz}
\stepcounter{section}

In the NJL model the free Dirac Lagrangian is supplemented by a 
chirally symmetric four quark contact interaction \cite{Na61}
\be
{\cal L}_{\rm NJL}= \bar q (i\dslash - \hat m^0 ) q +
      2G_{\rm NJL} \sum _{a=0}^{8}
\left( (\bar q {Q^a} q )^2
      +(\bar q {Q^a} i\gamma _5 q )^2 \right) .
\label{NJL}
\ee
Here $q$ denotes the quark field, which represents a vector 
in flavor space
\be
q=\pmatrix{u\cr d\cr s\cr}\ ,
\label{defqvec}
\ee
with each entry being a Dirac spinor. In the following isospin 
symmetry is assumed, {\it i.e.} the current quark mass matrix 
is taken to be $\hat m^0={\rm diag}(m^0,m^0,m^0_s)$. The flavor 
matrices $Q^a=(\frac{1}{3},\frac{\lambda^1}{2},\ldots,
\frac{\lambda^8}{2})$ are identical to those employed in the 
preceding chapters. The dimensionful coupling constant, 
$G_{\rm NJL}$ will shortly be determined from meson properties.
Except for the quark mass term this Lagrangian is invariant under 
the chiral transformation
\be
q_L\rightarrow Lq_L\ ,\quad
q_R\rightarrow Rq_R\ ,\quad
q_{R,L}=\frac{1}{2}\left(1\pm\gamma_5\right)q\ ,
\label{qchitrans2}
\ee
which has already been considered earlier (\ref{qchitrans}).

Using functional integral bosonization techniques the quark fields 
can be integrated out from the model Lagrangian (\ref{NJL}) in favor 
of composite meson fields \cite{Eg74,Eb86}. The resulting effective 
action is the sum, $\A_{NJL}=\A_F+\A_m$, of a fermion determinant
\be
\A_F=\Tr\log(i\Dslash)=\Tr\log_\Lambda\left\{i\dslash-
\frac{1}{2}\left[\left(1+\gamma_5\right)M+
\left(1-\gamma_5\right)M^{\dag})\right]\right\}
\label{fdet}
\ee
and a purely mesonic part
\be
\A_m=\int d^4x\left\{-\frac{1}{4G_{\rm NJL}}
\tr\left(M^{\dag}M-\hat m^0(M+M^{\dag})+(\hat m^0)^2\right)\right\}.
\label{ames}
\ee
The complex matrix $M=S+iP$ parametrizes the scalar and pseudo--scalar 
meson fields. From eq (\ref{fdet}) it can be deduced that the 
meson fields behave like
\be
M\rightarrow L M R^{\dag} 
\label{mchitr}
\ee
under the chiral transformation (\ref{qchitrans2}).
Employing the polar decomposition, $M=U\Phi$, the transformation 
property (\ref{uchitrans}) of the chiral field becomes apparent.
As the quark coupling constant is dimensionful, the 
functional trace (\ref{fdet}) is not renormalizable and requires 
regularization. This is indicated by the cut--off $\Lambda$, which 
apparently acquires a physical meaning. As in the original study 
\cite{Re88} of the NJL soliton, this cut--off will be introduced via 
the $O(4)$ invariant proper time regularization \cite{Sch51}, which 
is applied to the fermion determinant in Euclidean space. After 
Wick--rotating ($t=-ix_4=-i\tau$) to Euclidean space, the fermion 
determinant in general is a complex quantity, $\A_F=\A_R+\A_I$. The 
proper time regularization scheme consists of replacing the real part, 
$\A_R$ by a parameter integral with the short distance contribution 
cut off
\be
\A_R=\frac{1}{2}\Tr\log_\Lambda\left(\Dslash_E^{\dag}\Dslash_E\right)
=-\frac{1}{2}\int_{1/\Lambda ^2}^\infty
\frac{ds}s\Tr\exp\left(-s\Dslash_E^{\dag}\Dslash_E\right) .
\label{arreg}
\ee
Although the imaginary part
\be
\A_I=\frac{1}{2}\Tr\log\left((\Dslash_E^{\dag})^{-1}\Dslash_E\right)
\label{af3}
\ee
is finite it will be regularized\footnote{Regularization of the 
imaginary part is supposed to incorrectly reproduce the anomalous decay
$\pi^0\rightarrow\gamma\gamma$ \cite{Bl88}.  More recently it has been 
speculated \cite{Al93b} that this short--coming may be cured by higher 
order contributions to the ABJ triangle anomaly \cite{ABJ69} when the 
cut--off is finite.} as well. 

Although the starting point (\ref{NJL}) has been a quark theory, the 
bosonization procedure has led to an action ($\A_R+\A_I$), which 
essentially is a functional of the meson configuration $M$. Expanding 
this action in derivatives of $M$ yields Lagrangians which are 
similar to one of the Skyrme model \cite{Eb86}. However, models 
obtained in this fashion may not necessarily support soliton solutions. 
Therefore the complete fermion determinant will be retained and exactly
be computed for a given configuration $M$. In doing so, it will be 
recognized that the notion of quarks cannot be abandoned completely.

In order to determine the parameters of the model, the real part $\A_R$ 
is expanded up to quadratic order in the meson fields and subsequently 
continued back to Minkowski space. In this context a convenient 
parametrization for the scalar and pseudo--scalar fields is \cite{We93}
\be
M=\xi_0\xi_f\Sigma\xi_f\xi_0.
\label{defm}
\ee
The matrix $\Sigma$ is Hermitian whereas the matrices $\xi_0$ and
$\xi_f$ are unitary. The space--time dependent pseudo--scalar meson 
fluctuations $\eta_a(x)$ are contained in
$\xi_f(x)={\rm exp}(i\sum_{a=0}^8\eta_a(x)Q^a)$, while the quantity 
$\xi_0$ has been introduced to simplify the inclusion of the 
soliton. In the baryon number zero sector $\xi_0$
is replaced by the unit matrix. Varying the action with respect to
$\Sigma$ yields the Schwinger--Dyson or gap equation, which determines 
the vacuum expectation value 
$\langle \Sigma \rangle ={\rm diag}(m,m,m_s)$ 
\be
m_i & = & m_i^0+m_i^3\frac{N_C G_{\rm NJL}}{2\pi^2}
\Gamma\left(-1,(\frac{m_i}{\Lambda})^2\right)
=m_i^0-2G_{\rm NJL} \langle \bar q q \rangle _i .
\label{conmass}
\ee
Here $m_i=m,m_s$ denote the constituent masses of non--strange and 
strange quarks, respectively. The appearance of a non--vanishing quark 
condensate $\langle \bar q q \rangle _i$ is the outstanding feature of 
the spontaneous breaking of chiral symmetry \cite{Mi93}. Henceforth 
the scalar fields will be approximated by their vacuum expectation 
values, $\Sigma=\langle\Sigma\rangle$. By expanding the action in 
the pseudo--scalar fields $\eta^a$ their inverse propagator can be 
extracted \cite{We92,Ja92}. The on--shell condition requires this
meson propagator to possess a pole at the physical meson 
masses, $m_{\rm phys}$,
\be
\frac{(m_i^0+m_j^0)(m_i+m_j)}{2G_{\rm NJL}}+
\Pi_{ij}(m_{\rm phys}^2)=0\ .
\label{BSeq}
\ee
Here the polarization operator 
\be
\Pi_{ij}(q^2)=-2q^2f_{ij}^2(q^2)+2(m_i-m_j)^2f_{ij}^2(q^2)
-\frac{1}{2}(m_i^2-m_j^2)\Big(\frac{\langle\bar q q \rangle_i}{m_i}-
\frac{\langle\bar q q \rangle_j}{m_j}\Big)
\label{polar}
\ee
is given in terms of the quark condensates, $\langle\bar q q \rangle_i$
and the off--shell meson decay constants
\be
f^2_{ij}(q^2)=\frac{1}{4}(m_i+m_j)^2\frac{N_c}{4\pi^2}\int_0^1dx\
\Gamma\Big(0,[(1-x)m_i^2+xm_j^2-x(1-x)q^2]/\Lambda^2\Big).
\label{fq}
\ee
The Feynman parameter integral reflects the quark loop contained 
in the bosonized form (\ref{fdet}) of the NJL model action. From 
eq (\ref{fq}) the on--shell meson decay constants can be read off 
\be
f_\pi^2&=&m^2 \frac {N_c}{4\pi ^2} \int _0^1 dx
\Gamma\Big(0,[m^2-x(1-x)m_\pi^2]/\Lambda ^2\Big)
\label{fpi} \\
f_K^2&=&\frac 14 (m+m_s)^2 \frac {N_c}{4\pi ^2} \int _0^1 dx
\Gamma\Big(0,[xm^2+(1-x)m_s^2-x(1-x)m_K^2]/\Lambda ^2\Big) .
\label{fK}
\ee
Substituting in eq (\ref{polar}) the coupling constant, 
$G_{\rm NJL}$ by the quark condensate $\langle {\bar q}q\rangle$ 
via the gap--equation (\ref{conmass}), immediately leads to the 
Gell-Mann--Oakes--Renner (GMOR) relations, 
$f_\pi^2m_\pi^2=m^0\langle {\bar u}u +{\bar d}d\rangle$, and 
similarly for $f_K$  \cite{Ge68} .

The above eqs (\ref{conmass})--(\ref{fK}) relate the parameters of
the NJL model, $G_{\rm NJL}$, $\Lambda $ $m^0$ and $m_s^0$, to the
physical masses and decay constants. Using $m_\pi=135$MeV, $m_K=495$MeV 
and fixing the pion decay constant $f_\pi=93$MeV yields too small a 
value for the kaon decay constant\footnote{Determining $f_\pi$ or $f_K$ 
fixes the ratio $\Lambda /m$. This leaves one adjustable parameter, 
which for convenience is chosen to be the up constituent mass, $m$.}, 
see table \ref{ta_njlmes}. On the other hand, requiring $f_K=113$MeV 
gives too large a value for $f_\pi$.
\begin{table}
\caption{\label{ta_njlmes}\tenrm
The up and strange constituent and current masses, the cutoff
as well the pion and kaon decay constants in the NJL model. 
These results are taken from \protect\cite{We94a}.}
~
\newline
\centerline{\tenrm\smalllineskip
\begin{tabular}{ c c c c c c }
$m$ (MeV) & $m_s$ (MeV) & $m^0_s/m^0_u$
& $\Lambda$ (MeV) & $f_\pi$ (MeV) & $f_K$ (MeV) \\
\hline
350&577&23.5&641&93.0 &104.4\\
400&613&22.8&631&93.0 &100.3\\
450&650&22.4&633&93.0 & 97.4\\
500&687&22.3&642&93.0 & 95.5\\
\hline
350&575&24.3&698& 99.3&113.0\\
400&610&23.9&707&103.0&113.0\\
450&647&23.6&719&105.7&113.0\\
500&685&23.4&734&107.9&113.0\\
\end{tabular}}
\end{table}
In view of the discussion after eq (\ref{gasky}) the too small 
prediction for the ratio $f_K/f_\pi$ gives rise to the suspicion 
that the NJL model will underestimate the baryon mass differences.

For the discussion of static soliton configurations it is 
appropriate to introduce the Dirac Hamiltonian $h$
\be
i\beta\Dslash_E=-\partial_\tau-h\ ,
\label{de1}
\ee
which is time independent, {\it i.e.} $[\partial_\tau,h]=0$. 
Furthermore it will be assumed that $h$ is Hermitian. Denoting the 
eigenvalues of $h$ by $\epsilon_\nu$, the action $\A_F$ (\ref{fdet}) 
is found \cite{Da75,Ra82,Re89} to be decomposable into a vacuum part
\be
{\cal A}_0=\frac{1}{2}T\sum_\nu|\epsilon_\nu|
\label{avac}
\ee
and valence (anti--)quark parts
\be
{\cal A}_V^{\{\eta_\nu\}}=-T\sum_\nu \eta_\nu|\epsilon_\nu|
=-TE_V^{\{\eta_\nu\}} .
\label{aval}
\ee
Here $T$ denotes the Euclidean time interval under consideration.
The total action involves all possible sets of occupation numbers,
$\eta_\nu=0,1$, {\it i.e.} $\A_F=\A_0+{\rm log}\sum_{\{\eta_\nu\}}
{\cal A}_V^{\{\eta_\nu\}}$. These occupation numbers furthermore 
specify the baryon number
\be
B=\sum_\nu\left(\eta_\nu-\frac{1}{2}\right){\rm sgn}(\epsilon_\nu).
\label{bnonjl}
\ee
The soliton configuration uniquely determines the eigenvalues
$\epsilon_\nu$. Hence these occupation numbers functionally depend on 
the field configuration as well once the baryon number is fixed. The 
vacuum contribution, ${\cal A}_0$ is obviously divergent. The 
standard recipe to extract the vacuum part from the regularized 
determinant\footnote{The imaginary part vanishes as long as $h$ 
is Hermitian and time independent.} (\ref{arreg}) is to consider 
the limit $T\rightarrow\infty$. In that limit the eigenvalues of the 
differential operator $\partial_\tau$ lie densely and the sum over 
these eigenvalues is replaced by an integral
\be
\A_0=-\frac{T}{2}\sum_\nu\int_{-\infty}^{\infty}\frac{dz}{2\pi}
\int_{1/\Lambda^2}^\infty \frac{ds}{s}\
{\rm \exp}\left\{-s\left(z^2+\epsilon_\nu^2\right)\right\}\ .
\label{arstatic}
\ee
This expression allows one to read off the vacuum part of the 
energy, $E_0$, in the proper time regularization scheme because 
$A_0\rightarrow-TE_0$ when $T\rightarrow\infty$.

As in the purely mesonic models, the soliton configuration in the 
baryon number one sector is assumed to be of hedgehog structure
\be
\xi_0(\mbox{\boldmath $r$})=
{\rm exp}\left\{\frac{i}{2}
\mbox{\boldmath $\tau$}\cdot
\hat{\mbox{\boldmath $r$}}F(r)\right\}\ ,
\label{njlhedgehog}
\ee
(for the moment, the time dependent fluctuations are 
ignored ($\xi_f=1$)). Then the Dirac Hamiltonian becomes
\be
h=\mbox {\boldmath $\alpha$}\cdot\mbox{\boldmath $p$} +
\pmatrix{m\ {\rm exp}\left(i\gamma_5{\mbox{\boldmath $\tau$}}
\cdot\hat{\mbox{\boldmath $r$}} F(r)\right) 
&{| \atop |} & \hspace{-10pt} 0\cr
---------\hspace{-10pt}&-&\hspace{-10pt}-\cr 0 &|& 
\hspace{-8pt}m_s\cr} \ .
\label{h0}
\ee
As in the meson models the static soliton configuration does not 
effect the strange degrees of freedom.
In order to compute $E_V^{\{\eta_\nu\}}$ the occupation numbers
have to be determined from eq (\ref{bnonjl}) for a given profile 
function $F(r)$. In the case of $B=1$ the vacuum contribution 
$-(1/2)\sum_\nu {\rm sgn}(\epsilon_\nu)$, is either zero or one. 
Hence at most one of the $\eta_\nu$ can be different from zero. 
Minimization of the total energy enforces this to be the one with 
the lowest absolute value of the energy eigenvalue, $|\epsilon_\nu|$. 
This level is referred to as the valence quark state. The only 
non--vanishing occupation therefore is 
$\eta_{\rm val}=[1+{\rm sgn}(\epsilon_{\rm val})]$. The total 
energy functional is finally given by
\be
E_{\rm tot}[F]&=&E_0[F]-E_0[F\equiv0]
+\frac{N_C}{2}\eta_{\rm val}\epsilon_{\rm val}
+E_m\ , 
\label{etnjl} \\
E_0[F]&=&\frac{N_C}{2}\int^\infty_{1/\Lambda^2}
\frac{ds}{\sqrt{4\pi s^3}}\sum_\nu{\rm exp}
\left(-s\epsilon_\nu^2\right)\ , \quad
E_m=m_\pi^2 f_\pi^2\int d^3r  \left(1-\cF(r)\right),
\hspace{1.0cm}
\label{e0njl}
\ee
where the dependence on the number of colors has been made explicit
and a constant contribution, which is associated with the energy 
of the trivial meson configuration, has been subtracted. The mesonic 
part of the energy, $E_m$ originates from $A_m$ (\ref{ames}) after 
substituting $G_{\rm NJL}=m^0m/m_\pi^2f^2_\pi$ \cite{Eb86}, see also 
eqs (\ref{polar}) and (\ref{BSeq}).

Soliton configurations, which minimize the functional $E[F]$, indeed
exist for $m\ge350{\rm MeV}$ \cite{Re88}. These solitons are constructed 
by iteration, {\it i.e.} the Hamiltonian $h$ (\ref{h0}) is diagonalized 
for a test profile $F(r)$. The resulting eigenvalues and eigenvectors 
are substituted into the equation of motion $\delta E/\delta F(r)=0$ 
to update the profile function. This procedure is repeated until 
convergence is achieved\footnote{See ref \cite{Al94} for a detailed 
prescription of the numerical treatment. A parametrical solution to 
the minimization problem was considered in ref \cite{Di89}.}. This 
procedure, of course, provides the eigenvalues and eigenstates of the 
Dirac Hamiltonian, $\epsilon_\nu$ and $|\nu\rangle$, respectively. The 
resulting profile function, $F(r)$ is similar to the one displayed in 
figure \ref{fi_skyrme}, however, for $B=1$ the equation of motion only 
allows a solution with the reversed sign, $F(0)=-\pi$. This is in 
contrast to the purely mesonic models discussed in the previous 
chapters. The reason being that the quarks fields are maintained. This 
fixes the sign of the (anomalous) baryon current unambiguously whereas 
the sign of the Wess--Zumino (\ref{WZterm}) term could not be 
determined from meson properties. For the self--consistent soliton of 
the NJL model the various contributions to the energy are displayed in 
table \ref{ta_etheta}. By expanding the action in arbitrary fluctuations 
off the hedgehog configuration it can furthermore be shown, that the term 
linear in these fluctuations vanishes when $\delta E/\delta F(r)=0$
is satisfied \cite{We93}. Hence the hedgehog configuration represents 
a true solution to the full Euler--Lagrange equations.
\begin{table}
\caption{\label{ta_etheta}\tenrm
The soliton energy ${\Ss E_{\rm tot}}$ and its various contributions
according to the sum (\protect\ref{etnjl}) as functions of the
constituent quark mass ${\Ss m}$. All numbers are in MeV.}
~
\newline
\centerline{\tenrm\smalllineskip
\begin{tabular}{l|c c c c c c }
$m$           & 350 & 400 & 500 & 600 & 700 & 800 \\
\hline
$E_{\rm tot}$ &1236 &1239 &1221 &1193 &1161 &1130 \\
$E_V$         & 745 & 633 & 460 & 293 & 121 & -55 \\
$E_0$         & 459 & 571 & 728 & 869 &1012 &1103 \\
$E_m$         &  31 &  34 &  33 &  31 &  28 &  26 \\
\end{tabular}}
\end{table}
Obviously the total energy is quite insensitive to variations of the 
constituent quark mass. 

A major difference between the soliton in 
the NJL model and the soliton solutions considered in the previous 
chapters is the non--topological character of the NJL model soliton. 
As a consequence, this soliton does not necessarily represent the 
global minimum in the $B=1$ sector. Indeed, the soliton energy is 
larger than the three quark threshold as long as 
$m{_<\atop^\sim}420{\rm MeV}$ and a continuous deformation can
be constructed of the soliton to the configuration of three 
non--interacting quarks. Furthermore, the valence quark 
energy stays positive for a large range in the parameter 
space implying that the baryon number is carried by the explicit 
occupation of this orbit, {\it i.e.} $\eta_{\rm val}=1$. This
illustrates the above mentioned feature that the notion of quarks 
cannot be abandoned completely despite the model being a functional of 
meson fields. At this point it should be mentioned that this situation 
changes drastically when (axial)vector meson fields are added to the 
NJL model Lagrangian. For the corresponding self--consistent soliton 
the valence quark is strongly bound and the baryon number is actually 
carried by the polarized vacuum, {\it i.e.} $\eta_{\rm val}=0$ 
\cite{Al92,Do92}. This feature has been taken as a strong support 
for Witten's conjecture \cite{Wi83} that in effective meson theories 
the baryon number current is identical to the topological current, 
$B_\mu$, defined in eq (\ref{bnumber}). The argument goes as 
follows \cite{Al92}: The leading term in the gradient expansion of 
the baryon number current in the NJL model is indeed the topological 
current \cite{Go81}. However, the gradient expansion exclusively takes 
into account the vacuum part of the action, while explicitly occupied 
levels are ignored. Hence for these currents to be equivalent, the 
baryon number must be carried by the vacuum. This, apparently, is 
the case for the self--consistent soliton in the NJL model with 
(axial)vector meson fields. These results of the (axial)vector meson 
NJL model may furthermore be interpreted as an indication that explicit
valence quark and (axial)vector meson fields should not simultaneously 
be contained in a model. This conclusion that keeping both may cause 
double counting effects has also been reached from phenomenological 
studies on baryon properties in soliton models \cite{Ja89,Jo90}

\bigskip

\begin{subsatz}
\label{sec_yamasses}{\bf \hskip1cm
The Collective Approach to the NJL Model}
\addcontentsline{toc}{section}{\protect\ref{sec_yamasses}
The Collective Approach to the NJL Model}
\end{subsatz}
\stepcounter{section}

In this section the parameters in the collective Hamiltonian 
(\ref{collham}) will be derived within the context of the pseudo--scalar 
NJL model, which is defined in eq (\ref{NJL}). For this purpose the 
flavor rotating field configuration is parametrized as
\be
M(\mbox{\boldmath $r$},t)=A(t)\xi_0(\mbox{\boldmath $r$})A^{\dag}(t)
\langle\Sigma\rangle A(t)\xi_0(\mbox{\boldmath $r$})A^{\dag}(t)
\qquad A(t)\in SU(3) \ ,
\label{collansatz}
\ee
which represents the NJL model equivalent to eq (\ref{su3rot}).
Obviously, only the pseudo--scalar fields rotate in flavor space while 
the scalar fields are kept at their vacuum expectation values. In order 
to evaluate the fermion determinant (\ref{fdet}) for this field 
configuration, it is appropriate to transform the quark fields to the 
flavor rotating frame \cite{Re89}: $q=Rq^\prime$. This transformation 
eliminates the ``outer" rotations in (\ref{collansatz}) at the expense 
of an induced rotational part
\be
h_{rot}=\sum_{a=1}^8Q^a\Omega_a\ ,
\label{hrot}
\ee
with the angular velocities defined as in eq (\ref{defomega}).
In the rotating frame the Dirac operator reads
\be
i\beta\Dslash\ ^\prime=i\partial_t-h-h_{rot}-h_{SB}\ ,
\label{drot}
\ee
where $h$ is the static one--particle Hamiltonian (\ref{h0}). The 
symmetry breaking part of the Dirac operator is compactly displayed
by introducing 
${\cal T}=(\xi_0^{\dag}+\xi_0)/2+\gamma_5(\xi_0^{\dag}-\xi_0)/2$ 
\be
h_{SB} &=& \T\beta\left(A^{\dag}\langle\Sigma\rangle A
-\langle\Sigma\rangle\right)\T^{\dag}
\nonumber \\
&=&\frac{2(m-m_s)}{\sqrt3}\T\beta\left(
\sum_{i=1}^3 D_{8i}Q^i
+\sum_{\alpha=4}^7 D_{8\alpha}Q^\alpha
+(D_{88}-1)Q^8\right)\T^{\dag}.
\label{hsb}
\ee
The $SU(3)$ D--functions are those of eq (\ref{omtrans}), the 
argument being the collective rotation $A$. The $SU(2)$ invariant 
pieces have also been indicated. The computation proceeds by 
expanding the regularized action in powers of $h_{rot}$ and $h_{SB}$, 
which are approximated to be time independent. Furthermore, only 
expansions up to quadratic order have been considered in the 
literature \cite{We92a,We92,Bl92,Bl93}. As in the case of the 
classical energy, the action separates into valence quark and 
vacuum pieces. The valence quark piece is analyzed by treating the 
extended Dirac equation for the valence quark state
\be
\left(h+h_{rot}+h_{SB}\right)\Psi_{\rm val}=
\epsilon_{\rm val}\Psi_{\rm val}
\label{dirrotsb}
\ee
in stationary perturbation theory. The expansion of the vacuum part 
is more involved. Since regularization is mandatory, the detour over 
Euclidean space is unavoidable. In Euclidean space $h_{rot}$ is an 
anti--Hermitian quantity since it is linear in the  time derivative. 
Stated otherwise, $\Omega_a^{(E)}=i\Omega_a$ is considered to be a real 
quantity. Then the real and imaginary parts of the action are
\be
\A_R&\hspace{-0.3cm}=\hspace{-0.3cm}&
-\frac{1}{2}\int_{1/\Lambda^2}^\infty
\frac{ds}{s}{\rm Tr}\ {\rm exp}\left\{-s\left[
-\partial_\tau^2 + h^2+\{h,h_{SB}\}+h_{SB}^2
+[h,h_{rot}] - h_{rot}^2\right]\right\}\ ,
\hspace{1.0cm}
\label{arcoll} \\
\A_I&\hspace{-0.3cm}=\hspace{-0.3cm}&
\frac{1}{2} {\rm Tr}\ \left\{\left[
\left(\partial_\tau-h-h_{SB}\right)
\left(-\partial_\tau-h-h_{SB}\right)\right]^{-1}
\{h_{rot},h+h_{SB}\}\right\}+\ldots 
\label{aicoll0} \\
&\hspace{-0.3cm}\rightarrow\hspace{-0.3cm}&
\frac{1}{2}\int_{1/\Lambda^2}^\infty ds
{\rm Tr}\ \{h_{rot},h+h_{SB}\}
{\rm exp}\left\{-s\left[
-\partial_\tau^2 + h^2+\{h,h_{SB}\}+[h,h_{rot}]\right]\right\}
+\ldots \hspace{0.5cm}\ .
\label{aicoll}
\ee
Here the imaginary part has been regularized in a way consistent 
with the introduction of the cut--off, $\Lambda$ in eq (\ref{arcoll}).
Note also that the expression (\ref{arcoll}) is still exact, while 
eq (\ref{aicoll0}) already represents an expansion in $h_{rot}$ and 
$h_{SB}$. Abbreviating the arguments of the exponentials by $A_i$, 
where the subscript labels the order of the perturbation, these 
expressions are analyzed using the general formula for an expansion 
up to quadratic order
\be
e^{A_0+A_1+A_2}&=&e^{A_0}
+\int_0^1d\zeta e^{\zeta A_0}\left(A_1+A_2\right)
e^{\left(1-\zeta\right)A_0}
\nonumber \\ && \hspace{1cm}
+\int_0^1d\zeta\int_0^{1-\zeta}d\eta e^{\eta A_0}A_1
e^{\left(1-\zeta-\eta\right)A_0}A_1e^{\zeta A_0}+\ldots \ ,
\label{genexp}
\ee
which introduces Feynman parameter integrals. Since the perturbation
is taken to be static the temporal part of the functional trace is 
straightforwardly converted into Gaussian type integrals as in eq 
(\ref{arstatic}). The remainder of the trace is evaluated using the 
eigenstates, $|\mu\rangle$ of the Dirac Hamiltonian in the background 
of the chiral soliton (\ref{h0}). Having obtained the expanded 
Euclidean action, the coefficients of the collective Hamiltonian 
(\ref{collham}) are extracted after continuing back to Minkowski 
space ($\Omega_a^{(E)}\rightarrow\Omega_a=-i\Omega_a^{(E)}$).
These coefficients are listed in appendix D of ref \cite{Al95} and
do not need to be repeated here. For illustration, however, the 
matrix $\Theta_{ab}$, which defines the term of the collective 
Lagrangian bilinear in the angular velocities,
$L=(1/2)\Theta_{ab} \Omega_a\Omega_b+\ldots$ is shown because it 
contains the moments of inertia. According to the above mentioned 
decomposition of the action, $\Theta_{ab}$ is the sum
\be
\Theta_{ab}=\eta_{\rm val}\Theta^{\rm val}_{ab}
+\Theta^{\rm vac}_{ab}.
\label{thab1}
\ee
The valence quark piece is given by
\be
\Theta^{\rm val}_{ab}&=&2N_C\sum_{\mu\ne {\rm val}}
\frac{\langle{\rm val}|Q^a|\mu\rangle \langle\mu|Q^b|{\rm
val}\rangle} {\epsilon_\mu-\epsilon_{\rm val}}.
\label{thval}
\ee
In contrast to the purely mesonic models the cranking type structure 
\cite{In54,Ri80} of the inertial parameters is apparent. The vacuum 
contribution to the moment of inertia is obtained to be \cite{Re89,We92}
\be
\Theta^{\rm vac}_{ab}=2N_C\sum_{\mu\nu}
f_\Lambda(\epsilon_\mu,\epsilon_\nu)
\langle\mu|Q^a|\nu\rangle
\langle \nu|Q^b|\mu\rangle
\label{thvac}
\ee
where the cut-off function 
\be
f_\Lambda(\epsilon_\mu,\epsilon_\nu)=\frac{\Lambda}{\sqrt{\pi}}
\frac{e^{-(\epsilon_\mu/\Lambda)^2}-e^{-(\epsilon_\nu/\Lambda)^2}}
{\epsilon_\nu^2-\epsilon_\mu^2}
-\frac{{\rm sgn}(\epsilon_\nu)
{\rm erfc}\left(\left|\frac{\epsilon_\nu}{\Lambda}\right|\right)
-{\rm sgn}(\epsilon_\mu)
{\rm erfc}\left(\left|\frac{\epsilon_\mu}{\Lambda}\right|\right)}
{2(\epsilon_\mu-\epsilon_\nu)}
\label{thvacreg}
\ee
has the interesting property that it vanishes in the case
$\epsilon_\mu=\epsilon_\nu$ \cite{Re89}. This causes $\Theta_{ab}$
to vanish when the soliton is absent because the quark matrix elements 
of the flavor matrices $Q^a$ are diagonal for $F\equiv0$, {\it i.e.} 
$\langle \nu|Q^a|\mu\rangle\propto\delta_{\mu\nu}$. This feature just 
reflects rotational invariance, which is violated in the presence of 
the soliton\footnote{Special care has to be taken when choosing the
boundary conditions for the eigenstates diagonalizing $h$. Such 
boundary conditions may violate rotational invariance resulting 
in $\Theta_{ab}\ne0$ even for $F\equiv0$. Nevertheless a proper 
choice is possible, see appendix B of ref \cite{We92}.}. The moments 
of inertia for rotations into different directions correspond to 
certain choices of flavor matrices, {\it e.g.} $\alpha^2=\Theta_{33}$. 
For the strange moment of inertia, $\beta^2$ the contribution 
($\beta_I$) of the induced components (\ref{kaonind}) has been 
estimated in the gradient expansion, hence 
$\beta^2=\Theta_{44}+\beta_I$. These formulas for the moments of 
inertia have been obtained by expanding the action in $h_{rot}$. The 
expansion in $h_{rot}\times h_{SB}$ yields $\alpha_1$ and $\beta_1$. 
The distinction between these coefficients originates from the 
$SU(2)$ decomposition in eq (\ref{hsb}). Similarly, the expansion 
in $h_{SB}$ gives the contribution of the fermion determinant to
$\gamma$, $\gamma_T$, $\gamma_S$ and $\gamma_{TS}$, which acquire 
additional contributions by substituting the field configuration 
(\ref{collansatz}) into the mesonic part of the action $\A_m$ 
(\ref{ames}). After eliminating the coupling constant via
$G_{\rm NJL}=m^0m/m_\pi^2f^2_\pi$, the strange current quark
mass, $m^0_s$ only appears in form of the ratio $m_s^0/m^0$. Since 
this ratio is known to be insensitive on the regularization 
prescription \cite{Ha94} the above described expansion scheme avoids 
uncertainties stemming from the choice of the regularization 
prescription. On the contrary, the absolute values of the current 
quark masses are subject to major changes when the regularization 
scheme is altered.

The extraction of the coefficients of collective Hamiltonian from
the NJL model action is now completed and numerical results for 
the baryon mass differences will be discussed. Of course, these 
mass differences, which are displayed in table \ref{ta_njlcoll},
are obtained within the generalized Yabu--Ando approach, {\it i.e.} 
by exact diagonalization of the collective Hamiltonian according to 
the prescription outlined in appendix A.
\begin{table}
\caption{\label{ta_njlcoll}\tenrm
The mass differences of the low-lying ${\Ss \frac{1}{2}^+}$ and
${\Ss \frac{3}{2}^+}$ baryons with respect to the nucleon as obtained 
in the collective approach to the NJL model. The up-quark constituent 
mass ${\Ss m}$ is chosen such that the experimental 
${\Ss \Delta}$--nucleon mass difference is reproduced. The last 
column refers to the case when the symmetry breaker ${\Ss \gamma}$ 
is scaled by ${\Ss (f_K^{\rm expt.}/f_K^{\rm pred.})^2}$. All data 
(from ref \protect\cite{We93a}) are in MeV.}
~
\newline
\centerline{\tenrm\smalllineskip
\begin{tabular}{c|c c|c|c}
& $f_\pi$ fixed & $f_K$ fixed & Expt. & $f_K^{\rm corr.}$\\
\hline
$\Lambda$ & 105 & 109 & 177 & 175\\
$\Sigma$ & 148 & 151 & 254 & 248 \\
$\Xi$ & 236 & 243 & 379 & 396 \\
$\Delta$ & 293 & 293 & 293 & 291 \\
$\Sigma^*$ & 387 & 391 & 446 & 449 \\
$\Xi^*$ & 482 & 489 & 591 & 608 \\
$\Omega$ & 576 & 586 & 733 & 765 \\
\end{tabular}}
\end{table}
As already speculated earlier, the mass differences are underestimated 
in the collective approach to the NJL model. That this short--coming 
is due to the too small prediction for the ratio $f_K/f_\pi$ is 
also illustrated in table \ref{ta_njlcoll} by chosing two different 
set of parameters. First, $f_\pi=93$MeV is kept at its empirical value
and $m=407$MeV is chosen to reproduce the experimental
$\Delta$--nucleon mass difference yielding $f_K=99.8{\rm MeV}=
1.07f_\pi$. In the second set of parameters the cut--off $\Lambda$ 
is tuned to correctly give $f_K=114$MeV. Again $m=433$MeV is determined 
from the experimental $\Delta$--nucleon mass difference. Then 
$f_\pi=104.9$MeV is increased considerably, however, the ratio 
$f_K/f_\pi=1.09$ remains almost unaltered. Apparently the change in 
the mass differences for the two sets is not larger than the change 
for the ratio $f_K/f_\pi$. The sensitivity of the symmetry breaking 
parameters on $f_K$ has been traced in the gradient expansion \cite{We92}
\be
\gamma_{\rm grad.\ exp.}
&=&\frac{4}{3} \int d^3r \Bigg\{(m_K^2f_K^2-m_\pi^2f_\pi^2)(1-\cF)
\nonumber \\ && \hspace{2cm}
+\frac{f_K^2-f_\pi^2}{2}\cF\left(\fpt+\frac{2\sFt}{r^2}\right)
+\ldots\Bigg\} \ .
\label{grad1}
\ee
The dominating term in this expansion it the one involving 
$m_K^2f_K^2$. It is therefore intuitive to scale the vacuum
contribution to $\gamma$ by
$(f_K^{expt.}/f_K^{pred.})^2\approx\big(114/100\big)^2$ and 
re--evaluate the baryon spectrum. The resulting mass differences 
are in perfect agreement with the experimental data as can be 
observed from the last column of table \ref{ta_njlcoll}. Hence 
the problem of underestimating the mass differences in the collective 
approach is completely inherited from the meson sector of the model,
which is unable to properly reproduce the ratio $f_K/f_\pi$.

The expansion scheme employed above is somewhat different from the 
one used in refs \cite{Bl92,Bl93}. In the first place the meson
fields differ by a constant but flavor symmetry breaking amount,
$M\rightarrow M+{\hat m}^0$. Then the mesonic part, $\A_m$ vanishes 
and the current quark mass matrix is contained in the fermion 
determinant, $\A_F$. Of course, this re--parametrization does not make 
any difference when the Euler--Lagrange equations are solved exactly. 
However, in the collective approach the time dependent solutions are 
approximated by the rigidly rotation field configuration 
(\ref{collansatz}). Whence it is not obvious whether or not these 
parametrizations are equivalent. For simplicity the constituent quark 
masses have been set equal, {\it i.e.} $m=m_s$ in refs 
\cite{Bl92,Bl93}. This also implies the approximation $f_K=f_\pi$. In 
that treatment the remaining symmetry breaking term is of the structure 
(\ref{hsb}) too, but with the constituent masses replaced by the 
current masses. Furthermore a different regularization scheme has 
been used, which brings into the game one additional parameter. 
Nevertheless similar results for the baryon mass differences have been 
observed, in particular the symmetry breaking is underestimated when 
empirical values ($m^0\approx6{\rm MeV}$, $m_s^0\approx150{\rm MeV}$ 
\cite{PDG94}) are used. A mass splitting pattern similar to the one 
in the last column of table \ref{ta_njlcoll} was obtained when 
increasing $m_s^0$ by about 20\%. This is not surprising because the 
relation between $m_K$ and $m_s^0$ is almost linear as a consequence 
of the GMOR relations \cite{Ge68}. Thus the increase in $m_s^0$ is 
equivalent to a corresponding scaling of the symmetry breaking 
coefficient $\gamma$, {\it cf.} eq (\ref{grad1}).

Baryon properties can be computed similarly to the meson models. First,
the symmetry currents have to be obtained. The straightforward 
procedure is to extend the action by introducing external gauge fields,
$a_\mu^a(x)$, $i\partial_\mu\rightarrow i\partial_\mu+
a^a_\mu(x)\gamma^\mu\Gamma^a$, with $\Gamma^a$ being 
the generator of the symmetry under consideration. 
For example, $\Gamma^a\sim\mbox{\boldmath $\tau$}$ for the 
isovector vector current. The expressions linear in $a^a_\mu(x)$ 
are identified as the currents. These currents are again decomposed 
into a valence quark and vacuum piece
\be
j_\mu^a=N_C\left\{\eta_{\rm val}\ j^a_{{\rm val}\ \mu}+
j^a_{{\rm vac}\ \mu}\right\} \ .
\label{currnjl}
\ee
Denoting $\Psi_\mu(x)$ with the spatial representation of the 
eigenstates $|\mu\rangle$, a compact notation for the currents 
is possible
\be
j^a_{{\rm val}\ \mu}&=&
{\bar\Psi}_{\rm val}(x)A^{\dag}\gamma_\mu\Gamma^a A \Psi_{\rm val}(x)
+\sum_{\mu\ne{\rm val}}\left\{\Omega_a,
\frac{\langle{\rm val}|Q^a|\mu\rangle}{\epsilon_\mu-\epsilon_{\rm val}}
{\bar\Psi}_\mu(x)A^{\dag}\gamma_\mu\Gamma^a A\Psi_{\rm val}(x)\right\},
\hspace{1cm}
\label{jval} \\
j^a_{{\rm vac}\ \mu}&=&
-\frac{1}{2}\Bigg[\sum_\mu{\rm sgn}(\epsilon_\mu)
{\rm erfc}\left(\left|\frac{\epsilon_\mu}{\Lambda}\right|\right)
{\bar\Psi}_\mu(x)A^{\dag}\gamma_\mu\Gamma^a A \Psi_\mu(x)
\nonumber \\ &&\hspace{2cm}
-\sum_{\mu\nu}f_\Lambda(\epsilon_\mu,\epsilon_\mu)
\left\{\Omega_a,\langle\nu|Q^a|\mu\rangle
{\bar\Psi}_\mu(x)A^{\dag}\gamma_\mu\Gamma^a A\Psi_\nu(x)\right\}.
\label{jvac}
\ee
In order to obtain the vacuum part the detour over Euclidean space 
has to be taken again in order to be consistent with the proper time 
regularization scheme. The anti--commutators are understood between 
the collective coordinates, $A$, and the $SU(3)$ generators, $R_a$. 
The latter are related to the angular velocities, $\Omega_a$ via 
the quantization rule (\ref{Rgen}). The appearance of the regulator 
function, $f_\Lambda(\epsilon_\mu,\epsilon_\mu)$ is not accidental but 
rather guarantees the proper normalization of the baryon charges.
From the above expressions the radial functions like $V_i(r)$ in 
eq (\ref{spcurrps}) may be extracted when considering $\Gamma^a=Q^a$. 
Then the computation of static baryon properties proceeds along the 
lines outlined in section \ref{sec_masses}. This procedure has recently 
been employed \cite{Kim95} to compute the magnetic moments of the 
$\frac{1}{2}^+$ baryons in the NJL model. These results are summarized 
in table \ref{ta_magnjl}. As for all other treatments within the rigid 
rotator approach, the deviation from the $SU(3)$ symmetry relations
(\ref{su3mag}) is only moderate. The predictions compare with those 
obtained in the vector meson model, {\it cf.} table \ref{ta_emvm}.
\begin{table}
\caption{\label{ta_magnjl}\tenrm
The baryon magnetic moments in the NJL model. These results are 
taken from ref \protect\cite{Kim95} for the constituent quark mass 
${\Ss m=420{\rm MeV}}$. Data are given in nucleon magnetons
as well as ratios of the proton magnetic moment.}
~
\newline
\centerline{\tenrm\smalllineskip
\begin{tabular}{c | c c | c c}
& \multicolumn{2}{c|}{NJL} & \multicolumn{2}{c}{Expt.} \\
\hline
Baryon & $\mu_B$ & $\mu_B/\mu_p$ &
$\mu_B$ & $\mu_B/\mu_p$ \\
\hline
$p$        & 2.30 & 1.00 & 2.79 & 1.00 \\
$n$        &-1.66 &-0.72 &-1.91 &-0.68 \\
$\Lambda$  &-0.76 &-0.33 &-0.61 &-0.22 \\
$\Sigma^+$ & 2.34 & 1.02 & 2.42 & 0.87 \\
$\Sigma^0$ & 0.74 & 0.33 & ---  & ---  \\
$\Sigma^-$ &-0.85 &-0.37 &-1.16 &-0.42 \\
$\Xi^0   $ &-1.59 &-0.69 &-1.25 &-0.45 \\
$\Xi^-   $ &-0.67 &-0.29 &-0.69 &-0.25 \\
$\Sigma^0\rightarrow\Lambda$
&-1.44 &-0.62 &-1.61 &-0.58 \\
\end{tabular}}
\end{table}
However, a word of caution has to be added to the results obtained 
in  ref \cite{Kim95}. These contain $1/N_C$ corrections
(not included in eqs (\ref{currnjl}--\ref{jvac})),
which rely on a special ordering prescription for the collective
operators \cite{Go93}. Although this description is in agreement
with the generally expected form of the magnetic moment operator
\cite{Da94} it is not unambiguous either. In particular, the application
of this description to the axial current violates the PCAC relation 
by about 30\% \cite{Al94}. That is, this ordering does not 
respect one of the fundamental symmetries imposed in constructing 
the model Lagrangian. Moreover, the $1/N_C$ corrections
constitute the major contribution to the magnetic moments leading 
to the suspicion that the series has not yet converged. As a matter 
of fact the leading order (in $1/N_C$)  contribution to the isovector 
magnetic moment, $\mu_V=\mu_p-\mu_n$ is predicted to be about two 
\cite{Wa91,Kim95}, which is almost a factor three off the experimental 
value, which is $\mu_V=4.70$.

Substituting the generator $\Gamma^a=Q^0-2Q^8/\sqrt3$ into eqs 
(\ref{jval}) and (\ref{jvac}) gives the strange vector current in 
the NJL soliton model. The resulting predictions \cite{We95a} on 
the nucleon matrix element of the strange vector current have already
been presented in table \ref{ta_strange}.

Employing the generators $\Gamma^a=\gamma_5Q^a$ leads to the 
axial--vector currents. In analogy to the discussion before eq 
(\ref{naxff}), the components $a=3,8$ and $0$ may be used to extract 
the individual quark contributions $H_{u,d,s}$ to the axial--vector 
matrix element of the nucleon (\ref{paxmat}). For the constituent 
quark mass $m=420$ these are obtained to be 
\cite{Bl93b}\footnote{In these calculations 
the effective symmetry breaking has artificially been increased by 
tuning $m^0_s$, {\it cf.} the discussion after eq (\ref{grad1}).} 
\be
H_u(0)=0.64 (0.90)\quad H_d(0)=-0.24 (-0.48) \quad 
H_s(0)=-0.02 (-0.05) \ ,
\label{aflcont}
\ee
where the data in parentheses include the above mentioned PCAC 
violating $1/N_C$ corrections. In any event, the singlet contribution 
$H(0)=0.38$ is not effected by these corrections and turns out to be 
somewhat large. This presumably has to be interpreted as reminiscent 
of the quark model character of the NJL model. This statement is 
also supported by the fact that the valence contribution (\ref{jval}) 
to $H(0)$ absolutely dominates the vacuum part (\ref{jvac}). 
Nevertheless the prediction for the polarized nucleon structure 
function (\ref{EMCcomb}) $\Gamma_1(q^2=(10.7{\rm GeV})^2)=0.12 (0.16)$
is well within experimental data ($0.129\pm0.010$), at least when the 
$1/N_C$ corrections are ignored.

As a side remark the relation for the strangeness content fraction 
(\ref{xsskyrme}) will be explained to emerge from the NJL model. The 
strangeness content is defined as the expectation value of the 
associated bilinear
\be
\langle {\bar s}s\rangle&=&
\int D\bar q Dq \int d^4x
{\bar s}(x)s(x){\rm exp}\left(i {\cal A}_{\rm NJL}\right)
\nonumber \\
&=&\frac{\partial}{\partial \zeta}\int D\bar q Dq
{\rm exp}\left(i {\cal A}_{\rm NJL}
+\zeta\int d^4x{\bar s}(x)s(x)\right) \Bigg|_{\zeta=0} \ ,
\label{xsnjl1}
\ee
where ${\cal A}_{\rm NJL}=\int d^4x \cal L_{\rm NJL}$ denotes the 
action associated with the NJL Lagrangian (\ref{NJL}). Including the 
additional source term in the bosonization process and shifting the 
corresponding component of the meson fields, 
$M_{33}\rightarrow M_{33}-\zeta{\bar s}s$, 
moves the source into the mesonic part of the action. It is then 
straightforward to compute the derivative with respect to $\zeta$. Next 
the rotating hedgehog (\ref{collansatz}) is substituted and the baryon 
number zero contribution is subtracted. When normalizing with respect 
to the sum $\langle {\bar u}u+{\bar d}d+{\bar s}s\rangle$ the 
expression (\ref{xsskyrme}) is obtained.

In this section it has been shown that the rigid rotator approach to 
describe hyperons in the framework of chiral solitons can be applied 
to a microscopic theory of the quark flavor dynamics. The deficiencies 
of the numerical results are linked to an immanent problem of the NJL 
model, namely the too small prediction for the kaon decay constant. 
A further short--coming of this model, which has not been addressed 
here, is the presence of quark--antiquark thresholds, {\it i.e.} the 
model is not confining. This makes difficult the extension of the rigid 
rotator treatment to more sophisticated considerations like {\it e.g} 
the scaling approach of section \ref{sec_mixing}. As a consequence of 
non--confinement, the potential associated with the scaling variable 
($\mu$ in eq (\ref{su3breath})) possesses additional (unphysical) 
minima \cite{Ab95}, which makes a diagonalization as in section 
\ref{sec_mixing} unfeasible.

\bigskip

\begin{subsatz}
\label{sec_ckmasses}{\bf \hskip1cm Baryon Masses in the
Bound State Approach}
\addcontentsline{toc}{section}{\protect\ref{sec_ckmasses}
Baryon Masses in the Bound State Approach}
\end{subsatz}
\stepcounter{section}

In the present section the derivation of the parameters in the mass 
formula (\ref{bsmass}) within the NJL soliton model will be
described. According to the investigations in section \ref{sec_comp}, 
which concluded that the collective treatment gives smaller symmetry 
breaking in the baryon spectrum than the bound state approach, the
predictions of the NJL model on the baryon mass differences within the 
latter approach are expected to better agree with the experimental 
data. 

Similarly to the Skyrme model, the bound state approach to the NJL 
model is set up by substituting a P--wave {\it ansatz} with kaon 
quantum numbers for the fluctuating field
$\xi_f={\rm exp(i\sum_a\eta_a(x)Q^a)}$ in eq (\ref{defm}), see also 
eq (\ref{pwansatz})
\be
\eta_a(x)Q^a=\int \frac{d\omega}{2\pi}{\rm e}^{-i\omega t}
\pmatrix{0 & K(\mbox{\boldmath $r$},\omega)\cr
K^{\dag}(\mbox{\boldmath $r$},-\omega)&0\cr}\ ,
\quad K(\mbox{\boldmath $r$},\omega)=k_P(r,\omega)
\hat{\mbox{\boldmath $r$}}\cdot\mbox{\boldmath $\tau$}
\pmatrix{a_1(\omega)\cr a_2(\omega)\cr} \ .
\label{stft}
\ee
Here the Fourier transformation of the kaon isospinor, $K$ has 
been written explicitly. One proceeds along the standard path
and expands the NJL model action up to quadratic order in 
$k_P(r,\omega)$ in the background field of the chiral soliton,
$\xi_0={\rm exp}(i\mbox{\boldmath $\tau$}\cdot
\hat{\mbox{\boldmath $r$}}F(r)/2)$. In contrast to the collective 
approach the Dirac operator contains perturbative parts ($h_1$ 
and $h_2$) 
\be
i\beta\Dslash=i\partial_t-h-h_1-h_2
\label{direxp}
\ee 
which are not time independent but rather
\be
h_1&=&-\frac{m+m_s}{2}\int\frac{d\omega}{2\pi}{\rm e}^{-i\omega t}
\pmatrix{0 & u_0(\mbox{\boldmath $r$})\Omega(\omega)\cr
\Omega^{\dag}(-\omega) u_0(\mbox{\boldmath $r$})&0\cr}\ , 
\label{h1st} \\
\nonumber \\
h_2&=&\frac{m+m_s}{4}\int\frac{d\omega}{2\pi}
\frac{d\omega^\prime}{2\pi}{\rm e}^{-i(\omega+\omega^\prime)t}
\pmatrix{u_0(\mbox{\boldmath $r$})\beta\Omega(\omega)
\Omega^{\dag}(-\omega^\prime)u_0(\mbox{\boldmath $r$}) & 0 \cr
0&\hspace{-0.5cm}
-\beta\Omega^{\dag}(-\omega)\Omega(\omega^\prime)\cr}\ .
\hspace{0.5cm}
\label{h2st}
\ee
For simplicity the unitary, self--adjoint matrix 
\be
u_0(\mbox{\boldmath $r$})=\beta\left({\rm sin}\frac{\Theta}{2}
-i\gamma_5\hat{\mbox{\boldmath $r$}}\cdot{\mbox{\boldmath $\tau$}}
{\rm cos}\frac{\Theta}{2}\right)
\label{u0}
\ee
has been introduced and the spatial arguments of the isospinor
\be
\Omega(\omega)=k_P(r,\omega)\pmatrix{a_1(\omega)\cr a_2(\omega)\cr}
\label{bsOm}
\ee
have been omitted. By again taking the detour via Euclidean space 
to establish the proper time regularization, the expansion is 
performed with the formalism made available in eq (\ref{genexp}). 
The final expression is comprised in terms of local and bilocal
kernels $\Phi_1(r)$ and $\Phi_2(\omega;r,r^\prime)$, respectively
\cite{We93,We94a}
\be
\A^{(2)}[\Omega]&=&
\int_{-\infty}^{+\infty}\frac{d\omega}{2\pi}\Big\{\int drr^2
\int dr^\prime r^{\prime2}\ \Phi_2(\omega;r,r^\prime)
\Omega^\dagger(r,\omega)\Omega(r^\prime,\omega)
\nonumber \\
&&\qquad\qquad\qquad
+\int dr r^2\ \Phi_1(r)\Omega^\dagger(r,\omega)
\Omega(r,\omega)\Big\} \ .
\label{kernel}
\ee 
The explicit expressions of these kernels are summed up in appendix 
D. As a consequence of isospin invariance they are diagonal in the 
associated indices. As can be observed from eqs 
(\ref{phi1})--(\ref{phi2reg}) the kernels contain both, even and odd 
terms in the frequency $\omega$. The former stem from the real part of 
the Euclidean action, $\A_R$, while the latter are due to imaginary 
part, $\A_I$. Comparison with the bound state approach to the Skyrme 
model (section \ref{sec_fluc}) indicates that the imaginary part takes 
over the part of the Wess--Zumino term, $\Gamma_{\rm WZ}$ 
(\ref{WZterm}). Of course, this is expected rather than surprising 
because $\A_I$ is found to coincide with $\Gamma_{\rm WZ}$ 
in leading order of the gradient expansion \cite{Eb86}.

\begin{figure}[t]
\centerline{\hskip -2.0cm
\epsfig{figure=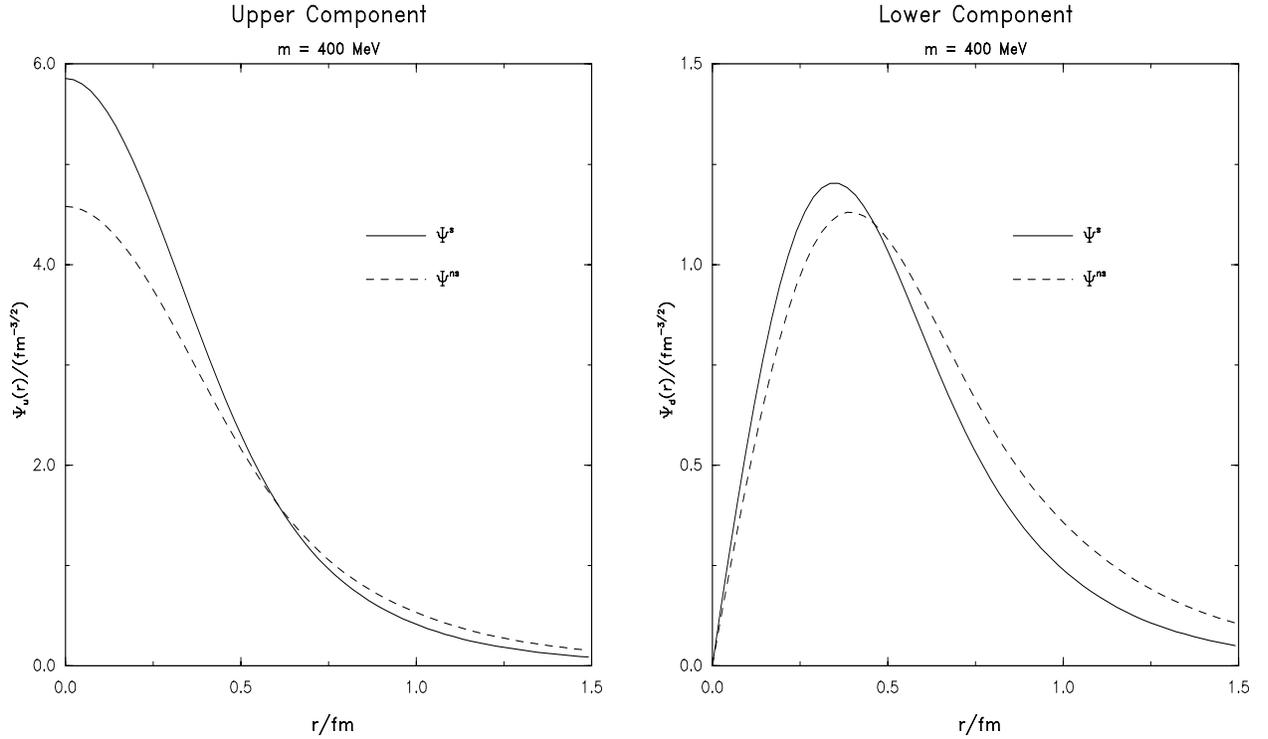,height=9.0cm,width=16.0cm}}
\caption{\label{fi_strq}\tenrm
Comparison of the radial dependencies of the induced strange valence 
quark (${\Ss s}$) and the non--strange valence quark (${\Ss ns}$) 
levels.}
\end{figure}

The NJL model analogue of the bound state equation (\ref{eqmkbs}) 
apparently is an integral equation rather than a differential 
equation
\be
r^2\left\{\int dr^\prime r^{\prime2}\Phi_2(\omega;r,r^\prime)
k_P(r^\prime,\omega)+\Phi_1(r)k_P(r,\omega)\right\}=0\ ,
\label{bssol}
\ee
which actually may be interpreted as the Bethe--Salpeter equation for 
the kaon field in the soliton background. In this equation, the 
frequency $\omega$ has to be adjusted to the bound state energy, 
$\omega_P$ such that a non--trivial solution, $k_P\ne0$ is 
obtained\footnote{See ref \cite{We94b} for a description of the 
numerical treatment of eq (\ref{bssol}).}. The resulting radial 
function is identified as the bound state wave--function. The explicit 
expressions (\ref{h1st}--\ref{bsOm}) for the perturbative parts of the 
Dirac Hamiltonian clearly illustrate that for small radii, $r$ the 
P--wave {\it ansatz} (\ref{stft}) is transmuted to an S--wave 
fluctuation as a consequence of the soliton background. Hence there is 
not centrifugal barrier for the bound state and therefore 
$k_P(r=0,\omega_P)\ne0$ is allowed. This, of course, is in no way 
different from the Skyrme model as can be inferred from figure 
\ref{fi_bound}. In the NJL model this bound state induces a strange 
($s$) valence quark field
\be
\Psi^s_{\rm val}(\omega_P)=
-\frac{m+m_s}{2}
\left(\epsilon_{\rm val}-\omega_P-h\right)^{-1}
\Omega^{\dag}(r,\omega_P)
\left(\sh-i\gamma_5\hat{\mbox{\boldmath $r$}}
\cdot\mbox{\boldmath $\tau$}\ch\right)
\Psi^{ns}_{\rm val}  \ ,
\label{sqind}
\ee
where $\Omega^{\dag}(r,\omega_P)$ contains the bound state 
wave--function $k_P(r,\omega_P)$.
This expression also illustrates how the isospin components of the 
up--down ($ns$) valence quark spinor are saturated. In figure 
\ref{fi_strq} the radial dependence of this induced strange valence 
quark is compared to that of the non--strange valence quark level. As 
for the kaon profile ({\it cf.} figure \ref{fi_bound}) one observes 
that the strange quark field is more strongly concentrated at the 
origin than the ``hedgehog" spinor. This shows that the strange degrees 
of freedom are bound by the soliton.
The fact that the upper component of the wave--function 
$\Psi^s_{\rm val}(\omega_P)$ is significantly larger than the lower 
one, indicates that the induced strange quark is of quark rather than 
antiquark character. Upon explicit computation it has been verified 
\cite{We94a} that the strangeness charge associated with this bound 
state indeed is negative. A bound state carrying positive strangeness 
was not detected. The occupation of such a state would corresponded 
to an exotic baryon with baryon number and strangeness both being $+1$. 
In the quark model language it would have the structure $qqq{\bar s}q$, 
a so--called pentaquark \cite{Oh94}.

In the next step the degeneracy between states of identical 
strangeness but different spin and isospin has to be removed.
Again collective coordinates are introduced for these symmetries
\be
M=A(t)\xi_0\xi_f\langle\Sigma\rangle\xi_f\xi_0A^{\dag}(t)
\qquad {\rm with}\qquad A(t)\in SU(2).
\label{cranknjl}
\ee
This parametrization can be formulated as in eq (\ref{bsrot2}) for
the pion and kaon fields. Hence all the relations between spin and 
isospin, which were discussed in section \ref{sec_fluc}, are recovered.
Nevertheless, it is interesting to consider the spin expectation 
value
\be
\langle\mbox{\boldmath $J$}\rangle =
\int D\bar q Dq \int d^3r \ q^{\dag}
\mbox{\boldmath $J$} q \ {\rm exp}\left(i {\cal A}_{\rm NJL}\right)
\label{spin1}
\ee
on the microscopic level. Here $\mbox{\boldmath $J$}$ is the spin
operator for a Dirac spinor
and ${\cal A}_{\rm NJL}=\int d^4x \cal L_{\rm NJL}$ denotes the action
associated with the NJL Lagrangian (\ref{NJL}). Since the spin
operator commutes with the iso-rotations $A(t)$ the transformation 
to the rotating frame $q=Aq^\prime$ is trivial
\be
\langle\mbox{\boldmath $J$}\rangle =
\int D\bar q^\prime Dq^\prime \int d^3r
\ q^{\prime{\dag} } \mbox{\boldmath $J$} q^\prime \ {\rm exp}
\left(i {\cal A}^\prime_{\rm NJL}\right).
\label{spin2}
\ee
${\cal A}^\prime_{\rm NJL}$ represents the NJL action in the
rotating frame, which also contains the Coriolis term
\be
{\cal A}^\prime_{\rm NJL}=\int d^4x \left({\cal L}_{\rm NJL}
-\frac{1}{2}q^{\prime{\dag}}{\mbox {\boldmath $\tau$}}\cdot
{\mbox {\boldmath $\Omega$}}q^\prime\right).
\label{spin3}
\ee
Substituting the definition of the grand spin ($\mbox{\boldmath $G$}$)
into eq (\ref{spin2}) yields
\be
\langle\mbox{\boldmath $J$}\rangle =
\int D\bar q^\prime Dq^\prime \int d^3r
\ q^{\prime{\dag} }\left(\mbox{\boldmath $G$} -
\frac{{\mbox{\boldmath $\tau$}}}{2}\right) q^\prime \ {\rm exp}
\left(i {\cal A}^\prime_{\rm NJL}\right).
\label{spin4}
\ee
The soliton contribution to the spin,
$\mbox{\boldmath $J$}^F$ is identified by differentiating
${\cal A}^\prime_{\rm NJL}$ with respect to the angular velocity
${\mbox{\boldmath $\Omega$}}$
\be
\langle\mbox{\boldmath $J$}\rangle =
\langle \mbox{\boldmath $G$} \rangle
+ \int D \bar q ^\prime D q^\prime \
\frac{1}{T}\frac{\partial {\cal A}^\prime_{\rm NJL}}
{\partial{\mbox{\boldmath $\Omega$}}} \
{\rm exp}\left(i A^\prime_{\rm NJL}\right)
=\langle\mbox{\boldmath $G$} \rangle+\mbox{\boldmath $J$}^F \ .
\label{spin5}
\ee
This immediately shows that the spin, $\mbox{\boldmath $J$}^K$
(\ref{jk}) associated with kaon bound state is identical to the 
grand spin in the flavor rotating frame. This could straightforwardly 
be shown on the microscopic level, while an elaborate chain of 
arguments (or tedious calculation)
was needed to establish this relation for the Skyrme model, see 
{\it e.g.} the discussion after eq (\ref{diagT2}). The identity 
$\mbox{\boldmath $J$}^K=\langle\mbox{\boldmath $G$}\rangle$
conisderably simplifies the computation of the spectral function 
$d(\omega)$ defined in eq (\ref{jk}) because the eigenstates, 
$|\mu\rangle$ of the one particle Hamiltonian (\ref{h0}) also 
diagonalize the grand spin projection, {\it i.e.}
$G_3|\mu\rangle=M_\mu\mu\rangle$. Hence, to extract the spectral 
function $d(\omega)$ one only has to repeat the expansion leading to 
eq (\ref{kernel}), with the projection quantum number, $M_\mu$, 
included when taking martix elements. To compute the other 
spectral function, $c(\omega)$ (\ref{lbscoll}) one first expands the 
action with respect to the angular velocity 
$\mbox{\boldmath $\Omega$}$
\be
\A_F=\A_F({\mbox{\boldmath $\Omega$}}=0)
+\sum_{a=1}^3\Omega_a\frac{\partial{\A_F}}{\partial\Omega_a}
\Big|_{{\mbox{\boldmath $\Omega$}}=0}
+{\cal O}({\mbox{\boldmath $\Omega$}}^2)
=\A_F^{(0)}+\A_F^{(1)}+
{\cal O}({\mbox{\boldmath $\Omega$}}^2)\ ,
\label{expomega}
\ee
where the contributions of order ${\mbox{\boldmath $\Omega$}}^2$ yield 
the spatial moment of inertia, $\alpha^2=\Theta_{33}$ (\ref{thab1}).
In the second step the linear term $\A_F^{(1)}$ is expanded up to
quadratic order in the kaon fluctuations. This painful exercise has 
been carried out in ref \cite{We94a} where the explicit expressions 
and numerical results for $c(\omega)$ and $d(\omega)$ may be traced. 
The numerical results for $\omega_P$ reflect that the kaon decay 
constant is underestimated in the NJL model. In case the cut--off, 
$\Lambda$ is fixed from $f_\pi$, $|\omega_P|$ decreases from about 
$200{\rm MeV}$ to $150{\rm MeV}$ when the constituent quark mass is 
varied from $350{\rm MeV}$ to $500{\rm MeV}$. These frequencies are 
smaller than the Skyrme model analogue ({\it cf.} table \ref{ta_bsdiff}) 
reflecting the fact that the kaon decay constant is underestimated in 
the NJL model. It truns out that $d_P$ is slightly smaller than unity. 
This difference with respect to the Skyrme model has been interpreted
\cite{We94a} as the polarization of the vaccum caused by the kaon 
bound state. Despite the denominator of $\chi=-c_P/d_P$ being smaller 
than in the Skyrme model, the numerical results for this ratio turn 
out to be similar. The baryon mass differences, which are obtained from 
eq (\ref{bsmass}), are exhibited in table \ref{ta_bsnjl}. The parameters
of the model are again obtained by first determining the ratio 
$\Lambda/m$ from either $f_\pi$ or $f_K$. Next the constituent 
quark mass is fitted to the $\Delta$--nucleon splitting. Then the 
remaining six mass differences are predicted as shown in table 
\ref{ta_njlcoll}.
\begin{table}
\caption{\label{ta_bsnjl}\tenrm
The mass differences of the low-lying ${\Ss \frac{1}{2}^+}$ and
${\Ss \frac{3}{2}^+}$ baryons with respect to the nucleon in the 
bound state approach to the NJL model. The up-quark constituent mass,
${\Ss m}$ is chosen such that the ${\Ss \Delta}$--nucleon mass 
difference is reproduced. All data (from ref.\protect\cite{We94a}) 
are in MeV.}
~
\newline
\centerline{
\begin{tabular}{c|c c|c}
& $f_\pi$ fixed & $f_K$ fixed & Expt. \\
\hline
$\Lambda$ & 132 & 137 &  177 \\
$\Sigma$ & 234 & 247 &  254 \\
$\Xi$ & 341 & 357 &  379 \\
$\Delta$ & 293 & 293 & 293 \\
$\Sigma^*$ & 374 &375 & 446 \\
$\Xi^*$ & 481 & 485 & 591  \\
$\Omega$ & 613 & 622 & 733 \\
\end{tabular}}
\end{table}
Comparing the results in table \ref{ta_bsnjl} with those in 
table \ref{ta_njlcoll} the prejudice from the beginning of this 
section is confirmed, that the bound state approach gives better 
agreement with the experimental mass differences. In particular 
this is the case for the $\frac{1}{2}^+$ baryons. Despite of this 
improvement the flavor symmetry breaking in the baryon spectrum
remains underestimated in the NJL model.

Finally this chapter will be concluded by also remarking that S--wave 
fluctuations off the NJL soliton have been investigated \cite{We94c}. 
As in the Skyrme model, the bound state in this channel is suited to 
describe the odd parity hyperon $\Lambda(1405)$. Here, however, 
the non--confining character of the NJL raises some problems since for 
small constituent quark masses the threshold for scattering the valence 
quark into the strange continuum is as small as the expected 
frequency of the bound state in the S--wave channel. Hence a bound 
state is only found when the constituent mass, $m$, of the up--quark 
(on which $m_s$ depends) exceeds a critical value. This value has 
been obtained to be $\sim425{\rm MeV}$ when the cut--off is linked 
to the experimental value of the pion decay constant. When $m$ is 
further increased the mass difference between the lowest odd parity 
hyperon and the nucleon shows almost no variation and is predicted to 
be about $415{\rm MeV}$. This is somewhat lower than the empirical 
value of $465{\rm MeV}$.

\vskip1cm
\begin{satz}
\label{chap_concl}{\large \bf \hskip1cm
Concluding Remarks}
\addcontentsline{toc}{chapter}{\protect\ref{chap_concl}
Concluding Remarks}
\end{satz}
\stepcounter{chapter}

In this article the incorporation of strange degrees of freedom 
in the soliton description of baryons has been surveyed. The models 
considered reach from the simple Skyrme model, which contains 
pseudo--scalar degrees of freedom only, to extended Skyrme type 
models containing vector meson degrees of freedom and finally
to a microscopic theory of the quark flavor dynamics. The latter 
has been specified to be the Nambu--Jona--Lasinio (NJL) model.
All these models are treated in a similar fashion. A chirally 
invariant Lagrangian is supplemented by appropriate flavor symmetry 
breaking terms with the model parameters being determined from meson 
properties as much as possible. Subsequently the soliton solutions 
corresponding to a unit baryon number are obtained. These (static) 
field configurations, which commonly are of hedgehog structure, do
not possess the quantum numbers of physical baryons. In order to 
generate states with good spin and flavor quantum numbers approximations
to the time--dependent solutions of the Euler--Lagrange equations are 
considered. In general these approximations introduce time--dependent
coordinates or fields, which in turn are quantized canonically. Here two 
seemingly opposite procedures to gain hyperon wave--functions has 
been reflected about: (i) the introduction of collective coordinates in 
the whole $SU(3)$ flavor space and (ii) the construction of hyperon 
states such as a kaon cloud bound by the soliton background field. 
In the former approach strange degrees of freedom are considered as 
large amplitude fluctuations off the soliton, while in the latter 
the strange fields are limited to the harmonic approximation. Although 
these two treatments seem to coincide in the large symmetry breaking 
limit, they apparently do not in the case that flavor symmetry breaking 
is small. This has been argued from the simple fact that the bound state 
approach does not reproduce the Gell--Mann Okubo mass relation. For 
model parameters, which are adjusted to the meson properties, the 
Skyrme model examinations have revealed that the bound state approach 
overestimates the flavor symmetry breaking in the baryon spectrum. The 
collective treatment appears to be superior in this respect.

Despite of the progress achieved so far, one should acknowledge that 
the collective treatment has not yet been brought to complete 
fruition. For example, the treatment of the induced strange fields, 
which are mandatory to maintain the proper divergence of the currents, 
may undergo some changes when double counting effects are accounted for. 
As argued, these changes are supposedly small, in particular for the 
baryon properties. Also fields induced by the symmetry breaking have 
not yet been considered in meson models. The studies in the NJL model,
however, indicate that these effects are negligible. This may be 
argued from the fact that those contributions to the baryon masses 
are small, which are quadratic in the symmetry breaking part of 
the Dirac Hamiltonian. These terms essentially correspond to the 
inclusion of quark excitations due to symmetry breaking.

The collective approach has been employed to extensively study the 
static properties of the low--lying $\frac{1}{2}^+$ baryons. In 
particular, the role of virtual strange quark--antiquark excitations 
in the nucleon has been examined. It has been established 
that the inclusion of symmetry breaking effects in the nucleon 
wave--functions significantly reduces the predictions on the 
amount of strangeness in the nucleon. Furthermore, it has been 
explained how the nucleon matrix elements of various operators,
which are bilinear in the strange quark spinors, evolve differently 
with symmetry breaking when the collective Hamiltonian (including 
flavor symmetry breaking terms) is diagonalized exactly. These effects 
go together with strong deviations from flavor covariant baryon 
wave--functions. Nevertheless a reasonable agreement with the Cabibbo 
scheme of the semi--leptonic hyperon decays is achieved. 

The investigation of the baryon magnetic moments demonstrates that even 
the exact diagonalization within rigid rotator approach apparently 
contains symmetries which are not reflected in nature. This feature 
not only indicates the demand for refined quantization procedures, it 
even effects the classical soliton configuration because the strange 
fields do not acquire the asymptotic form expected for a kaon
wave--function. This happens to be the case in both, the rigid rotator 
as well as the bound state approach. Hence one is tempted to ask the 
question whether flavor symmetry breaking could cause deviations from 
the hedgehog shape of the classical field configuration. For related 
studies, space dependent kaon fields need to be considered. The harmonic 
expansion of the bound state approach will certainly not be sufficient.

The soliton description of baryons has proven especially successful 
in explaining the data for polarized muon scattering. In all 
models the polarized structure function 
$\Gamma_1^p(q^2=(10.7{\rm GeV})^2)$ was predicted to be only slightly 
above 0.1, which compares favorably with the experimental value,
$0.129\pm0.010$. As a reminder it should be noted that the 
na{\"\i}vely expected matrix element of the axial singlet current, 
$H(0)$=1, together with $g_A=1.25$ and the assumption of flavor 
symmetry yields $0.22$. As verified in the case of the NJL model, 
chiral solitons not only exist in effective meson theories but they 
may also occur in microscopic models of the quark flavor dynamics. 
These quark model solitons similarly explain the smallness of 
$\Gamma_1^p$, even for configurations which are dominated by the 
valence contribution. Thus, one is tempted to consider the smallness 
of $\Gamma_1^p$ as manifestation for the soliton structure of the 
baryons. However, as has been indicated, in order to account for 
the detailed structure of the baryons the pure pseudo--scalar solitons 
need to be extended; {\it e.g.} by vector mesons or explicit quark 
fields.

\vskip2cm

{\bf \hskip1cm Acknowledgements}
\smallskip

The author is indebted to J. Schechter, B. Schwesinger, N. W. Park, 
A. Subbaraman, R. Alkofer and H. Reinhardt for many helpful 
contributions and to L. Gamberg for carefully reading the manuscript. 
Furthermore numerous illuminating discussions on the subjects covered 
in this article with G. Holzwarth, H. Walliser, Ulf.--G. Mei{\ss}ner, 
P. Jain, U. Z\"uckert and A. Abada are gratefully acknowledged.

\vskip2cm

\vskip2cm

\appendix

\stepcounter{chapter}
{\large \bf \hskip1cm Appendix A}
\addcontentsline{toc}{chapter}{Appendix A}

\smallskip

In this appendix the explicit forms of the right $SU(3)$ 
generators $R_a\ (a=1,..,8)$ are displayed in terms of 
differential operators with respect to $SU(3)$ ``Euler angles"
\cite{Ya88}. An appropriate definition of these angles is given by
parametrizing the collective flavor rotations (\ref{su3rot}) via
\be
A={\rm e}^{-i(\alpha/2)\lambda_3}
{\rm e}^{-i(\beta/2)\lambda_2}
{\rm e}^{-i(\gamma/2)\lambda_3}{\rm e}^{-i\nu\lambda_4}
{\rm e}^{-i(\alpha^\prime/2)\lambda_3}
{\rm e}^{-i(\beta^\prime/2)\lambda_2}
{\rm e}^{-i(\gamma^\prime/2)\lambda_3}
{\rm e}^{-i(\rho/\sqrt{3})\lambda_8}\ .
\label{Apara}
\ee
The group manifold is completely covered by varying the 
angles $\alpha,\beta,...,\rho$ according to
\be
0\le\alpha,\gamma,\alpha^\prime,\gamma^\prime<2\pi,\
0\le\beta,\beta^\prime<\pi,\ 0\le\nu<\pi/2,\ 
0\le\rho<3\pi.
\label{Vangle}
\ee
Since the $SU(3)$ generators are linear operators they may 
in general be written as linear combinations of differential 
operators \cite{Ne67}
\be
R_a=id_{ba}(\mbox{\boldmath $\alpha$)}
\frac{\partial}{\partial\alpha_b},
\label{Rgenapp}
\ee
where $\mbox{\boldmath $\alpha$}=
(\alpha_1,\alpha_2,...,\alpha_8)=
(\alpha,\beta,...,\rho)$ compactly refers to the eight 
``Euler angles". The coefficient functions 
$d_{ab}(\mbox{\boldmath $\alpha$})$ are 
extracted from the defining equation of the $SU(3)$ algebra
\be
AR_aA^{\dag}=\frac{1}{2}A\lambda_aA^{\dag}
=\frac{1}{2}\lambda_bD_{ba}(\mbox{\boldmath $\alpha$}),
\label{Deqn}
\ee
where $D_{ab}$ denotes the adjoint representation of the 
rotation matrix $A$, see eq (\ref{omtrans}). Explicit computation of 
the {\it LHS} of eq (\ref{Deqn}) provides the quantities $M_{ab}$, 
which are defined by
\be
ŠAR_aA^{\dag}=\lambda_bM_{bc}(\mbox{\boldmath $\alpha$})
d_{ca}(\mbox{\boldmath $\alpha$}).
\label{defmab}
\ee
From this one may read off
\be
d_{ab}(\mbox{\boldmath $\alpha$})
=\left(M^{-1}(\mbox{\boldmath $\alpha$})\right)_{ac}
D_{cb}(\mbox{\boldmath $\alpha$}).
\label{Cdab}
\ee
The explicit expressions are
\be
R_1&=&i\frac{{\rm cos}\gamma^\prime}{{\rm sin}\beta^\prime}
\frac{\partial}{\partial\alpha^\prime}
-i{\rm sin}\gamma^\prime
\frac{\partial}{\partial\beta^\prime}
-i{\rm cos}\gamma^\prime{\rm cot}\beta^\prime
\frac{\partial}{\partial\gamma^\prime}\ ,
\nonumber \\
R_2&=&-i\frac{{\rm sin}\gamma^\prime}{{\rm sin}\beta^\prime}
\frac{\partial}{\partial\alpha^\prime}
-i{\rm cos}\gamma^\prime
\frac{\partial}{\partial\beta^\prime}
+i{\rm sin}\gamma^\prime{\rm cot}\beta^\prime
\frac{\partial}{\partial\gamma^\prime}\ ,
\nonumber \\
R_3&=&-i\frac{\partial}{\partial\gamma^\prime}\ ,
\nonumber \\
R_4&=&-i{\rm sin}
\left(\gamma-\rho+\frac{\alpha^\prime-\gamma^\prime}{2}\right)
\frac{{\rm sin}\frac{\beta^\prime}{2}}
{{\rm sin}\beta{\rm sin}\nu}
\frac{\partial}{\partial\alpha}
-i{\rm cos}
\left(\gamma-\rho+\frac{\alpha^\prime-\gamma^\prime}{2}\right)
\frac{{\rm sin}\frac{\beta^\prime}{2}} {{\rm sin}\nu}
\frac{\partial}{\partial\beta}
\nonumber \\
&&-i\left[2{\rm sin}
\left(\rho+\frac{\alpha^\prime+\gamma^\prime}{2}\right)
\frac{{\rm cos}\frac{\beta^\prime}{2}} {{\rm sin}2\nu}
-{\rm sin}
\left(\gamma-\rho+\frac{\alpha^\prime-\gamma^\prime}{2}\right)
{\rm cot}\beta\frac{{\rm sin}\frac{\beta^\prime}{2}} 
{{\rm sin}\nu}\right]
\frac{\partial}{\partial\gamma}
\nonumber \\
&&-\frac{i}{2}{\rm cos}
\left(\rho+\frac{\alpha^\prime+\gamma^\prime}{2}\right)
{\rm cos}\frac{\beta^\prime}{2}\frac{\partial}{\partial\nu}
-\frac{3i}{4}{\rm sin}
\left(\rho+\frac{\alpha^\prime+\gamma^\prime}{2}\right)
{\rm tan}\nu {\rm cos}\frac{\beta^\prime}{2}
\frac{\partial}{\partial\rho}
Š\nonumber \\
&&+\frac{i}{2}{\rm sin}
\left(\rho+\frac{\alpha^\prime+\gamma^\prime}{2}\right)
\left[{\rm cos}\frac{\beta^\prime}{2}{\rm tan}\nu
+\frac{{\rm cot}\nu}{{\rm cos}\frac{\beta^\prime}{2}}\right]
\frac{\partial}{\partial\alpha^\prime}
\nonumber \\
&&+i{\rm cos}
\left(\rho+\frac{\alpha^\prime+\gamma^\prime}{2}\right)
{\rm cot}\nu{\rm sin}\frac{\beta^\prime}{2}
\frac{\partial}{\partial\beta^\prime}
+\frac{i}{2}{\rm sin}
\left(\rho+\frac{\alpha^\prime+\gamma^\prime}{2}\right)
\frac{{\rm cot}\nu}{{\rm cos}\frac{\beta^\prime}{2}}
\frac{\partial}{\partial\gamma^\prime}\ ,
\nonumber \\
R_5&=&i{\rm cos}
\left(\gamma-\rho+\frac{\alpha^\prime-\gamma^\prime}{2}\right)
\frac{{\rm sin}\frac{\beta^\prime}{2}}
{{\rm sin}\beta{\rm sin}\nu}
\frac{\partial}{\partial\alpha}
-i{\rm sin}
\left(\gamma-\rho+\frac{\alpha^\prime-\gamma^\prime}{2}\right)
\frac{{\rm sin}\frac{\beta^\prime}{2}} {{\rm sin}\nu}
\frac{\partial}{\partial\beta}
\nonumber \\
&&-i\left[2{\rm cos}
\left(\rho+\frac{\alpha^\prime+\gamma^\prime}{2}\right)
\frac{{\rm cos}\frac{\beta^\prime}{2}} {{\rm sin}2\nu}
+{\rm cos}
\left(\gamma-\rho+\frac{\alpha^\prime-\gamma^\prime}{2}\right)
{\rm cot}\beta\frac{{\rm sin}\frac{\beta^\prime}{2}} 
{{\rm sin}\nu}\right]
\frac{\partial}{\partial\gamma}
\nonumber \\
&&+\frac{i}{2}{\rm sin}
\left(\rho+\frac{\alpha^\prime+\gamma^\prime}{2}\right)
{\rm cos}\frac{\beta^\prime}{2}\frac{\partial}{\partial\nu}
-\frac{3i}{4}{\rm cos}
\left(\rho+\frac{\alpha^\prime+\gamma^\prime}{2}\right)
{\rm tan}\nu {\rm cos}\frac{\beta^\prime}{2}
\frac{\partial}{\partial\rho}
\nonumber \\
&&+\frac{i}{2}{\rm cos}
\left(\rho+\frac{\alpha^\prime+\gamma^\prime}{2}\right)
\left[{\rm cos}\frac{\beta^\prime}{2}{\rm tan}\nu
+\frac{{\rm cot}\nu}{{\rm cos}\frac{\beta^\prime}{2}}\right]
\frac{\partial}{\partial\alpha^\prime}
\nonumber \\
&&-i{\rm sin}
\left(\rho+\frac{\alpha^\prime+\gamma^\prime}{2}\right)
{\rm cot}\nu{\rm sin}\frac{\beta^\prime}{2}
\frac{\partial}{\partial\beta^\prime}
+\frac{i}{2}{\rm cos}
\left(\rho+\frac{\alpha^\prime+\gamma^\prime}{2}\right)
Š\frac{{\rm cot}\nu}{{\rm cos}\frac{\beta^\prime}{2}}
\frac{\partial}{\partial\gamma^\prime}\ ,
\nonumber \\
R_6&=&-i{\rm sin}
\left(\gamma-\rho+\frac{\alpha^\prime+\gamma^\prime}{2}\right)
\frac{{\rm cos}\frac{\beta^\prime}{2}}
{{\rm sin}\beta{\rm sin}\nu}
\frac{\partial}{\partial\alpha}
-i{\rm cos}
\left(\gamma-\rho+\frac{\alpha^\prime+\gamma^\prime}{2}\right)
\frac{{\rm cos}\frac{\beta^\prime}{2}} {{\rm sin}\nu}
\frac{\partial}{\partial\beta}
\nonumber \\
&&+i\left[2{\rm sin}
\left(\rho+\frac{\alpha^\prime-\gamma^\prime}{2}\right)
\frac{{\rm sin}\frac{\beta^\prime}{2}} {{\rm sin}2\nu}
+{\rm sin}
\left(\gamma-\rho+\frac{\alpha^\prime+\gamma^\prime}{2}\right)
{\rm cot}\beta\frac{{\rm cos}\frac{\beta^\prime}{2}} 
{{\rm sin}\nu}\right]
\frac{\partial}{\partial\gamma}
\nonumber \\
&&+\frac{i}{2}{\rm cos}
\left(\rho+\frac{\alpha^\prime-\gamma^\prime}{2}\right)
{\rm sin}\frac{\beta^\prime}{2}\frac{\partial}{\partial\nu}
+\frac{3i}{4}{\rm sin}
\left(\rho+\frac{\alpha^\prime-\gamma^\prime}{2}\right)
{\rm tan}\nu {\rm sin}\frac{\beta^\prime}{2}
\frac{\partial}{\partial\rho}
\nonumber \\
&&-\frac{i}{2}{\rm sin}
\left(\rho+\frac{\alpha^\prime-\gamma^\prime}{2}\right)
\left[{\rm sin}\frac{\beta^\prime}{2}{\rm tan}\nu
+\frac{{\rm cot}\nu}{{\rm sin}\frac{\beta^\prime}{2}}\right]
\frac{\partial}{\partial\alpha^\prime}
\nonumber \\
&&+i{\rm cos}
\left(\rho+\frac{\alpha^\prime-\gamma^\prime}{2}\right)
{\rm cot}\nu{\rm cos}\frac{\beta^\prime}{2}
\frac{\partial}{\partial\beta^\prime}
+\frac{i}{2}{\rm sin}
\left(\rho+\frac{\alpha^\prime-\gamma^\prime}{2}\right)
\frac{{\rm cot}\nu}{{\rm sin}\frac{\beta^\prime}{2}}
\frac{\partial}{\partial\gamma^\prime}\ ,
\nonumber \\
R_7&=&i{\rm cos}
\left(\gamma-\rho+\frac{\alpha^\prime+\gamma^\prime}{2}\right)
\frac{{\rm cos}\frac{\beta^\prime}{2}}
{{\rm sin}\beta{\rm sin}\nu}
\frac{\partial}{\partial\alpha}
-i{\rm sin}
\left(\gamma-\rho+\frac{\alpha^\prime+\gamma^\prime}{2}\right)
\frac{{\rm cos}\frac{\beta^\prime}{2}} {{\rm sin}\nu}
\frac{\partial}{\partial\beta}
\nonumber \\
Š&&+i\left[2{\rm cos}
\left(\rho+\frac{\alpha^\prime-\gamma^\prime}{2}\right)
\frac{{\rm sin}\frac{\beta^\prime}{2}} {{\rm sin}2\nu}
-{\rm cos}
\left(\gamma-\rho+\frac{\alpha^\prime+\gamma^\prime}{2}\right)
{\rm cot}\beta\frac{{\rm cos}\frac{\beta^\prime}{2}} 
{{\rm sin}\nu}\right]
\frac{\partial}{\partial\gamma}
\nonumber \\
&&-\frac{i}{2}{\rm sin}
\left(\rho+\frac{\alpha^\prime-\gamma^\prime}{2}\right)
{\rm sin}\frac{\beta^\prime}{2}\frac{\partial}{\partial\nu}
+\frac{3i}{4}{\rm cos}
\left(\rho+\frac{\alpha^\prime-\gamma^\prime}{2}\right)
{\rm tan}\nu {\rm sin}\frac{\beta^\prime}{2}
\frac{\partial}{\partial\rho}
\nonumber \\
&&-\frac{i}{2}{\rm cos}
\left(\rho+\frac{\alpha^\prime-\gamma^\prime}{2}\right)
\left[{\rm sin}\frac{\beta^\prime}{2}{\rm tan}\nu
+\frac{{\rm cot}\nu}{{\rm sin}\frac{\beta^\prime}{2}}\right]
\frac{\partial}{\partial\alpha^\prime}
\nonumber \\
&&-i{\rm sin}
\left(\rho+\frac{\alpha^\prime-\gamma^\prime}{2}\right)
{\rm cot}\nu{\rm cos}\frac{\beta^\prime}{2}
\frac{\partial}{\partial\beta^\prime}
+\frac{i}{2}{\rm cos}
\left(\rho+\frac{\alpha^\prime-\gamma^\prime}{2}\right)
\frac{{\rm cot}\nu}{{\rm sin}\frac{\beta^\prime}{2}}
\frac{\partial}{\partial\gamma^\prime}\ ,
\nonumber \\
R_8&=&-\frac{i\sqrt{3}}{2}
\frac{\partial}{\partial\rho}\ .
\label{Rexpl}
\ee

Here we also want to outline how the eigenvalue problem for the 
collective Hamiltonian (\ref{collham}) reduces to coupled differential 
equations for functions, which only depend on the strangeness changing
angle $\nu$. Up to the normalization a suitable decomposition of the 
baryon wave--functions is given by \cite{Ya88}
\be
\Psi(I,I_3,Y;J,J_3,Y_R)=
\sum_{M_L,M_R}
D^{(I)*}_{I_3,M_L}(\alpha,\beta,\gamma)
f_{M_L,M_R}^{(I,Y;J,Y_R)}(\nu){\rm e}^{iY_R\rho}
D^{(J)*}_{M_R,-J_3}(\alpha^\prime,\beta^\prime,\gamma^\prime).
\label{Dpsi}
\ee
The $D$--functions refer to $SU(2)$ Wigner functions.
It is important to note that the sums over the intrinsic spins 
($M_R=-J,-J+1,...,J$) and isospins ($M_L=-I,-I+1,...,I)$ are subject 
to the constraint $M_L-M_R=(Y-Y_R)/2$. Using the explicit 
forms for the $SU(3)$ generators (\ref{Rexpl}) the action of the 
Šquadratic Casimir operator $C_2=\sum_{a=1}^8R_a^2$ on the 
baryon wave--function (\ref{Dpsi}) is found to be
\be
&&C_2\Psi(I,I_3,Y;J,J_3,Y_R)=
\sum_{M_L,M_R}
D^{(I)*}_{I_3,M_L}(\alpha,\beta,\gamma){\rm e}^{iY_R\rho}
D^{(J)*}_{J_3,M_R}(\alpha^\prime,\beta^\prime,\gamma^\prime)
\nonumber \\
&&\hspace{-0.5cm} 
\times \Bigg\{
-\frac{1}{4}\left[\frac{d^2}{d\nu^2}
+\left(3{\rm cot}\nu-{\rm tan}\nu\right)\frac{d}{d\nu}\right]
+\frac{I^2+J^2}{{\rm sin}^2\nu}
+\frac{M_L^2}{{\rm cos}^2\nu}
+\frac{M_R^2}{4}\left(3+\frac{1}{{\rm cos}^2\nu}\right)
\nonumber \\ && 
-\frac{1+{\rm cos}^2\nu}{{\rm sin}^2\nu{\rm cos}^2\nu}M_LM_R
+\frac{3Y_RM_L}{2{\rm cos}^2\nu}
-3\frac{1+{\rm cos}^2\nu}{4{\rm cos}^2\nu}Y_RM_R
+\left(\frac{3}{4}+\frac{9}{16}{\rm tan}^2\nu\right)\Bigg\}
f_{M_L,M_R}^{(I,Y;J,Y_R)}(\nu)
\nonumber \\
&&-\frac{{\rm cos}\nu}{{\rm sin}^2\nu}
\sqrt{\left(I+M_L+1\right)\left(I-M_L\right)
\left(J+M_R+1\right)\left(J-M_R\right)}\
f_{M_L+1,M_R+1}^{(I,Y;J,Y_R)}(\nu)
\nonumber \\
&&-\frac{{\rm cos}\nu}{{\rm sin}^2\nu}
\sqrt{\left(I-M_L+1\right)\left(I+M_L\right)
\left(J-M_R+1\right)\left(J+M_R\right)}\
f_{M_L-1,M_R-1}^{(I,Y;J,Y_R)}(\nu).
\label{C2psi}
\ee
Obviously the dependence on the angles other than $\nu$ 
can be factorized leaving a set of coupled ordinary 
differential equations in the variable $\nu$. This becomes 
even more transparent by displaying the $\nu$ dependence 
of the dominating symmetry breaking term in the collective 
Hamiltonian (\ref{collham})
\be
1-D_{88}=\frac{3}{2}{\rm sin}^2\nu .
\label{D88}
\ee
Eq. (\ref{C2psi}) also illustrates how the intrinsic functions 
$f_{M_L,M_R}^{(I,Y;J,Y_R)}(\nu)$ depend on the spin 
and isospin quantum numbers. 

The eigenvalue equation $C_2\Psi=\mu\Psi$ yields the flavor symmetric 
$SU(3)$ D--functions, which correspond to irreducible representations. 
As an example we display the non--vanishing intrinsic functions for 
the baryon octet. The constraint $M_L-M_R=(Y-1)/2$ has to be taken 
into consideration.
\be
\hspace{-0.5cm}
N&:&\
f_{\frac{1}{2},\frac{1}{2}}^{\frac{1}{2},1;
\frac{1}{2},1}(\nu)={\rm cos}^2\nu\ ,
Šf_{-\frac{1}{2},-\frac{1}{2}}^{\frac{1}{2},1;
\frac{1}{2},1}(\nu)= {\rm cos}\nu\ ;\hspace{1.52cm}
\Lambda:\
f_{0,\frac{1}{2}}^{0,0;\frac{1}{2},1}(\nu)
={\rm sin}\nu{\rm cos}\nu \  ;
\nonumber \\ \hspace{-0.5cm}
\Sigma&:&\
f_{0,\frac{1}{2}}^{1,0;\frac{1}{2},1}(\nu)=
\frac{1}{\sqrt{2}}{\rm cos}\nu{\rm sin}\nu \ ,
f_{-1,-\frac{1}{2}}^{1,0;\frac{1}{2},1}(\nu)=
{\rm sin}\nu\ ;\quad
\Xi:\
f_{-\frac{1}{2},\frac{1}{2}}^{\frac{1}{2},-1;\frac{1}{2},1}(\nu)
={\rm sin}^2\nu\ . 
\label{su3dexap}
\ee
Obviously non of these wave--functions vanishes execpt at the 
boundaries $\nu=0,\pi/2$. This is, of course, a special feature 
of the ground states, which reside in the {\bf 8} representation. 
The isoscalar wave--functions associated with baryons in higher 
dimensional representations, which carry the same physical quantum 
numbers ($I,J,Y$), may well develop nodes.

\begin{figure}[t]
\centerline{\hskip -1.5cm
\epsfig{figure=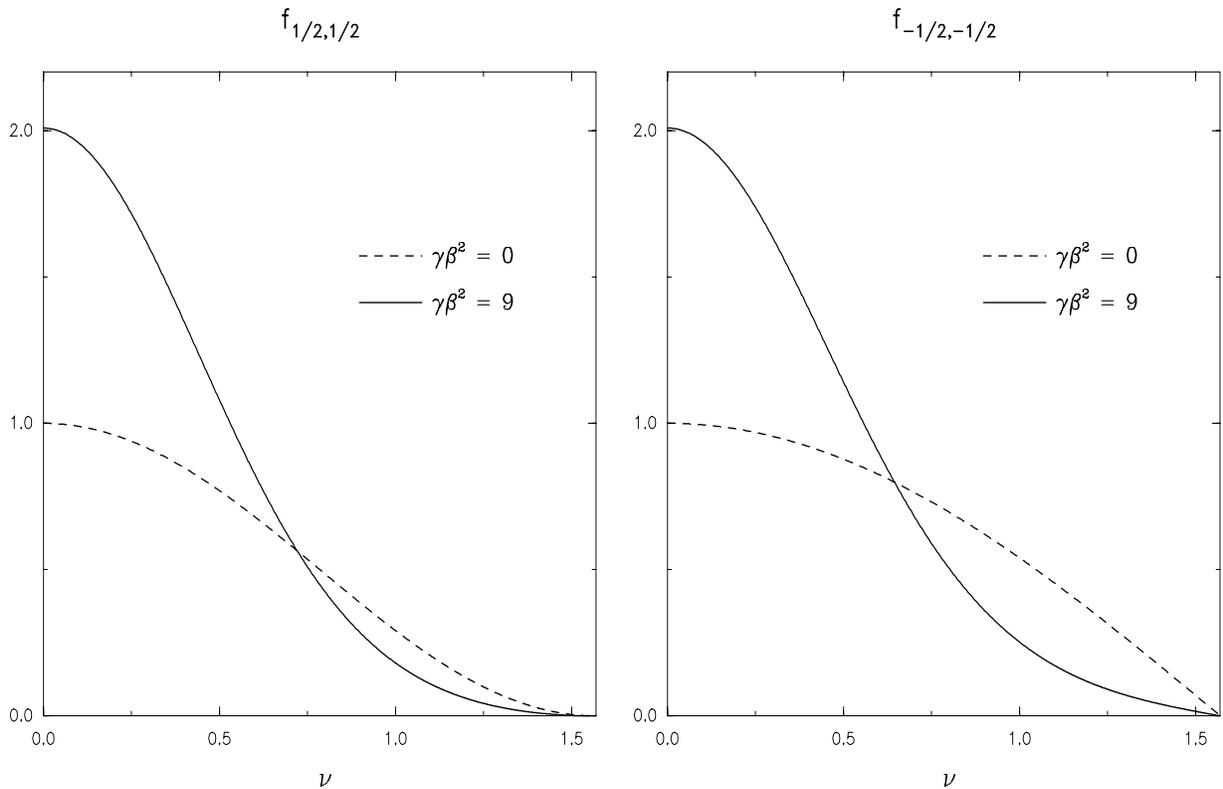,height=10.0cm,width=16.0cm}}
\caption{\label{fi_yabu}\tenrm
The dependencies of the nucleon scalar functions on the 
strangeness changing angle ${\Ss \nu}$ for two values of the 
symmetry breaking. The case ${\Ss \gamma\beta^2=0}$ should be 
compared with the expressions for (${\Ss N}$) in eq 
(\protect\ref{su3dexap}).}
\end{figure}

When the eigenvalue problem is augmented by symmetry breaking 
terms the intrinsic function deviate from (\ref{su3dexap}) such 
that they get more pronounced at small $\nu$, {\it i.e.} 
rotations into the direction of strangeness are suppressed. This can 
also be deduced from figure \ref{fi_yabu} where the dependencies 
of the nucleon isoscalar functions 
$f_{\pm\frac{1}{2},\pm\frac{1}{2}}^{\frac{1}{2},1;\frac{1}{2},1}(\nu)$
are displayed for the symmetric case, $\gamma\beta^2=0$ as well as 
for large symmetry breaking $\gamma\beta^2=9$. All other 
symmetry breakers have been ignored.

It should be noted that this procedure to compute the eigenvalues of
the collective Hamiltonian (\ref{collham}) is completely equivalent to 
diagonalizing it in a basis built from $SU(3)$ representations.
In addition to the constraint $Y_R=1$ for the allowed representations 
one more condition has to be satisfied, which stems from the 
grand spin symmetry of the hedgehog {\it ansatz}. For $\nu=0$ 
the wave--function must be invariant under a combined spin and isospin 
transformation $T$. Denoting the coordinate and isospin space ``Euler 
angles" by $R_J=(\alpha^\prime,\beta^\prime,\gamma^\prime)$ and
$R_I=(\alpha,\beta,\gamma)$, this implies
\be
&&\sum_{M_L,M_R} D^{(I)*}_{I_3,M_L}(R_I T)
f_{M_L,M_R}^{(I,Y;J,Y_R)}(0)D^{(J)*}_{M_R,-J_3}(T^{-1}R_J) 
\nonumber \\ &&\hspace{2cm}
=\sum_{M_L,M_R} D^{(I)*}_{I_3,M_L}(R_I)
f_{M_L,M_R}^{(I,Y;J,Y_R)}(0) D^{(J)*}_{M_R,-J_3}(R_J)
\label{nu0a}
\ee
which can only be solved if 
\be
f_{M_L,M_R}^{(I,Y;J,Y_R)}(0)={\cal C}(I;Y,Y_R)
\delta_{M_L,M_R}\delta_{I,J}. 
\label{nu0b}
\ee
Because of completeness, at least one of ${\cal C}(I;Y,Y_R)$
must not vanish within a specified representation. Hence in each 
of the $SU(3)$ representations, which constitute the basis for 
diagonalizating the collective Hamiltonian, a state with 
$I=J$ must exist \cite{Ma84}. Since the static field configuration 
(\ref{su3hedgehog}) commutes with $\lambda^8$ this state must 
furthermore satisfy $Y=Y_R$. The fact that ${\cal C}(1/2;1,1)$
does not depend on the intrinsic projections $M_{L,R}$ can also 
be seen in figure \ref{fi_yabu}. In the limit of large 
symmetry breaking the isoscalar functions can be approximated 
by $f_{M_L,M_R}(\nu)\approx f_{M_L,M_R}(0)\delta(\nu)$. It is 
therefore obvious from eq (\ref{nu0b}) that this limit 
corresponds to the two flavor model.

Here also a few comments on the treatment of the slow rotator 
discussed in section \ref{sec_slowrot} will be added. In a first 
step the explicit form (\ref{C2psi}) for the Casimir operator is used
in order to express the Hamiltonian (\ref{slowham}) as a second 
order differential equation for the isoscalar functions 
$f_{M_L,M_R}^{(I,Y;J,Y_R)}(\nu)$, which are defined in eq (\ref{Dpsi}).
These coupled differential equations are then integrated by standard 
means. In order to evaluate matrix elements like (\ref{magslow}) one 
again employs the decomposition (\ref{Dpsi}) to reduce these to 
expressions which only contain functions of the strangeness changing 
angle $\nu$ and $f_{M_L,M_R}^{(I,Y;J,Y_R)}(\nu)$. The 
final result is obtained by integrating with respect to the 
measure 
\be
\int_0^{\frac{\pi}{2}}\ d\nu\ {\rm sin} 2\nu\ {\rm sin}^2\nu\
\Big\{\ldots\Big\} \ .
\label{numeasure}
\ee
For example, the contribution of the first term in eq (\ref{magslow}) 
to the magnetic moment of the proton is found to be
\be
\hspace{-0.2cm}
\mu_p&\hspace{-0.2cm}=&\hspace{-0.2cm}-\frac{8\pi}{3}M_N
\int_0^{\frac{\pi}{2}}\ d\nu\ {\rm sin} 2\nu\ {\rm sin}^2\nu\
m_1(\nu) \Bigg\{\frac{2}{3}{\rm sin}^2\nu
\left(\left(f_{\frac{1}{2},\frac{1}{2}}
^{\frac{1}{2},1;\frac{1}{2},1}(\nu)\right)^2
\hspace{-0.1cm} -\left(f_{-\frac{1}{2},-\frac{1}{2}}
^{\frac{1}{2},1;\frac{1}{2},1}(\nu)\right)^2\right)
\label{examslow} \\ && \hspace{-0.5cm}
-\frac{2}{9}\left[\left(1+{\rm cos}^2\nu\right)
\left(\left(f_{\frac{1}{2},\frac{1}{2}}
^{\frac{1}{2},1;\frac{1}{2},1}(\nu)\right)^2
+\left(f_{-\frac{1}{2},-\frac{1}{2}}Š^{\frac{1}{2},1;
\frac{1}{2},1}(\nu)\right)^2\right)
+8{\rm cos}\nu f_{\frac{1}{2},\frac{1}{2}}
^{\frac{1}{2},1;\frac{1}{2},1}(\nu)
f_{-\frac{1}{2},-\frac{1}{2}}
^{\frac{1}{2},1;\frac{1}{2},1}(\nu)\right]
\Bigg\} \ ,
\nonumber
\ee
where the isoscalar functions $f(\nu)$ are obtained by constructing
the eigenstates of the Hamiltonian (\ref{slowham}). Their normalization
is determined by the one of $\Psi(I,I_3,Y;J,J_3,Y_R)$.
The $\nu$ dependence of $m_1$ is purely due to the implicit 
dependence of the chiral angle $F=F(r,\nu)$ in the slow rotator 
approach \cite{Sch92b}
\be
\hspace{-0.5cm}
m_1(\nu)=\int_0^\infty \hspace{-0.2cm}
dr r^2  \sFt\left[f_\pi^2+
\frac{1}{e^2}\left(F^{\prime 2}+\frac{\sFt}{r^2}\right)+
\frac{\epsilon^2_6}{4\pi^4}\frac{F^{\prime 2}\sFt}{r^2}+
\frac{2}{3}(f_K^2-f_\pi^2)\cF\right] .
\label{m1examp}
\ee
A prime indicates a derivative with respect to the radial 
coordinate, {\it i.e.} $F^\prime=\partial F(r,\nu)/\partial r$.

\bigskip

\stepcounter{chapter}
{\large \bf \hskip1cm Appendix B}
\addcontentsline{toc}{chapter}{Appendix B}

\smallskip

In this appendix we will present the quantities relevant for the 
collective Hamiltonian in the vector meson model, which is discussed 
in Section \ref{sec_inertia}. Also a few typographical errors are 
corrected, which happened to occur in the corresponding expressions of 
ref \cite{Pa92}. For convenience the notation $\alpha^{\prime\prime}=
x\alpha^\prime$, etc. will be introduced, where $x$ measures the 
strangeness symmetry breaking (\ref{qmratio}).

The classical mass is given by
\be
E&=&4\pi\int dr\Big[{\ft\over 2}(\fpt r^2+2\sFt)
-{r^2\over{2g^2}}(\wpt+m_{\rho}^2\wo^2)
+{1\over{g^2}}[\gpt+{{G^2}\over{2r^2}}(G+2)^2]
\nonumber \\
&&+{{m_V^2}\over{g^2}}(1+G-\cF)^2+{\go\over g}\fp \wo \sFt-
{{2\gw}\over g}\gp\wo \sF
\nonumber \\
&&+{\gt\over g}\fp\wo G(G+2)+{1\over g}(\gw+\gt)\fp\wo
[1-2(G+1)\cF+\cFt]
\nonumber \\
&&+(1-\cF)\Big\{4\delta^\prime r^2+2(2\bep-{\alp\over g^2})
(\fpt r^2 +2\sFt)
\nonumber \\
Š&&-{{2\alp}\over{g^2}}\big[\wo^2r^2-2(G+1-\cF)^2-4(1+\cF)(1+G-\cF)
\big]\Big\}\Big].
\label{mcl}
\ee
For the moment of inertia for rotations in coordinate space, $\alpha^2=
\alpha_S^2+\alpha_{SB}^2$ we get
\be
\alpha_S^2&=&
{8\pi\over 3}\int dr \Big[\ft r^2
\sFt-{4\over{g^2}}(\phi^{\prime2}+2{\phi^2\over r^2}+m_\rho^2\phi^2)
\nonumber \\
&&+{{m_\rho^2}\over{2g^2}}r^2[(\xi_1+\xi_2)^2+2(\xi_1-1+\cF)^2]
\nonumber \\
&&+{1\over{2g^2}}[(3\xi_1^{\prime2} +2\xi_1^\prime\xi_2^\prime+
\xi_2^{\prime2})r^2+2(G^2+2G+2)\xi_2^2+4G^2(\xi_1^2+\xi_1\xi_2-2\xi_2
-\xi_2+1)]
\nonumber \\
&&+{4\over g}\go \phi\fp \sFt+{4\over g}\gt\phi\fp[(G-\xi_1)(1-\cF)
+(1-\cF)^2-G\xi_1]
\nonumber \\
&&+{{2\gw}\over g}\{\phi^\prime \sF(G-\xi_1+2-2\cF)
+\phi \sF(\xi_1^\prime-G^\prime)
\nonumber \\
&&+\phi\fp[
2+2\sFt+(\xi_1-G-2)\cF-2(\xi_1+\xi_2)]\}
\nonumber \\
&&\qquad
-{1\over2}[\eta^{\prime2}r^2+2\eta^2
+\tilde m_\eta^2r^2\eta^2]
\nonumber \\
&&-{\go\over{3gf_\pi}}[\eta^\prime
(\xi_1+\xi_2)\sFt+2\eta\fp(G+\xi_1)\sF]
-{{3\gt}\over{gf_\pi}}\eta^\prime(G+1-\cF)^2(\xi_1+\xi_2)
\nonumber \\
&&-{\gw\over{gf_\pi}}\big(\eta^\prime \big[(G+\xi_1)G+(\xi_1+\xi_2)
[(1-\cF)^2-2G\cF]\big]+\eta(G\xi_1^\prime-\gp\xi_1)\big)
\nonumber \\
&&+{{\gw g}\over{2f_\pi}}[\eta(\phi\wop -\wo\phi^\prime)-\eta^\prime
\phi\wo]\Big],
\\ \nonumber \\
\alpha_{SB}^2&=&
-{{4\pi}\over3}\int dr\Big[
8(\beta^\prime-{{\alp}\over{2g^2}})\Big\{2r^2\sFt\cF-
{1\over{f_\pi^2}}\big[\big(\eta^{\prime2}r^2+2\eta^2
\nonumber \\
&&\qquad\qquad-
{1\over2}(\fpt r^2+2\sFt)\eta^2\big)\cF
-2r^2\sF\fp\eta\eta^\prime\big]\Big\}
+{{\delta^\prime}\over{4f_\pi^2}}r^2\cF\eta^2
\nonumber \\
&&+{{4\alp}\over{g^2}}r^2\big[\cF[2(\xi_1-1+\cF)^2+(\xi_1+\xi_2)^2
-{8\over{r^2}}\phi^2]+4\sFt(\xi_1-1+\cF)\big]
\nonumber \\
&&-{{2\alp}\over{g^2f_\pi^2}}\big[(1+G-\cF)\sFt\eta^2
+[\wo^2r^2-2(1+G-\cF)^2]\cF\eta^2Š\nonumber \\
&&\qquad+2f_\pi\wo \sF\eta(\xi_1+\xi_2)\big]\Big].
\label{alpha}
\ee

The expression for the moment of inertia associated with rotations 
into the direction of strangeness is slightly lengthy. We therefore 
split it into pieces according to which part of the action 
they originate (flavor symmetric, anomalous and flavor symmetry
breaking)
\be
\beta^2= \beta^2_S+\beta^2_{an}+\beta^2_{SB}
\label{splitbeta}
\ee
and list them separately.
\be
\beta^2_S&=&\pi\int dr\Bigg\{2f_\pi^2\Big[r^2(1-\cF)-
2r^2(1+\ch)^2\Wpt-2r^2\sh\fp W\Wp
\nonumber \\
&&\quad+[\sF(3\sF+4\sh)-(1+2\ch+\cF)^2+{1\over2}r^2\fpt]W^2\Big]
\nonumber \\ &&\hspace{-1cm}
+{1\over{g^2}}\Big[2G^\prime DE+4r^2(S^\prime+{{\wo}\over4}E)^2
+2[{{\wo}\over2}D+G(S-1)]^2
\nonumber \\
&&\quad-2(D^\prime+{E\over2}G)^2+2r^2\wop E(1-S)
-{2\over{r^2}}(G+1)(G+2)D^2\Big]
\nonumber \\ && \hspace{-1cm}
+2{{m_V^2}\over{g^2}}\Big[2r^2\wo W\sh-(1+4\ch+3\cF)(1+G-\cF)W^2
+2r^2\left(S-1+\ch\right)^2
\nonumber \\
&&\quad-{{r^2}\over2}\left(E+2\ch\Wp-\fp W\right)^2
-[D+2W\sh(1+\ch)]^2\Big]\Bigg\}
\label{betas}
\\ \nonumber \\
\beta^2_{an}&=&{{\pi}\over{g}}\int dr\Bigg\{{g\over{\pi^2}}
\Big\{\fp\big[3\sFt\sh
+4\sF{\rm cos}^2\frac{F}{2}(1+\ch)\big]W
\nonumber \\
&&\qquad\qquad
+2\ch \sFt(1+\ch)\Wp\Big\}
\nonumber \\
&&-(2\go+{3\gw})\Big\{2\fp\sh \sFt W+2\wo\sh \sFt W\Wp
\nonumber \\
&&\quad -\wo\fp({7\over2}\sF+4\sh)\sF W^2
\nonumber \\
&&\quad-{8\over3}\sh\sF(1+\ch)(1+G-\cF)\Wp
\nonumber \\
&&\quad+{1\over3}\wo(1+2\ch+\cF)
[\fp(1+2\ch+\ch)W^2+4\sF(1+\ch)W\Wp]
\nonumber \\
&&\quad-{2\over3}\sFt[\sh(E+2\sh\Wp-\fp W)
-2(1+\ch)S\Wp+(1-\cF)\Wp]
\nonumber \\
&&\quad+{4\over3}\fp\sF[(1+2\ch+\cF)(S-1+\ch)W
\nonumber \\
&&\qquad\qquad +\sh(D+2W\sh(1+\ch))]\Big\}Š\nonumber \\
&&+\gw\Big\{\wop[(1+G-\cF)(3\sF+4\sh)-\sF(1+4\ch+3\cF)]W^2
\nonumber \\
&&\quad-2\sF(1+G-\cF)(1-S)E+2\fp G(G+2)\sh W
+{\wo\over2}\fp D^2-\wo\sF DE
\nonumber \\
&&\quad+\wo G(G+2)(2\sh W\Wp-{\fp\over2}W^2)
+4\sF\sh\gp W\hspace{-0.1cm}
-\wo\gp(3\sF+4\sh)W^2
\nonumber \\
&&\quad-\wop(1+2\ch+\cF)[D+2W\sh(1+\ch)]W
\nonumber \\
&&\quad+\wo[(1+\ch)(G+2)D\Wp+(1+2\ch+\cF)(\Dp+{E\over2}G)W]
\nonumber \\
&&\quad+2\gp\sh[D+2W\sh(1+\ch)]
+\hspace{-0.1cm}2\gp(1+2\ch+\cF)(S-1+\ch)W
\nonumber \\
&&\quad+G(G+2)\big[2(1+\ch)(S-1+\ch)\Wp-\sh(E+2\sh\Wp-\fp W)]
\nonumber \\
&&\quad-(1+G-\cF)\Big[2(1+2\ch+\cF)(\Sp+{\wo\over4}E)W
\nonumber \\
&&\quad+(1+\ch)[\wo D+2G(1-S)]\Wp+2\sh(\Dp+{E\over2}G)\Big]
\nonumber \\
&&\quad+\fp\Big[[{\wo\over2}D+G(1-S)]
[D+2W\sh(1+\ch)]+(G+2)(S-1+\ch)D\Big]
\nonumber \\
&&\quad+2\sF(S-1+\ch)(\Dp+{E\over2}G)\hspace{-0.1cm}
-\hspace{-0.1cm}2\sF\big[(\Sp+{\wo\over4}E)[D+2W\sh(1+\ch)]
\nonumber \\
&&\quad+{1\over4}[\wo D+2G(1-S)](E+2\sh\Wp-\fp W)\big]\Big\}
\nonumber \\
&&-(\gw+2\gt)\Bigg[2\sh(1+G-\cF)^2(\fp+\wo\Wp)W
\nonumber \\
&&\quad +\wo\fp[D+2W\sh(1+\ch)]^2
\nonumber \\
&&\quad+{\wo\over2}\fp(1+G-\cF)(1-G+8\ch+7\cF)W^2
\nonumber \\
&&\quad-2\wo\sF(E+2\sh\Wp-\fp W)[D+2W\sh(1+\ch)]
\nonumber \\
&&\quad-2(1+G-\cF)^2\Big[\sh(E+2\sh\Wp-\fp W)
\nonumber \\
&&\qquad\qquad-2(1+\ch)(S-1+\ch)\Wp\Big]
\nonumber \\
&&\quad+4\sF(1+G-\cF)(S-1+\ch)(E+2\sh\Wp-\fp W)\Bigg]
\Bigg\},
\label{betaan}
\\ \nonumber \\
\hspace{-0.1cm}
\beta^2_{SB}&=&4\pi\int\Bigg\{(2\bep-{\alp\over{g^2}})
\Big[\cF[\cF-1+2(1+\ch)^2(\Wpt+{2\over{r^2}}
{\rm cos}^2{F\over2}W^2)
\nonumber \\
&&\quad+4\sh\fp\Wp W-\fpt W^2-{2\over{r^2}}\sF(3\sF+4\sh)W^2]
\nonumber \\
&&\quad+(\fpt+{2\over{r^2}}\sin^2F)(1+\ch)(1-2\ch)W^2Š\nonumber \\
&&\quad-2(1+\ch)(\sF+\sh)(\fp\Wp+{2\over{r^2}}\sF\ch W)W\Big]
\nonumber \\
&&+(2\bpp-{\app\over{g^2}})(1+\ch)\Big[2(\ch-1)
+2(1+\ch)(\Wpt+{2\over{r^2}}{\rm cos}^2{F\over2}W^2)
\nonumber \\
&&\quad-2\sh(\fp\Wp W+{2\over{r^2}}\sF\ch W^2)
-\ch(\fpt+{2\over{r^2}}sin^2F)W^2\Big]
\nonumber \\
&&-\alp\cF\Big[4\wo\sh W-{2\over{r^2}}(1+G-\cF)(1+4\ch+3\cF)W^2\Big]
\nonumber \\
&&-2(\alp\cF+\app)\Big[(S-1+\ch)^2-{1\over4}(E+2\sh\Wp-\fp W)^2
\nonumber \\
&&\qquad\qquad-{1\over{2r^2}}[D+2W\sh(1+\ch)]^2\Big]
\nonumber \\
&&+(1+\ch)[\alp(2\ch-1)+\app\ch][\wo^2-{2\over{r^2}}(1+G-\cF)^2]W^2
\nonumber \\
&&+2[\alp(\sF+\sh)+\app\sh]\Big[\wo(S-1+\ch)W
\nonumber \\
&&\qquad-{1\over{r^2}}(1+G-\cF)[D+2W\sh(1+\ch)]W\Big]
\nonumber \\
&&+2[\alp(\cF+\ch)+\app(1+\ch)]\Big[r^2\wo\sh
+{{r^2}\over2}\fp(E+2\sh\Wp-\fp W)
\nonumber \\
&&\qquad-\sF[D+2W\sh(1+\ch)]-
\hspace{-0.1cm}(1+2\ch+\cF)(1+G-\cF)W\Big]W
\nonumber \\
&&+2[2\alp(\sF+\sh)+\app(\sF+2\sh)]\sF(1+G-\cF)W^2
\nonumber \\
&&-2\alp\sF\Big[2r^2\sh(S-1+\ch)+r^2(1+\ch)(E+2\sh\Wp-\fp W)\Wp
\nonumber \\
&&\qquad+(1+2\ch+\cF)[D+2W\sh(1+\ch)]W
\nonumber \\
&&\qquad-[(1+G-\cF)(3\sF+4\sh)-\sF(1+4\ch+3\cF)]W^2\Big]
\nonumber \\
&&-[\dep(1+4\ch+3\cF)+\ddp(3+4\ch+\cF)]W^2\Bigg\}. 
\label{betasb}
\ee
Finally the symmetry breaking parameters, $\gamma$, $\alpha_1$ and
$\beta_1$ are integrals over the classical profile functions only
\be
\gamma&=&{16\pi\over 3}\int dr\Bigg\{2(\ddp-\dep)r^2(1-\cF)
+2(\bep-\bpp-{{\alp-\app}\over{2g^2}})\cF(\fpt r^2+2\sFt)
\nonumber \\
&&\quad+{{\app-\alp}\over{g^2}}\Big[\cF[\wo^2 r^2-2(1+G-\cF)^2]
+4\sFt(G+1-\cF)\Big]
\Bigg\}, \label{gammavm}\\
\alpha_1&=&2\beta_1={{16\pi}\over{\sqrt3}}{{\app-\alp}\over{g^2}}
\int drr^2\wo \cF.
\label{ab1vm}
\ee

For completeness we also display the Skyrme model expression for 
the moment of inertia $\beta^2=\beta^2_S+\beta^2_{SB}$ which Šenters
the computations of Section \ref{sec_quant}. The contribution 
from the Wess--Zumino term is included in $\beta^2_S$.
\be
\beta^2_S&=&\pi\int dr \Bigg\{
2\ft r^2\left(1-\cF\right)
+\frac{1}{2e^2}\left(1-\cF\right)\left(\fpt r^2+2\sFt\right)
\nonumber \\
&&\qquad\qquad
-2\left(2\ft r^2+\frac{1}{e^2}\sFt\right)
\left(1+\ch\right)^2\Wpt
\nonumber \\
&&\qquad
-2\fp\left[2\ft r^2\sh+\frac{4}{e^2}\sFt\sh
+\frac{6}{e^2}\sF\ch\left(1+\ch\right)^2\right]W\Wp
\nonumber \\
&& \hspace{-0.5cm}
+\Bigg[\fpt\left(\ft r^2+\frac{8}{e^2}\sFt
+\frac{8}{e^2}\sF\sh\right)
+\frac{2}{r^2}\sF\left(\ft r^2+\frac{\sFt}{e^2}\right)
\left(3\sF+4\sh\right)
\nonumber \\
&&\qquad
-\frac{1}{r^2}\left(2\ft r^2+\frac{1}{2e^2}\fpt r^2
+\frac{2}{e^2}\sFt\right)
\left(1+2\ch+\cFt\right)^2\Bigg]W^2
\label{betasks} \\
&& \hspace{-0.5cm}
+\frac{\fp}{\pi^2}\left[3\sh\sFt
+4{\rm cos}^2\frac{F}{2}\sF\left(1+\ch\right)\right]W
+\frac{2}{\pi^2}\ch\sFt\left(1+\ch\right)\Wp
\Bigg\},
\nonumber \\
\nonumber \\
\beta_{SB}^2&=&8\pi\int dr r^2\Bigg\{
\bep\Bigg[\cF\Big[\cF-1+
2\left(1+\ch\right)^2\left(\Wpt+
\frac{2}{r^2}{\rm cos}^2\frac{F}{2}W^2\right)
\nonumber \\
&&\qquad\qquad\qquad
+4\fp\sh\Wp W-\fpt W^2-\frac{2}{r^2}\sF\left(3\sF+4\sh\right)W^2\Big]
\nonumber \\
&&\qquad\qquad\qquad 
+\left(\fpt+\frac{2}{r^2}\sFt\right)
\left(1+\ch\right)\left(1-2\ch\right)W^2
\nonumber \\
&&\qquad\qquad\qquad 
-2\left(1+\ch\right)\left(\sh+\sF\right)
\left(\fp\Wp+\frac{2}{r^2}\sF\ch W\right)W \Bigg]
\nonumber \\ &&
+\bpp\left(1+\ch\right)\Bigg[
2\left(\ch-1\right)+2\left(1+\ch\right)\left(\Wpt+
\frac{2}{r^2}{\rm cos}^2\frac{F}{2}W^2\right)
\nonumber \\
&&\qquad\qquad\qquad
-\ch\left(\fpt+\frac{2}{r^2}\sFt\right)W^2
Š-2\sh\left(\fp\Wp+\frac{2}{r^2}\sF\ch W\right)\Bigg]
\nonumber \\ &&
-\frac{1}{2}\Bigg[
\dep\left(3\cF+4\ch+1\right)+\ddp\left(\cF+4\ch+3\right)
\Bigg]W^2
\Bigg\} \ .
\label{betasksb}
\ee

\bigskip

\stepcounter{chapter}
{\large \bf \hskip1cm Appendix C}
\addcontentsline{toc}{chapter}{Appendix C}

\smallskip

In this appendix a short remark on the conservation of the 
axial current and the role of the induced fields is added. 
For this illustrative purpose it suffices to only consider the 
non--linear $\sigma$--model (\ref{Lnls}) supplemented by the 
Wess--Zumino term (\ref{WZterm}). Although no solution to the 
corresponding equation of motion 
\be
r^2 F^{\prime\prime}+2rF^\prime={\rm sin}2F
\label{eqmnls}
\ee
for the chiral angle exists, it may be used formally since the 
Skyrme term (\ref{Skterm}) is omitted for convenience only.
Here we are interested in those parts of the axial--vector 
current, which contain the angular velocities (\ref{defomega})
,$\Omega_4,\ldots,\Omega_7$, for the rotations into the direction 
of strangeness. In the notation of refs \cite{Pa91,Pa92} these parts 
are given by
\be
A_i^a=\left[A_3(r)\delta_{ik}+A_4(r){\hat r}_i{\hat r}_k\right]
d_{k\alpha\beta}D_{a\alpha}\Omega_\beta \ .
\label{ax34}
\ee
As usual the convention $i,j,k=1,2,3$ and $\alpha,\beta=4,\ldots,7$ is 
adopted. The contribution \cite{Ka87} from the Wess--Zumino term is 
purely due to the rotation of the classical fields (\ref{su3rot}) 
\be
A_3^{\rm WZ}&=&-\frac{1}{2\pi^2 r^2}F^\prime \sF\ \sh \ , 
\nonumber \\
A_4^{\rm WZ}&=&-A_3^{\rm WZ}+\frac{1}{2\pi^2 r^2}\sFt\
{\rm sin}^2\frac{F}{2} \ .
\label{ax34wz}
\ee
A straightforward computation shows that this piece has 
a non--vanishing divergence
\be
\partial_i A_i^{a\ {\rm WZ}}={\hat r}_a
\frac{3}{4\pi^2}\fp \frac{{\rm sin}^3F}{r^2} \ .
\label{divawz}
\ee
ŠAs long as the field configuration is restricted to the rotating 
hedgehog (\ref{su3rot}) there will be no other contribution to 
$A_3$ and/or $A_4$. Hence the (partial) conservation of the 
axial current is violated. On the other hand the non--linear 
$\sigma$--model contributes via the induced components 
(\ref{kaonind},\ref{kaonansatz}) 
\be
A_3^{nl\sigma}&=&\frac{1}{r}\ch\left(1+\ch\right)\cF W
\nonumber \\
A_4^{nl\sigma}&=&-A_3^{nl\sigma}+\frac{1}{2}\fp \sh\ W
+\ch\left(1+\ch\right)\Wp
\label{ax34nls}
\ee
Using the truncated equation of motion (\ref{eqmnls}) the divergence 
of this part of the axial current is obtained to be 
\be
\partial_i A_i^{a\ nl\sigma}&=&{\hat r}_a\frac{\ch}{r}
\Bigg\{r^2\left(1+\ch\right)\Wpt-r^2\sh\Wp+2r\left(1+\ch\right)\Wp
\nonumber \\ && \hspace{1cm}
+W\left[\frac{r^2}{4}\fpt-2\ch\cF\left(1+\ch\right)\right]\Bigg\}\ .
\label{divanls}
\ee

Upon some integrations by parts and application of eq (\ref{eqmnls}) 
the relevant parts of the moment of inertia for rotations into the 
direction of strangeness (\ref{betasks}) can be cast into the form
\be
\beta^2&=&\frac{1}{\pi}\int_0^\infty dr \Bigg\{
\fp\left[3\sFt\sh+4\sF{\rm cos}^2\frac{F}{2}
\left(1+\ch\right)\right]W
\nonumber \\ && \hspace{1cm}
+2\ch\sFt\left(1+\ch\right)\Wp\Bigg\}
\nonumber \\ &&
+\pi\int_0^\infty dr \Bigg\{
-2r^2\left(1+\ch\right)^2W^{\prime 2}
\nonumber \\ && \hspace{1cm}
+\left[\frac{r^2}{2}F^{\prime 2}
-4\ch\cF\left(1+\ch\right)\right]\left(1+\ch\right)W^2\Bigg\}
+\ldots\ ,
\label{betawznls}
\ee
where the ellipsis indicate terms which are neither due to the 
non--linear $\sigma$ model nor the Wess--Zumino term.
Straightforward calculation shows that the variation of 
$\beta^2$ yields the proper conservation of the axial current
\be
\partial_i A_i^a={\hat r}_a\frac{1} {4\pi^2r^2}
\ch\left(1+\ch\right)^{-1}
\frac{\delta \beta^2}{\delta W}\ .
\label{diva34}
\ee
Stated otherwise, the proper divergence of the axial current 
can only be obtained when the induced fields are taken into 
account. The inclusion of a Lagrange multiplier such 
that the overlap (\ref{overlap}) vanishes will spoil the 
Šrelation (\ref{diva34}) unless the radial functions $A_3$ and 
$A_4$ are altered accordingly. Unfortunately the equation of 
motion $\delta \beta^2/\delta W$ does not have a solution in 
the flavor symmetric case, even when the Skyrme term is included.
The reason being the existence of a zero mode, which represents 
a solution to the homogeneous part of the differential equation 
without being necessarily orthogonal to the inhomogeneity. However, 
this point does not give too much of a worry as for the physical 
situation no such zero mode exists because the flavor symmetric is
broken explicitly.

\bigskip

\stepcounter{chapter}
{\large \bf \hskip1cm Appendix D}
\addcontentsline{toc}{chapter}{Appendix D}

\smallskip

In this appendix some formulas are summarized which are helpful 
for studying the bound state approach introduced in chapter 
\ref{chap_bound}. These expressions are taken from refs 
\cite{Oh91,Rho92,Ku89,Sch95a,We94a}.

After re--expressing the parameters in the symmetry breaking parts 
of the Lagrangian (\ref{LSB}) in terms of physical quantities, the 
Lagrangian for the fluctuating kaon field $K$ defined in eq 
(\ref{bsansatz}) is given by
\be
{\cal L}_K&=&\left(D_\mu K\right)^{\dag}D^\mu K 
- m_K^2 K^{\dag}K
-\frac{f_\pi^2m_\pi^2}{4f_K^2}K^{\dag}
\left(\xi_\pi+\xi_\pi^{\dag}-2\right)K
\nonumber \\ &&\hspace{0.5cm}
-\frac{1}{8}K^{\dag}K{\rm tr}\left(
\partial_\mu\xi_\pi^{\dag}\partial^\mu\xi_\pi
+\frac{1}{4e^2f_K^2}\left[\xi_\pi^{\dag}\partial_\mu\xi_\pi,
\xi_\pi^{\dag}\partial_\nu\xi_\pi\right]^2\right)
\nonumber \\ &&\hspace{0.5cm}
-\frac{1}{8e^2f_K^2}\Bigg\{
\left(D_\mu K\right)^{\dag}D_\nu K
{\rm tr}\left(p^\mu p^\nu\right)
+\left(D_\mu K\right)^{\dag}D^\mu K
{\rm tr}\left(\partial_\nu\xi_\pi^{\dag}\partial^\nu\xi_\pi\right)
\nonumber \\ &&\hspace{1.5cm}
-3\left(D_\mu K\right)^{\dag}\left[p^\mu,p^\nu\right]D_\nu K\Bigg\}
-\frac{iN_C}{4f_K^2}B_\mu\left[K^{\dag}D^\mu K -
\left(D_\mu K\right)^{\dag}K\right]\ ,
\label{L2CK}
\ee
where use has been made of the covariant derivative 
$D_\mu=\partial_\mu+(\xi_\pi^{\dag}\partial_\mu\xi_\pi
+\xi_\pi\partial_\mu\xi_\pi^{\dag})/2$. Furthermore 
$p_\mu=\partial_\mu\xi_\pi\xi_\pi^{\dag}+
\xi_\pi^{\dag}\partial_\mu\xi_\pi$ refers to the induced axial--vector 
field, see also eq (\ref{defpR}). Finally $B_\mu$ is the topological 
current (\ref{bnumber}). The radial functions in the 
differential equation (\ref{eqmkbs}) are \cite{Ca85,Sch92a}
\be
h(r)&=&1+\frac{1}{2e^2f_K^2}\frac{\sFt}{r^2}\ ,
\label{hrbst} \\
f(r)&=&1+\frac{1}{4e^2f_K^2}\left(\fpt+2\frac{\sFt}{r^2}\right)\ ,
\label{frbst} \\
\lambda(r)&=&\frac{N_C}{8\pi^2f_K^2}\fp\frac{\sFt}{r^2}\ ,
\label{lrbst} \\
V_P(r)&=&\frac{2}{r^2}{\rm sin}^4\frac{F}{2}
\left[1+\frac{1}{4e^2f_K^2}\left(\fpt+\frac{\sFt}{r^2}\right)\right]
-\frac{1}{4}\left(\fpt+2\frac{\sFt}{r^2}\right)
\nonumber \\ && \hspace{0.2cm}
-\frac{1}{4e^2f_K^2}\Bigg\{
\frac{2}{r^2}\sFt\left(2\fpt+\frac{\sFt}{r^2}\right)
-\frac{6}{r^2}\left({\rm sin}^4\frac{F}{2}\frac{\sFt}{r^2}
+\frac{d}{dr}\fp\sFt{\rm sin}^2\frac{F}{2}\right)\Bigg\}
\nonumber \\ && \hspace{0.2cm}
-\frac{1}{r^2}\Bigg\{{\rm sin}^2\frac{F}{2}
\left[4+\frac{1}{e^2f_K^2}
\left(\fpt+\frac{\sFt}{r^2}\right)\right]
-\frac{3}{2e^2f_K^2}
\left(\cF\frac{\sFt}{r^2}-\frac{d}{dr}\fp\cF\right)\Bigg\}
\nonumber \\ && \hspace{0.2cm}
+\frac{2}{r^2}\left(\fpt+\frac{\sFt}{r^2}\right)
+\frac{m_\pi^2f_\pi^2}{2f_K^2}\left(\cF-1\right)
\label{vrbst}
\ee
The spectral functions $c(\omega)$ and $d(\omega)$ may by decomposed 
in terms of the grand spin channels according to eq (\ref{kaondecom}). 
For the present discussion the presentation of the bound state channel 
($L=1$, $G=1/2$) suffices \cite{Ca88,Oh91}
\be
c_P(\omega)+1&=&2\omega\int dr\  k_P^*(r,\omega)
\Bigg\{\frac{4}{3}r^2f(r){\rm cos}^2\frac{F}{2}
\nonumber \\ && \hspace{2cm}
+\frac{1}{2e^2f_K^2}\left[\frac{4}{3}\sFt{\rm cos}^2\frac{F}{2}
-\frac{d}{dr}r^2\sF\fp\right]\Bigg\}k_P(r,\omega)\ .
\label{cpara} 
\ee
Since one has $\omega=0$ for the zero--mode, it is apparent that 
$c_P$ is (minus) unity in the flavor symmetric case.

The coefficients in the magnetic moment operator (\ref{magbs}) read
\cite{Oh91}, see also \cite{Ku89}
\be
\mu_{S,0}&=&-\frac{2M_N}{3\pi\alpha^2}\int dr r^2 \fp \sFt\ , \quad
\mu_{V,0}=\frac{1}{2}M_N\alpha^2 \ ,
\nonumber \\ 
\mu_{S,1}&=&\chi\mu_{S,0}-\frac{M_N}{3}\int dr\Bigg\{
4r^2k_P^2{\rm cos}^2\frac{F}{2}
\nonumber \\ && \hspace{2cm}
+\frac{1}{e^2f_K^2}
\left[4k_P^2\sFt{\rm cos}^2\frac{F}{2}
+r^2k_P^2\fpt{\rm cos}^2\frac{F}{2}
+3r^2k_Pk_P^\prime\fp\sF\right]\Bigg\}\ ,
\nonumber \\
\mu_{V,1}&=&\frac{M_N}{3}\int dr\Bigg\{
r^2k_P^2{\rm cos}^2\frac{F}{2}\left(1-{\rm sin}^2\frac{F}{2}\right)
+\frac{N_CM_N}{24\pi^2f_K^2}\omega_Pr^2k_P^2\fp\sFt
\nonumber \\ && \hspace{0.5cm}
+\frac{1}{4e^2f_K^2}\Bigg[4k_P^2\sFt{\rm cos}^2\frac{F}{2}
\left(3-8{\rm sin}^2\frac{F}{2}\right)
+r^2k_P^2\fpt{\rm cos}^2\frac{F}{2}
\left(1-18{\rm sin}^2\frac{F}{2}\right)
\nonumber \\ && \hspace{1.5cm}
+2k_P^{\prime2}\sFt+3k_Pk_P^\prime\fp\sF
\left(3-4{\rm sin}^2\frac{F}{2}\right)\Bigg]\Bigg\}\ ,
\label{magbspara}
\ee
where the arguments of the bound state wave--function 
$k_P=k_P(r,\omega_P)$ have been omitted.

Finally the integral kernels $\Phi_{1,2}$, which show up in the bound
state equation of the NJL soliton model will be presented. The local 
kernel $\Phi_1(r)$ does not depend on the eigen energy $\omega$. 
It acquires contributions from the meson part of the action as well as 
those terms involving $h_1$
\be
&&\hspace{-2cm}\Phi_1(r)=
-\frac{\pi}{2}m_\pi^2f_\pi^2\left(1+\frac{m_s}{m}\right)
\left({\rm cos}\Theta+\frac{m_s^0}{m^0}\right)
\nonumber \\ &&
-\frac{N_C}{4}\eta_{\rm val}(m+m_s)
\int \frac{d\Omega}{4\pi}\psi_{\rm val}^{\dag}(\mbox{\boldmath $r$})
u_0(\mbox{\boldmath $r$})\beta u_0(\mbox{\boldmath $r$})
\psi_{\rm val}(\mbox{\boldmath $r$})
\label{phi1} \\ &&
-\frac{N_C}{4}(m+m_s) \int_{1/\Lambda^2}^\infty
\frac{ds}{\sqrt{4\pi s}}\int \frac{d\Omega}{4\pi}
\Big\{\sum_{\mu=ns}\epsilon_\mu e^{-s\epsilon_\mu^2}
\psi_\mu^{\dag}(\mbox{\boldmath $r$})u_0(\mbox{\boldmath $r$})
\beta u_0(\mbox{\boldmath $r$}) \psi_\mu(\mbox{\boldmath $r$})
\nonumber \\ && \hspace{7cm}
+\sum_{\rho=s}\epsilon_\rho e^{-s\epsilon_\rho^2}
\psi_\rho^{\dag}(\mbox{\boldmath $r$})\beta
\psi_\rho(\mbox{\boldmath $r$})\Big\}.
\nonumber
\ee
The unitary matrix $u_0$ has been defined in eq (\ref{u0}), while the 
grand spinors $\psi_\mu(\mbox{\boldmath $r$})$ denote the spatial 
representations of the eigenstates of the one--particle Dirac 
Hamiltonian (\ref{h0}). The integral $\int (d\Omega/4\pi)$ indicates 
that the average with regard to the internal degrees of freedom has 
been taken. The bilocal kernel
$\Phi_2(\omega;r,r^\prime)$ originates from the terms quadratic
in $h_1$ and is symmetric in $r$ and $r^\prime$
\be
&&\hspace{-1cm} \Phi_2(\omega;r,r^\prime)=
-\frac{N_C}{4}(m+m_s)^2 \int \frac{d\Omega}{4\pi}
\int \frac{d\Omega^\prime}{4\pi} \Bigg\{\eta_{\rm val} \sum_{\rho=s}
\frac{\psi_{\rm val}^{\dag}(\mbox{\boldmath $r$})
u_0(\mbox{\boldmath $r$})\psi_\rho(\mbox{\boldmath $r$})
\psi_\rho^{\dag}(\mbox{\boldmath $r$}^\prime)
u_0(\mbox{\boldmath $r$}^\prime)
\psi_{\rm val}(\mbox{\boldmath $r$}^\prime)}
{\epsilon_{\rm val}-\omega-\epsilon_\rho}
\nonumber \\ && \hspace{3cm}
-\sum_{\mu=ns\atop\rho=s}
\psi_\mu^{\dag}(\mbox{\boldmath $r$})
u_0(\mbox{\boldmath $r$})\psi_\rho(\mbox{\boldmath $r$})
\psi_\rho^{\dag}(\mbox{\boldmath $r$}^\prime)
u_0(\mbox{\boldmath $r$}^\prime)
\psi_\mu(\mbox{\boldmath $r$}^\prime)
{\cal R}_{\mu,\rho}(\omega)\Bigg\}\ .
\label{phi2}
\ee
The regulator function appearing eq (\ref{phi2}) is obtained to be
\be
&&\hspace{-2.5cm}{\cal R}_{\mu,\rho}(\omega)=
\int_{1/\Lambda^2}^\infty ds \sqrt{\frac{s}{\pi}}
\Big\{\frac{{\rm e}^{-s\epsilon_\mu^2}
+{\rm e}^{-s\epsilon_\rho^2}}{s}
+[\omega^2-(\epsilon_\mu+\epsilon_\rho)^2]
R_0(s;\omega,\epsilon_\mu,\epsilon_\rho)
\nonumber \\ &&\hspace{2.5cm}
-2\omega\epsilon_\rho R_1(s;\omega,\epsilon_\mu,\epsilon_\rho)
+2\omega\epsilon_\mu R_1(s;\omega,\epsilon_\rho,\epsilon_\mu)
\Big\},
\label{phi2reg}
\ee
wherein the Feynman parameter integrals 
\be
R_i(s;\omega,\epsilon_\mu,\epsilon_\nu)
=\int_0^1 x^i dx\ {\rm exp}\left(-s[(1-x)\epsilon_\mu^2
+x\epsilon_\nu^2-x(1-x)\omega^2]\right)
\label{regfctapp}
\ee
reflect the quark loops in the soliton background. The regulator 
function (\ref{regfctapp}) is nothing but the analogue of the 
incomplete $\Gamma$--functions in eqs (\ref{fq}-\ref{fK}) when 
the soliton background field is present.

\vfill\eject


\end{document}